\documentclass[DIV13,12pt,a4paper,headinclude]{scrartcl}
\usepackage[english]{babel}
\usepackage{amsmath}
\usepackage{amssymb}
\usepackage{bbm}
\usepackage{empheq}
\usepackage{mathrsfs}
\usepackage{mathtools}
\usepackage{array}
\usepackage{bbold}
\usepackage{dsfont}
\usepackage{graphicx}
\usepackage{psfrag}
\usepackage{pstricks}
\usepackage{color}
\usepackage{booktabs}
\usepackage[normal,margin=10pt,font=small,labelfont=bf,
labelsep=period]{caption}
\usepackage[hang]{subfigure}
\usepackage[onehalfspacing]{setspace}
\usepackage{placeins}
\usepackage[bottom]{footmisc} %footnotes always at bottom of page
\usepackage{indentfirst}
\usepackage{overpic}

\numberwithin{equation}{section}

\newcommand{\WHlike}{WH-like}
\newcommand{\Ofourloc}{\ensuremath{{\sf O}(4)_{\rm loc}}}
\newcommand{\Ofour}{\ensuremath{{\sf O}(4)}}
\newcommand{\wrt}{w.\,r.\,t.\ }
\newcommand{\ie}{i.\,e. }
\newcommand{\eg}{e.\,g. }
\newcommand{\ea}{\ensuremath{e^{a}_{\ \mu}}}
\newcommand{\oab}{\ensuremath{\omega^{ab}_{\ \ \mu}}}
\newcommand{\ZNk}{\ensuremath{Z_{{\rm N}\,k}}}
\newcommand{\STr}{{\rm{STr}}\,}
\newcommand{\tr}{{\rm{tr}}\,}
\newcommand{\Tr}{{\rm{Tr}}\,}
\newcommand{\rank}{{\rm{rank}}\,}
\newcommand{\bD}{\beta_{\rm D}}
\newcommand{\aD}{\alpha_{\rm D}}
\newcommand{\aL}{\alpha_{\rm L}'}
\newcommand{\bm}{\bar{\mu}}
\newcommand{\hP}{\hat{\rm P}}
\newcommand{\D}[2]{\left[\mathcal{D} #1 \right]_{#2}}

\begin{document}
\renewcommand{\baselinestretch}{1.1}
\begin{titlepage}
\enlargethispage{2\baselineskip}
\title{
\vspace{-1.5cm}
\begin{flushright}
\rm
\normalsize{MITP/14-061}
\bigskip
\vspace{0.5cm}
\end{flushright}
\rm A new functional flow equation\\ for Einstein-Cartan quantum gravity}
\date{}
\author{{\large U. Harst and M. Reuter}\\
{\small Institute of Physics, University of Mainz}\\[-0.2cm]
{\small Staudingerweg 7, D-55099 Mainz, Germany}}
\maketitle
\thispagestyle{empty}
\begin{abstract} 
\vspace{-0.5cm}
\renewcommand{\baselinestretch}{1.0}\selectfont
\noindent We construct a special-purpose functional flow equation which facilitates 
non-perturbative renormalization group (RG) studies on theory spaces involving a 
large number of independent field components that are prohibitively complicated 
using standard methods. Its main motivation are quantum gravity theories in which 
the gravitational degrees of freedom are carried by a complex system of tensor 
fields, a prime example being Einstein-Cartan theory, possibly coupled to matter. 
We describe a sequence of approximation steps leading from the functional RG 
equation of the Effective Average Action to the new flow equation which, as a 
consequence, is no longer fully exact on the untruncated theory space. However, 
it is by far more ``user friendly'' when it comes to projecting the abstract 
equation on a concrete (truncated) theory space and computing explicit 
beta-functions. The necessary amount of (tensor) algebra reduces drastically, and 
the usually very hard problem of diagonalizing the pertinent Hessian operator is 
sidestepped completely. In this paper we demonstrate the reliability of the 
simplified equation by applying it to a truncation of the Einstein-Cartan theory 
space. It is parametrized by a scale dependent Holst action, depending on a ${\sf O}(4)$ 
spin-connection and the tetrad as the independent field variables. We compute the 
resulting RG flow, focusing in particular on the running of the Immirzi parameter, 
and compare it to the results of an earlier computation where the exact equation 
had been applied to the same truncation. We find consistency between the two 
approaches and provide further evidence for the conjectured non-perturbative 
renormalizability (asymptotic safety) of quantum Einstein-Cartan gravity. We also 
investigate a duality symmetry relating small and large values of the Immirzi 
parameter ($\gamma \rightarrow 1/\gamma$) which is displayed by the beta-functions 
in absence of a cosmological constant.
\end{abstract}
\end{titlepage}
\newpage 

\tableofcontents
\section{Introduction}
Searching for a fundamental quantum theory of gravity is a long-standing and on-going quest of modern physics. Since the perturbative non-renormalizability of a quantum field theory of metric gravity was shown \cite{tHooft-Velt,goroff-sag,vandeVen} various different approaches as e.g. string theory, loop quantum gravity (LQG) or Asymptotic Safety have been pursued in order to find a solution to this problem. These approaches differ in which of the constituents are accounted for the failure of the perturbative renormalizability of metric gravity, i.e. quantum field theory, perturbation theory or the choice of variables that serve as a carrier of the fundamental gravitational degrees of freedom. While string theory leaves the framework of quantum field theory by choosing strings of finite length as fundamental degrees of freedoms, it essentially sticks to perturbative methods. LQG, in contrast, is constructed non-perturbatively relying on a Hamiltonian formalism that uses a special choice of variables to parametrize phase space, which are the quantized canonically.

The Asymptotic Safety scenario for gravity is the most conservative approach among these as it attributes the theory's perturbative non-renormalizability only to the unjustified use of perturbation theory. It can be seen as a specific, particularly natural ultraviolet (UV) completion of the effective field theory framework pioneered by Donoghue \cite{donoghue1, donoghue2}. As conjectured by Weinberg \cite{wein} metric gravity in four dimensions could be well-defined in the UV at a non-Gaussian fixed point (NGFP) in the space of all actions, i.e. a fixed point of the renormalization group flow, that corresponds to an interacting theory. If this conjectured NGFP exists but is located at large couplings outside the realm of perturbation theory it cannot be found or examined by perturbative methods, even though the description of gravity as a conventional quantum field theory remains valid at all scales.

Thus, by definition, the investigation of the Asymptotic Safety scenario in gravity can only be carried out using non-perturbative methods in quantum field theory. An appropriate tool for this investigation is the functional renormalization group equation (FRGE) for the effective average action (EAA) which was first derived for scalar \cite{avact} and Yang-Mills theory \cite{ym1,ym2,nonabavact,ym3}, and later on for gravity \cite{mr}. This exact functional equation for a running effective action functional typically cannot be solved in full generality but allows for non-perturbative approximations by choosing an ansatz for the form of the action functional, a so-called truncation. The FRGE rendered possible an investigation of the renormalization group (RG) flow of metric gravity in arbitrary dimensions of spacetime. Since then numerous studies of different approximations have been carried out\footnote{For an overview of this field of research we refer to the review articles \cite{livrev,percrev,frankrev,martinrev}.}, all of which indicate the existence of a NGFP for metric gravity in $d=4$ spacetime dimensions. Also the inclusion of matter fields, that causes divergences in the perturbative approach already at the one-loop level, has been explored in some detail \cite{percadou,perper1a, perper1b,vacca}, and it was found that Asymptotic Safety of gravity is compatible with the matter content of the Standard Model, although general bounds on the number of matter fields were found to exist. Taken together an impressive amount of evidence for Asymptotic Safety of metric gravity at a NGFP in the space of diffeomorphism invariant action functionals has been collected, such that it is likely to be a true feature of this theory space and not merely an artifact of the approximations applied. Theories of metric gravity whose continuum limit is taken at this NGFP are called Quantum Einstein Gravity (QEG).

Despite its success in metric gravity, it is important to keep in mind that the concept of Asymptotic Safety is neither linked nor restricted to using the metric as the carrier field for the gravitational degrees of freedom. In the search for an asymptotically safe theory of gravity the only restriction to be obeyed is that the space of action functionals on which the RG flow is considered, the ``theory space'', contains an action functional that gives rise to Einstein's equation. Otherwise the resulting quantum theory will lack the classical regime of general relativity. As there is a remarkable number of Lagrangian variational principles that all give rise to Einstein's equation, we find a variety of different theory spaces, that are equally plausible to be investigated. For instance, a ``tetrad only'' theory space, containing the Einstein-Hilbert action re-expressed in terms of the tetrad field has been examined in \cite{Harst2012}. 

The focus of the present work is on the Einstein-Cartan theory space \cite{Hehl-Book}. This theory space consists of all action functionals constructed from the tetrad, $e^a{}_\mu$ and an independent spin connection $\omega^{ab}{}_\mu$, which serve as carrier fields for the gravitational degrees of freedom. In particular it contains the Hilbert-Palatini \cite{Ortin2007, Peldan} and the Holst \cite{Holst-Paper} action that are well known from the classical theory.

It should be noted that all these theory spaces do not have very much in common, except for the fact that they contain points which give rise to equivalent field equations: In particular, they are of different ``size'' (i.e. the action functionals contained are not in one-to-one correspondence to each other), the constituent fields differ in the number of independent field components and in the symmetry group of gauge transformations. For this reason the RG flows on these spaces and their fixed point structure are a priori completely unrelated from each other such that {\it an investigation of the Asymptotic Safety scenario of quantum gravity has to be carried out separately for each of these spaces}. 

Even if two or more theory spaces were found to contain a fixed point suitable for the Asymptotic Safety construction, we cannot expect the quantum theories defined at these fixed points to be similar or even equivalent. Therefore the investigation of the Einstein-Cartan theory space might lead to finding new quantum theories of gravity, that are inequivalent to QEG. Nonetheless, such an inequivalent theory may give rise to certain similar predictions. However such predictions cannot be assumed from the outset but have to be assessed on the level of observable quantities, i.e. after both theories of gravity are found to be well-defined.

Up to now, there is only little literature on the RG flow in Einstein-Cartan theory space \cite{je:longpaper,je:proc,je:lett,bene-speziale, bene-speziale2}. This is mainly due to the fact, that an exact evaluation of the FRGE for the EAA is extremely involved, even for comparatively small truncations such as the Holst truncation containing only Newton's constant, the cosmological constant and the Immirzi parameter \cite{Immirzi, Barbero} as its running couplings. The complexity of the task is brought about mainly by the quadruplication of the number of independent carrier fields, namely 40 in case of the $(e, \omega)$-setting compared to 10 in the metric formulation. Due to the increased complexity of the calculation no exact evaluation of the FRGE for the Einstein-Cartan theory space has been carried out so far. The results in \cite{je:longpaper,je:proc,je:lett} are based on the RG flow of the Holst truncation evaluated using a (non-perturbative) proper-time approximation of the FRGE for the EAA.  The investigation presented in \cite{bene-speziale,bene-speziale2} constitutes a purely perturbative calculation for an even smaller truncation corresponding to a vanishing cosmological constant ($\lambda\equiv 0$).

For more thorough investigations of the RG flow in Einstein-Cartan theory space using more involved truncations, approximative RG equations are needed in order to reduce the mathematical complexity of the task. However, before such an approximation can be used its reliability has to be confirmed.

Therefore, the purpose of this paper is twofold:

\noindent{\bf(i)} We develop a new type of approximate RG equation for the EAA being structurally similar to the Wegner-Houghton equation \cite{Wegner1973}, which is by far easier to apply to enlarged truncations of the Einstein-Cartan theory space.

\noindent{\bf(ii)} We apply this new approximate RG equation to the Holst truncation using the same setting as investigated in \cite{je:longpaper}. By comparing our results with the results obtained using a structurally exact equation \cite{je:longpaper} we carry out a first assessment of the reliability of the new Wegner-Houghton-like flow equation.

In the remainder of this Introduction we recall some aspects of the RG methods used to investigate Asymptotic Safety in gravity before, and at the end of this section, we outline the structure of the present work.

\subsection{RG flows and Asymptotic Safety}

The central idea at the very heart of the renormalization group is that the quantum fluctuations $\varphi$ and ghost fluctuations $\{\hat\Xi, \hat{\bar{\Xi}}\}$, if any, which are integrated over in the path integral for the generating functional,

\begin{equation}
Z[J;\bar\Phi]=\int{\cal D}\varphi\,{\cal D}\hat\Xi\,{\cal D}\hat{\bar{\Xi}} \,\text{exp}\big[-S_\Lambda[\bar\Phi+\varphi]-S_{\rm gf}[\varphi;\bar\Phi]-S_{\rm gh}[\varphi,\hat\Xi,\hat{\bar\Xi};\bar\Phi]+\textstyle \int \displaystyle \!J\cdot (\varphi,\hat\Xi,\hat{\bar{\Xi}})\big]
\end{equation}

\noindent can be taken into account in a piecewise manner. This approach was pioneered by Wilson \cite{wilson1,wilson2}, who applied this idea directly at the level of the path integral, dividing it into several integrations each corresponding to a certain ``momentum shell'' of quantum fluctuations. By considering an infinitesimal momentum shell integration of the Wilson type one can derive the Wegner-Houghton renormalization group equation \cite{Wegner1973} for the bare action $S_\Lambda$, that expresses the scale derivative $\partial_\Lambda S_\Lambda$ in terms of functional derivatives $\delta S_\Lambda/\delta \varphi$ of the action at that scale. Instead of using a sharp momentum cutoff splitting the functional integration into two domains one might also use smooth regulator functions, that essentially have the same effect. Different choices of these functions lead to different renormalization group equations \cite{BagnulsRev, Sumi2000}, as e.\,g. the Polchinski equation, all of which have in common that their solutions relate {\it bare} actions describing the same quantum system with different UV cutoff scales. Thus, in all these cases a further functional integration of the low momentum quantum fluctuations has to be carried out in order to obtain the corresponding Green's functions, from which all physical observables are to be calculated.

\paragraph{Effective average action (EAA).} It is possible to implement similar ideas directly at the level of the {\it effective} action $\Gamma$, leading to the concept of the effective average action (EAA) $\Gamma_k$ and the related functional RG equation \cite{avactrev1}. In this case we achieve the coarse graining effect by the addition of a (scale-de\-pen\-dent) mode suppression term $\Delta S_k[\varphi]=\frac{1}{2} \int\!\varphi {\cal R}_k(p^2)\, \varphi$ to the bare action $S[\varphi]$ with ${\cal R}_k(p^2) \propto k^2 R^{(0)}(p^2/k^2)$, where the shape function $R^{(0)}$ is dimensionless and interpolates smoothly between $R^{(0)}(0)=1$ and $\lim_{z\rightarrow\infty}R^{(0)}(z)=0$. Thus the mass scale $k$ acts as a variable infrared cutoff: the contributions of field modes with $p \lesssim k$ are suppressed in the functional integral, while all others are integrated out as usual. The scale-dependent generalization of the effective action, the EAA, $\Gamma_k[\phi]$, depending on the classical field $\phi=\langle \varphi \rangle$, satisfies the exact FRGE
\begin{equation}\label{scalarFRGE}
 \partial_t \Gamma_k=\frac{1}{2}\,\STr\bigg[\Big[\Gamma^{(2)}_k+{\cal R}_k\Big]^{-1} \partial_t {\cal R}_k\bigg]\:.
\end{equation}

\noindent The EAA, compared to the running bare action, is more closely related to the physics at scale $k$ since the functional $\Gamma_k$ directly corresponds to an effective action at the scale $k$ such that no further functional integral has to be carried out in order to connect it with physical observables.  

\paragraph{Asymptotic Safety.} The existence of non-perturbatively renormalizable fundamental theories is closely related to the existence of complete RG trajectories of the EAA. Using the functional RG approach, we are able to examine if the space of all action functionals, the theory space ${\cal T}$, contains RG fixed points and complete trajectories emanating from them. Each of them defines a candidate for a well-defined fundamental quantum field theory. The set of all trajectories pulled into a certain fixed point under the inverse RG flow is called the UV attractive hypersurface $\mathscr{S}_{\rm UV}$ of the fixed point. If $\mathscr{S}_{\rm UV}$ is finite dimensional, with $s={\rm dim}(\mathscr{S}_{\rm UV})$, we can pin down the one trajectory ``realized in Nature'' by measuring $s$ independent observables at a fixed scale $k_0$. Once a trajectory in $\mathscr{S}_{\rm UV}$ is selected, all other observables are predictions of the theory. Moreover, also the bare action is a prediction of the RG flow; it is obtained from the fixed point condition and does {\it not} serve as an input in this setting.

This generalized notion of renormalizability depends crucially on the existence of suitable fixed points in theory space and the gravitational EAA is a suitable tool to search for them: It is Background Independent by construction, and it comes with a natural non-perturbative approximation scheme, namely the truncation of theory space.

\subsection{Exploring Asymptotic Safety on truncated theory spaces}\label{InvestigatingAS}
\paragraph{Theory space.}
The theory space, $\cal T$, is the most important ingredient to any RG study as it indirectly defines the system under consideration. Its defining properties are the field content $\Phi$ of which all its points, \ie all action functionals $A: \Phi\mapsto A[\Phi]$ are built from, and the symmetry group ${\bf G}$ under whose action these functionals are left invariant:

\begin{equation}\label{GenTheorySpace}
 {\cal T}=\{A:\Phi\mapsto A[\Phi]\,|\, A \text{ invariant under }{\bf G}\}
\end{equation}
We assume that $\cal T$ allows for a set $\{P_\alpha[\Phi]\}$ of basis functionals, such that any point in theory space can be expanded as $A[\Phi]=\sum_{\alpha=1}^\infty \bar{u}_\alpha P_\alpha[\Phi]$. The generalized couplings $\bar{u}_\alpha$ serve as coordinates in theory space and any trajectory in theory space is therefore parametrized by an infinite set of coordinate functions, or ``running couplings'', $\bar{u}_\alpha(k)$. The EAA assumes the form $\Gamma_k[\Phi]=\sum_{\alpha=1}^\infty \bar{u}_\alpha(k) P_\alpha[\Phi]$. The FRGE defines a vector field on $\cal T$ whose integral curves are the RG trajectories solving the FRGE. By expanding it in terms of the basis functionals, and replacing the couplings by their dimensionless counterparts $u_\alpha(k)=\bar{u}_\alpha(k) k^{-d_\alpha}$, we obtain the component form of the FRGE, an equivalent system of autonomous differential equations:
\begin{equation}
 \partial_t u_\alpha(k)=\beta_\alpha(u_1,u_2,\cdots)\:.\nonumber
\end{equation}
Here $t \equiv \ln k$ is the ``RG time'', and the dimensionless $\beta$-functions $\beta_\alpha$ do not depend on $k$ explicitly.

\paragraph{Truncations.} Usually it is not possible to solve or even derive this complete system of differential equations for a given theory space. Thus, one has to approximate the exact flow (the exact FRGE gives rise to) in a non-perturbative way, which can be done by reducing the basis $\{P_\alpha[\Phi]\}$ to a finite (or infinite) subset $\{P_i[\Phi]\}\subset\{P_\alpha[\Phi]\}$, in order to render the calculation technically feasible. This kind of approximation is known as a {\it truncation of theory space}. Geometrically {\it a truncation of theory space} corresponds to neglecting all components of the vector field generated by the FRGE that are not tangent to the subspace of the truncation. As the vector field is first projected onto the subspace of the truncation and then its integral curves are determined, the resulting trajectories do not coincide with the exact trajectories projected onto the subspace under consideration: The difference is exactly the error in the predictions of the running couplings that we pick up by truncating the theory space.

Obviously, the magnitude of the truncation error will depend very much on the truncation chosen. We should expect that extending a given truncation by adding additional invariants will decrease the error of the truncation. However, truncations of the same number of invariants will differ largely in reliability depending on the extent to which they cover the physics content of the full theory. Thus, in any functional RG study it is a major challenge to find a truncation large enough to reliably describe the physics of the model while at the same time being small enough to allow for an explicit computation.

In gravity even truncated RG studies are particularly complicated: In metric gravity already the two dimensional truncation $\{\int\! \sqrt g, \int\! \sqrt g R\}$ and its three dimensional extension $\{\int\! \sqrt g, \int\! \sqrt g R, \int\! \sqrt g R^2\}$ amount to demanding calculations \cite{mr,oliver2}. With the help of computer algebra systems it has been possible to extent these truncations up to $\int\! \sqrt g R^8$ \cite{codellob,codelloc,Machado2008} and only recently first studies of the infinite dimensional $f(R)$-truncation have been published \cite{BeneCara,Demmel2012}. In the even more involved setting of Einstein-Cartan gravity, as yet only up to three dimensional truncations could be analyzed \cite{je:longpaper}.

\paragraph{Symmetries of theory space and choice of the FRGE.} It is desirable that the RG equation employed to analyze the theory space \eqref{GenTheorySpace} is invariant under the symmetry group ${\bf G}$, \ie it should retain the symmetries, as otherwise the flow generated by the FRGE will leave ${\cal T}$. This consideration is particularly important when gauge theories like gravity are considered. Here, the theory space can be reduced to gauge invariant\footnote{To be precise, {\it background} gauge invariant functionals, see below.} functionals of the fields, if it is possible to construct a FRGE that preserves gauge invariance. We will show below how the FRGE \eqref{scalarFRGE} can be generalized to a gauge invariant setting by the use of the background field method. 

Any theory space whose functionals exhibit a certain symmetry can be seen as embedded in, and hence a truncation of, the larger theory space constructed from the same field content but without the additional symmetry requirement. Using an exact FRGE that retains the symmetry on this larger space will lead to an RG flow in which the symmetric subspace forms a self-consistent truncation, \ie the resulting trajectories lie entirely either inside the subspace or outside of it. Thus, if we find a fixed point in the symmetric subspace it gives rise to a complete RG trajectory whose IR endpoint is, as well, symmetric. Using a different exact FRGE in the larger space that, however, does not retain the symmetry, we should expect to find the same FP as its existence should be a property of the underlying theory space rather than relying on the exact form of the RG equation. If we take an asymptotically safe trajectory emanating from this FP and follow it to the IR, it will leave the symmetric subspace at some point, but its IR endpoint, eventually, will be symmetric again, as we then have integrated out all fluctuation modes of a symmetric theory, albeit in a non-covariant manner. Thus, it should be possible to describe and analyze the RG flow of a symmetric theory, using both types of FRGE, one that retains the symmetry throughout and one that restores it only at the level of the effective action $\Gamma_{k=0}$.

On the other hand, when considering practical calculations that always involve a truncation of theory space, the symmetry preserving approach becomes even more advantageous. Using a truncation of a fixed number of invariants we can always approximate the symmetric subspace better than the larger total theory space. Thus, the quality of the truncation can be increased the more, the larger the symmetry group is, such that with a comparable calculational effort the covariant approach is expected to lead to more accurate results.

Nonetheless it should be kept in mind that using a non-covariant approach for a symmetric theory is comparable to choosing a worse truncation, but should not be despised as being inappropriate from the outset.

\paragraph{Interrelations of different EAA-based RG settings.} Using the gravitational effective average action and its FRGE to find and investigate asymptotically safe theories consists of the following steps:

\begin{enumerate}
\item[\bf(i)] Define a theory space of action functionals.
\item[\bf(ii)] For practical reasons: Choose a truncation of theory space to be analyzed.
\item[\bf(iii)] Choose an appropriate FRGE and compute the resulting RG flow in this truncation.
\item[\bf(iv)] Analyze the structure of the resulting flow, search for UV attractive fixed points and find complete RG trajectories.
\end{enumerate}
Within this program different theory spaces should lead to different quantum theories, as their fixed point structures are generally not related to each other. Even if two theory spaces are motivated by the same classical theory, it could well be that one of them exhibits a FP suitable for the Asymptotic Safety construction while the other does not, indicating that the sought-for fundamental QFT can only be formulated with the symmetries and field content of one of the two theory spaces. Hence, we keep in mind  that {\it the existence of NGFPs has to be investigated independently for every theory space under consideration.}

On a {\it given theory space}, the application of {\it different FRGEs} to the {\it same truncation} should lead, however, to comparable results. The most trustable results should be obtained by the FRGE \eqref{scalarFRGE} and its generalization to gauge theories, as it is structurally exact on the untruncated level. Often ``structurally approximate'' FRGEs are applied which do not yield the exact answer even if no truncation is made. Their approximations usually can only be justified in comparison to the exact flow equation or to other well-approved approximations. In particular, {\it any new ``structurally approximate'' RG equation should be tested this way}. 

Furthermore, for a {\it given FRGE} we expect that {\it different truncations} of the {\it same theory space} lead to similar results if the essential physics is captured by both truncations. Extending a given truncation should obviously lead to a better approximation of the exact, untruncated flow, such that the stability of fundamental properties, as \eg the critical exponents of a given fixed point, under extension of the truncation can be considered a measure for its quality. Usually this kind of quality check requires a second complete RG analysis of a more involved truncation. In section \ref{QECG:Projection} we introduce a method that allows for a similar test of the truncation {\it within a single computation}.

\subsection[Theory spaces of metric, tetrad, and Einstein-Cartan gravity]{Theory spaces of metric, tetrad, and \hspace{3cm} Einstein-Cartan gravity}\label{prelim:TheorySpaces}

In classical General Relativity there exists a remarkably rich variety of 
different variational principles which give rise to Einstein's equation, or 
equations equivalent to it but expressed in terms of different field variables. 
The best known examples are the Einstein-Hilbert action expressed in terms of 
the metric, $S_{\text{EH}}[g_{\mu\nu}]$, or the tetrad, respectively, 
$S_{\text{EH}}[e^a_{\ \mu}]$. The latter action functional is obtained by 
inserting the representation of the metric in terms of vielbeins into the 
former: $g_{\mu\nu}=\eta_{ab} e^a_{\ \mu} e^b_{\ \nu}$. 

Another classically equivalent formulation, at least in absence of spinning 
matter, is provided by the first order Hilbert-Palatini action 
$S_{\text{HP}}[e^a_{\ \mu}, \omega^{ab}_{\ \ \mu}]$ which, besides the tetrad, 
depends on the spin connection $\omega^{ab}_{\ \ \mu}$ assuming values in the 
Lie algebra of ${\sf O(}1,3)$. Variation of $S_{\text{HP}}$ with respect to 
$\omega^{ab}_{\ \ \mu}$ leads, in vacuo, to an equation of motion which 
expresses that this connection has vanishing torsion. It can be solved 
algebraically as $\omega=\omega(e)$ which, when inserted into $S_{\text{HP}}$, 
brings us back to $S_{\text{EH}}[e]\equiv S_{\text{HP}}[e,\omega(e)]$. 

Still another equivalent formulation is based upon the self-dual Hilbert-Palatini 
action $S_{\text{HP}}^{\text{sd}}[e^a_{\ \mu}, \omega^{(+)\, ab}_{\ \ \ \mu}]$, 
which only depends on the (complex, in the Lorentzian case) self-dual 
projection of the spin connection, $\omega^{(+)\,ab}_{\ \ \ \mu}$ 
\cite{A1,A2,R,T,Kiefer}. This action in turn is closely related to the Plebanski 
action \cite{Pleb}, containing additional 2-form fields, and to the 
Capovilla-Dell-Jacobson action \cite{CDJ1,CDJ2} which involves essentially only a 
self-dual connection. Similarly, Krasnov's diffeomorphism invariant Yang-Mills 
theories \cite{Krasnov1,Krasnov2} allow for a ``pure connection'' reformulation of 
General Relativity as well as deformations thereof.

The above variational principles are Lagrangian in nature; the fields employed 
provide a parametrization of configuration space. The corresponding Legendre 
transformation yields a Hamiltonian description in which the ``carrier fields'' 
of the gravitational interaction parametrize a phase-space now. In this way the 
ADM-Hamiltonian \cite{ADM} and Ashtekar's Hamiltonian \cite{Ash-Ham}, for 
instance, make their appearance.

Regarding the ongoing search for a quantum (field) theory of gravity this 
multitude of classical formalisms offers many equally plausible possibilities 
to explore. A priori it is not clear which one of the above hamiltonian 
systems, if any, is linked to the as yet unknown fundamental quantum theory 
in the simplest or most easy to guess way.

The most prominent advantage of the functional RG approach to quantum gravity compared to other approaches is now, that we do not have to specify the classical dynamical system we are going to quantize from the outset. In this sense, it depends on the classical input data to a lesser extent. The only inspiration we draw from the classical system is its field content $\Phi$ and its group of symmetry transformations ${\bf G}$ that together form the theory space, the respective system gives rise to. The RG analysis then allows us to search for fixed points in this space, which, if found, predicts the fundamental action of the system being quantized.

The above examples of classical actions for gravity motivate the following theory spaces:

\noindent{\bf (i) Einstein gravity.} In case of the common metric description of gravity we have $\Phi=g_{\mu\nu}$, and the gauge group is given by the diffeomorphisms of the manifold ${\cal M}$, ${\bf G}={\sf Diff}({\cal M})$. We denote the corresponding theory space by
\begin{equation}
{\cal T}_{\rm E}=\big\{A[g_{\mu\nu},\cdots]\,|\text{ inv. under } {\bf G}={\sf Diff}({\cal M})\big\}\;.
\end{equation}
Most of the investigations on asymptotically safe gravity have been carried out within this theory space or extensions thereof, when additional matter fields are coupled to gravity. All of these studies found a NGFP suitable for the Asymptotic Safety construction.

\noindent{\bf (ii) ``Tetrad only'' gravity.} Expressing the metric in terms of the tetrad does not merely amount to a change of field variables on the level of theory space. As (in Euclidean spacetimes, that we consider throughout) there exists an ${\sf O}(d)$ manifold of tetrads that correspond to the same metric, this ambiguity is usually treated as an additional ${\sf O}(d)_{\rm loc}$ gauge freedom. Thus, the theory space of ``tetrad only'' gravity is given by 
\begin{equation}
{\cal T}_{\rm tet}=\big\{A[e^a{}_{\mu},\cdots]\,| \text{ inv. under } {\bf G}={\sf Diff}({\cal M})\ltimes {\sf O}(d)_{\rm loc}\big\}\;.
\end{equation}
This theory space ${\cal T}_{\rm tet}$ was first explored in \cite{Harst2012}. It is intermediate between ${\cal T}_{\rm E}$ and the Einstein-Cartan theory space to be introduced next, in the sense that the gauge group is already enlarged compared to metric gravity, while the connection is still fixed to the Levi-Civita choice. Exploring ${\cal T}_{\rm tet}$ thus can help to understand the cause of differences found between RG studies of metric and Einstein-Cartan gravity \cite{Harst2012}.

\noindent{\bf (iii) Einstein-Cartan gravity.} In comparison to the ``tetrad only'' case in Einstein-Cartan gravity we introduce the spin connection $\omega^{ab}_{\ \ \mu}$ as an additional independent field variable. The corresponding theory space is, hence, defined by
\begin{equation}
{\cal T}_{\rm EC}=\big\{A[e^a{}_{\mu},\omega^{ab}_{\ \mu},\cdots] \,| \text{ inv. under } {\bf G}={\sf Diff}({\cal M})\ltimes {\sf O}(d)_{\rm loc}\big\}\;.
\end{equation}
The first fully non-perturbative investigation of this space with $d=4$ has been published recently \cite{je:lett,je:proc,je:longpaper}. 

It is this theory space ${\cal T}_{\rm EC}$ that we shall consider in the present work, using the same truncation and $d=4$ as in the previous study, but employing a new, ``structurally approximative'' RG equation whose reliability thus can be tested by direct comparison to the findings of \cite{je:longpaper}.

\noindent{\bf (iv) Chiral gravity.} Also the restriction of the Hilbert-Palatini action to spin connections of a defined chirality gives rise to Einstein's equation. The corresponding theory space is
\begin{equation}
{\cal T}^\pm_{\rm EC}=\big\{A[e^a{}_{\mu},(\omega^{(\pm)})^{ab}_{\ \mu},\cdots]\,| \text{ inv. under } {\bf G}={\sf Diff}({\cal M})\ltimes {\sf O}(4)_{\rm loc}\big\}
\end{equation}
A first publication about the RG flow of ${\cal T}^\pm_{\rm EC}$ is in preparation \cite{ChiralLetter}.

\subsection{QECG in Holst truncation}

The main difference to ``tetrad only'' gravity lies in the fact, that in ${\cal T}_{\rm EC}$ the spin connection $\omega^{ab}{}_\mu$ is considered an independent field variable. Thus, the spacetime connection $\Gamma^\lambda_{\mu\nu}$ is no longer restricted to the Levi-Civita choice, but is allowed to carry torsion \cite{Hehl-1,Hehl-2, Mielke-newvar}. A fundamental QFT of gravity whose UV limit is taken at a NGFP in this theory space is called `Quantum Einstein Cartan Gravity' (QECG). First evidence for the existence of a suitable NGFP has been reported in \cite{je:lett,je:proc,je:longpaper}.

The truncation of ${\cal T}_{\rm EC}$ which we are going to employ is motivated by the Holst action 
\begin{equation}\label{Ho-Action}
 S_{\rm Ho}[e,\omega]=-\frac{1}{16 \pi \hat{G}}\int \!{\rm d}^4x\: e \Big[e_a{}^\mu e_b{}^\nu\Big(F^{ab}{}_{\mu\nu}-\frac{1}{2 \gamma}\varepsilon^{ab}{}_{cd} F^{cd}{}_{\mu\nu}\Big)-2\Lambda\Big].
\end{equation}
The Holst action \cite{Holst1996} generalizes the Hilbert-Palatini action of classical Einstein-Cartan gravity by the addition of the Immirzi term (with the Immirzi parameter $\gamma$ as its coupling), that vanishes on spacetimes without torsion and hence does not have a counterpart in ${\cal T}_{\rm tet}$. Thus, by the presence of this term, our truncation explicitly reflects the fact that ${\cal T}_{\rm EC}$ is a generalization of ${\cal T}_{\rm tet}$. Note, however, that also the other two terms in $S_{\rm Ho}$ differ from the torsionless case $\omega=\omega(e)$, see also the Appendix \ref{Ho-ActioninTorsion}.

There are several motivations for the inclusion of the Immirzi term. 

\noindent The first one is that the Holst action is the starting point for several other approaches to the quantization of gravity, for instance canonical quantum gravity in Ashtekar's variables \cite{A1,A2,R}, LQG \cite{T}, or spin foam models \cite{Perez2003}. Thus, the RG analysis of the Holst action may help to find and understand relations between the Asymptotic Safety scenario and these alternative approaches to quantum gravity. In LQG the dimensionless Immirzi parameter $\gamma$ determines the spectrum of area and volume operators and thus enters the expression for the entropy of black holes \cite{R}. Usually, in LQG, $\gamma$ is considered a parameter of fixed value parameterizing different quantum theories of gravity. In the RG approach, however, $\gamma$ has the status of a running coupling constant. In order to compare the two approaches to Quantum Gravity it is important therefore to relate these two quite different conceptions of the Immirzi parameter in some way.

A second motivation for including the Immirzi term is that when we include fermionic matter in the truncation so that the fermions act as a source of torsion in the field equations, the presence of the Immirzi term then gives rise to a ${\sf CP}$ violating four-fermion interaction, whose coupling constant depends on $\gamma$. Thus in this case the Immirzi parameter leads to an observable effect \cite{Perez_Rovelli:Immirzi_Parameter}, at least in principle, that may have had consequences for the evolution of the early universe \cite{Freidel:Immirzi1}. 

\subsection{Scope of the present paper}
The study of pure gravity, that we present here, can be seen as a first step towards the technically hard problem of fermions (or other matter fields) coupled to Einstein-Cartan gravity. To make computationally demanding investigations of this kind feasible we try to simplify the calculational scheme as far as possible by constructing and testing a new ``structurally approximate'' flow equation. By identifying the domain of validity of this \WHlike\ equation we therefore, besides performing an in-depth analysis of the pure gravity case, provide substantial preliminary work for future projects on matter coupled gravity.

The study presented in the following is the second fully non-perturbative RG analysis of Einstein-Cartan gravity, while the first RG study on ${\cal T}_{\rm EC}$ has been carried out in \cite{je:longpaper,je:proc,je:lett}. For the present calculation we deliberately chose a similar setting, using the same truncation with comparable gauge-fixing and ghost terms, so that, except for minor details, the only difference between the two calculations lies in the RG equation used to study the flow. While in \cite{je:longpaper} a well-tested RG equation which makes use of the proper-time approximation has been employed, we present here the first application of the new \WHlike\ flow equation. Due to the similar framework we are able to directly compare both RG equations; as neither of them is an exact equation we cannot judge the absolute validity of the respective approximations they make use of, but searching for common features of the two resulting RG flows both calculations may support each other, such that the credibility of these results is strengthened. 

The rest of this paper is organized as follows: In {\it section 2} we introduce the truncation studied and give details about the gauge-fixing conditions used as well as the resulting ghost action. {\it Section 3} introduces the new WH-like flow equation used to analyze the RG flow. {\it Section 4} contains the main part of the calculation, namely the computation of $\Gamma^{(2)}$ and the corresponding traces on the RHS of the WH-like flow equation up to second order in the spin connection $\omega$, evaluated for constant background fields. Before we can derive the exact form of the resulting $\beta$-functions we discuss in {\it section 5} different possibilities for projecting the RHS of the Flow equation onto the invariants under consideration. {\it Section 6} contains a detailed analysis of the resulting RG flow as well as a comparison to the similar study carried out in \cite{je:longpaper}. A summary and the conclusions can be found in {\it section 7}. Various technical issues of gravity theories with torsion and the new flow equation are relegated to the appendices A and B, respectively.

\vspace{1cm}
\section{The truncated Einstein-Cartan theory space}\label{QECG:truncation}
In this paper we are concerned with quantum field theories of gravity constructed, in $d=4$, from the vielbein and the spin-connection, the EAA depends on $\hat \Phi=\{\hat{e}^a{}_{\mu}, \hat{\omega}^{ab}{}_{\mu}\}$ therefore. Moreover, we impose the same symmetries as in classical Einstein-Cartan gravity, namely invariance under diffeomorphisms of the spacetime manifold ${\cal M}$, ${\sf Diff}({\cal M})$, and local frame rotations, ${\sf O}(4)_{\rm loc}$. Therefore the total group of gauge transformations has the semi-direct product structure ${\bf G}={\sf Diff}({\cal M})\ltimes{\sf O}(4)_{\rm loc}$. The proper treatment of the gauge freedom demands for the introduction of corresponding ghost fields; we shall write $\hat\Xi=\{{\cal C}^\mu, \Sigma^{ab}\}$ and $\hat{\bar \Xi}=\{{\bar{\cal C}}_\mu, \bar{\Sigma}_{ab}\}$ for the diffeomorphism and ${\sf O}(4)$ ghosts and anti-ghosts, respectively.

\vspace{0.2cm}
\paragraph{(A) Gauge transformations.} Under ${\bf G}$-transformations the quantum fields $\hat\Phi$ transform according to
\vspace{0.5cm}
\begin{equation}\label{gauge_transf}
\begin{aligned}
 \delta_{\rm D}(v)\hat{e}^a{}_{\mu}&= {\cal L}_v\hat{e}^a{}_{\mu}   & \delta_{\rm D}(v)\hat{\omega}^{ab}{}_{\mu}&= {\cal L}_v\hat{\omega}^{ab}{}_{\mu},\\
 \delta_{\rm L}(\lambda)\hat{e}^a{}_{\mu}&=\lambda^a{}_{b}\hat{e}^b{}_{\mu}, & \delta_{\rm L}(\lambda)\hat{\omega}^{ab}{}_{\mu}&=-\hat\nabla_\mu \lambda^{ab},
\end{aligned}
\end{equation}
\vspace{0.2cm}

\noindent
for diffeomorphisms and Lorentz transformations, respectively. Here, ${\cal L}_v$ denotes the Lie derivative \wrt\ the vector field $v$ and $\hat\nabla_\mu$ is the covariant derivative formed from the spin-connection $\hat\omega^{ab}{}_\mu$. The ghosts transform as diffeomorphism and ${\sf O}(4)$ tensors of the rank their index structure implies, respectively, \eg $\delta_{\rm D}(v){\cal C}^{\mu}= {\cal L}_v{\cal C}^{\mu}$, $\delta_{\rm L}(\lambda){\cal C}^{\mu}=0$.

Up to this point, the theory space of quantum actions in our setting is
\begin{equation}
 {\cal T}^{\rm quant}_{\rm EC} = \big\{A[\hat{\Phi},\hat{\Xi},\hat{\bar\Xi}]\,\big|\, \text{inv. under } {\bf G}={\sf Diff}({\cal M})\ltimes{\sf O}(4)_{\rm loc}\big\}.
\end{equation}
This space is still too small though. In order to implement Background Independence \cite{daniel2} and to carry over gauge invariance to the level of effective actions, we make use of the background field method \cite{back1, back2}. Thus, we split the quantum fields arbitrarily into a classical background field and a quantum fluctuation, $\hat \Phi=\bar\Phi+\varphi$, with 
$\bar\Phi= \{\bar{e}^a{}_\mu,\bar{\omega}^{ab}{}_\mu\}$ and $\varphi=\{\varepsilon^a{}_\mu,\tau^{ab}{}_\mu\}$. Having performed this split there exist two possibilities to distribute the gauge transformations of the full quantum fields over their constituents: 

\vspace{0.5cm}
\noindent {\bf (i)} Under {\it background gauge transformations $\delta^{\rm B}$} the background fields transform in the same manner as the respective full quantum fields, while the fluctuations transform as tensors. In particular we have

\begin{equation}
\delta^{\rm B}_{\rm D}(v) \bar \Phi={\cal L}_v \bar{\Phi} \qquad \text{and} \qquad \delta^{\rm B}_{\rm D}(v) \varphi={\cal L}_v \varphi  
\end{equation}
for the diffeomorphisms, while for ${\sf O}(d)_{\rm loc}$ transformations 
\begin{equation}
\begin{aligned}
\delta^{\rm B}_{\rm L}(\lambda)\bar{\omega}^{ab}{}_\mu&= -\bar\nabla_\mu \lambda^{ab} ,& \delta^{\rm B}_{\rm L}(\lambda)\tau^{ab}{}_\mu&=\lambda^{a}{}_c\tau^{cb}{}_\mu+\lambda^{b}{}_c\tau^{ac}{}_{\mu}\\
\delta^{\rm B}_{\rm L}(\lambda)\bar{e}^a{}_\mu&= \lambda^{a}{}_{b}\bar{e}^b_{\ \mu}, & \delta^{\rm B}_{\rm L}(\lambda)\varepsilon^a{}_\mu&=\lambda^{a}{}_{b}\varepsilon^b_{\ \mu}
\end{aligned}
\end{equation}

\vspace{0.5cm}
\noindent {\bf (ii)} Under {\it true gauge transformations $\delta^{\rm G}$}, however, the background fields stay invariant, while the fluctuations carry the transformation of the whole quantum field, \ie
\begin{equation}
\delta^{\rm G}_{\rm D}(v)\bar\Phi=\delta^{\rm G}_{\rm L}(\lambda)\bar\Phi=0,
\end{equation}
\begin{equation}
\delta^{\rm G}_{\rm D}(v)\varphi=\delta_{\rm D}(v)\hat\Phi\qquad {\rm and} \qquad \delta^{\rm G}_{\rm L}(\lambda)\varphi=\delta_{\rm L}(\lambda)\hat\Phi.
\end{equation}

If we use a gauge fixing action $S_{\rm gf}[\varphi;\bar\Phi]$ that is invariant under background gauge transformations, while it breaks the true gauge invariance, the propagator of the fluctuation fields is well-defined, and the functional integral in question
\begin{equation}\label{2ndPartFunct}
\begin{split}
Z_k[J;\bar\Phi]\equiv e^{W_k[J;\bar\Phi]}\equiv \int{\cal D}\varphi\,{\cal D}\hat\Xi\,{\cal D}\hat{\bar{\Xi}} \,{\rm exp}\Big[-S[\bar\Phi&+\varphi]-\Delta S_k[\varphi;\bar\Phi]-S_{\rm gf}[\varphi;\bar\Phi]\\ &-S_{\rm gh}[\varphi,\hat\Xi,\hat{\bar\Xi};\bar\Phi]+{\textstyle \int}\!J\cdot (\varphi,\hat\Xi,\hat{\bar{\Xi}})\Big]
\end{split}
\end{equation}
is invariant under background gauge transformations, provided that the ghost action is constructed by employing the Faddeev-Popov method and using a covariant reparametrization of the symmetry group ${\bf G}$ \cite{dhr1,je:longpaper}.

In \eqref{2ndPartFunct} we also introduced the familiar mode suppression term $\Delta S_k$, that depends on the IR cutoff scale $k$ and suppresses the contributions of $\varphi$-modes with momenta smaller than $k$ by giving a mass to them. It is quadratic in the fluctuations and vanishes for $k=0$. It is of the general form
\begin{equation}\label{IR-cutoff}
\begin{aligned}
\Delta S_k &= \frac{1}{2} \int\!{\rm d}^4 x\:\sqrt{\bar{g}}\:\big(\varphi,\hat\Xi,\hat{\bar{\Xi}} \big){{\cal R}_k [\bar\Phi]}\begin{pmatrix} \varphi \\ \hat\Xi\\ \hat{\bar\Xi} \end{pmatrix}\\&=\frac{1}{2} \int\!{\rm d}^4 x\,\sqrt{\bar{g}}\:\varphi\,{\breve{{\cal R}}_k [\bar\Phi]}\,\varphi+ \int\!{\rm d}^4 x\,\sqrt{\bar{g}}\ \hat{\bar\Xi}\,{{\cal R}^{\rm gh}_k [\bar\Phi]}\,\hat\Xi\:.
\end{aligned}
\end{equation}
Further properties of the mode suppression kernel ${\cal R}_k$ will be specified below.

From \eqref{2ndPartFunct} we can construct the one parameter family of effective actions $\tilde{\Gamma}_k$ corresponding to the bare actions $S+\Delta S_k$ as the Legendre transform of $W_k$. The effective average action, defined as $\Gamma_k[\bar\varphi,\Xi,\bar\Xi;\bar\Phi]=\tilde\Gamma_k[\bar\varphi,\Xi,\bar\Xi;\bar\Phi]-\Delta S_k[\bar\varphi,\Xi,\bar\Xi;\bar\Phi]$ with the expectation values $\bar\varphi=\langle\varphi\rangle=\{\bar{\epsilon}^a{}_\mu,\bar{\tau}^{ab}{}_\mu\}$ and $\Xi=\langle\hat\Xi\rangle=\{\xi^\mu,\Upsilon^{ab}\}$ as its arguments, can then be shown to satisfy the exact FRGE \cite{avact,nonabavact,mr,oliverbook1}
\begin{equation}\label{gaugeFRGEgen}
\partial_t \Gamma_k = \frac{1}{2}\,{\rm STr} \Big[\Big(\Gamma_k^{(2)} + {\cal R}_k\big(\Delta\big)\Big)^{-1}\partial_t{\cal R}_k\big(\Delta\big)\Big]\;.
\end{equation}
The RHS of \eqref{gaugeFRGEgen} provides a vector field on the theory space of Quantum Einstein-Cartan Gravity, whose final definition reads
\begin{equation}
 {\cal T}_{\rm EC} = \big\{A[\Phi,\bar{\Phi},\Xi,\bar\Xi]\,\big|\, \text{inv. under background transf. } {\bf G}={\sf Diff}({\cal M})\ltimes{\sf O}(4)_{\rm loc}\big\}.
\end{equation}

Any complete trajectory we may find on ${\cal T}_{\rm EC}$, \ie an integral curve of this vector field, emanating from a fixed point $\Gamma_\ast$ of finite dimensional critical hypersurface in the UV ($k\rightarrow\infty$) running towards the ordinary effective action $\Gamma=\Gamma_{k=0}$, amounts to a (non-perturbatively) renormalizable field theory of gravity with the tetrad and the spin-connection as fundamental field variables.

In practice, however, only approximations of the exact equation \eqref{2ndPartFunct} can be dealt with easily which are obtained by making an ansatz for the functional form of $\Gamma_k$ containing only a finite number of invariants, thus truncating the theory space to a finite dimensional subspace. As the exact character of the flow equation is spoiled in all explicit functional RG studies anyhow, it may be legitimate to already {\it start from an approximative FRGE}, that retains its non-perturbative character and background gauge invariance, but is an approximation already at the untruncated level. Equations of this type often have been proven useful in the past already, and in the next section we introduce a new Wegner-Houghton(WH)-like flow equation, tailored to our special needs.

\paragraph{(B) A Holst type truncation.} In order to find a basis of the theory space before the background split we must specify a complete set of linearly independent field monomials composed from the fields $\{e^a,\omega^{ab}\}$ and invariant under the action of the total symmetry group ${\bf G}$. This problem can be reformulated in a more geometrical fashion: In our setting we deal with a four dimensional Euclidean manifold ${\cal M}$ without boundary ($\partial {\cal M}=0$). Thus we may ask equivalently: How many different 4-forms can be constructed from a generic (not necessarily invertible) tetrad one-form $e^a$ and the spin connection one-form $\omega^{ab}$ that form a (pseudo-)scalar w.\,r.\,t. the \Ofourloc-transformations? It is well-known that up to the topological Euler, Pontryagin and Nieh-Yan invariants, there are only three four-forms satisfying this condition:
\begin{equation}\label{threeMonomials}
\begin{aligned}
 {\rm(a)}& & \varepsilon_{a b c d} \,e^{a}\wedge e^{b}\wedge e^{c}\wedge e^{d}&  \\
 {\rm(b)}& &\varepsilon_{a b c d}\, e^{a}\wedge e^{b}\wedge F^{cd}&\\
 {\rm(c)}& & e^{a}\wedge e^{b}\wedge F_{ab}&. 
\end{aligned}
\end{equation}
 Moreover, on spacetimes with vanishing Nieh-Yan invariant \cite{Nieh-Yan, Zan-Chan} the last form (c) is equal to the square of the torsion form $T^a= {\rm d}e^a + \omega^a{}_b \wedge e^b$:
 \begin{equation}\label{totalDivergence}
  {\rm d}(e^{a}\wedge T_a)=0= T^a \wedge T_a - e^{a}\wedge e^{b}\wedge F_{ab}\: .
\end{equation}
Here, $F^{ab}$ denotes the field strength $F^{ab}= {\rm d} \omega^{ab} + \omega^a{}_c\wedge\omega^{cb}$. The monomials (a), (b) that are contracted by the $\varepsilon$-tensor correspond to \Ofour-scalars, while only (c) forms an \Ofour-pseudo-scalar.

The first two terms (a), (b) form the Hilbert-Palatini action of classical gravity in the first order formalism, and the third monomial (c) is the Immirzi term. It is parity-odd, and from relation (\ref{totalDivergence}) we find that it vanishes on torsionless manifolds. If we restrict ourselves to local invariants, the Holst action i.\,e. the sum of the three monomials (\ref{threeMonomials}) amounts to the most general ansatz possible: its three invariants form a basis of the theory space {\it when the invertibility of the tetrad is not assumed and no background fiels are employed}. Remarkably enough, this space is finite dimensional.

Unfortunately, the formalism we use to compute the RG flow requires us to add a gauge-fixing term $S_{\rm gf}$ and a cutoff action $\Delta S_k$ to the truncation ansatz which makes the background field formalism indispensable. Both $S_{\rm gf}$ and $\Delta S_k$ can only be constructed if we assume the invertibility of the background vielbein $\bar{e}^a$. As a result the RG flow unavoidably mixes the above 3 invariants to the infinitely many additional invariants that can be constructed using the inverse background vielbein. Thus, the three-dimensional theory space described by an ansatz of the form of the Holst action
\begin{equation}\label{Ho-trunc}
\begin{aligned}
 \Gamma_{{\rm Ho}\, k}[e,\omega]&=-\frac{1}{16 \pi G_k}\left(\frac{1}{2}\int \varepsilon_{a b c d}\, e^{a}\wedge e^{b}\wedge F^{cd}- \frac{1}{\gamma_k}\int e^{a}\wedge e^{b}\wedge F_{ab}\right.\\
 &\hspace{6.5 cm}\left.-\frac{\Lambda_k}{12} \int \varepsilon_{a b c d} \,e^{a}\wedge e^{b}\wedge e^{c}\wedge e^{d} \right)\\
&= -\frac{1}{16 \pi G_k}\int {\rm d}^4x\, e\left[e_a^{\ \mu} e_b^{\ \nu} \left( F^{ab}_{\ \ \mu\nu}-\frac{1}{2\gamma_k} \varepsilon^{ab}_{\ \ cd}F^{cd}_{\ \ \mu\nu}\right)- 2\Lambda_k\right]
\end{aligned}
\end{equation}
necessarily must be regarded a three-dimensional {\it truncation} of an actually infinite dimensional theory space, when its RG flow is calculated using the exact FRGE \eqref{gaugeFRGEgen} or related functional RG techniques. We choose the truncation \eqref{Ho-trunc} as the starting point of our investigations of the RG flow of QECG. It remains to specify our choice of the gauge fixing and ghost action.

\paragraph{(C) Covariant derivatives.} In this paper we distinguish three kinds of covariant derivatives, namely the derivative $D_\mu$ which is covariant with respect to diffeomorphisms, symbolically $D= \partial + \Gamma$, the derivative $\nabla_\mu$ covariant with respect to ${\sf O}(4)$-transformations, $\nabla=\partial + \omega$, and the fully covariant derivative ${\cal D}= \partial + \Gamma+\omega$. In addition, their background counterparts $\bar{D}_\mu,\bar{\nabla}_\mu, \bar{\cal D}_{\mu}$ constructed from the corresponding background connections appear.

\paragraph{(D) Gauge fixing conditions.} As we are dealing with a gauge theory we need to fix a gauge for both, diffeomorphisms and Lorentz transformations, and add the corresponding ghost action to the truncation ansatz \eqref{Ho-trunc}. From this point on we need to assume the presence of an  invertible background vielbein $\bar{e}^a{}_\mu$. As in \cite{je:longpaper} we choose a gauge fixing action of the form 
\begin{equation}\label{QuadForm_gf}
\Gamma^{\rm gf}_{k}[\bar{e},\bar{\omega};\bar{\varepsilon}] =\frac{1}{2} \frac{1}{16 \pi G_k}\int\! {\rm d}^4 x \,\bar{e}\, \bar{g}^{\mu\nu} {\cal F}_\mu {\cal F}_\nu+\frac{1}{2} \int\! {\rm d}^4 x\, \bar{e}\,  {\cal G}^{ab} {\cal G}_{ab}\:,
\end{equation}
where ${\cal F}_\mu$ is the diffeomorphism and ${\cal G}_{ab}$ the Lorentz/\Ofour-gauge fixing condition. Explicitly, we have chosen the gauge fixing conditions
\begin{align}\label{Diff_gf}
 {\cal F}_{\mu}&= \frac{1}{\sqrt{\aD}}\bar{e}_a^{\ \nu} \left( \bar{\cal D}_\nu \bar{\varepsilon}^{a}_{\ \mu}+ \beta_{\rm D}\bar{\cal D}_\mu \bar{\varepsilon}^{a}_{\ \nu}\right)\\
{\cal G}^{ab}&= \frac{1}{\sqrt{\alpha_{\rm L}}}\bar{g}^{\mu\nu}\left( \bar{e}^b_{\ \nu} \bar{\varepsilon}^a_{\ \mu} - \bar{e}^a_{\ \nu} \bar{\varepsilon}^b_{\ \mu}\right)\:.\label{O4_gf}
\end{align}
Here $\bar{\cal D}_\mu$ is the covariant derivative constructed from both the (background) space-time and the spin-connection. 

Several comments are in order here.

\vspace{0.5cm}
\noindent{\bf (i)} The prefactors occurring in $\Gamma^{\rm gf}_k$ containing the running Newton constant $G_k$ but no gauge fixing parameters may seem unusual at first sight: First, note that the usual gauge fixing parameters for the diffeomorphisms, $\aD$, and the \Ofour-transformations, $\alpha_{\rm L}$, are shifted to the definition of the gauge fixing condition. In consequence, the ghost action derived from these gauge conditions will depend on the gauge parameters and we will see later on that only this choice allows us to discuss the limit of Landau gauge, $\alpha_{\rm L},\aD\rightarrow0$. As it can be argued that this limit is a fixed point of the generically running gauge fixing parameters, this gauge is considered a specifically reliable choice in truncations that do not treat $\aD$ and $\alpha_{\rm L}$ as running couplings.

\vspace{0.5cm}
\noindent{\bf (ii)} The prefactor in the diffeomorphism part containing the Newton constant is convenient as it simplifies the combination of contributions to the Hessian from the action and the gauge fixing term, while it renders $\aD$ dimensionless at the same time.

\vspace{0.5cm}
\noindent{\bf (iii)} For the \Ofour-part we start with a gauge fixing action  of the standard form. Note that $\alpha_{\rm L}$ has mass dimension $-4$. In order to establish an overall prefactor of $(16 \pi G_k)^{-1}$ of the gauge fixing action we redefine
\begin{equation} 
\alpha_{\rm L}=16 \pi G_k \aL 
\end{equation}
with $\aL$ having a remaining mass dimension of $-2$.

\vspace{0.5cm}
\noindent{\bf (iv)} In the diffeomorphism gauge condition \eqref{Diff_gf} we encounter an additional gauge parameter $\beta_{\rm D}$; it has been shown in \cite{je:lett,je:proc,je:longpaper} that this choice of gauge conditions indeed fixes the gauge of the 10 dimensional total gauge group ${\bf G}= {\sf Diff} \ltimes \Ofourloc$ completely for all $\beta_{\rm D}\neq -1$. It is hence confirmed that the gauge fixing term breaks the invariance of the action under true gauge transformations, $\delta^{\rm G}(\lambda,v)$, such that the resulting propagator is well-defined.

\vspace{0.5cm}
\noindent{\bf (v)} By construction, the above gauge fixing action enjoys full background gauge invariance. Thus, the above set of gauge conditions satisfies the requirements set by the general background field method.

\paragraph{(E) Construction of the ghost action.} A certain complication arises since the usual Faddeev-Popov construction for the ghost action leads to a background gauge invariant ghost action only provided that we first {\it reparameterize the gauge transformations in a covariant fashion} \cite{dhr1,je:longpaper}. This is rooted in the non-trivial structure of the gauge group ${\bf G}= {\sf Diff} \ltimes \Ofourloc$ of the system: Pure diffeomorphisms do not map \Ofour\ tensors onto \Ofour\ tensors. For that reason we have to augment the pure diffeomorphisms by an \Ofourloc-transformation that \Ofour-covariantizes the Lie derivative that generates the diffeomorphisms. We denote the reparameterized diffeomorphisms by $\widetilde{\widetilde{\delta^{\rm G}_{\rm D}}}(w)=\delta^{\rm G}_{\rm D}(w)+\delta^{\rm G}_{\rm L}(w \cdot \bar\omega)$. For the vielbein fluctuation we then obtain explicitly
\begin{equation}
 \begin{aligned}
  \widetilde{\widetilde{\delta^{\rm G}_{\rm D}}}(w)\varepsilon^{a}_{\ \mu}&=\left(\delta^{\rm G}_{\rm D}(w)+\delta^{\rm G}_{\rm L}(w \cdot \bar{\omega})\right)\varepsilon^{a}_{\ \mu}\\
&=\mathcal{L}_w e^{a}_{\ \mu} + w^\rho \bar{\omega}^a_{\ b \rho} e^b_{\ \mu}\\
&=  e^a_{\ \rho} \bar{D}_\mu w^\rho +w^\rho \bar{\cal D}_\rho e^a_{\ \mu} + w^{\rho}\bar{T}^\nu_{\rho \mu} e^a_{\ \nu}\:.
 \end{aligned}
\end{equation}
From the last line the background covariant tensor character of the expression becomes obvious. 

With this covariant reparametrization of the group of gauge transformations ${\bf G}$ at hand the ghost action is constructed, as usual, by replacing the transformation parameters of diffeomorphisms and \Ofourloc-transformations with the corresponding diffeomorphism and \Ofour-ghost fields, $(w^\mu,\lambda^{ab})\mapsto ({\xi}^\mu, \Upsilon^{ab})$. As the ghost sector is treated classically, \ie all renormalization effects of the ghost couplings are neglected, we can set all tetrad and spin connection fluctuations to zero, even before the Hessian is computed. The resulting ghost action (where we already have replaced the quantum fields by their expectation values) reads then explicitly
\begin{equation}\label{Ghost-action}
 \begin{aligned}
  S_{\rm gh}[\bar{\varepsilon},\bar{\tau},&\bar{\xi},{\xi},\bar{\Upsilon},\Upsilon;\bar{e},\bar{\omega}]\bigg|_{\genfrac{}{}{0pt}{}{\bar{\varepsilon}=0}{\bar{\tau}=0}}=\\
&=-\!\int\!{\rm d}^4 x \,\bar{e} \left(\bar{\xi}_\mu \bar{g}^{\mu\rho} \frac{\partial {\cal F}_\rho}{\partial \varepsilon^a_{\ \nu}} \widetilde{\widetilde{\delta^{\rm G}_{\rm D}}}({\xi})\varepsilon^a_{\ \nu}
+
\bar{\xi}_\mu \bar{g}^{\mu\rho} \frac{\partial {\cal F}_\rho}{\partial \varepsilon^a_{\ \nu}} \delta^{\rm G}_{\rm L}( \Upsilon)\varepsilon^a_{\ \nu}\right.
\\&\hspace{3.5cm}\left.+
\bar{\Upsilon}_{ab} \frac{\partial {\cal G}^{ab}}{\partial \varepsilon^c_{\ \nu}} \widetilde{\widetilde{\delta^{\rm G}_{\rm D}}}({\xi})\varepsilon^c_{\ \nu}
+
\bar{\Upsilon}_{ab} \frac{\partial {\cal G}^{ab}}{\partial \varepsilon^c_{\ \nu}} \delta^{\rm G}_{\rm L}(\Upsilon)\varepsilon^a_{\ \nu}\right)\bigg|_{\genfrac{}{}{0pt}{}{\varepsilon=0}{\tau=0}}\\
&=
-\!\!\int \!\!{\rm d}^4 x \, \bar{e}\left[ \bar{\xi}_\mu \bar{g}^{\mu\nu}\! \left(\bar{D}_\sigma\bar{D}_\nu\! +\! \beta_{\rm D} \bar{D}_\nu \bar{D}_\sigma\! -\! \bar{D}_\alpha \bar{T}^\alpha_{\nu\sigma}\!-\!\beta_{\rm D} \bar{D}_{\nu} \bar{T}^\alpha_{\alpha \sigma}\right){\xi}^\sigma\right.\\
&\hspace{2.1cm} + \bar{\xi}_\mu (\bar{e}_b^{\ \mu}\bar{e}_a^{\ \rho} \bar{\nabla}_\rho)\Upsilon^{ab}\\
&\hspace{2.1cm}+ \bar{\Upsilon}_{ab} \, 2 \,\bar{g}^{\mu\nu} \bar{e}^b_{\ \nu} \bar{e}^a_{\ \alpha}(\delta^{\alpha}_{\ \sigma} \bar{D}_\mu-\bar{T}^\alpha_{\mu \sigma}){\xi}^\sigma\\
&\left.\hspace{2.1cm}+ \bar{\Upsilon}_{ab} (2\,\delta^a_{\ c}\delta^{b}_{\ d})\Upsilon^{cd}\right]\,.
 \end{aligned}
\end{equation}
The different covariant derivatives appearing here act on all objects to their right.

\vspace{0.5cm}
\section{The new functional flow equation}\label{prelim:flow_equations}
The derivation of the new Wegner-Houghton(WH)-like flow equation that we present in the following consists of three successive approximative steps, but clearly has the logical status of a motivation only, 
since the validity of each of these steps can eventually be proven 
only by a comparison to results obtained from the exact equation \eqref{gaugeFRGEgen} or another reliable approximation of it. 

Our main reason for using an approximative instead of the exact RG equation is that the computational effort needed to obtain the RG flow in a given truncation can be greatly reduced in this way. This is especially useful when considering the Einstein-Cartan theory space ${\cal T}_{\rm EC}$: RG studies on this space show a drastically increased complexity mainly due to the quadruplication of independent field components compared to the metric theory space ${\cal T}_{\rm E}$. If we further want to include fermionic matter to the model in future calculations, the exact treatment is rendered a hopelessly involved task. Here, a well-tested simplified approximative FRGE would be more than welcome.

\paragraph{(A) Derivation of the new equation.} We start out from the FRGE for the effective average action, \eqref{gaugeFRGEgen},
\begin{equation}\label{startingPoint}
 \partial_{t} \Gamma_k= 
 \frac{1}{2}\;\textrm{STr}\Bigg[ 
 \frac{\partial_t \mathcal{R}_k( \Delta)}{\Gamma^{(2)}_k + 
 \mathcal{R}_k(\Delta)}\Bigg] \quad \textrm{with} \quad \mathcal{R}_k(\Delta)=  
 \mathcal{Z}_k k^2 R^{(0)}\left(\frac{\Delta}{k^2}\right)\:.
\end{equation}
The dimensionless shape function, $R^{(0)}(x)$ interpolates 
smoothly between $1$ and $0$ according to $ R^{(0)}(0)=1 \quad \textrm{and} \quad \lim_{x \rightarrow \infty} R^{(0)}(x)=0.$ The argument $\Delta$ is called cutoff operator since the 
discrimination between ``low'' and ``high'' momentum fluctuations is obtained with 
respect to its eigenmodes whose eigenvalues satisfy $\lambda<k^2$ and $\lambda>k^2$, 
respectively. The factor ${\cal Z}_k$, being a matrix in field space, is adapted to a given truncation by the so-called ${\cal Z}_k\!=\!\zeta_k$-rule: ${\cal Z}_k$ should be chosen such that for any eigenmode of $\Gamma^{(2)}_k$ with eigenvalue $\zeta_k p^2$ the sum $\Gamma^{(2)}_k + {\cal R}_k$ has eigenvalue $\zeta_k(p^2 + k^2 R^{(0)}(p^2/k^2))$. 

In a first step of approximation we choose a ``spectrally adjusted'' or ``type 
III'' \cite{codelloc} cutoff operator, namely $\Delta= \mathcal{Z}^{-1}_k 
\Gamma^{(2)}_k[\bar \Phi, \bar \Phi]$, depending on the background fields 
$\bar \Phi$ only (i.\,e. with the fluctuations $\bar\varphi$ set to zero). 
At least within the truncation we are going to consider, this choice can be seen as an approximation to the 
standard choice, the background covariant Laplacian $\Delta=-\bar{\cal D}^2$. Denoting the background field independent part of $\Gamma^{(2)}_k$ by $\Gamma^{(2)}_{0\,k}$, the $\mathcal{Z}_k\!=\!\zeta_k$-rule entails the identity $\mathcal{Z}^{-1}_k \Gamma^{(2)}_{0\,k}=-\partial^2\: \mathds{1}$. Using the spectrally adjusted cutoff, the supertrace 
of the flow equation allows for a simple spectral representation as it only 
depends on a single differential operator. We obtain

\begin{equation}\label{WExpression}
\begin{aligned}
\partial_t\Gamma_k 
 &=
 \frac{1}{2}\; \textrm{STr}\!\!\left.\left[ \frac{2 
 \left(R^{(0)}\!\left(\frac{x}{k^2}\!\right)\!-\!\frac{x}{k^2}R^{(0)}{}'\!
 \left(\frac{x}{k^2}\!\right)\right) - \eta R^{(0)}\!\left(\frac{x}{k^2}\!\right) 
 +\frac{\partial_t x}{k^2}R^{(0)}{}'\!\left(\frac{x}{k^2}\!\right)}{\frac{x}{k^2} 
 + R^{(0)}\!\left(\frac{x}{k^2}\!\right)}\right]
 \right|_{x=\mathcal{Z}^{-1}_k \Gamma^{(2)}_k} \\
 &= 
 \frac{1}{2}\;\textrm{STr}\!\left.\left[W_1(x)-\eta W_2(x) 
 + \frac{\partial_t \big(\mathcal{Z}^{-1}_k \Gamma^{(2)}_k\big)}{k^2} W_3(x) 
 \right]\right|_{x=\mathcal{Z}^{-1}_k \Gamma^{(2)}_k},
\end{aligned}
\end{equation}

\noindent
where $\eta=-\mathcal{Z}^{-1}_k(\partial_t\mathcal{Z}_k)$. The last term in the square
brackets reflects the newly introduced scale dependence of the cutoff operator. 

As a next step we represent the three functions $W_{\rm 1,2,3}(x)$ defined in the second line of (\ref{WExpression}) in terms of their respective Laplace transforms, $\widetilde{W}_{\rm 1,2,3} (s)$. This results in

\begin{equation}
\begin{aligned}\label{PTFE1}
 \partial_t\Gamma_k 
 =
 \frac{1}{2}\; \textrm{STr}\bigg[\int_0^\infty \!\!\!\! \textrm{d}s
 \bigg( \widetilde{W}_1(s)- \eta \widetilde{W}_2(s) &+ \frac{\partial_t 
 \big(\mathcal{Z}^{-1}_k \Gamma^{(2)}_k\big)}{k^2} \widetilde{W}_3(s)\!\bigg) 
 e^{-s \mathcal{Z}^{-1}_k \Gamma^{(2)}_k }\bigg]\\
 =
 \frac{1}{2}\int_0^\infty\!\!\!{\rm d} s\Bigg[ \widetilde{W}_1(s) \; \textrm{STr}\bigg[e^{-s \mathcal{Z}^{-1}_k \Gamma^{(2)}_k }\bigg]&-\widetilde{W}_2(s) \; \textrm{STr}\bigg[ \eta\, e^{-s \mathcal{Z}^{-1}_k \Gamma^{(2)}_k }\bigg]\\
&+\frac{\widetilde{W}_3(s)}{k^2}\; \textrm{STr}\bigg[\partial_t 
 \big(\mathcal{Z}^{-1}_k \Gamma^{(2)}_k\big)e^{-s \mathcal{Z}^{-1}_k \Gamma^{(2)}_k }\bigg]\!\Bigg].
\end{aligned}
\end{equation}
\vspace{0.2cm}

\noindent
The three different contributions in \eqref{PTFE1} correspond to different levels of approximation: If we only take into account the first term $\propto \widetilde{W}_1(s)$, the computation amounts to a usual 1-loop calculation with a non-standard regulator. Including the second one $\propto \widetilde{W}_2(s)$ is a first step of RG improvement: The matrix $\eta$ corresponds to the running couplings being fed back into the RHS of the flow equation. Finally, the third term $\propto \widetilde{W}_3(s)$ takes into account the additional $k$-dependence of the RHS, due to the running of the couplings in the type III cutoff operator; it can be seen as a second step of RG improvement.

It will be helpful to rewrite \eqref{PTFE1} in yet another form:

\begin{equation}
\begin{aligned}\label{PTFE2}
 \partial_t\Gamma_k 
 &=
 \frac{1}{2}\; \textrm{STr}\!\left[\int_0^\infty \! \textrm{d}s
 \left( \widetilde{W}_1(s)- \eta \widetilde{W}_2(s) + \frac{\partial_t 
 \big(\mathcal{Z}^{-1}_k \Gamma^{(2)}_k\big)}{k^2} \widetilde{W}_3(s)\right) 
 e^{-s \mathcal{Z}^{-1}_k \Gamma^{(2)}_k }\right]\\
 &=
 \frac{1}{2}\; \textrm{STr}\!\left[\int_0^\infty \! \textrm{d}s 
 \left(-\frac{\partial_t f_k(s)}{s}\right) e^{-s  \mathcal{Z}^{-1}_k 
 \Gamma^{(2)}_k }\right]\\
 &=
 \frac{1}{2}\; \textrm{STr}\!\left[\int_0^\infty \! \textrm{d}s 
 \left(-\frac{D_t f_k(\mathcal{Z}_k s)}{s}\right) e^{-s  \Gamma^{(2)}_k } 
 \right].
\end{aligned}
\end{equation}
\vspace{0.2cm}

\noindent
Here the second equality sign is meant as a {\it definition} for the matrix valued 
function $f_k(s)$. Furthermore in the last line of \eqref{PTFE2} the scale derivative $D_t$ only acts on the {\it explicit}
$k$-dependence of the function $f_k$, i.\,e. not on its argument, $\mathcal{Z}_k s$.

In a second step of approximation we replace $f_k(s)$ in the integrand by a simpler function of $s$, which is chosen such that the above definition nonetheless qualitatively reproduces the desired properties of ${\cal R}_k$. In particular, the chosen function $f_k(s)$ should regularize the integral in \eqref{PTFE2} both in the UV and the IR. Thus, it should fall off to zero quickly for arguments $s>k^{-2}$ and $s<\Lambda^{-2}$, where $\Lambda$ denotes some UV cutoff scale. A choice frequently adopted in the literature is the one parameter set of functions $f^m_k(s)$ given by \cite{prop} 
\begin{equation}\label{SmearingFunctions}
 f^m_k(s)= \frac{\Gamma(m+1, s k^2)-\Gamma(m+1, s \Lambda^2)}{\Gamma(m+1)},
\end{equation}
where $m$ is an arbitrary real, positive parameter, and $\Gamma(\alpha, x)= 
\int_x^\infty \textrm{d}t\; t^{\alpha-1} e^{-t}$ denotes the incomplete Gamma 
function. Note that $\Lambda$, thanks to the scale 
derivative in the integrand, can safely be taken to infinity at the level of the flow equation. As a result this yields a FRGE of the form
\begin{equation}\label{ProperTimeFE}
 \partial_t\Gamma_k=-\frac{1}{2}\; \textrm{STr}\!\left[\int_0^\infty \! 
 \frac{\textrm{d}s}{s} \;D_t f^m_k(\mathcal{Z}_k s)\; e^{-s \Gamma^{(2)}_k } 
 \right],
\end{equation}
which is known as the ``proper-time'' (PT) flow equation \cite{mrpt}. 

A flow equation of this type has widely been used in the literature \cite{PT1,PT2,PT5}; it is well known to give accurate results for several scalar theories \cite{PTaccurate} and has also been applied to metric gravity \cite{prop}.

The first non-perturbative study of the theory space ${\cal T}_{\rm EC}$ has also been carried out using a flow equation of the proper-time type \cite{je:lett,je:proc,je:longpaper}. Therein, first evidence for the existence of asymptotically safe trajectories in this space was found, and we will compare our findings on QECG carefully precisely to these results.

Let us come back to the proper-time FRGE \eqref{ProperTimeFE} and perform the last step of 
approximation which is the new one. The main idea here is to use different smearing functions $f^m_k$ 
depending on the momentum dependence of the exponentiated operator. At this 
point we specialize for the case of constant background fields $\bar{\Phi}$, such 
that $\Gamma^{(2)}_k$ is diagonal in momentum space. Then we can represent the 
functional trace by a momentum integral according to
\begin{equation}
 \begin{aligned}
  \partial_t\Gamma_k 
  &=
  -\frac{1}{2}\, \textrm{STr}\!\left[\int_0^\infty \! \frac{\textrm{d}s}{s} 
  \partial_t f^m_k(s) e^{-s  \mathcal{Z}^{-1}_k \Gamma^{(2)}_k }\right]\\
  &=
  -\frac{1}{2}\,\textrm{str}\left[\!\int_0^\infty \! 
  \frac{\textrm{d}s}{s}\partial_t f^m_k(s) \int \! \textrm{d}^d x\; 
  \textrm{d}^d p \;  \langle x |p \rangle \langle p | 
  e^{-s \mathcal{Z}_k^{-1}\Gamma^{(2)}_k} |x\rangle\right]\\
  &\stackrel{\bar{\Phi}=
  \textrm{const}}{=}-\frac{1}{2}\int_0^\infty \! 
  \frac{\textrm{d}s}{s}\partial_t f^m_k(s) \int \! \textrm{d}^d x 
  \int \!\frac{\textrm{d}^d p}{(2\pi)^d} \; \textrm{str}
  \left[e^{-s \mathcal{Z}_k^{-1}\Gamma^{(2)}_k(p)}\right].
 \end{aligned}
\end{equation}
Here, $\textrm{str}$ denotes the remaining {\it algebraic} supertrace. Recalling that 
$\mathcal{Z}_k$ is adapted to $\Gamma^{(2)}_{0\;k}$ by the $\mathcal{Z}_k\!=\!\zeta_k$-rule which implies $\mathcal{Z}^{-1}_k \Gamma^{(2)}_{0\,k}=-\partial^2 \:\mathds{1}$ we can pull out the dominant quadratic momentum dependence in the exponent 
and expand the remainder as a power series in the momentum variable $p$:
\begin{equation}
\begin{aligned}
  \mathcal{Z}^{-1}_k\Gamma^{(2)}_k
  &=
  \mathcal{Z}^{-1}_k\Gamma^{(2)}_{0\;k} + 
  \mathcal{Z}^{-1}_k\Gamma^{(2)}_{\bar{\Phi}\;k} 
  = 
  p^2 \mathds{1} + \mathcal{Z}^{-1}_k\Gamma^{(2)}_{\bar{\Phi}\;k} \\ 
  \Rightarrow 
  e^{-s\mathcal{Z}^{-1}_k\Gamma^{(2)}_k}
  &=
  e^{-sp^2}e^{-s\mathcal{Z}^{-1}_k\Gamma^{(2)}_{\bar{\Phi}\;k}}
  = e^{-sp^2}\left(\sum_{n=0}^{\infty} p^n A_n\right).
\end{aligned}
\end{equation}
In this series expansion the quantities $A_n$ carry the matrix character of the 
expression. In principle we have to sum over different possible momentum 
dependences in each order (e.g. $p^2$ and $p_\mu p_\nu$ in second order), but 
due to the symmetric momentum integration those will combine to a single 
contribution for every even power $p^{2n}$, while the odd powers vanish. 
Therefore we have

\begin{equation}
 \begin{aligned}\label{pExpansion}
  \partial_t\Gamma_k 
  &=
  -\frac{1}{2}\int_0^\infty \! \frac{\textrm{d}s}{s} \partial_t f^m_k(s) 
  \int\!\textrm{d}^d x\int \!\frac{\textrm{d}^d p}{(2\pi)^d} \; 
  \textrm{str}\left[ e^{-sp^2} \sum_{n=0}^{\infty} p^{2n} A_{2n}\right]\\
  &=
  -\frac{1}{2}\,\textrm{str}\left[\int_0^\infty \! \frac{\textrm{d}s}{s} 
  \int \! \textrm{d}^d x \; \sum_{n=0}^{\infty} \partial_t\left( f^m_k(s) 
  \int \!\frac{\textrm{d}^d p}{(2\pi)^d} \;  e^{-sp^2}  p^{2n}\right) 
  A_{2n}\right].
 \end{aligned}
\end{equation}
\vspace{0.2cm}

\noindent
At this point we employ our last approximation. It consists in choosing the parameter $m$
of the smearing function {\it differently} in each term of the sum $\sum_{n=0}^{\infty}$ in \eqref{pExpansion}. We pick the value 
$m=m(n)\equiv n+d/2-1$. We will see in a moment how this specific choice of $m$ changes the character of the cutoff procedure \cite{PT1}, and why this is advantageous. 

Inserting the explicit form of the family $f^m_k(s)$ of smearing functions \eqref{SmearingFunctions} we obtain

\begin{equation}
 \begin{aligned}\label{CutoffRel}
  f^{n+d/2-1}_k(s) \int \!\frac{\textrm{d}^d p}{(2\pi)^d} \; e^{-sp^2} p^{2n} 
  &=
  \frac{\Gamma(n+d/2, s k^2)-\Gamma(n+d/2, s \Lambda^2)}{\Gamma(n+d/2)}
  \cdot \\
  &\hspace{3.9 cm}\cdot 2 v_d \int_{0}^\infty \textrm{d}y\; e^{-sy}\; y^{n+d/2-1}\\
  &=
  \frac{2\,v_d}{s^{d/2+n}} \left(\Gamma(n+d/2, s k^2)-\Gamma(n+d/2, s \Lambda^2)\right)\\
  &=
  \int_{|p|=k}^{\Lambda} \!\frac{\textrm{d}^d p}{(2\pi)^d} \; e^{-sp^2}  p^{2n},
 \end{aligned}
\end{equation}

\noindent
where $v_d=(2^{d+1}\pi^{d/2} \Gamma(d/2))^{-1}$. From \eqref{CutoffRel} we observe that this last step of approximation actually corresponds to a {\it transition from proper-time 
regularization to momentum cutoff regularization} \cite{PT1}: Using the proper-time cutoff with a single smearing function $f^m$, a specific power of the momentum 
$p^{2m-d +2}$ is cut off sharply, while the others are cut off smoothly in such a way that background gauge invariance is retained. Adopting different smearing functions for each power of momentum $p^{2n}$ we obtain a {\it sharp momentum cutoff at all orders in $p$}.  

Thus the principal drawback of this step of approximation is that we spoil the background gauge invariance of the exact FRGE since we now cut off the gauge dependent momentum $p$ instead of the gauge invariant proper-time parameter $s$. As we shall argue in more detail later on, however, breaking the covariance of the FRGE should have a similar effect as choosing a less good truncation of the theory space. Hence, we will have to assess the reliability of this approximative step at the level of the resulting RG flow and its universal properties.

Inserting the relation \eqref{CutoffRel} into the flow equation \eqref{pExpansion} 
and resumming the series to an exponential we obtain

\begin{equation}
 \begin{aligned}
  \partial_t\Gamma_k 
  &=
  -\frac{1}{2}\textrm{str}\left[\int_0^\infty \! \frac{\textrm{d}s}{s} 
  \int \! \textrm{d}^d x \; \sum_{n=0}^{\infty} \partial_t\left( 
  \int_{|p|=k}^{\Lambda} \!\frac{\textrm{d}^d p}{(2\pi)^d} \; 
  e^{-sp^2}  p^{2n}\right) A_{2n}\right]\\
  &=
  -\frac{1}{2}\textrm{str}\left[\int_0^\infty \! \frac{\textrm{d}s}{s} 
  \int \! \textrm{d}^d x \; D_t \int_{|p|=k}^{\Lambda} \!
  \frac{\textrm{d}^d p}{(2\pi)^d} \;  e^{-s\mathcal{Z}_k^{-1}\Gamma^{(2)}_k} 
  \right]\,.
 \end{aligned}
\end{equation}
\vspace{0.2cm}

\noindent
In a slightly symbolic notation, this brings us to the final form of the new flow equation:
\begin{equation}\label{newWHlikeFE}
\boxed{\partial_t \Gamma_k=\left.\frac{1}{2}\; D_t\; \textrm{STr}\right|_k \ln \left(\Gamma^{(2)}_k\right).}
\end{equation}
Here STr$|_k$ denotes the infrared (IR) regularization of the trace by a sharp cutoff of the momentum integral, and from now on the scale derivative $D_t$ acts only on the explicit $k$-dependence due to this IR cutoff. Furthermore, at this level we can safely take the limit $\Lambda\rightarrow \infty$ as the expression is UV finite due to the scale derivative $D_t$. It is a FRGE of this form that we use to study the RG flow of QECG in the next section.

\paragraph{(B) Summary: the approximations leading from (3.1) to (3.12).} The derivation of the new flow equation, \eqref{newWHlikeFE}, from the exact FRGE \eqref{startingPoint} involves the following three approximative steps

\noindent{\bf (i)} Use the complete Hessian $\Gamma^{(2)}_k$ as the cutoff operator $\Delta$ in the exact FRGE \eqref{gaugeFRGEgen} in order to obtain a spectral representation of its RHS by a Laplace transformation.

\noindent{\bf (ii)} Replace the exact Laplace transform of the RHS by one member (fixed $m$) of the set of functions $f^m_k(s)/s$ which are chosen in such a way that they reproduce certain convergence properties of the exact Laplace transform.

\noindent{\bf (iii)} Choose a different value for $m$ for the terms of different momentum dependence in the integrand. Explicitly we choose $m=n+d/2-1$ for the terms proportional to $p^{2n}$.

Since these approximations, unavoidably, lack a strict mathematical 
justification we shall demonstrate their validity later on by a comparison to the result 
obtained from the proper-time equation for pure gravity \cite{je:longpaper}. If it turns out reliable, the new FRGE \eqref{newWHlikeFE} suggests itself for the analysis of even more involved systems, like first order gravity coupled to fermionic matter, in future investigations.

\paragraph{(C) Relation to the Wegner-Houghton equation.} Our approximation to the exact FRGE, \eqref{newWHlikeFE}, for 
the effective average action $\Gamma_k$ is formally equivalent to the 
Wegner-Houghton equation (for the running bare action $S_k$) \cite{Wegner1973} in the 
special case of constant background fields. In this case only the term 
corresponding to 1\,PI contributions on the RHS of the WH equation is 
non-vanishing. Therefore, it is plausible that the same equation may serve as an 
approximation for both $S_k$ and $\Gamma_k$. Due to this formal equivalence we will refer to the new FRGE \eqref{newWHlikeFE} also as a ``\WHlike\ flow equation''.

\section{Application of the new FRGE to the Holst action}\label{Evaluation_FE1}
In this section the \WHlike\ flow equation \eqref{newWHlikeFE} is applied to the Holst truncation. We will thereby develop general methods to evaluate the traces on the RHS, and we discuss a number of conceptual difficulties that arise in comparison to metric gravity.

The calculation described in this section consists of the following major steps. Starting point is the \WHlike\ equation \eqref{newWHlikeFE},
\begin{equation}\label{WHlike-allg}
 \partial_{t} \Gamma_k[\bar{e},\bar{\omega}] =\frac{1}{2} D_t \, \STr \bigg|_k \ln\Big( \Gamma^{(2)}_k[\bar e,\bar \omega]\Big)\, .
\end{equation}
In {\it subsection (\ref{Evaluation_FE})} we discuss how this rather formal equation has to be interpreted concretely and how its RHS will be expanded in the constant background fields $\bar{e}$ and $\bar{\omega}$. Subsequently, in {\it subsection (\ref{HessianET})}, we compute the main basic ingredient to the equation, the Hessian $\Gamma^{(2)}$, from our general truncation ansatz 
\begin{equation}\label{HoTruncation}
 \Gamma_k= \Gamma^{\rm Ho}_k+\Gamma^{\rm gf}_k+S_{\rm gh}\equiv\breve{\Gamma}_k+ S_{\rm gh}\:,
\end{equation}
whose constituents we introduced already. In order to calculate the trace in \eqref{WHlike-allg} we will introduce in {\it subsection (\ref{QECG:FE:Decomposition})} a decomposition of the fluctuation fields $(\bar{\varepsilon}^a_{\ \mu}, \bar{\tau}^{ab}_{\ \ \mu})$ that retains the value of the trace. An expression for $\Gamma^{(2)}_k$ in the corresponding decomposed basis is obtained in {\it subsection (\ref{QECG:FE:HessDecFieldBasis})}. In {\it subsection (\ref{EVFLowEq})} we discuss the structure of the result obtained, when the RHS is evaluated up to second order in the background field $\bar{\omega}$, but before it is projected onto the invariants of the Holst truncation.

\subsection{The general strategy for the trace calculations}\label{Evaluation_FE}
\paragraph{(A) Structure of the Hessian.} The Hessian $\Gamma^{(2)}_k$ in \eqref{WHlike-allg} is defined as the second variation of the effective action $\Gamma_k$ \wrt the (generalized) fluctuation fields $\bar{\varphi}_i$ evaluated on the background field configuration $\bar{\Phi}_i=\Phi_i\big|_{\bar{\varphi}_j=0}=\bar{\Phi}_i+\bar{\varphi}_i|_{\bar{\varphi}_j=0}$:

\begin{equation}
 \Big[\Gamma^{(2)}_{ij}(x,y)\Big]^{a_ib_j}=\frac{1}{\bar{e}}\frac{\delta}{\delta \bar{\varphi}^{b_j}_j(y)}\frac{1}{\bar{e}}\frac{\delta}{\delta \bar{\varphi}^{a_i}_i(x)} \Gamma\bigg|_{\varphi_k=0} = \langle x, i, a_i|\Gamma^{(2)}|y,j,b_j\rangle
\label{DefOfHessian}
\end{equation}

\noindent
Here, the indices $i, j,\cdots$ label different field variables and the indices $a_i, b_j, \cdots$ denote their respective index structure. Thus $\bar{\varphi}$ is the set of all fluctuation fields considered; as long as the fluctuations are not further decomposed we simply have
\begin{equation}
 \bar{\varphi}=\{\bar\varepsilon,\bar\tau\}, \quad \bar{\Phi}=\{\bar{e},\bar{\omega}\} \quad \text{and} \quad \Phi=\bar{\Phi}+\bar{\varphi} =\{e,\omega\}
\end{equation}
in the Grassmann even sector. If the value of $i$ is fixed, $\bar{\varphi}_i$ denotes a specific fluctuation field. Similarly $\bar{\varphi}^{a_i}_i$ denotes a specific component of this field. (We have suppressed the dependence of $\Gamma$ on the ``RG-time'' $t$ in \eqref{DefOfHessian} in order to avoid unnecessary notational complexity.)

The last equality in \eqref{DefOfHessian} introduces a convenient bra-ket notation. The Hessian is a Hermitian operator that can be represented by its matrix elements in any basis. Here it is given in a basis labeled by the field types $i, j$, their components $a_i, b_j$ and the spacetime points $x, y$. Later on we will see how a basis transformation to momentum space and decomposed field variables leads to the most simple form of this Hessian operator.

For the further evaluation of \eqref{WHlike-allg} it is crucial that the Hessian is computed for an {\it $x$-independent} background field configuration $\{\bar{e}^a_{\ \mu},\bar{\omega}^{ab}_{\ \ \mu}\}$. Only in this case the position dependence of the operator can be separated off, leading to the following form of the matrix elements in field space (with the component indices $a_i,b_j$ suppressed):
\begin{equation}
 \Gamma^{(2)}_{ij}(x,y)=\Gamma^{(2)}_{ij}(\partial_{x^\mu})\frac{\delta^{(4)}(x-y)}{\bar{e}}\:.
\end{equation}
As we will see below explicitly, this structure corresponds to the fact, that the operator is diagonal in a generalized momentum space. Since this special form is preserved if the operator is applied multiple times, functions of this operator decompose according to

\begin{equation}
 \Big[f\big(\Gamma^{(2)}\big)\Big]_{ij}(x,y)=\Big[f\big(\Gamma^{(2)}(\partial_{x^\mu})\big)\Big]_{ij}\frac{\delta^{(4)}(x-y)}{\bar{e}}\:,
\end{equation}

\noindent
where the function $f$ is understood as a power series in its (matrix-valued) argument.

\paragraph{(B) Cutoff operators $\bar{\cal D}^2$ and $\Box$.}
For the evaluation of the trace it is convenient to choose a basis that is adapted to the form of the Hessian operator and that at the same time allows for a simple IR cut-off procedure for modes of momentum smaller than the cutoff scale $k$. The notion of a `momentum' is defined here by a generalized covariant Laplacian $-\bar{\cal D}^2$: Its eigenfunctions with eigenvalue $p^2$ correspond to fluctuations carrying momentum $p^2$. The (IR regulated) trace is then obtained by summing over its eigenfunctions with $p^2\geq k^2$. 

Using the \WHlike\ flow equation we give up background gauge covariance to some extent in order to obtain a structurally simpler form of the RHS, where all integrations can be carried out explicitly, leading to $\beta$-functions that are rational functions in all couplings (except for logarithmic contributions to the running of $\Lambda_k$). To this end we take the trace by integrating over the complete set of eigenfunctions of the flat Laplacian $-\Box=-\bar{g}^{\mu\nu}\partial_\mu\partial_\nu$ and using an IR cutoff \wrt its eigenvalues $p^2\geq k^2$. The precise form of its eigenfunctions depends, of course, on the flat background spacetime $\bar{g}^{\mu\nu}$: As already mentioned, it is crucial to our approach that we choose constant background fields $\{\bar e, \bar\omega\}$, resulting in a flat manifold with non-vanishing torsion. As we want to consider the case of a boundaryless manifold, we choose the flat torus $T^4$ as background manifold, or equivalently a 4-cube in $\mathds{R}^4$ of volume $L^4$ with periodic boundary conditions. 

On a flat torus the spacetime connection is given by
\begin{equation}
\bar{K}^\lambda{}_{\mu\nu}=\bar{\Gamma}^\lambda_{\mu\nu}=\bar{e}_a{}^\lambda\bar{\omega}^a{}_{b\mu}\bar{e}^b{}_{\nu}\:,
\end{equation}
which is obtained from the standard split of the connection into its Levi-Civita part and the contorsion $K^{\lambda}{}_{\mu\nu}$ according to
\begin{equation}
 \bar{\Gamma}^\lambda_{\mu\nu}\equiv \big(\Gamma_{\rm LC}(\bar{e})\big)^\lambda_{\mu\nu}+ \bar{K}^\lambda{}_{\mu\nu}
\end{equation}
since $\Gamma_{\rm LC}(\bar{e})=0$ in the case at hand. Consequently, the torsion on the torus is given by the antisymmetric part of the contorsion, $\bar{T}^\lambda{}_{\mu\nu}=\bar{K}^\lambda{}_{[\mu\nu]}$.

Now we are in the position to express the action of the covariant Laplacian $-\bar{\cal D}^2$ in terms of the flat Laplacian $-\Box\equiv -\partial^\mu\partial_\mu$ and contorsion terms. Acting on a scalar we obtain
\begin{equation}
 -\bar{\cal D}^2\phi=-\Box\phi+\bar{g}^{\mu\nu} \bar{K}^\tau{}_{\mu\nu}\partial_\tau\phi\:,
\end{equation}
while for a vector field we find
\begin{equation}
 -\bar{\cal D}^2 v^\rho\!=\!-\Box v^\rho\!-\!2\bar{g}^{\mu\nu}\bar{K}^\rho{}_{\!\mu\tau}\partial_\nu v^\tau\!+\bar{g}^{\mu\nu} \bar{K}^\tau{}_{\!\mu\nu}\partial_\tau v^\rho\!+\bar{g}^{\mu\nu}\bar{K}^\alpha{}_{\!\mu\nu}\bar{K}^\rho{}_{\!\alpha\tau}v^\tau\!-\bar{g}^{\mu\nu}\bar{K}^\rho{}_{\!\mu\alpha}\bar{K}^\alpha{}_{\!\nu\tau}v^\tau\!.
\end{equation}
Thus, we see explicitly how the two candidates for the cutoff operator, $-\bar{\cal D}^2$ and $-\Box$, differ by contorsion terms. We observe that the additional terms are at most first order in derivatives, while in the highest (second) order of derivatives both choices coincide. Hence, we might expect that both cutoff operators will lead to comparable results in the high momentum regime. In this respect the change of cutoff operator, $\bar{\cal D}^2\rightarrow\Box$, is similar to switching between a type I and a type II or III cutoff operator \cite{codelloc}. However, the latter all amount to manifestly covariant choices, while we have to face the additional approximation that lies in the abandoning of background gauge invariance.

On $T^4$, the eigen-basis of $-\Box$ consists of ``plane waves''
\begin{equation}
 \psi_n(x)=\langle x|n\rangle=\frac{1}{L^{2}}\exp \bigg[i \frac{2\pi}{L} n_a \bar{e}^a_{\ \mu} x^\mu\bigg]\:,
\end{equation}
where $n=(n_a)$ is a four-component vector of (dimensionless) integers. All eigenvalues of the $-\Box$-operator are thus of the form $p^2= (2\pi/L)^2 n_a n^a$.

If we now take the functional trace of an operator $\hat{O}$ with matrix elements
\begin{equation}\label{Oform}
 \langle x |\hat{O}|y\rangle=O (\partial_{x^\mu}) \frac{\delta^{(4)}(x-y)}{\bar{e}},
\end{equation}
thereby imposing a UV cutoff $N(L)=\frac{L}{2 \pi}  P$ on the components $n_a$, we find
\begin{equation}
 \begin{aligned}
  \Tr \hat{O}\Big|^P&= \sum_{n_1,\cdots,n_4=-N}^N \langle n|\hat{O} |n\rangle = \sum_{n_1,\cdots,n_4=-N}^N \int {\rm d}^4 x\, {\rm d}^4 y \,\bar{e}^2 \langle n|x\rangle \langle x|\hat{O}|y\rangle \langle y|n\rangle\\
&=\frac{1}{(2\pi)^4}\int {\rm d}^4 x\, \bar{e}  \left[\left(\frac{2\pi}{L}\right)^4 \sum_{n_1,\cdots,n_4=-N}^N O\left( i \frac{2 \pi}{L} n_a \bar{e}^a_{\ \mu}\right)\right]\\
&= \int {\rm d}^4 x \,\bar{e} \frac{1}{(2\pi)^4} \left[\int^P_{-P} {\rm d}^4 p \, O(i p_\mu)+{\cal O}\Big(\frac{1}{L}\Big)\right]\, .
 \end{aligned}
\end{equation}
In the last step we have approximated the sum over equidistant sampling points by the corresponding integral over $p_\mu$. This approximation gets exact in the limit of infinitely many sampling points, \ie $L\rightarrow \infty$. (This approximation is equivalent to employing the Euler-MacLaurin formula and retaining its first term only.) 

In our calculation we shall read off the prefactor of the various momentum integrals in order to identify the running of the couplings $(\Lambda_k,\gamma_k,G_k)$ under consideration: Since the corresponding invariants evaluated at constant background fields are proportional to the spacetime volume $\int {\rm d}^4 x$ in front and therefore scale with $L^4$, we know that, even in the case of a finite spacetime volume $L^4$, the correction terms ${\cal O}(1/L)$, that scale at most like $L^3$, do not contribute to the running of the couplings. We thus conclude that the RG flow of the three couplings $(\Lambda_k,\gamma_k,G_k)$ on a finite torus is equivalent to the limit of an infinite torus, although the value of the invariants diverges in this (formal) limit. In the following we therefore choose to work in this infinite volume limit. In this limit the basis functions of $-\Box$ are given by
\begin{equation}
  \psi_p(x)=\langle x|p\rangle=\frac{1}{{\sqrt{2\pi}\,}^{4}}\exp \Big[ip_\mu x^\mu\Big],
\end{equation}
with $p_\mu \in \mathds{R}^4$, and the functional trace is then evaluated using formula
\begin{equation}
 \Tr \hat{O}= \int\! {\rm d}^4 x\, \bar{e} \int\! \frac{{\rm d}^4 p}{(2\pi)^4} \tr O(i p_\mu)
\end{equation}
for any operator of the form \eqref{Oform}. Here, tr denotes the remaining trace over the algebraic part of the operator $\hat{O}$.

\paragraph{(C) Irreducible basis in field space.}
After having specified how the functional part of the trace should be evaluated, we now turn to its algebraic part. Here, we have to be careful not to overcount the number of independent field components of $\varphi_i^{a_i}$, that may be less than the range of the index $a_i$ due to imposed symmetry or transversality conditions. For this reason we introduce a new basis for the algebraic part of the operator, given by $I_i$ basis fields $\Phi^{I_i}$ for each type of field $i$. The transformation matrices $v_{i\ a_i}^{\: I_i}(p)$, that depend on the direction of the momentum variable $p^\mu$ for transversal fields $\varphi^{a_i}_i$, connect the two bases. The new index $I_i$ runs from 1 up to the number of independent field components of the field $\varphi_i$. We will refer to this new basis as ``irreducible''. 

As an example consider the symmetric metric fluctuation field $h_{\mu\nu}$. The above reparametrization of field space corresponds to introducing a new field $H_I$ with $I=1,\cdots,10$, that represents the 10 independent components of the metric fluctuation. The transformation between the field types is then given by the relation
\begin{equation}
 h_{\mu\nu}=\sum_{I=1}^{10} (v_h)^I{}_{\mu\nu}\,H_I.
\end{equation}
 
The irreducible basis fields are chosen orthonormal such that the transformation matrices satisfy
\begin{equation}
 \sum_{a_i} v_{i\ a_i}^{\:I_i}\,v_{iJ_i}^{\ \ a_i}=\delta^{I_i}_{\ J_i}\quad \text{or, in matrix notation,}\quad v_i\,v_i^T= \mathds{1}.
\end{equation}
Moreover, we construct the operators
\begin{equation}\label{projector}
 \left(P_i\right)^{a_i}{}_{b_i}=\sum_{I_i}v_{iI_i}^{\ \ a_i}v_{i\ b_i}^{\:I_i}\quad \text{or} \quad \widetilde{P}_i= v_i^T\,v_i\:,
\end{equation}
that are defined on the space of tensors with index structure $a_i$. They are projectors onto the subspace spanned by the basis fields $\Phi^{I_i}$, \ie project a tensor onto its part exhibiting the symmetry and transversality properties of the field $\varphi_i$.

Of course, we can transform any matrix that shows the symmetry and transversality properties of the fields $\varphi_i$ from one basis to the other. When the indices are suppressed we will denote the matrix expressed in the ``reducible'' basis by a tilde on top. This also explains the notation in \eqref{projector}. For a general matrix $M$ we thus have:
\begin{equation}
\begin{aligned}
 (M_{ij})_{\ J_j}^{I_i}&=\sum_{a_i,b_j} (v_i)^{I_i}_{\ \ a_i}\, (M_{ij})_{\ b_j}^{a_i}\,(v_j)^{\ b_j}_{J_j} & &\Leftrightarrow & &M_{ij}=v_i \widetilde{M}_{ij}v_j^T\,,\\
 (M_{ij})_{\ b_j}^{a_i}&= \sum_{I_i,J_j}(v_i)_{I_i}^{\ \ a_i}\,(M_{ij})_{\ J_j}^{I_i} \,(v_j)_{\ b_j}^{J_j} & &\Leftrightarrow & &\widetilde{M}_{ij}=v^T_i M_{ij}v_j\,.
\end{aligned}
\end{equation}
The trace of the algebraic part of an operator, denoted \tr, is easily seen to be independent of the basis then:
\begin{equation}
\begin{aligned}
 \tr M=\sum_{i}(M_{ii})^{I_i}_{\ I_i}&=\sum_{i}\big(v_i \widetilde{M}_{ii}v_i^T\big)^{I_i}_{\ I_i}=\sum_{i}\big(v_i^Tv_i \widetilde{M}_{ii}\big)^{a_i}_{\ a_i}\\&=\sum_{i} \big(\widetilde{P}_i\widetilde{M}_{ii}\big)^{a_i}_{\ \  a_i}\hspace{0.2cm}=\sum_{i}\Big(\widetilde{M}_{ii}\Big)^{a_i}_{\ \ a_i}\hspace{0.5cm}=\tr \widetilde{M}\,.
 \end{aligned}
\end{equation}
Here the symmetry properties of $\widetilde{M}$ amount to the fact that $\widetilde{P}_i\widetilde{M}_{ii}=\widetilde{M}_{ii}=\widetilde{M}_{ii}\widetilde{P}_i$.

The new basis of the total field space is thus given by the set of basis elements that are characterized by the 4-momentum $p$ and the index $I_i$ and will be abstractly denoted by $|p,I_i\rangle$. Their representation in the previous ``position space'' basis is given by
\begin{equation}\label{PlaneWaveRel}
 \langle x, i,a_i|p,j,I_j\rangle= \delta_{ij} \:v_{i\ a_i}^{\:I_j}(p) \:\frac{e^{i p_\mu x^\mu }}{{\sqrt{2\pi}\,}^{4}}\:.
\end{equation}

\paragraph{(D) Computation of the trace.} Now we are able to evaluate the functional trace of the FRGE \eqref{WHlike-allg} in the generalized momentum space basis $|p, i, I_i\rangle$ introduced above, by first expressing the matrix elements of the Hessian in terms of the position space basis $|x,i,a_i\rangle$ and applying \eqref{PlaneWaveRel} then. We discuss the Grassmann-even part of the trace in detail and include the ghost sector, that should be treated in complete analogy, only at the very end:

\begin{equation}
\begin{aligned}
 \Tr\left[\ln\breve{\Gamma}^{(2)}\right]&=\sum_{i,I_i}\int\! {\rm d}^4 p\,\langle p,i,I_i|\ln \breve{\Gamma}^{(2)}|p,i,I_i\rangle\\
&=\sum_{\genfrac{}{}{0pt}{}{i,j,k,}{I_i,a_j,b_k}}\!\!\!\int\! {\rm d}^{4}x \, {\rm d}^{4}y\, \bar{e}^2{\rm d}^4 p\, \langle p,i,I_i|x,j,a_j\rangle \langle x,j,a_j |\ln \breve{\Gamma}^{(2)}|y,k,b_k\rangle\\[-0.3cm] & \hspace{8.7cm}\cdot\langle y,k,b_k|p,i,I_i\rangle\\[0.3cm]
&= \sum_{\genfrac{}{}{0pt}{}{i,j,k,}{I_i,a_j,b_k}}\!\!\!\int\! {\rm d}^{4}x \, {\rm d}^{4}y\, {\rm d}^4 p\,\bar{e}^2\ \delta_{ij} v_{i\ a_j}^{\:I_i} \frac{e^{-i p_\mu x^\mu }}{{\sqrt{2\pi}\,}^{4}}\bigg[\ln \breve{\Gamma}^{(2)}(\partial_{x^\mu})\bigg]^{a_j}_{jk\:b_k}\!\!\!\frac{\delta^{(4)}(x\!-\!y)}{\bar{e}}\\
& \hspace{8.7cm} \cdot \delta_{ki} v_{i I_i}^{\ \  b_k} \frac{e^{i p_\mu y^\mu }}{{\sqrt{2\pi}\,}^{4}}\\
&=\int\!{\rm d}^{4}x\,\bar{e}\int\! \frac{{\rm d}^4 p}{(2\pi)^4}\sum_{i,I_i,a_i,b_i} v_{i I_i}^{\ \  b_i}v_{i\ a_i}^{\:I_i}\bigg[\ln \breve{\Gamma}^{(2)}(i p_\mu)\bigg]^{a_i}_{ii\:b_i}\\
&= \int\!{\rm d}^{4}x\,\bar{e}\int\! \frac{{\rm d}^4 p}{(2\pi)^4} \sum_{i,a_i,b_i}(P_i)^{b_i}_{\ a_i} \big[\ln \breve{\Gamma}^{(2)} (i p_\mu)\big]^{a_i}_{ii\:b_i}\:.
\end{aligned}
\end{equation}

\noindent
Next we rewrite the momentum integral in spherical coordinates, splitting it into a radial and an angular part, where the former is cut off at the IR momentum scale $k$:

\begin{equation}
 \STr\bigg|_k\!\!\left[\ln\breve{\Gamma}^{(2)}\right]=\frac{1}{{(2\pi)^4}}\!\int\!{\rm d}^{4}x\,\bar{e}\int_k^\infty\!{\rm d} p\, (p^2)^{3/2} \!\int\!{\rm d}\Omega_p \! \sum_{i,a_i,b_i}(P_i)^{b_i}_{\ a_i} \big[\ln \breve{\Gamma}^{(2)} (i p_\mu)\big]^{a_i}_{ii\:b_i}.
\end{equation}

\noindent
The angular part of the momentum integration can be carried out using the rules of symmetric integration, such that odd powers of $p_\mu$ in the integrand vanish and even ones are replaced by certain combinations of the metric, for instance, 

\begin{align}
 \int {\rm d}\Omega_p\, p_{\mu}p_\nu f(p^2)&= \frac{\pi^2}{2} \bar{g}_{\mu\nu} p^2f(p^2)\\
\int {\rm d}\Omega_p\, p_{\mu}p_\nu p_{\rho}p_\sigma f(p^2)&= \frac{\pi^2}{12} \big[\bar{g}_{\mu\nu}\bar{g}_{\rho\sigma}+\bar{g}_{\mu\rho}\bar{g}_{\nu\sigma}+\bar{g}_{\mu\sigma}\bar{g}_{\nu\rho}\big] p^4 f(p^2)\:.
\end{align}

\noindent
Performing this integration is obviously a simple algebraic task such that only a radial integral with an integrand depending on $p^2$ remains. This remaining integration is removed, however, when we apply the scale derivative $D_t$ to it leaving us with minus the integrand evaluated at the lower boundary $p=k$.

If we finally include the ghost sector with the usual minus sign the full \WHlike\ flow equation \eqref{WHlike-allg} can hence be written in the form

\begin{equation}\label{WHlike_konkret}
\boxed{\begin{aligned}
 \partial_t \Gamma_k[\bar{e},\bar{\omega}]&=-\frac{k^4}{(2\pi)^4} \int\!{\rm d}^{4}x\, \bar{e}\Bigg[\int\!{\rm d}\Omega_p \sum_{i,a_i,b_i}\frac{1}{2}(P_i)^{b_i}_{\ a_i} \big[\ln \breve{\Gamma}^{(2)}_k (i p_\mu)\big]^{a_i}_{ii\:b_i}\\
 &\hspace{4.1cm}-\int\!{\rm d}\Omega_p \!\!\!\sum_{{\genfrac{}{}{0pt}{}{i\in}{{\rm ghosts}}},a_i,b_i}(P_i)^{b_i}_{\ a_i} \big[\ln S^{(2)}_{{\rm gh}\,k} (i p_\mu)\big]^{a_i}_{ii\:b_i}\Bigg]_{p=k}
\end{aligned}}
\end{equation}

At this point only the logarithm of the Hessian $\breve{\Gamma}^{(2)}_k$ is left awaiting evaluation. First of all, we should mention that the logarithm $\ln M$ is only defined for a matrix $M$ given in the ``irreducible'' basis, \ie with capital indices $I_i,J_j$. This is due to the fact that the determinant of any matrix fulfilling \eg transversality conditions is zero, $\det \widetilde{M}=0$, as its rows are not independent. Hence it is not invertible in the larger matrix space with index structure $a_i,b_j$ and thus its logarithm $\ln \widetilde{M}$ is not defined.

As we want to project the RHS onto invariants that are at most quadratic in the (constant) background field $\bar{\omega}^{ab}_{\ \ \mu}$, we now expand the logarithm up to second order in this field. To this end we split the Hessian into a part independent of $\bar{\omega}$ and a part containing all $\bar{\omega}$-dependence,

\begin{equation}
 \breve{\Gamma}^{(2)}= H_0 + V(\bar{\omega})\: ,
\end{equation}

\noindent
with $V$ being a matrix whose elements are at least linear in $\bar{\omega}$. Thus, the logarithm can be expanded in $V$ according to

\begin{equation}
\begin{aligned}
 \ln \breve{\Gamma}^{(2)}&=\ln\! \big[ H_0\! +\! V(\bar{\omega})\big]=\ln \!\big[ H_0\big(\mathds{1}\! +\! H_0^{-1}V(\bar{\omega})\big)\big]=\ln\! \big[ H_0\big]\! +\! \ln\! \big[\mathds{1}\!+\! H_0^{-1}V(\bar{\omega})\big]\\&= \ln \big[ H_0\big]+H_0^{-1}V(\bar{\omega})-\frac{1}{2}\big(H_0^{-1}V(\bar{\omega})\big)^2+{\cal O}(\bar{\omega}^3)\:.
\end{aligned}
\end{equation}

\noindent
While the parts depending on $\bar\omega$ can be computed by simple matrix multiplication, the logarithm of $H_0$ is rewritten using $\tr \ln H_0=\ln \det H_0$. We thus find in the two bases 

\begin{equation}\label{logarithm}
 \boxed{\begin{aligned}
  \tr\ln \breve{\Gamma}^{(2)}&=\sum_{i,a_i,b_i}(P_i)^{b_i}_{\ a_i} \big[\ln \breve{\Gamma}^{(2)} (i p_\mu)\big]^{a_i}_{ii\:b_i}\\
&= \ln \det H_0+\tr H_0^{-1}V-\frac{1}{2}\, \tr(H_0^{-1}V)^2+{\cal O}(\bar\omega^3)\\
&= \ln\det v \widetilde{H_0} v^T+\tr \widetilde{P}\widetilde{H_0^{-1}}\widetilde{V}-\frac{1}{2} \,\tr\widetilde{P}(\widetilde{H_0^{-1}}\widetilde{V})^2+{\cal O}(\bar\omega^3)\:.
 \end{aligned}}
\end{equation}

The last hurdle to take is the computation of the inverse of $H_0$. It turns out that we will have to perform a complete transverse-traceless decomposition of the fluctuation fields $\bar{\varepsilon}, \bar{\tau}$ to achieve this goal. In this field basis, $H_0$ has a very simple structure then: It is block diagonal in the field space (indices $i,j$) and diagonal in the space of field components (indices $a_i, b_j$), \ie we can write

\begin{equation}
\begin{aligned}\label{H0inv1}
(H_0)^{\ \: a_i}_{ij\ b_j}&=(H_0)_{ij} \otimes P_{i\ b_j}^{\; a_i} & \text{for}\qquad a_i &\sim b_j\:,\\
(H_0)^{\ \: a_i}_{ij\ b_j}&=0 & \text{for}\qquad a_i &\nsim b_j\:.
 \end{aligned}
\end{equation}

\noindent
Here, the block-diagonality is reflected in the fact, that the elements of $H_0$ are only non-zero if the corresponding fields $\varphi_i$ and $\varphi_j$ have an index structure of the same type ($a_i \sim b_j$), \eg both are scalars, divergence-free vectors etc. Hence, in this basis $H_0$ can be inverted easily by inverting the first factor of the tensor product:

\begin{equation}\label{H0inv2}
 (H_0^{-1})^{\ \: a_i}_{ij\ b_j}=(H_0^{-1})_{ij} \otimes P_{i\ b_j}^{\; a_i}
\end{equation}

\noindent
Correspondingly the determinant in \eqref{logarithm} simplifies to

\begin{equation}
 \det v \widetilde{H_0} v^T=\det\big[ (H_0)_{ij} \otimes v^{I_i}_{\ a_i} P_{i\ b_j}^{\; a_i} v_{J_j}^ {\ \,b_j}\big]=\det\big[ (H_0)_{ij} \otimes \mathds{1}^{I_i}_{\ J_j}\big]\:.
\end{equation}

\noindent
Upon exploiting the identity $\det \big[ M\otimes N\big] = (\det M)^{\rank N}\cdot (\det N)^{\rank M}$ we arrive at
\begin{equation}\label{detH0}
\boxed{  \det v \widetilde{H_0} v^T= \prod \big(\det (H_0)_{ij}\big)^{\delta^{I_i}_{J_j}}\:,}
\end{equation}
Here the symbolic product sign denotes the product over the different blocks of the matrix $(H_0)_{ij}$ in field space. In each block the fields have the same index structure, and $\delta^{I_i}_{\ J_j}$ therefore results in the number of independent field components, the fields in each block have.

\paragraph{(E) Summary of the formal trace evaluation.} Taken together eqns. {\bf \eqref{WHlike_konkret}}, {\bf \eqref{logarithm}} and {\bf \eqref{detH0}} give a concrete meaning to the RHS of the \WHlike\ flow equation {\bf \eqref{WHlike-allg}} and show up the way we are going to evaluate it. Note that, using the \WHlike\ flow equation, we have reduced this task to (comparatively) easy manipulations of matrices in field space, which amounts to a great simplification compared to the use of the FRGE or a proper-time flow equation. Moreover, we note that using the last line of \eqref{logarithm} together with \eqref{detH0}, we never have to construct the basis of independent field components or the transformation matrices $v_i$ explicitly as we were able to reformulate all algebraic matrix contributions in terms of the ``reducible'' basis using only the projectors to the irreducible subspaces.

\paragraph{(F) Limitations of the approach.} The approach outlined above has a number of limitations. First, as our technique requires the use of constant background fields, we can only project out invariants that do not vanish in this case and thus have at least some part that stays non-zero for algebraic reasons (in our case the $\bar{\omega}^2$-term of the field strength $\bar{F}$). Nonetheless, even if this is the case it might not be possible to uniquely map the remaining constant parts onto the invariants one started with. Projecting out the flow of couplings that correspond to certain invariants from the RHS then becomes ambiguous, a problem which we will address in more detail in section \ref{QECG:Projection}.

Second, our approach does not retain background gauge invariance. Thus, we do not quantize the system on all possible background spacetimes, but the flat background plays a distinct role. Note that the loss of background gauge invariance is not due to the very use of a constant background field to evaluate the RHS; when employing the FRGE, for example, it is also possible to choose a specific background spacetime to project out the invariants one is interested in. The important difference here is, that we use a cutoff operator $-\Box$ to evaluate and cut off the trace, that does not correspond to the covariant momentum operator evaluated on that background, \ie $-\bar{\cal D}^2 \big|_{\bar{e},\bar{\omega}={\rm const}}\neq -\Box$.

However, as pointed out before, the loss of background gauge invariance is not too much a drawback. At the level of the effective action $\Gamma=\Gamma_{k=0}$, where all fluctuations have been integrated out and from which all observables are to be calculated, it does not matter which background was used to classify the fluctuation fields according to their momentum. Hence, the exact untruncated flow in the background invariant approach leads to the same effective action as the one using a distinct background spacetime. The difference is the theory space the trajectory lies in: In the background invariant approach the fields $\{\bar e, \bar \omega\}$ arrange themselves to gauge invariant field monomials, that amount to a subspace of the theory space spanned by all possible field monomials of $\{\bar e, \bar \omega\}$, where the trajectory in our case lives in. As we should expect that the existence of a UV fixed point does not depend on the details of the cutoff procedure chosen, both approaches should in principle be suitable to investigate the UV behavior of the theory. Seen in this light, the loss of background invariance is equivalent to an additional approximation, as we truncate a larger theory space to the same number of invariants of a given truncation ansatz.

\subsection[The Hessian $\breve{\Gamma}^{(2)}_k$ in the $(\bar{\varepsilon}, \bar{\tau})$-basis]{The Hessian $\boldsymbol{\breve{\Gamma}^{(2)}_k}$ in the $\boldsymbol{(\bar{\varepsilon}, \bar{\tau})}$-basis}\label{HessianET}
In this subsection we present the result of the second variation of the truncation ansatz with respect to the fields $\ea$ and \oab, where $\delta\ea=\bar{\varepsilon}^a_{\ \mu}$ and $\delta \oab =\bar{\tau}^{ab}_{\mu}$. Suppressing all indices, the pertinent quadratic action $\breve{\Gamma}^{\text{quad}}_k$ is connected to the Hessian $\breve{\Gamma}^{(2)}_k$ according to
\begin{equation}
 \breve{\Gamma}^{\rm quad}_k=\frac{1}{2} \int{\rm d}^4x\,{\rm d}^4y\, \bar{e}^2\:\begin{pmatrix}\bar{\varepsilon}(x),\bar{\tau}(x)\end{pmatrix}\breve{\Gamma}^{(2)}_k(x,y) \begin{pmatrix} \bar{\varepsilon}(y)\\\bar{\tau}(y) \end{pmatrix}\:.
\end{equation}
Thus, the Hessian can be obtained from $\breve{\Gamma}^{\rm quad}_k$ by first symmetrizing the quadratic form \wrt both fluctuation fields and their index symmetries, and then canceling the $\frac{1}{2}$ factor, the integrations and the leftmost fluctuation field, while replacing the rightmost fluctuation field by $\delta^{(4)}(x-y)/\bar{e}$.

We will present $\breve{\Gamma}^{\rm quad}_k=\Gamma^{\rm quad}_{{\rm Ho}\,k}+\Gamma^{\rm quad}_{{\rm gf}\,k}$ as a sum of the contributions stemming from the Holst action and those from the gauge fixing terms. Note that the latter only contributes to the $(\bar{\varepsilon},\bar{\varepsilon})$-block of the quadratic form and that $\Gamma_{{\rm gf}\,k}$ is already quadratic in the fluctuations such that $\Gamma_{{\rm gf}\,k}^{\rm quad}=\Gamma_{{\rm gf}\,k}$. The evaluation of $\Gamma^{\rm quad}_{{\rm Ho}\,k}$ is straightforward, yielding
\begin{equation}
\begin{aligned}\label{QuadForm_Ho}
 \Gamma^{\rm quad}_{{\rm Ho}\,k}\!=\!-\frac{1}{2} \frac{Z_{{\rm N}k}}{16 \pi \hat{G}}\!\int\!\! {\rm d}^4 x\bigg[\bar{\varepsilon}^c_{\ \rho}\! \Big[\varepsilon^{\mu\nu\rho\sigma}\varepsilon_{abcd}\Big(\frac{1}{4}\Big(\!\delta^{a}_{\, [e }\delta^{b}_{\, f]}\!-\!\frac{1}{\gamma_k}\varepsilon^{ab}_{\ \ ef}\!\Big)\bar{F}^{ef}_{\ \ \mu\nu}\! -\! \bar{e}^a_{\ \mu}\bar{e}^b_{\ \nu}\Lambda_k \Big)\!\Big]\bar{\varepsilon}^{d}_{\ \sigma}\\
+\frac{1}{2}\,\bar{\varepsilon}^e_{\ \rho} \varepsilon^{\mu\nu\rho\sigma}\varepsilon_{abef}\bar{e}^f_{\ \sigma} \Big(\delta^a_{\ [c}\delta^b_{\ d]}-\frac{1}{\gamma_k}\varepsilon^{ab}_{\ \ cd}\Big) \bar{\nabla}_\mu \bar{\tau}^{cd}_{\ \ \nu}\\
+2\,\bar{\tau}^{ce}_{\ \ \mu}\left[\bar{e}\, \bar{e}_a^{\ \mu}\bar{e}_b^{\ \nu}\Big(\delta^a_{\ [c}\delta^b_{\ d]}-\frac{1}{\gamma_k}\varepsilon^{ab}_{\ \ cd}\Big)\eta_{ef}\right]\bar{\tau}^{fd}_{\ \ \nu}\bigg]\:.
\end{aligned}
\end{equation}
The ghost action $S_{\rm gh}$, eq. \eqref{Ghost-action}, too, is already quadratic in $\{{\xi}_\mu,\Upsilon^{ab}\}$ so that the Hessian can be directly read off from $S_{\rm gh}\equiv S^{\rm quad}_{\rm gh}$.

\subsection{Irreducible components of the dynamical fields}\label{QECG:FE:Decomposition}

\paragraph{Decomposition of $\bar{\boldsymbol\varepsilon}$ and $\bar{\boldsymbol\tau}$.}
In the above undecomposed form it turns out technically not feasible to directly invert the total $\bar{\omega}$-independent part of the bosonic Hessian operator $H_0= \breve{\Gamma}^{(2)}_k|_{\bar{\omega}=0}$ even for constant background fields $\bar{e}$. This is owed to the fact, that $H_0$ is a 40\,$\times$\,40-matrix operator (16+24 independent components of $\bar{\varepsilon}^a_{\ \mu}$ and $\bar{\tau}^{ab}_{\ \ \mu}$) with a complicated matrix structure. 

To facilitate the diagonalization we decompose the fluctuation fields into transverse and longitudinal parts \wrt the generalized ``plane wave'' basis introduced before which we are going to employ in order to evaluate the trace later on. For the vielbein fluctuation we write

\begin{equation}\label{tetrad_decomp}
\bar{\mu}^{\frac{1}{2}} \bar{\varepsilon}^{a}_{ \ \mu}(x)= \frac{\partial^a\partial_\mu}{-\Box}a(x)+ \frac{\partial_\mu}{\sqrt{-\Box}}b^a(x) + \frac{\partial^a}{\sqrt{-\Box}}c_\mu(x) + d^a_{\ \mu}(x)\:,
\end{equation}

{\noindent}while the spin connection fluctuation is decomposed according to 

\begin{equation}\label{connection_decomp}
 \bar{\tau}^{ab}_{~\mu} (x)\! =\! \frac{\bar{\mu}^{ \frac{1}{2}} }{\sqrt{2}}\!\left[\!\frac{\partial_\mu \partial^{[a}}{ - \Box}A^{b]} (x) \!+\! \frac{\partial^{[a}}{\sqrt{\! - \Box}} B^{b]}_{~\,\mu} (x) \!+\! \varepsilon^{ab}_{~~cd}\frac{\partial_\mu \partial^c}{- \Box} C^d (x) \!+\! \varepsilon^{ab}_{~~cd} \frac{\partial^c}{\sqrt{\! - \Box}} D^d_{~\mu} (x)\!\right].
\end{equation}

\noindent
Several comments on this decomposition are in order:

\vspace{0.2cm}
\noindent{\bf (A)} All fields occurring in the decomposition have a vanishing divergence in all their indices, \ie
\begin{equation}
 \partial_a b^a = 0 = \partial^\mu c_\mu\:,~~ \partial_a d^a_{~\mu} = 0\:,~~\partial^\mu d^a_{~\mu} = 0
\end{equation}
and
\begin{equation}
 \partial_a A^a = 0 = \partial_a C^a\:,~~\partial_a B^a_{~\mu} = 0 = \partial_a D^a_{~\mu}\:,~~ \partial^\mu B^a_{~\mu} = 0 = \partial^\mu D^a_{~\mu}\:.
\end{equation}
Here, partial derivatives with an \Ofour-index contain the background vielbein implicitly, $\partial_a =\bar{e}_a^{\ \mu} \partial_{\mu}$, $\partial^{a}=\bar{e}^a_{\ \mu}\partial^{\mu}=\bar{e}^a_{\ \mu}\bar{g}^{\mu\nu}\partial_{\nu}$.

\vspace{0.2cm}
\noindent{\bf (B)} We have introduced a rescaling of the fluctuations using the positive definite operator $-\Box=-\bar{g}^{\mu\nu}\partial_\mu \partial_\nu$; the powers of $\sqrt{-\Box}$ present in the different terms hereby correspond to the number of partial derivatives acting on the respective component field. Thus, we achieve that, first, all component fields have the same mass dimension and, second, all $\Box$ operators appearing in the Hessian that are only due to the decomposition will be canceled by the denominators, such that they cannot be confused with the original kinetic terms. Hence, the Hessian operator $\Gamma^{(2)}_{\rm Ho}$ will be first order in the derivatives, before and after the decomposition. Note that when passing to the generalized plane wave basis, we find the simple replacement rules $\partial_\mu\rightarrow i p_\mu$, $\sqrt{-\Box}\rightarrow \sqrt{p^2}$, for derivatives acting to the right. 

\vspace{0.2cm}
\noindent{\bf (C)} In the decompositions \eqref{tetrad_decomp}, \eqref{connection_decomp} we have introduced a mass scale $\bar{\mu}$, that may appear artificial at first sight, but is crucial in order to define a Hessian operator of definite mass dimension from the quadratic form $\Gamma^{\rm quad}$: As the tetrad is dimensionless but the spin connection has mass dimension 1, the different blocks of the matrix resulting from splitting off the fluctuations differ in dimension. Such a matrix, however, does not give rise to an operator whose spectrum and trace are well defined, as we then have to sum up quantities of different mass dimension. By introducing the additional mass scale $\bar{\mu}$ we equalize the mass dimensions of all component fields. In principle, this can be done in an arbitrary way, rescaling for example only the vielbein fluctuation by $\bar{\mu}^1$, but as in \cite{je:longpaper} we opted for the symmetric scheme.

\vspace{0.2cm}
\noindent{\bf (D)} In comparison to the conventions of ref. \cite{je:longpaper} we used a different numerical prefactor in the decomposition \eqref{connection_decomp}. Only with this choice the trace of an operator is invariant under the decomposition. Otherwise, the component fields contribute with different weights to the trace. As a simple check of the decomposition the following equation should hold without any additional numerical prefactors in front of the component fields:

\begin{equation}
 \int{\rm d}^4 x \,\bar{\tau}_{ab}^{\ \ \mu}\bar{\tau}^{ab}_{\ \ \mu}=\int{\rm d}^4 x \, \big(A^a A_a+ B_a^{\ \mu} B^a_{\ \mu}+C^a C_a+ D_a^{\ \mu} D^a_{\ \mu}\big)
\end{equation}

\noindent
For the decomposition of the tetrad an analogous relation holds.

In a second step we further decompose all tensorial component fields in the above decompositions, $\{d,B,D\}$, into trace, antisymmetric and symmetric-traceless part. Using the $d^a_{\ \mu}$-field as an example the decomposition reads
\begin{equation}\label{d_decomp}
 d^a_{\ \mu}= -\frac{1}{\sqrt{3}}\left(\bar{e}^a_{\ \mu}+\frac{\partial^a \partial_\mu}{(-\Box)} \right)d +\frac{1}{\sqrt{2}}\,\varepsilon^{a}_{\ bcd}\bar{e}^b_{\ \mu} \frac{\partial^c}{\sqrt{-\Box}}\, d^d+ \widehat{d}^a_{\ \mu}\,.
\end{equation}
Here, $d$ describes the scalar trace mode, $d^a$ is a divergence-free vector field representing the antisymmetric part of $d^a_{\ \mu}$ and $\widehat{d}^a_{\ \mu}$ is a traceless symmetric tensor ($\bar{e}_a^{\ \mu}\widehat{d}^a_{\ \mu}=0$, $\bar{e}^{b \mu}\widehat{d}^a_{\ \mu}=\bar{e}^{a \mu}\widehat{d}^b_{\ \mu}$), with vanishing divergence in both indices separately, $\partial_a \widehat{d}^a{}_\mu=0=\partial^\mu\widehat{d}^a{}_\mu$.

Remarks analogous to points {\bf (B)} and {\bf (D)} above apply here, concerning the rescaling with the operator $-\Box$ and the numerical prefactors chosen.

For the other tensor fields, $\{B,D\}$, we proceed in complete analogy. As we will work only in the completely decomposed setting later on, the tensor fields $\{d,B,D\}$ will not be of importance any more. For that reason, we will drop from now on the hat on the last component field, $\{\widehat{d},\widehat{B},\widehat{D}\}\rightarrow\{d,B,D\}$, such that the latter denote divergence-less, symmetric and traceless tensors, from now on.

\paragraph{Decomposition of the ghost fields.} The diffeomorphism ghost $\xi_\mu$  can be split into a transverse part $g_\mu$ and a longitudinal component $f$ according to
\begin{equation}\label{DiffGh_decomp}
 \xi_\mu= \frac{\partial_\mu}{\sqrt{-\Box}}f+ g_\mu\quad\text{with}\quad \partial^\mu g_\mu=0
\end{equation}
The \Ofour-ghost $\Upsilon^{ab}$ consists of two parts $\{F,G\}$ that are dual to each other and are rescaled by a factor of $\bar{\mu}$:
\begin{equation}
\label{O4Gh_decomp}
 \bm^{-1}\Upsilon^{ab}=\frac{1}{\sqrt{2}}\left( \frac{\partial^{[a} }{\sqrt{-\Box}}F^{b]}+\varepsilon^{ab}_{\ \ cd} \frac{\partial^{c}}{\sqrt{-\Box}} G^{d}\right)\quad \text{with} \quad\partial_a F^a=0=\partial_a G^a.
\end{equation}
The decomposition of the anti-ghosts $\bar{\xi}$ and $\bar{\Upsilon}$ is analogous and can be obtained by putting a bar on every field appearing in eqns. \eqref{DiffGh_decomp} and \eqref{O4Gh_decomp}.

\subsection{The Hessian in the decomposed field basis}\label{QECG:FE:HessDecFieldBasis}
By substituting the decompositions of the fluctuations introduced in the previous subsection into the expression of the quadratic forms \eqref{QuadForm_Ho}, \eqref{QuadForm_gf} and for the ghost sector \eqref{Ghost-action} we can transform the quadratic forms to the component field basis. After this substitution we use integration by parts until all derivatives act to the right; as the background manifold is a flat torus no boundary terms arise. A large fraction of the occurring terms vanishes due to the transversality conditions that all component fields satisfy. Nonetheless, the decomposition down to the level of the component fields generates a huge number of terms: The Grassmann-even sector consists in total of 14 component fields, so that the corresponding Hessian is a 14\,$\times$\,14 matrix in field space whose 196 entries are, due to their remaining index structure, still complicated expressions involving $\bar{e}^a_{\ \mu}$, $\bar{\omega}^{ab}_{\ \ \mu}$ and the partial derivative operator $\partial_\mu$.

For the evaluation of the flow equation we split the Hessian according to $\breve{\Gamma}^{(2)}=H_0+V(\bar{\omega})$ and $S_{\rm gh}^{(2)}=H^{\rm gh}_0+V^{\rm gh}(\bar{\omega})$ into a ``free'' part $H_0$ and an interaction part $V(\bar{\omega})$. The next two paragraphs will be devoted to these building blocks.

\paragraph{The $H_0$ part of the Hessian.}\enlargethispage{1cm}
While the total Hessian is an extremely complicated object in the decomposed field basis, its free part $H_0$ is comparatively simple although it can not be written in an as compact form as in \eqref{QuadForm_Ho}. The free part of the total quadratic form stemming from the Holst action \eqref{QuadForm_Ho} and the gauge fixing term \eqref{QuadForm_gf} in the component field basis, reads explicitly
\begin{equation}\label{H0_decomposed}
\begin{aligned}
 &\breve{\Gamma}^{\rm quad}_k\Big|_{\bar{e}={\rm const}, \bar{\omega}=0}= \frac{1}{2}\frac{\ZNk}{16 \pi \hat{G}}\int\! {\rm d}^4 x\,\bar{e}\,\times\\[0.2cm]
&\times \left[ 
\begin{pmatrix}
a\\ d\\ B \\ D
\end{pmatrix}^{\rm \!\! T}
\left(\begin{smallmatrix}
\frac{(1+\bD)^2}{\aD \bm} \hP^2& \frac{\sqrt{3}}{\bm}\left( 2 \Lambda_k+ \frac{\bD(1+\bD)}{\aD}\hP^2\right)&0&0\\
\frac{\sqrt{3}}{\bm}\left( 2 \Lambda_k+ \frac{\bD(1+\bD)}{\aD}\hP^2\right)& \frac{1}{\bm}\left(4\Lambda_k+3 \frac{\bD^2}{\aD}\hP^2\right) &2\sqrt{2}\hP&-\frac{2\sqrt{2}}{\gamma_k}\hP\\
0& 2\sqrt{2}\hP& 2\bm &-\frac{2}{\gamma_k}\bm\\
0&-\frac{2\sqrt{2}}{\gamma_k}\hP&-\frac{2}{\gamma_k}\bm& 2 \bm\\
\end{smallmatrix}\right)
\begin{pmatrix}
 a\\ d\\ B \\ D
\end{pmatrix}\right.\\[0.2cm]
&+
\begin{pmatrix}
b_a\\c_a\\d_a\\A_a\\B_a\\C_a\\D_a
\end{pmatrix}^{\rm \!\!\! T}
\!\!\!\left(\begin{smallmatrix}
\frac{2}{\aL\bm}& -\frac{2}{\bm}\left(\Lambda_k+ \frac{1}{\aL}\right)&0&0&0&0&0\\
-\frac{2}{\bm}\left(\Lambda_k+\frac{1}{\aL}\right)& \frac{1}{\bm}\left(\frac{\hP^2}{\aD}+\frac{2}{\aL}\right)&0&0&-\frac{2}{\gamma_k}\hP&0&2\hP\\
0&0&\frac{1}{\bm}\left(2\Lambda_k+\frac{4}{\aL}\right)&0&\sqrt{2}\hP&0&-\frac{\sqrt{2}}{\gamma_k}\hP\\
0&0&0&0&-\frac{\sqrt{2}}{\gamma_k}\bm&0&\sqrt{2}\bm\\
0&-\frac{2}{\gamma_k}\hP&\sqrt{2}\hP&-\frac{\sqrt{2}}{\gamma_k}\bm&\bm&\sqrt{2}\bm&-\frac{1}{\gamma_k}\bm\\
0&0&0&0&\sqrt{2}\bm&0&-\frac{\sqrt{2}}{\gamma_k}\bm\\
0&2\hP&-\frac{\sqrt{2}}{\gamma_k}\hP&\sqrt{2}\bm&-\frac{1}{\gamma_k}\bm&-\frac{\sqrt{2}}{\gamma_k}\bm&\bm
\end{smallmatrix}\right)\!\begin{pmatrix}
 b^a\\c^a\\d^a\\A^a\\B^a\\C^a\\D^a
\end{pmatrix}\\[0.2cm]
&+ \left.
\begin{pmatrix}
d_a^{\ \mu}\\B_a^{\ \mu}\\D_a^{\ \mu}
\end{pmatrix}^{\rm \!\! T}
\left(\begin{smallmatrix}
-\frac{2}{\bm}\Lambda_k&-\sqrt{2}\hP&\frac{\sqrt{2}}{\gamma_k}\hP\\
-\sqrt{2}\hP&-\bm&\frac{\bm}{\gamma_k}\\
\frac{\sqrt{2}}{\gamma_k}\hP&\frac{\bm}{\gamma_k} &-\bm
\end{smallmatrix}\right)
\begin{pmatrix}
 d^a_{\ \mu}\\B^a_{\ \mu}\\D^a_{\ \mu}
\end{pmatrix}\right]\;.
\end{aligned}
\end{equation}
Here, we introduced the notations $c^a=\bar{e}^a{}_\mu c^\mu$ and $\hP\equiv\sqrt{-\Box}$.

The free part of the Hessian, $H_0$, can be read off from this result, being the 14\,$\times$\,14 block diagonal matrix operator in field space, whose blocks are given by the above matrices, including the overall prefactor $\ZNk/(16 \pi \hat G)$.

Let us comment on the general structure of $H_0$:

\vspace{0.2cm}
\noindent{\bf (A)} While the above form of $H_0$ is notationally more complicated than in the undecomposed basis, structurally it is simpler: We observe that it is diagonal in the remaining index structure and block diagonal in field space, such that scalar, vector and tensor components only couple to themselves. 

This simple form was to be expected from the outset, as all component fields are transverse and the partial derivative $\partial_\mu$ in the free part $H_0$ is the only object carrying an index (the constant tetrads $\bar{e}^a_{\ \mu}$ are being absorbed by partial derivatives changing the type of their index). Thus, they can only combine to $-\Box$-operators in $H_0$ as all free indices give a vanishing contribution due to the transversality of the fields. For the same reason, no coupling between component fields of different tensor type can occur in $H_0$.

Due to this simplicity $H_0$ can be inverted by inverting its matrix structure in field space and taking the tensor product with the identity operators on the respective field space \ie scalars, transverse vectors and transverse symmetric traceless tensors (cf. eqns. \eqref{H0inv1}, \eqref{H0inv2}). Thus, by the decomposition of the fields we have reached our main goal, namely the explicit invertibility of $H_0$.

\vspace{0.2cm}
\noindent{\bf (B)} One can check the plausibility of the result \eqref{H0_decomposed} by several consistency checks \cite{je:longpaper}. As the component fields with small/capital letters correspond to the fluctuations $\bar{\varepsilon}$/$\bar{\tau}$ we can easily compare with the $\bar{\varepsilon}$/$\bar{\tau}$-block structure in the undecomposed basis. 

\vspace{0.2cm}
\noindent{\bf (i)} We see that all elements stemming from the $\bar{\varepsilon}\bar{\varepsilon}$-block are proportional to $\bm^{-1}$, while those from the $\bar{\tau}\bar{\tau}$-block come with $\bm$ and the $\bar{\varepsilon}\bar{\tau}$-blocks are not affected. Thus the $\bm$-rescaling has the desired effect of giving $H_0$ a well defined mass dimension.

\vspace{0.2cm}
\noindent{\bf (ii)} The first order derivatives of the Holst action appearing in the $\bar{\varepsilon}\bar{\tau}$-block are here reflected in the $\hP$-operators in the corresponding blocks. All $\hP^2$-operators in $H_0$ appear in the $\bar{\varepsilon}\bar{\varepsilon}$-block and correspond to the second order derivatives that are part of the diffeomorphism gauge fixing action.

\vspace{0.2cm}
\noindent{\bf (iii)} We can divide the component fields into tensors and pseudo tensors. Since the complete fluctuation $\bar{\tau}$ transforms as a tensor under parity, the tensor components come with an even number and the pseudo-tensor components with an odd number of $\varepsilon_{abcd}$ pseudo-tensors as prefactors in the field decomposition \eqref{connection_decomp}, \eqref{d_decomp}. As noted before, the Immirzi term in the Holst action corresponds to its pseudo scalar part; thus, all pseudo scalar contributions to $\breve{\Gamma}^{\rm quad}$ are proportional to $\frac{1}{\gamma_k}$, while the scalar contributions do not contain the Immirzi parameter. As one can check, the Immirzi parameter occurs in all matrix elements coupling tensors with pseudo tensors (and only there) forming pseudo scalar contributions.

\vspace{0.2cm}
\noindent{\bf(C)} We have cross-checked our result for $H_0$ with the corresponding result from \cite{je:longpaper} and were able to verify that all differences occurring are due to the different numerical prefactors in the decomposition of the fields and a factor of 2 traced back to a different definition of the \Ofour-gauge condition.

\vspace{0.5cm}
In the ghost sector the analogous decomposition is much simpler. For the free part $H_0^{\rm gh}$ of the quadratic form $S_{\rm gh}$ we find:

\begin{equation}\label{Sgh_decomposed}
 S_{\rm gh}\Big|_{\genfrac{}{}{0pt}{}{\bar{e}={\rm const},}{\bar{\omega}=0}}=\int\! {\rm d}^4 x\,\bar{e}\,
\begin{pmatrix}
 \bar{f}\\\bar{g}_a\\\bar{F}_a\\\bar{G}_a
\end{pmatrix}^{\!\!\rm T}
\begin{pmatrix}
 -\frac{(1+\bD)}{\sqrt{\aD}}\hP^2 &0&0&0\\
0&0&-\frac{\bm}{\sqrt{2\aD}}\hP&0\\
0&-\frac{\sqrt{2}\bm}{\sqrt{\alpha_{\rm L}}}\hP&\frac{2 \bm^2}{\sqrt{\alpha_{\rm L}}}&0\\
0&0&0&\frac{2\bm^2}{\sqrt{\alpha_{\rm L}}}
\end{pmatrix}
\begin{pmatrix}
 f\\g^a\\F^a\\G^a
\end{pmatrix}
\end{equation}

\noindent
From \eqref{Sgh_decomposed} we see explicitly that for $\bD=-1$ the diffeomorphism gauge fixing condition breaks down and the ghost operator develops a zero mode in the scalar sector, while it is invertible for all other values of $\bD$.

\paragraph{The interaction part $V(\bar{\omega})$.} 
For the interaction part $V(\bar{\omega})$ of the Hessian the index structure of its elements does not simplify in a comparable manner: Here we encounter the objects $\bar{\omega}^{ab}_{\ \ \mu}$, $\varepsilon_{abcd}$ and $\partial_\mu$ which carry indices that can be contracted in various ways using the background vielbein $\bar{e}^a_{\ \mu}$ and its inverse. As long as the free indices of the resulting expressions do not belong to partial derivatives, they may contribute to $V(\bar\omega)$. For this reason, most of the matrix elements of $V(\bar\omega)$ are so complicated expressions that it is not instructive to write down $V(\bar\omega)$ in the decomposed basis. (The total expression would fill many pages.)

Let us mention, however, that if we split $V(\bar\omega)=V^1(\bar\omega)+V^2(\bar\omega)$, where $V^1$ is linear and $V^2$ quadratic in $\bar{\omega}$ (note that there are no higher order terms in the action), these matrices have a certain block structure in field space, that can already be read off from the $(\bar{\varepsilon},\bar{\tau})$ representation \eqref{QuadForm_Ho} and \eqref{QuadForm_gf}. For constant background fields $\{\bar{e},\bar{\omega}\}$ we find schematically:
\begin{equation}
 V^1(\bar\omega)=
\left.\begin{pmatrix}
 \Big(\Gamma^{(2)}_{\rm gf}\Big)_{\bar{\varepsilon} \bar{\varepsilon}}&\Big(\Gamma^{(2)}_{\rm Ho}\Big)_{\bar{\varepsilon} \bar{\tau}}\\[0.2cm]
\Big(\Gamma^{(2)}_{\rm Ho}\Big)_{\bar{\tau} \bar{\varepsilon}}&0
\end{pmatrix}\right|_{\genfrac{}{}{0pt}{}{\text{linear}}{\text{in }\bar{\omega}}}\!\!,\ 
V^2(\bar\omega)=
\left.\begin{pmatrix}
 \Big(\Gamma^{(2)}_{\rm gf}+\Gamma^{(2)}_{\rm Ho}\Big)_{\bar{\varepsilon} \bar{\varepsilon}}&0\\[0.2cm]
0&0
\end{pmatrix}\right|_{\genfrac{}{}{0pt}{}{\text{quadratic} }{\text{in }\bar{\omega}}}
\end{equation}
This structure can be exploited to simplify the calculation later on.

\subsection{Evaluation of the flow equation}\label{EVFLowEq}
At this point, after having defined all ingredients to the flow equation 
\begin{multline}\label{WHlike_konkret2}
 \partial_t \Gamma_k[\bar{e},\bar{\omega}]=-\frac{k^4}{(2\pi)^4} \int\!{\rm d}^{4}x\, \bar{e}\bigg[\int\!{\rm d}\Omega_p \sum_{i,a_i,b_i}\frac{1}{2}(P_i)^{b_i}_{\ a_i} \big[\ln \breve{\Gamma}^{(2)} (i p_\mu)\big]^{a_i}_{ii\:b_i}\\
 -\int\!{\rm d}\Omega_p \sum_{{\genfrac{}{}{0pt}{}{i\in}{{\rm ghosts}}},a_i,b_i}(P_i)^{b_i}_{\ a_i} \big[\ln S^{(2)}_{\rm gh} (i p_\mu)\big]^{a_i}_{ii\:b_i}\bigg]_{p=k}
\end{multline}
we are now able to evaluate both of its sides. If we substitute the Holst truncation \eqref{Ho-trunc} into its left hand side and switch to dimensionless couplings,
\begin{equation}
  g_k=k^{2}G_k,\qquad \lambda_k=k^{-2} \Lambda_k,
\end{equation}
in order to obtain an autonomous system of $\beta$-functions, we find

\begin{equation} 
\begin{aligned}\label{LHS}
 \partial_t\Gamma_k[\bar{e},\bar{\omega}]=&-\frac{k^2}{16 \pi g_k}\left(2-\frac{\partial_t g_k}{g_k}\right) & &\!\!\!\!\!\cdot\int\!{\rm d}^4 x\, \bar{e}\, \bar{e}_a^{\ \mu}\bar{e}_b^{\ \nu} \bar{F}^{ab}_{\ \ \mu\nu}\\
&+\frac{k^2}{16 \pi g_k}\left(2-\frac{\partial_t g_k}{g_k}-\frac{\partial_t \gamma_k}{\gamma_k}\right) \frac{1}{\gamma_k}& &\!\!\!\!\!\cdot\int\!{\rm d}^4 x\, \bar{e}\, \frac{1}{2} \bar{e}_a^{\ \mu}\bar{e}_b^{\ \nu} \varepsilon^{ab}_{\ \ cd} \bar{F}^{cd}_{\ \ \mu\nu}\\
&+\frac{k^2}{16 \pi g_k}\left(2-\frac{\partial_t g_k}{g_k}+2+\frac{\partial_t \lambda_k}{\lambda_k}\right) 2 \lambda_k k^2& &\!\!\!\!\!\cdot\int\!{\rm d}^4 x\, \bar{e}\:.
\end{aligned}
\end{equation}
We may further simplify the field monomials by inserting the constant background fields $\bar{e}$ and $\bar{\omega}$, leading to 

\begin{equation}
  \bar{e}\, \bar{e}_a^{\ \mu}\bar{e}_b^{\ \nu} \bar{F}^{ab}_{\ \ \mu\nu}=\bar{e}\,(\bar{\omega}_{abc}\bar{\omega}^{acb}-\bar{\omega}^{a}_{\ ca}\bar{\omega}^{bc}_{\ \ b}), \quad 
\frac{1}{2}\,\bar{e}\, \bar{e}_a^{\ \mu}\bar{e}_b^{\ \nu} \varepsilon^{ab}_{\ \ cd} \bar{F}^{cd}_{\ \ \mu\nu}=\bar{e}\varepsilon_{abcd}\bar{\omega}_{e}^{\ ab}\bar{\omega}^{ecd}\;.
\end{equation}

{\noindent}Here, we have used the background vielbein to formally change the spacetime index of $\bar{\omega}$ to an \Ofour-index. This is only done for notational simplicity; if needed, the tetrads can be restored at any point of the calculation.

Now, let us turn to the right hand side of \eqref{WHlike_konkret2}. We have to extract all terms from the logarithms that are independent of or second order in $\bar{\omega}$, as the invariants we want to project on are either of zeroth or of second order in $\bar{\omega}$. For this case the expansion of the logarithms \eqref{logarithm} simplifies to
\begin{equation}\label{logarithm_simpl}
\tr\ln \breve{\Gamma}^{(2)} = \frac{1}{2}\ln \bigg[\prod \big(\det (H_0)_{ij}\big)^{\delta^{I_i}_{J_j}}\bigg]^2+\tr \widetilde{P}\widetilde{H_0^{-1}}\widetilde{V^2}-\frac{1}{2} \tr\widetilde{P}(\widetilde{H_0^{-1}}\widetilde{V^1})^2+\cdots
\end{equation}
(and analogous for the ghost sector). The dots stand for terms linear in $\bar{\omega}$ and ${\cal O}(\bar{\omega}^3)$-terms. Note, that all terms linear in $\bar\omega$ vanish anyway when the momentum integration is carried out, as no field monomials exist that have a linear part for a constant field $\bar{\omega}$. Thus, \eqref{logarithm_simpl} still represents a full expansion of the right hand side up to second order in $\bar{\omega}$. In the $\bar{\omega}$-independent part of \eqref{logarithm_simpl} we have replaced $\ln\det H_0\rightarrow \frac{1}{2}\ln \det H_0^2$, as $H_0$ may have negative eigenvalues.

As a next step we now substitute $H_0$, $V^1$ and $V^2$ in the decomposed field basis into \eqref{logarithm_simpl} and the result into \eqref{WHlike_konkret2}. Then we are left with an expression containing all possible monomials that are quadratic in $\bar{\omega}$; in order to prepare for a comparison of their coefficients on both sides of the flow equation \eqref{WHlike_konkret2} we would like to cast it into the form
\begin{equation}
\boxed{\begin{aligned}
 \partial_t \Gamma_k= \text{rhsF}\, \cdot k^2 &\int\!{\rm d}^4 x\, \bar{e}\,(\bar{\omega}_{abc}\bar{\omega}^{acb}\!-\!\bar{\omega}^{a}_{\ ca}\bar{\omega}^{bc}_{\ \ b}) +\,\text{rhsF}^* \cdot k^2 \int\!{\rm d}^4 x\, \bar{e}\,\varepsilon_{abcd}\bar{\omega}_{e}^{\ ab}\bar{\omega}^{ecd}\\+ \,\text{rhs}\Lambda\, \cdot k^4  &\int\!{\rm d}^4 x\, \bar{e} + k^2 \int\!{\rm d}^4 x\, \bar{e} \,(\text{further, independent $\bar{\omega}^2$-monomials})\:.
\end{aligned}}\label{RHS}
\end{equation}
Here, rhsF, rhsF${}^*$, and rhs$\Lambda$ denote, in an obvious notation, the dimensionless functions of the dimensionless couplings $(\lambda, \gamma, g)$ and the gauge fixing parameters $(\aD,\aL,\bD)$ which arise when we compute the terms of interest delivered by the supertrace on the right hand side of the flow equation.

\paragraph{Properties of rhsF and rhsF${}^*$.} If we analyze the expression \eqref{RHS} in more detail, we find that the exact form of the coefficients rhsF and rhsF${}^*$ depends on the {\it basis of independent field monomials quadratic in $\bar{\omega}$} which we choose; only after we have fixed the form of the ``further independent $\bar{\omega}^2$-monomials'' in \eqref{RHS}, thus selecting a particular basis in theory space, the coefficients of the three invariants we are actually interested in assume uniquely defined values! The details of this {\it projection ambiguity} are discussed in detail in the next section.\footnote{The explicit form of the complete RHS containing all independent $\bar{\omega}^2$-monomials does not depend on the basis of theory space chosen, but only on the gauge parameters. Since the result for general gauge parameters fills about 20 pages without being very illuminating, we only display this unambiguous result for the $(\aD,\aL,\bD)\!=\!(0,0,0)$ gauge in Appendix \ref{Appendix:UnambiguousRHS}.} 

While the detailed expressions for the functions rhsF and rhsF${}^*$ are basis dependent, we can, however, discuss the general form of the result, \ie how these functions structurally depend on the couplings. This is possible as the generic form of all prefactors of the $\bar\omega^2$-expressions corresponding to the same parity is equal. In any basis chosen the functions rhsF and rhsF${}^*$ are linear combinations of these prefactors and, thus, of the same form. 

We find that neither rhsF nor rhsF${}^*$ depends on the Newton constant $g$, and their functional form in terms of the Immirzi parameter is very simple. This can be understood as $\breve{\Gamma}^{\rm quad}$ contains the Newton constant only as a global prefactor that drops out in the expansion of the logarithm and is only present in the $\bar{\omega}$-independent determinant \ie in rhs$\Lambda$. On the other hand $\breve{\Gamma}^{\rm quad}$ contains a factor $\gamma^{-1}$ in every parity-odd element, leading to a simple $\gamma$-dependence of its inverse. We therefore expect, that the parity-even prefactor rhsF only contains even powers and, in contrast, rhsF${}^*$ only odd powers of $\gamma$. The dependence on the cosmological constant is, however, very involved and comprises a complicated dependence on the gauge fixing parameters as well, whose details depend on the basis chosen. Explicitly the coefficient functions possess the structure
\begin{equation}\label{rhsForm}
\boxed{
\begin{aligned}
  {\rm rhsF}(\lambda,\gamma)&=\frac{1}{\gamma^2} \frac{P_8(\lambda)}{N(\lambda)}+\frac{P_{10}(\lambda)}{N(\lambda)}\,,\\
 {\rm rhsF}^*(\lambda, \gamma)&=\frac{1}{\gamma} \frac{P_9(\lambda)}{N(\lambda)}\:.
\end{aligned}
}
\end{equation}
Here, $N(\lambda)$ is a common denominator, which is a polynomial in $\lambda$ of degree 10 given by
\begin{multline}\label{Denominator}
 N(\lambda)=(\lambda-1)^2(\aL\lambda+2)^2\big(3\aD\lambda^2+(2\bD^2+\bD-1)\lambda+\bD^2+2\bD+1\big)\cdot\\\cdot\big(2\aD\aL\lambda^2+4\aD\lambda-1\big)^2\big(1+\bD\big)^2
\end{multline}
Remarkably, $N(\lambda)$ turns out basis independent. The functions $P_n(\lambda)$ are polynomials in $\lambda$ of degree $n$. Their explicit form is extremely complicated and basis dependent. Each of them fills several pages as long as the gauge parameters are not set to specific values.

\paragraph{Properties of rhs$\Lambda$.} The coefficient function rhs$\Lambda$ is basis independent as no other invariants constructed from the tetrad only exist. It is given by
\begin{multline}\label{rhsL}
 {\rm rhs}\Lambda(\lambda,\gamma,g)=-\frac{1}{32\pi^2} \bigg[\ln\Big(\frac{\gamma^2-1}{\gamma^2}\Big)^{24}+\ln\big[(\lambda-1)^{10}(2+\aL \lambda)^6\big]\\ +\ln\big[(3\aD\lambda^2\!+\!(2\bD^2\!+\!\bD\!-\!1)\lambda\!+\!\bD^2\!+\!2\bD\!+\!1)^2(2\aD\aL\lambda^2\!+\!4\aD\lambda\!-\!1)^6\big]\\
-\ln\big[ g^{68}\pi^{68}M^{160}\mu^{32}(1+\bD)^4\big]-\ln{\cal N}\bigg]
\end{multline}
Here, $\mu$ and $M$ are dimensionless mass parameters: $\mu\equiv\bar{\mu}/k$ stems from the rescaling of the fluctuation fields performed earlier and $M\equiv\bar{M}/k$ is a mass parameter, that has to be introduced in order to render the argument of the logarithm dimensionless. As both parameters only occur in the combination $\mu M^5$ we can substitute them by a single parameter $m=M=\mu$ in the following. ${\cal N}$ is a purely numerical factor with $\ln{\cal N}\approx 167.74$. 

Note that in the above expression for ${\rm rhs}\Lambda$ the limit of $(\aL,\aD)\rightarrow0$ is well-defined as no argument of the logarithm vanishes completely. If we, however, decompose ${\rm rhs}\Lambda= {\rm rhs}\Lambda_{\rm grav} + {\rm rhs}\Lambda_{\rm gh}$ into graviton and ghost contributions we find that both parts contain logarithmic terms that diverge in this limit, but cancel each other in the sum. Here we see explicitly that this cancelation of divergences is only obtained due to the incorporation of the gauge parameters into the gauge conditions; otherwise the ghost contributions would have been $(\aL,\aD)$-independent and the limit of the $(\aL,\aD)=(0,0)$-gauge would not be well defined.

\paragraph{General structure of the RG equations.} Independent of the detailed expressions for the coefficients in \eqref{rhsForm} we can derive the form of the resulting $\beta$-functions for the couplings $(\lambda,\gamma,g)$ by equating the coefficients in  \eqref{LHS} and \eqref{RHS}, leading to

\begin{equation}\label{BetaForm}
\boxed{ \begin{aligned}
  \partial_t g=\beta_g&= g \big[2+ 16 \pi g \text{ rhsF}(\lambda,\gamma)\big]\\
  \partial_t \gamma=\beta_\gamma&= -16 \pi g \gamma\big[\gamma \text{ rhsF}^*(\lambda,\gamma) + \text{ rhsF}(\lambda,\gamma)\big]\\
  \partial_t \lambda=\beta_\lambda&= 16 \pi g \lambda \text{ rhsF}(\lambda,\gamma)+8 \pi g \text{ rhs}\Lambda(\lambda,\gamma,g) - 2 \lambda
 \end{aligned}}
\end{equation}
This system of coupled RG differential equations is one of our main results. In the rest of this paper we shall study the resulting RG flow for various gauge choices and projection schemes that will be introduced in section \ref{QECG:Projection}. The properties of the RG flow are discussed then in detail in section \ref{QECG:Results}.

\section{Projection schemes in theory space}\label{QECG:Projection}
In this section we discuss the influence of the choice of projection scheme in theory space on the form of the resulting $\beta$-functions. Since typically only projections of RG flows to a subspace of theory space can be studied, the conclusions of this section are of general importance beyond the present application. 

In {\bf subsection (\ref{ChoiceOfBasis})} we demonstrate that {\it defining a projection scheme requires not only to specify the basis invariants spanning the subspace of theory space to be considered, but also, either explicitly or implicitly, to specify the basis invariants to be discarded}. 

In {\bf subsections (\ref{ParityOddSubSp}) and (\ref{ParityEvenSubSp})} we apply these general findings to the present truncation and discuss {\it how the freedom of choice of basis invariants to be discarded can be used to optimize the reliability of the projected flow.}

\subsection{Role of the basis in the exact theory space}\label{ChoiceOfBasis}
It is clear from the outset that $\beta$-functions generically depend on the basis chosen in theory space, since couplings serve as coordinates in theory space and the $\beta$-functions obtain as their scale derivatives. Thus, $\beta$-functions themselves and the running couplings they give rise to can not be considered physical quantities. All observable quantities must depend on them in such a way that their basis dependence drops out.

\begin{figure}[htb]
 \includegraphics[height=2.8cm]{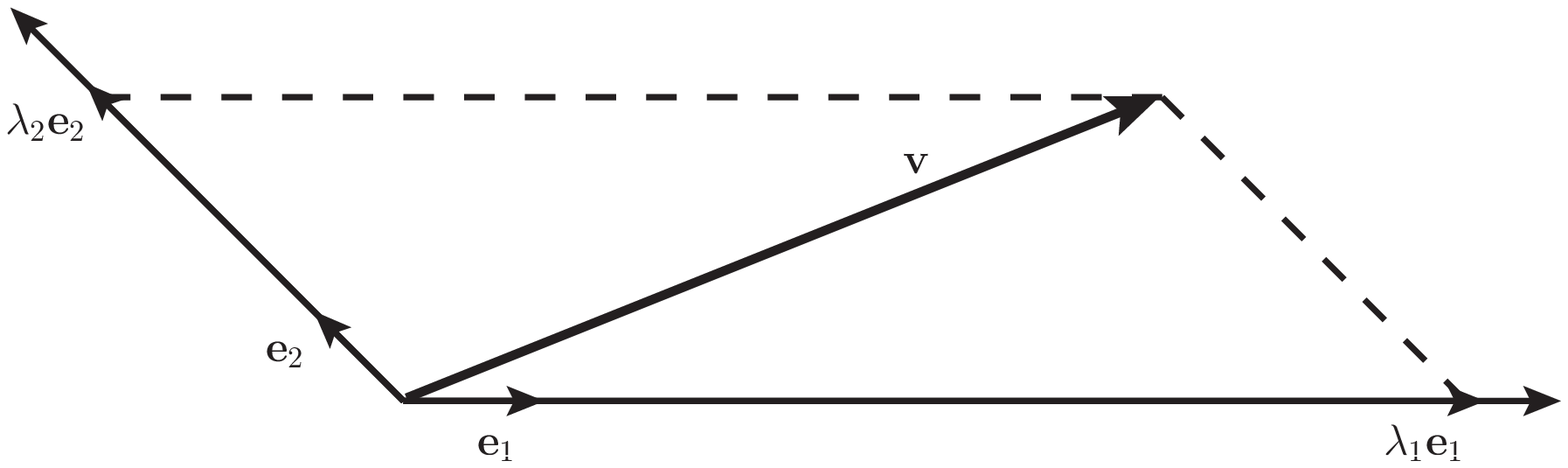}
\includegraphics[height=2.8cm]{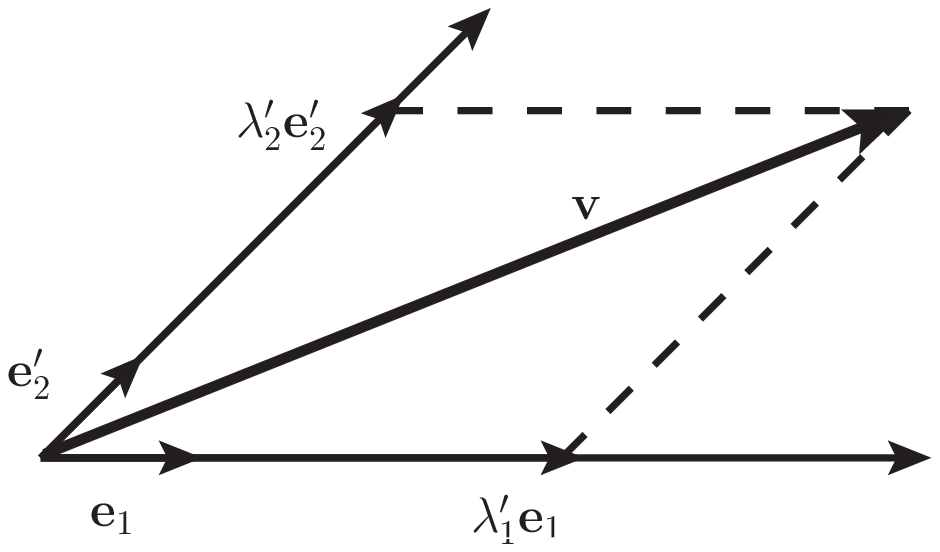}
\caption{Illustrative example: The vector ${\bf v}$ is decomposed \wrt the two bases $({\bf e}_1,{\bf e}_2)$ and $({\bf e}_1,{\bf e}_2')$. Although the basis vector ${\bf e}_1$ does not change under the basis transformation, the corresponding coordinate $\lambda_1\rightarrow \lambda_1'$ transforms non-trivially. This is because the spaces spanned by ${\bf e_2}$ and ${\bf e_2'}$ that form the kernel of the projection do not coincide.}\label{projection}
\end{figure}

A more subtle consequence of the basis dependence is that $\beta$-functions of a given invariant may transform non-trivially even if we perform a change of basis involving only {\it the other} basis invariants. 

In a general vector space the coordinate of a fixed basis element only stays constant if the space spanned by the other basis elements is invariant under the basis transformation (cf. Fig. \ref{projection}). Thus, in order to fix a coordinate of some vector \wrt\ a given basis element we also have to specify {\it the space spanned by all other basis elements}. This corresponds to the definition of a projection operator onto the basis element: Its kernel defines the space spanned by the other basis elements while its range is spanned by the element we want to project on. Note that only if the vector space is equipped with a scalar product, orthogonal projections can be defined, that allow the construction of the kernel from the range of the projector. Thus, for the definition of a projection scheme in a vector space lacking the notion of orthogonality, like theory space, {\it we not only need to specify the invariants we want to project on, but also those which should be discarded}.

If we use a systematic expansion of the invariants in theory space, as \eg the derivative expansion, the $\beta$-functions of couplings corresponding to invariants up to a given order will not change any more once all basis invariants up to that order are fixed. From the above general considerations this is easily understood since in this case the kernel of the projection scheme used is implicitly defined as the space of higher derivative invariants. As this kernel does not change no matter which invariants are used as its basis, the $\beta$-functions of the lower derivative couplings are defined unambiguously once a basis in the range of the projection scheme has been chosen.  

Within a subspace of a fixed order of derivatives, however, we have to define the kernel of the projection scheme explicitly before the $\beta$-function of a given coupling is determined and computed concretely. 

\paragraph{Metric gravity.} This effect can be illustrated by a well-known example from metric gravity: In $d\neq 4$ we find 3 independent field monomials ($R_{\mu\nu\rho\sigma}R^{\mu\nu\rho\sigma}$, $R_{\mu\nu}R^{\mu\nu}$, $R^2$) in the curvature-squared subspace, that may serve as its basis. Even if we only want to compute the $\beta$-function of the $R^2$ coupling, we have to define the space spanned by the other two basis elements and thus give some information about the choice of basis in the 3-dimensional subspace before it can be determined. 

Note that a projection scheme can also be defined implicitly by using a specific background spacetime: In \cite{oliver2} a truncation of metric gravity has been considered that only includes the $R^2$ term and none of the other curvature-squared invariants. In order to determine the $\beta$-function of its coupling a maximally symmetric background spacetime has been used with
\begin{equation}
R_{\mu\nu}=\frac{1}{d}g_{\mu\nu} R \quad \text{and} \quad R_{\mu\nu\rho\sigma}=\frac{1}{d(d-1)}(g_{\mu\rho}g_{\nu\sigma}-g_{\mu\sigma}g_{\nu\rho})R\;.
\end{equation}
Using these relations one can see immediately that the combinations
\begin{equation}
R_{\mu\nu} R^{\mu\nu}-\frac{1}{d}R^2 \quad \text{and} \quad R_{\mu\nu\rho\sigma}R^{\mu\nu\rho\sigma}-\frac{2}{d(d-1)}R^2
\end{equation}
of the curvature-squared invariants (and all linear combinations of them) are mapped to zero once we employ a maximally symmetric spacetime. Thus they define the kernel of the projection scheme that is implicitly defined by the choice of the maximally symmetric background spacetime.

In $d=4$ the three curvature squared monomials from above can be combined to a topological term, the Euler invariant $\chi_{\rm E}$ of the manifold using the Gauss-Bonnet theorem (see \eg \cite{Ortin2007}). Thus, on spacetimes with $\chi_{\rm E}=0$ the three invariants form an overcomplete basis of the curvature squared subspace. Clearly, the $\beta$-functions will depend on our choice which of the three monomials is considered as linearly dependent on the other two and thus is excluded from the basis. Reducing an overcomplete basis to a complete one hence also amounts to the definition of a projection scheme and can been seen as an example of the projection ambiguity of the $\beta$-functions, albeit a more trivial one.

\paragraph{Projection ambiguity in Einstein-Cartan gravity.} In the case of Einstein-Cartan gravity, we find an analogous ambiguity: We expanded the right hand side of the flow equation up to second order in the spin connection and from this expression we want to extract the coefficient of the terms that correspond to the Immirzi term. On any manifold of the same topology as the flat torus we work on, the value of the Nieh-Yan invariant is zero and therefore $\int e_a\wedge e_b\wedge F^{ab}=\int T^a\wedge T_a$. Hence, the Immirzi term lies within the subspace of parity-odd torsion squared monomials. In Appendix \ref{torsion} we show that {\it of the four different contractions of two torsion tensors with the $\varepsilon$-symbol, $\{T^{2(-)}_1, T^{2(-)}_2,T^{2(-)}_3, T^{2(-)}_4\}$, only two are linearly independent}. Thus, the subspace is two dimensional and in order to determine the prefactor of the Immirzi term, rhsF${}^\ast(\lambda,\gamma)$, we have to first choose the second basis monomial in this space.

Up this point, all observations apply even to an ideal untruncated calculation and are thus independent of the details of our calculation. Working with the \WHlike\ flow equation \eqref{WHlike-allg}, however, requires us to use constant background fields $\{\bar{e},\bar{\omega}\}$. For this reason, a second ambiguity of the above kind arises for the curvature term: As from the field strength tensor evaluated on a constant background only the $\bar{\omega}^2$-part remains ($\bar{F}^{ab}{}_{\mu\nu}=\bar{\omega}^{a}{}_{c[\mu}\bar{\omega}^{cb}{}_{\nu]}$), {\it the curvature term in the action is indistinguishable from a certain combination of parity-even torsion squared monomials $\{T^{2(+)}_1, T^{2(+)}_2, T^{2(+)}_3\}$} evaluated on the same background. This space is three dimensional (cf. Appendix \ref{torsion}). Hence we have to first specify the space spanned by the other two basis vectors before we can extract the corresponding prefactor rhsF($\lambda,\gamma$).

In the remaining two subsections we describe which choices of basis have been considered in our study, and how the inherent ambiguity can be exploited to find an optimized choice of basis as well as an optimized gauge condition. 

\subsection[The subspace of parity-odd $T^2$-invariants]{The subspace of parity-odd $\boldsymbol{T^2}$-invariants}\label{ParityOddSubSp}
In the following we will make use of the decomposition of the torsion tensor in $d=4$ into irreducible components (cf. \cite{Baekler2011a,Shapiro2002}) according to
\begin{equation}\label{Torsion_decomp}
 T^{\lambda}_{\ \mu\nu}= \frac{1}{3}\big(\delta^{\lambda}_{\ \nu}T_{\mu}-\delta^{\lambda}_{\ \mu}T_{\nu} \big)+\frac{1}{6 e}\varepsilon^{\lambda}_{\ \mu\nu\sigma}S^{\sigma} + q^{\lambda}_{\ \mu\nu}
\end{equation}
with $q^{\lambda}_{\ \mu \lambda}=0$, $q^{\lambda}_{\ \mu \nu}=-q^{\lambda}_{\ \nu \mu}$ and $\varepsilon^{\mu\nu\rho\sigma}q_{\nu\rho\sigma}=0$. The details of this decomposition are further explained in Appendix \ref{Torsion:Prelim} to which the reader could turn at this point.

As already mentioned before, the subspace of parity-odd torsion squared monomials is two dimensional. Making use of the irreducible decomposition \eqref{Torsion_decomp} of the torsion tensor a quite natural choice of basis of this space arises, which is given by the two monomials 
\begin{equation}
I_4= S_\mu T^\mu\quad \text{and}\quad I_5= e^{-1} \varepsilon_{\alpha\beta\gamma\delta}q^{\alpha\beta\mu}q^{\gamma\delta}{}_\mu\;.
\end{equation}
If one wanted to carry out a more general calculation including all the subspace in the truncation considered, one would probably choose to work with this basis, as the choice $\{I_4$, $I_5\}$ seems less arbitrary than any choice of 2 invariants among the $\{T^{2(-)}_i$, $i=1,\cdots,4\}$. However, as our truncation only contains the Immirzi term, which is given as the linear combination 
\begin{equation}
\int e_a\wedge e_b\wedge F^{ab}=\frac{1}{4}\int\!{\rm d}^4 x\, e \,T^{2(-)}_1=\int \!{\rm d}^4 x\, e \left(-\frac{1}{3} I_4+I_5\right),
\end{equation}
we are bound to choose this combination as one of our basis elements. For the second basis element we have a free choice among all other linearly independent combinations of $I_4$ and $I_5$. In the following we will demonstrate how this freedom can be used in order to optimize the truncation and the choice of gauge parameters. 

\paragraph{Strategy for optimizing the truncation.} The very idea of choosing a good truncation is that all RG trajectories of the exact, untruncated flow that start in the subspace defined by the truncation invariants lie almost completely within this subspace of theory space anyway, such that the truncation captures all essential features of the flow. This requires that the components of the exact RG flow causing the departure from the subspace are considered in some sense ``small''. A truncation becomes exact only when the corresponding subspace in theory space is mapped onto itself under RG transformations. 

Usually it is not possible to judge the reliability of a given truncation by this criterion without performing a completely independent second RG analysis considering a more general truncation. This is because the right hand side of the flow equation is usually projected directly onto the invariants contained in the truncation such that all information about other invariants generated by the flow is lost at that point.

In our calculation we have projected the right hand side onto the space of torsion squared invariants and by doing that, we have computed how the RG flow starting within the Holst truncation ``leaks'' from the truncation into this larger part of theory space. If the Holst truncation was an exact truncation, the right hand side would contain the invariants $I_4$ and $I_5$ only in the linear combination corresponding to the Immirzi term. In this case the choice of a second basis invariant would be obsolete, as the coordinate of the Immirzi invariant would not depend on it, while this second coordinate stays zero. 

This consideration suggests that in an inexact truncation the ratio of the two coordinates may serve as an indicator for the reliability of the truncation. Pictorially speaking, this ratio of the coordinates, which generically depends on all couplings, is related to the angle at which the RG flow departs from the subspace of the truncation at this point in theory space. Unfortunately, one would need to define a scalar product in theory space in order to give this illustrative interpretation of the ratio a quantitative meaning: As long as we cannot normalize the basis elements, we can rescale one of them resulting in an inverse scaling of the respective coordinate which alters the ratio of the two coordinates and as long as we cannot choose the basis orthogonally, the ratio will also depend on the angle between the basis elements. Thus, without a scalar product, we can only draw qualitative conclusions from the ratio of coordinates. These conclusions will implicitly assume that if we were to introduce a scalar product in theory space, the norm of the most natural basis elements $\{I_4, I_5\}$ should be of the same order of magnitude and they should not turn out practically collinear. Taken together, we assume that the elements of theory space parametrized by an angle $\varphi$ according to
\begin{equation}
 {\bf v}(\varphi)=\sin(\varphi) I_4+ \cos(\varphi)I_5
\end{equation}
should all have a norm within the same order of magnitude. If we choose the monomial ${\bf v}(\varphi)$, besides the Immirzi term, as a second basis element, we can discuss a continuous class of bases which we will denote by ${\cal B}^{(-)}_{{\bf v}(\varphi)}$.

Besides this family of bases, we introduce a discrete set of further bases according to 
\begin{equation}
{\cal B}^{(-)}_i\equiv\bigg\{\frac{1}{4}\int\!{\rm d}^4 x\, e\, T^{2(-)}_1,\frac{1}{2} \int \!{\rm d}^4 x\,e\, T^{2(-)}_i\bigg\}\quad\text{with}\quad i=2,3,4
\end{equation}
and 
\begin{equation}
{\cal B}^{(-)}_{I_5}\equiv\bigg\{\frac{1}{4}\int\!{\rm d}^4 x\, e\, T^{2(-)}_1,\frac{3}{2} \int \!{\rm d}^4 x\,e\, I_5\bigg\}.
\end{equation}
Here, $T^{2(-)}_i$ denote the four different parity-odd torsion squared monomials, that result from contracting two torsion tensors with the $\varepsilon$-symbol. Their explicit form is given in Appendix \ref{torsion}.

As ${\bf v}(\varphi)$ describes a rotation in the $(I_4,I_5)$-plane, the basis ${\cal B}^{(-)}_{{\bf v}(\varphi)}$ is, up to a rescaling of the second basis element, equivalent to each of the discrete bases for a certain value of $\varphi$. Explicitly these values are given by $\tan(\varphi)=\{\infty,\frac{1}{6},\frac{2}{3},0\}$ for the bases $\{{\cal B}^{(-)}_i,{\cal B}^{(-)}_{I_5}\}$.

\paragraph{Optimizing the choice of basis in theory space.} With these bases defined we can discuss the coordinate ratios as a measure for the reliability of our truncation explicitly. To this end we decompose the result for the RHS of the flow equation \eqref{RHS} first into its parity-even and -odd parts
\begin{equation}
\text{rhs}= \text{rhs}^{(+)}+\text{rhs}^{(-)}
\end{equation}
and the parity-odd part $\text{rhs}^{(-)}$ is further decomposed into a linear combination of the basis invariants. In the basis ${\cal B}^{(-)}_i$ we introduce the notation
\begin{equation}
\text{rhs}^{(-)}= \big(\text{rhs*F}\big)_{{\cal B}^{(-)}_i} \cdot \frac{1}{4}\int\!{\rm d}^4 x\, e \,T^{2(-)}_1+ \text{rhsT}^{2(-)}_i \cdot \frac{1}{2} \int\!{\rm d}^4 x\, e\, T^{2(-)}_i\,,
\end{equation}
for $i=2,3,4$. Thus, $\big(\text{rhs*F}\big)_{{\cal B}^{(-)}_i}$ and $\text{rhsT}^{2(-)}_i$ play the role of coordinate functions in the basis ${\cal B}^{(-)}_i$. From the discussion of the general form of the RHS (cf. eq. \eqref{rhsForm}) we know that these are functions of the couplings $\lambda$ and $\gamma$, but that the $\gamma$-dependence drops out when we compute the ratio of both functions. Thus we are led to a function of the form
\begin{equation}
R^{(-)}(\lambda)=\frac{\big(\text{rhs*F}\big)_{{\cal B}^{(-)}_i}}{\text{rhsT}^{2(-)}_i}=\frac{P^{{\cal B}^{(-)}_i}_9(\lambda)}{P^{T^{2(-)}_i}_9(\lambda)}\;.
\end{equation}
As it is the ratio of two polynomials of degree 9 and a polynomial of degree $n$ generally has $n$ zeros, not all of which have to be real, we expect these functions to exhibit up to 9 zeros and 9 poles on the real axis. 

\begin{figure}[t]
{\small
\subfigure[Basis ${\cal B}^{(-)}_2$]{
\begin{psfrags}
 \input{Pictures/Ratio46-psfrag.tex}
 \includegraphics[width=0.45\linewidth]{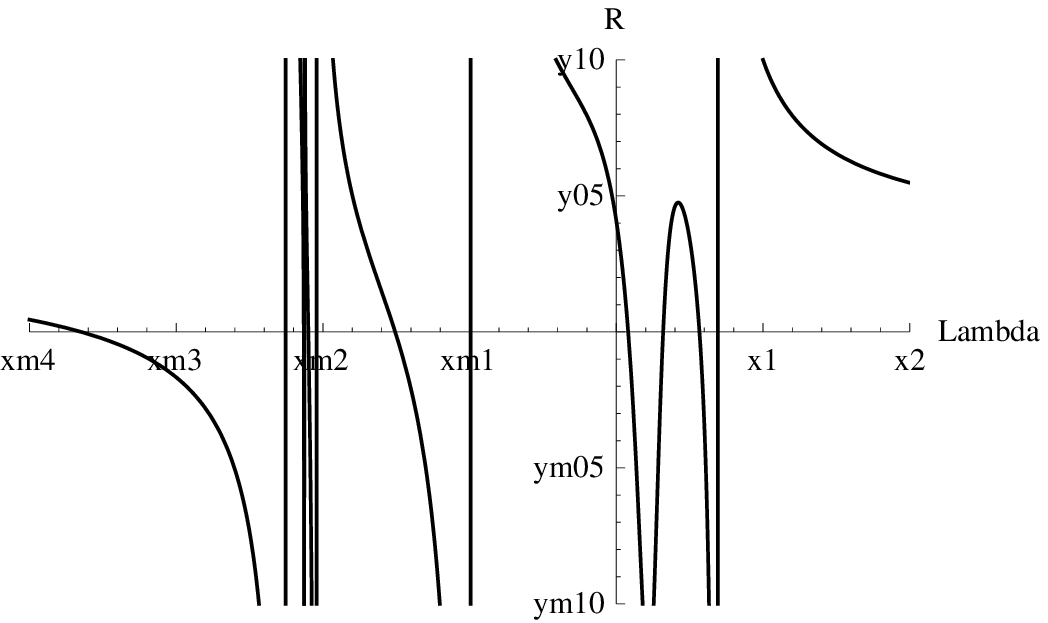}
\end{psfrags}}
\subfigure[Basis ${\cal B}^{(-)}_3$]{
\begin{psfrags}
 \input{Pictures/Ratio56-psfrag.tex}
 \includegraphics[width=0.45\linewidth]{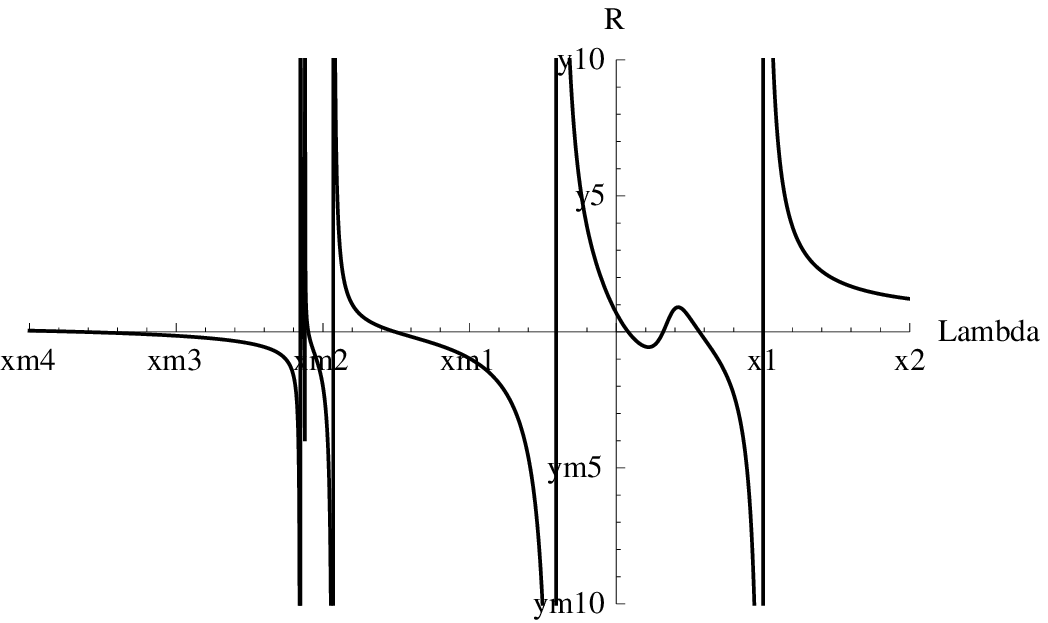}
\end{psfrags}}}
\caption{Typical plots of the coordinate ratio $R^{(-)}(\lambda)$ for two different bases as a function of the cosmological constant $\lambda$, shown here for the $(\aD,\aL,\bD)=(1,1,0)$ gauge.}
\label{RatioPlot}
\end{figure}

In Fig. \ref{RatioPlot} we have plotted the coordinate ratio as a function of $\lambda$ for the bases ${\cal B}^{(-)}_2$ and ${\cal B}^{(-)}_3$. From the figure we observe that many of these poles and zeros generically occur in the most interesting part of the phase diagram, namely at small (positive or negative) cosmological constant $\lambda$. For that reason the ratio function is wildly fluctuating at small $\lambda$ indicating that the reliability of the truncation strongly depends on the value of $\lambda$. Taking this into account it does not make sense to simply choose one basis and further discuss specific properties of the resulting RG flow in different gauges, as the reliability of these results would be highly questionable.

Instead we try to find a specific gauge that improves the situation. As most of the zeros of the polynomials $P_9(\lambda)$ dependent on the gauge parameters, we can try to move them such that they are situated outside the region of small $\lambda$. It turns out that almost all movable zeros tend to larger values of $\lambda$ for small values of $\aL,\aD$. In the limit $\aL,\aD \rightarrow 0$ the polynomials simplify: Because the prefactors of all higher orders in $\lambda$ vanish, only polynomials of degree two remain. Stated differently, all but two zeros are removed by moving them to infinity. If we, in addition, send $\bD\rightarrow0$ the ratio of the polynomials further simplifies as in this limit a common factor $(1-\lambda)$ of numerator and denominator can be canceled, such that we are left with a single pole.

\begin{figure}[t]
{\small
\subfigure[Basis ${\cal B}^{(-)}_2$]{
\begin{psfrags}
 \input{Pictures/Ratio46000-psfrag.tex}
 \includegraphics[width=0.45\linewidth]{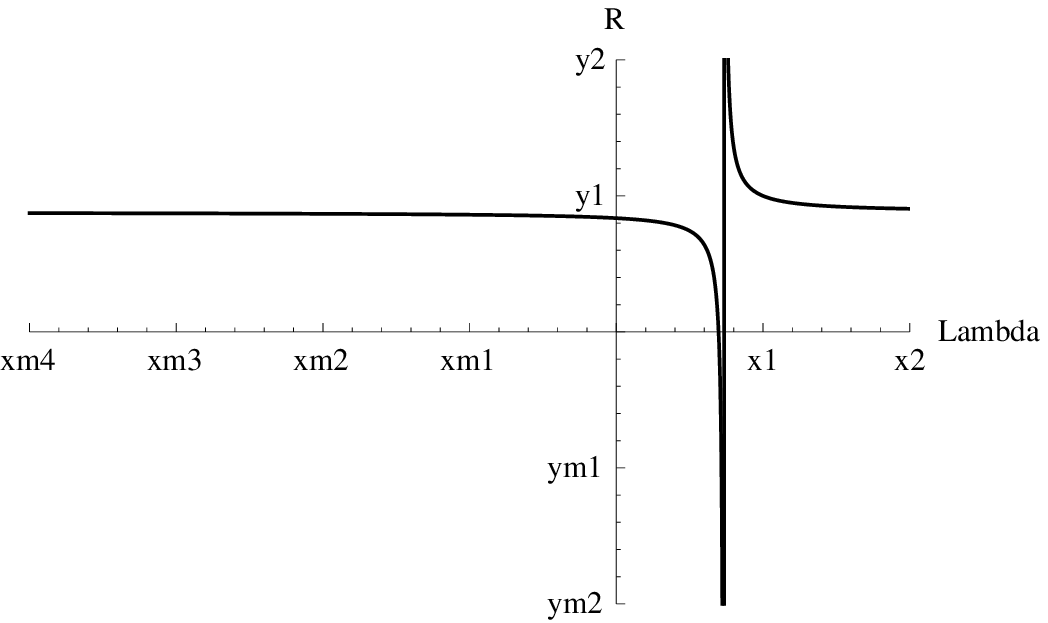}
\end{psfrags}}
\subfigure[Basis ${\cal B}^{(-)}_3$]{
\begin{psfrags}
 \input{Pictures/Ratio56000-psfrag.tex}
 \includegraphics[width=0.45\linewidth]{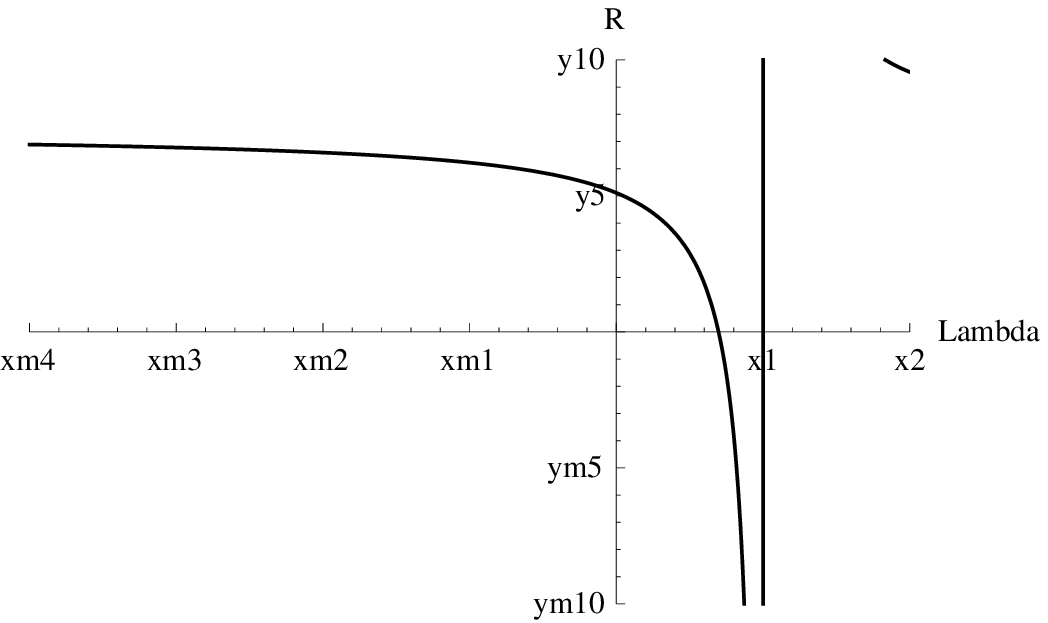}
\end{psfrags}}}
\caption{The coordinate ratio $R^{(-)}(\lambda)$ for the same bases as in Fig. \ref{RatioPlot} as a function of $\lambda$ in the $(\aD,\aL,\bD)=(0,0,0)$ gauge.}
\label{RatioPlot2}
\end{figure}

In Fig. \ref{RatioPlot2} the coordinate ratios are shown for the same bases as in Fig. \ref{RatioPlot} but for the $(\aD,\aL,\bD)=(0,0,0)$-gauge. We observe a tremendous simplification in the graphs and find that the ratio now is a constant function except for the vicinity of the one remaining pole, which, unfortunately, cannot be removed within our gauge freedom. Moreover, we find that the asymptotic value and the width of the pole remain basis depend quantities (note the different scales of the two plots). Nonetheless, the comparison between Figs. \ref{RatioPlot} and \ref{RatioPlot2} shows strikingly that {\it the $(0,0,0)$-gauge should be preferred compared to other gauges, in any basis}. Besides the fact that the $(0,0,0)$-point in gauge parameter space can be argued to be a fixed point of the RG flow in this space, we find here a new argument in favor of this gauge, namely that {\it it optimizes the consistency of a given truncation}.

\begin{figure}[ht]
\centering
{\small
\begin{psfrags}
 \input{Pictures/OddBasisOpt-psfrag.tex}
 \includegraphics[width=0.75\linewidth]{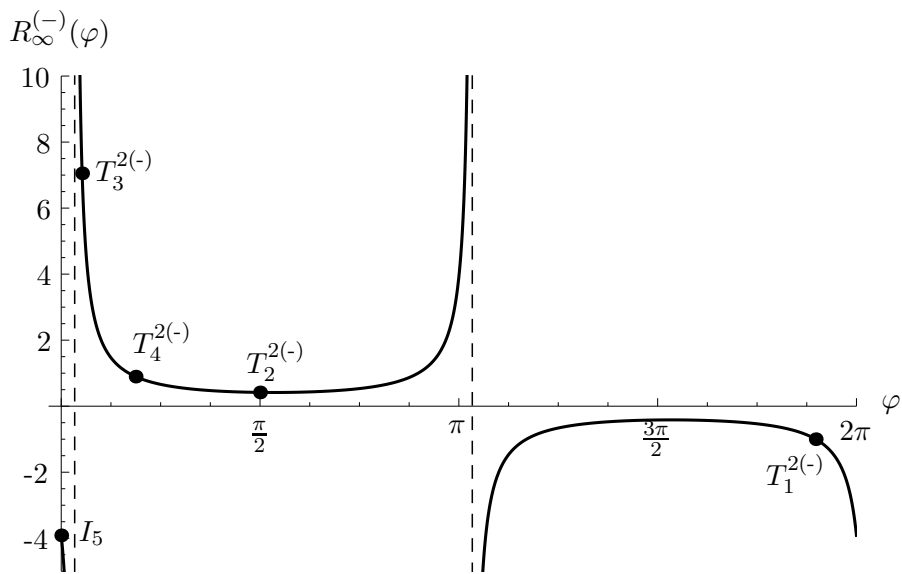}
\end{psfrags}}
\caption{Asymptotic value $R^{(-)}_{\infty}=\lim_{\lambda\rightarrow\pm \infty}R^{(-)}(\lambda)$ of the coordinate ratio for the continuous set of bases ${\cal B}^{(-)}_{{\bf v}(\varphi)}$ in $(\aD,\aL,\bD)=(0,0,0)$ gauge as a function of the parameter $\varphi$.}
\label{AsympValue}
\end{figure}

In a last step of basis optimization we consider the asymptotic value of the ratio functions $R^{(-)}_{\infty}=\lim_{\lambda\rightarrow\pm \infty}R^{(-)}(\lambda)$. As we can freely scale each of the corresponding basis functionals, which leads to an inverse scaling of the ratio, it is difficult to compare the different discrete bases. For that reason, we now turn over to the continuous set of bases ${\cal B}^{(-)}_{{\bf v}(\varphi)}$ and discuss the asymptotic value as a function of its parameter $\varphi$. The corresponding graph is shown in Fig. \ref{AsympValue}.

The resulting function is of the expected form: If we decompose a fixed vector {\bf v} \wrt different bases whose first element is held fixed and the second rotates, we should find a $2\pi$-periodic function for the coordinate ratio. It has poles at those points where the second basis element points into the direction of the vector {\bf v}, such that the first coordinate (the denominator) vanishes. The minima of the absolute value of the function occur when the second basis element is chosen orthogonal to the vector {\bf v} we wish to decompose. At the angle at which both basis elements are collinear, both coordinates diverge in opposite direction to $\pm \infty$ resulting in a ratio of $-1$; this angle corresponds to the point labeled by $T^{2(-)}_1$ in Fig. \ref{AsympValue}.

Under the assumption that the basis elements ${\bf v}(\varphi)$ are of approximately the same norm all these observations also apply to the graph in Fig. \ref{AsympValue}. Here we have also marked the points that correspond to the directions of the discrete torsion squared monomials $T^{2(-)}_i$ and $I_5$ as second basis elements. We find that the monomial $T^{2(-)}_3$ is almost collinear with our result for right hand side of the flow equation, as it lies very close to the pole of the function. On the other hand, $T^{2(-)}_2$, lies in the vicinity of the minimum of the function, indicating that it is ``almost orthogonal'' to our right hand side expression. For this reason, we have chosen the basis ${\cal B}^{(-)}_2$ in the parity-odd subspace for the further detailed analysis of the resulting RG flow. 

As a last remark let us mention that the width of the pole, given by the angle difference between the point $T^{2(-)}_1$ and the pole, may serve as a measure for the reliability of the truncation. In our case we find that it is of order $\pi/8$ and thus not particularly small, indicating that an extension of the Holst truncation to the full torsion squared subspace could be sensible in order to improve the stability of the results.

\subsection[The subspace of parity-even $T^2$-invariants]{The subspace of parity-even $\boldsymbol{T^2}$-invariants}\label{ParityEvenSubSp}
In the subspace of parity-even $T^2$-invariants the situation is more complicated as the space is 3-dimensional. Let us stress again, that the basis ambiguity here is only due to the choice of constant background fields, which, however is inevitable if one wants to make use of the \WHlike\ flow equation. In an RG analysis that uses a general background the curvature term could be identified unambiguously from the terms that are first order in derivatives and the spin connection. Evaluated on a constant background, however, we cannot distinguish the curvature term from a certain combination of torsion squared invariants. This kind of approximation is well known and often used to make the most complicated computations feasible; choosing \eg maximally symmetric spacetimes as a background in metric gravity results in an indistinguishability of the curvature squared monomials \cite{oliver2}. In contrast to this example from metric gravity, here {\it the choice of a constant background does not completely fix the projection scheme} that maps onto our truncation space, but it only maps the curvature term into the torsion squared subspace. Thus an additional basis ambiguity arises here, that has to be fixed by completing the projection scheme.

As for the parity-odd case the basis monomials constructed from the irreducible torsion components,

\begin{equation}
I_1=T_\mu T^\mu, \quad I_2= S_\mu S^\mu \quad \text{and} \quad I_3=q^{\mu\nu\rho}q_{\mu\nu\rho},
\end{equation}

\vspace{0.2cm}
\noindent
appear as the most natural choice for a basis in the parity-even subspace. Alternatively, one could choose the three torsion squares $T^{2(+)}_i$, $i=1,2,3$, defined in Appendix \ref{Torsion:Quad_Inv}. 
However, if we evaluate the curvature term on the constant background $(\bar{e},\bar{\omega})$ we find:

\begin{equation}
\begin{aligned}
\bar{e}_a\wedge \bar{e}_b \wedge *\bar{F}^{ab}&=\big(\bar{\omega}_{abc}\bar{\omega}^{acb}-\bar{\omega}_{ab}{}^a\bar{\omega}_{c}{}^{bc}\big)\bar{e}\,{\rm d}^4 x\\&=\frac{1}{4} \Big(\bar{T}^{2(+)}_1 +2 \bar{T}^{2(+)}_2 - 4 \bar{T}^{2(+)}_3\Big)\bar{e}\,{\rm d}^4 x\\ &=\bigg(\!-\frac{2}{3}\bar{I}_1 -\frac{1}{24}\bar{I}_2 + \frac{1}{2} \bar{I}_3\bigg)\bar{e}\,{\rm d}^4 x\,.
\end{aligned}
\end{equation}

\vspace{0.2cm}
\noindent
Thus, it does not coincide with any of the basis vectors but is a combination of all three basis elements in both natural bases. As this combination has to be held fixed as the first basis element that we want to project on, arbitrary choices of the other two basis invariants among the $T^{2(+)}_i$ and the $I_i$, but also among the spin connection squares $\bar{\omega}_{abc}\bar{\omega}^{abc}, \bar{\omega}_{abc}\bar{\omega}^{acb}, \bar{\omega}_{ab}{}^a\bar{\omega}_{c}{}^{bc}$ are equally plausible.

Unfortunately also the discussion of the coordinate ratios for the different bases does not lead to a unique preference of one specific basis: Although also in this subspace an enormous simplification of the ratio functions can be reached by using the (0,0,0)-gauge, the structure of the right hand side expression \eqref{rhsForm} causes the ratios to be functions of the two couplings $(\lambda,\gamma)$, which complicates a systematic analysis.

Since we were not able to single out one specific basis as being optimal \wrt the consistency of the truncation, we analyzed the RG flow for four different, more or less arbitrarily chosen bases, which are defined by their elements according to

\begin{equation}
 \begin{aligned}
  {\cal B}^{(+)}_1&=
   \Big\{
    \int \bar{e} (\bar{\omega}_{abc}\bar{\omega}^{acb}-\bar{\omega}_{ab}{}^{a}\bar{\omega}^{cb}{}_{c}),
    \int \bar{e} (\bar{\omega}_{abc}\bar{\omega}^{acb}+\bar{\omega}_{ab}{}^{a}\bar{\omega}^{cb}{}_{c}),
    2 \int \bar{e} \,\bar{\omega}_{abc}\bar{\omega}^{abc}\Big\},\\
 {\cal B}^{(+)}_2&= \Big\{    \int \bar{e} (\bar{\omega}_{abc}\bar{\omega}^{acb}-\bar{\omega}_{ab}{}^{a}\bar{\omega}^{cb}{}_{c}),
    2 \int \bar{e} \,\bar{\omega}_{ab}{}^{a}\bar{\omega}^{cb}{}_{c},
    2 \int \bar{e} \, \bar{\omega}_{abc}\bar{\omega}^{abc}\Big\}, \\
   {\cal B}^{(+)}_3&=
\Big\{ \int \bar{e} (\bar{\omega}_{abc}\bar{\omega}^{acb}-\bar{\omega}_{ab}{}^{a}\bar{\omega}^{cb}{}_{c}),
    2 \int \bar{e} \, \bar{\omega}_{abc}\bar{\omega}^{acb},
    2 \int \bar{e} \, \bar{\omega}_{abc}\bar{\omega}^{abc}\Big\},\\
 {\cal B}^{(+)}_4&=\Big\{\int \bar{e} (\bar{\omega}_{abc}\bar{\omega}^{acb}-\bar{\omega}_{ab}{}^{a}\bar{\omega}^{cb}{}_{c}),
    -\frac{1}{4} \int \bar{e} \, I_2,
    \phantom{-}\frac{3}{2} \int \bar{e} \, I_3 \Big\}\:. \end{aligned}
\end{equation}

\vspace{0.5cm}
\noindent
Here $\int \equiv\int \!{\rm d}^4x$, and the prefactors of the various invariants are chosen such that the bases are related to each other by transformation matrices of unit determinant.

With these bases specified we are now able to discuss the properties of the RG flow of the Holst truncation in the next section.

\vspace{0.5cm}
\section{Analysis of the RG flow}\label{QECG:Results}
After having introduced the different projection schemes we are going to analyze, we are now able to write down the explicit form of the $\beta$-functions for the three couplings of the Holst action. From \eqref{BetaForm}, \eqref{rhsForm}, and \eqref{rhsL} we find that for all the bases the $\beta$-functions are of the general form
\begin{subequations}\label{Beta_System}
\begin{empheq}[box=\fbox]{align}
  \beta_g(\lambda,\gamma,g)&=g\left[2+\eta_N(\lambda,\gamma,g)\right],\label{Beta_g}\\
 \beta_\gamma(\lambda,\gamma,g)&= -\frac{16 \pi g \gamma}{N(\lambda)} \left[P_9(\lambda)+\frac{1}{\gamma^2}P_8(\lambda)+P_{10}(\lambda)\right],\label{Beta_gamma}\\
 \beta_\lambda(\lambda,\gamma,g)&= -2\lambda + \frac{16 \pi g \lambda}{N(\lambda)}\left[\frac{1}{\gamma^2}P_8(\lambda)+P_{10}(\lambda)\right]\nonumber\\
&\hspace{0.5cm}-\frac{g}{4\pi}\!\left[\!12\ln\left(\!\frac{\gamma^2-1}{\gamma^2}\!\right)^{\!\!2}\!\!+\!5\ln(1-\lambda)^{2}\!-\!96\ln m^{2}\!-\!34\ln g^{2}\!-\!\ln {\cal N}'\right]\!,\label{Beta_lambda}
 \end{empheq}
\end{subequations}
with the anomalous dimension of Newton's constant
\begin{equation}
\boxed{\eta_N(\lambda,\gamma,g)=\frac{16 \pi g}{N(\lambda)}\left(\frac{1}{\gamma^2}P_8(\lambda)+P_{10}(\lambda)\right)\:.}
\end{equation}
Here, we have already specialized the logarithmic terms in $\beta_\lambda$ to the case of the preferred $(0,0,0)$-gauge, as we will restrict the discussion to this case. ${\cal N}'$ is a numerical constant given by
\begin{equation} 
\ln {\cal N}' =\ln {\cal N} + 68\ln \pi -6 \ln 2\approx 241.42\:. 
\end{equation}
In the (0,0,0)-gauge also the $P_n(\lambda)$ polynomials simplify and their degree is then smaller than $n$, but for notational consistency we will stick to this notation. In addition it reminds us of true complexity of the $\beta$-functions in terms of $\lambda$ for a general choice of gauge.

As we have discussed before, the virtue of choosing $\bD=0$ in addition to $(\aD,\aL)=(0,0)$ lies in the fact, that the polynomials $P_n(\lambda)$ then contain a factor of $(1-\lambda)$. In the $\beta$-functions every polynomial $P_n(\lambda)$ is divided by the denominator $N(\lambda)$ such that this factor is canceled. In the $(0,0,0)$-gauge the new common denominator is thus given by $N(\lambda)/(1-\lambda)$ and the numerators are composed of the polynomials $P_n(\lambda)/(1-\lambda)$. Since the following discussion is restricted to this preferred gauge, we will present only the explicit expressions of these ``rescaled'' quantities.

The denominator $N(\lambda)$ simplifies considerably in the $(0,0,0)$-gauge and assumes the explicit form
\begin{equation}
 \frac{N(\lambda)}{1-\lambda}\stackrel{(0,0,0)}{=}4 (\lambda-1)^2\,.
\end{equation}

For the rest of this section we will consider four distinct bases in the subspace of torsion-squared invariants. As we have shown in the last section the optimal basis in the parity-odd sector is ${\cal B}^{(-)}_2$, while for the parity-even part we have introduced the four bases ${\cal B}^{(+)}_{\{1,2,3,4\}}$. As a shorthand notation for the bases of the complete torsion squared subspace we will denote to the combinations of these bases by ${\cal B}_i=({\cal B}^{(-)}_2,{\cal B}^{(+)}_i)$, $i=1,2,3,4$.

For each of these bases ${\cal B}_i$ the polynomials $P_n(\lambda)$ assume different explicit forms. As the polynomial $P_9(\lambda)$ only depends on the basis in the parity-odd sector, it is equal for all four bases considered. In the $(0,0,0)$-gauge it assumes the simple linear form
\begin{equation}
 \frac{P_9(\lambda)}{1-\lambda}=-\frac{5}{64 \pi^2}(15\lambda-11)\:.
\end{equation}
The polynomials $P_8(\lambda)$ and $P_{10}(\lambda)$ take on different forms in the different bases. Their explicit form for the (0,0,0)-gauge is given in Table \ref{Poly:Table}. We observe that basis ${\cal B}_1$ plays a special role as the polynomial $P_8$ vanishes in this case. We will investigate the deep implications of this fact during the rest of this section.

\begin{table}[htb]
\centering \renewcommand{\arraystretch}{1.15}
 \begin{tabular}{lcccc}\toprule
     \phantom{asdfsadkf}           &          ${\cal B}_1$           &          ${\cal B}_2$        & ${\cal B}_3$ & ${\cal B}_4$\\ \midrule\\[-0.4cm]
 $ \frac{P_8(\lambda)}{1-\lambda}$& 0 & $\frac{5}{96 \pi^2}$ & $-\frac{5}{96 \pi^2}$ & $-\frac{5}{32 \pi^2}$\\[0.4cm]
  $\frac{P_{10}(\lambda)}{1-\lambda}$& $\frac{55\lambda^2-42\lambda-18}{128 \pi^2}$ & $-\frac{43\lambda^2-186\lambda+178}{384 \pi^2}$ & $\frac{373\lambda^2-438\lambda+70}{384 \pi^2}$ & $\frac{(\lambda-1)(52\lambda-25)}{32 \pi^2}$\\[.15cm] \bottomrule
 \end{tabular}
\caption{Explicit forms of the polynomials $P_8(\lambda)$ and $P_{10}(\lambda)$ in $(\aD,\aL,\bD)=(0,0,0)$-gauge for the four bases ${\cal B}_i$, which we will explore in detail.}\label{Poly:Table}
\end{table}

\paragraph{Structure of this section.} The analysis of the system of flow equations \eqref{Beta_System} in this section is organized as follows: In the first two subsections we will explore the RG flow in two two-dimensional truncations beginning with the $(\lambda,g)$-truncation for a fixed value of the Immirzi parameter $\gamma$ followed by the analysis of the $(\gamma,g)$-system for a fixed cosmological constant $\lambda$. In the third subsection we discuss the full three dimensional RG flow of the Holst truncation, including its fixed point structure and its phase portrait. In all of these subsections we will first compare the results for the four distinct bases ${\cal B}_i$, $i=1,\cdots,4$, before we evaluate the qualitative similarities with the analysis carried out in \cite{je:longpaper}. We thereby keep in mind that the ultimate justification for the applicability of our new, structurally simplified \WHlike\ flow equation can only lie in the accordance of its predictions with other---either exact or at least well approved approximate---RG flow equations. 

\subsection[The $(\lambda,g)$-subsystem]{The $\boldsymbol{(\lambda,g)}$-subsystem}\label{gl-system}

\paragraph{(A) Differential equations.} The system of flow equations we are going to discuss in this subsection is given by the two $\beta$-functions \eqref{Beta_g} and \eqref{Beta_lambda},

\begin{equation}\label{glsubsystem}
 \begin{aligned}
  \partial_t g=\beta_g(\lambda,\gamma,g)&=g\left[2+\eta_N(\lambda,\gamma,g)\right],\\[0.5cm]
  \partial_t \lambda= \beta_\lambda(\lambda,\gamma,g)&= -2\lambda + \frac{16 \pi g \lambda}{N(\lambda)}\left[\frac{1}{\gamma^2}P_8(\lambda)+P_{10}(\lambda)\right]\\
&\hspace{0.5cm}-\frac{g}{4\pi}\!\!\left[\!12\ln\!\left(\!\!\frac{\gamma^2-1}{\gamma^2}\!\!\right)^{\!\!2}\!\!+\!5\ln(1-\lambda)^{2}\!-\!96\ln m^{2}\!-\!34\ln g^{2}\!-\!\ln {\cal N}'\right]\!.
 \end{aligned}
\end{equation}

{\noindent}Throughout this subsection we treat $\gamma$ as an external parameter that does not run and can be set to an arbitrary fixed value; in particular we will discuss how the RG flow depends on the value of $\gamma$. 

Let us first address the limitations of this truncation due to divergences inherent in the $\beta$-functions. We find that the $\beta$-functions are not well defined for the values $\gamma=0$ and $\gamma=\pm1$. While the latter divergence was expected from the outset, as the action depends in this limit only on one chiral component of the spin connection, the reason for the former remains somewhat unclear. It should be noted that the quadratic divergence in both $\beta$-functions at $\gamma\rightarrow0$ is basis dependent since it ceases to exist in basis ${\cal B}_1$ where $P_8=0$. Nevertheless a logarithmic divergence of $\beta_\lambda$ at $\gamma\rightarrow0$ still remains even in this basis. The limit $\gamma\rightarrow \pm \infty$ is, however, perfectly well defined in all bases. When we discuss properties of the RG flow that apply for any value of $\gamma$ in the following, we implicitly exclude these pathological cases of $\gamma=0, \pm 1$.

Furthermore, we find that the 2-dimensional theory space, the $(\lambda,g)$-plane, has a boundary at $\lambda=1$ for any value of $\gamma$. As not only the logarithmic term in $\beta_\lambda$ diverges on this line, but also the denominator $N(\lambda)$, the ratio $\beta_\lambda/\beta_g$ stays finite in the limit $\lambda\rightarrow1$. Hence, the RG flow simply stops at this line, similar to the situation at $\lambda=0.5$ in metric gravity \cite{mr,frank1}. In contrast to the metric case the singularity here is a pole of even order, such that the flow does not change direction at this line.

\begin{figure}[tb]
\centering
{\small
\begin{psfrags}
 \input{Pictures/glSketch-psfrag.tex}
 \includegraphics[width=0.45\linewidth]{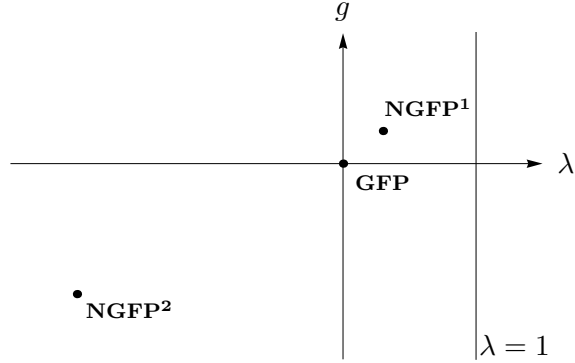}
\end{psfrags}
}
\caption{Sketch of the fixed point structure of the $(\lambda,g)$-system. Besides the position of the three fixed points, the barrier of the flow at $\lambda=1$ is depicted.}
\label{glSketch}
\end{figure}

\paragraph{(B) Fixed point structure.}
As we will see shortly the $(\lambda,g)$-system exhibits in total three FPs, a Gaussian FP, denoted {\bf GFP}, and two non-Gaussian fixed points ${\bf NGFP^1}$ and ${\bf NGFP^2}$, respectively. Their approximate location in theory space is depicted in Fig. \ref{glSketch} in order to illustrate the situation, before the details of their properties are discussed.

\paragraph{(C) The Gaussian fixed point.}
First we observe that the system \eqref{glsubsystem} allows for a Gaussian fixed point at $(\lambda,g)=(0,0)$. However, the critical exponents, defined as the negative eigenvalues of the stability matrix at the fixed point
\begin{equation}
 \left.\mathscr{B}\right|_{g=g^*,\lambda=\lambda^*} = \left.
\begin{pmatrix}
 \partial_g \beta_g & \partial_\lambda \beta_g\\
 \partial_g \beta_\lambda & \partial_\lambda \beta_\lambda
\end{pmatrix}
\right|_{g=g^*,\lambda=\lambda^*}
\end{equation}
cannot be determined as the flow cannot be expanded at the {\bf GFP}: Although $\beta_\lambda$ is well defined at the {\bf GFP} its partial derivative \wrt $g$ is not. Thus, it is not possible to linearize the RG flow in the vicinity of the {\bf GFP} in the usual way. Nonetheless it possible to approximate the system in this region by taking into account only the terms linear in the couplings and the logarithmic term that causes the divergence of the derivative at the {\bf GFP}. The resulting approximate system of flow equations reads 
\begin{equation}
 \begin{aligned}
  \partial_t g=\beta_g&=2g\\
  \partial_t \lambda=\beta_\lambda&=-2\lambda+ \frac{17}{2\pi}g \ln g^2 - C g\:,
 \end{aligned}
\end{equation}
where $C=(4\pi)^{-1}[12\ln(1-1/\gamma^2)^{2}- 96\ln m^{2} -\ln{\cal N}']$ is a constant. It is easy to integrate this system of differential equations; its solution subject to the initial condition $(\lambda_0,g_0)$ imposed at the scale $k=k_0$ is given by
\begin{equation}
 \begin{aligned}
  g(k)&= g_0 \big({\textstyle\frac{k}{k_0}}\big)^{\!\!2}\\
  \lambda(k)&= \lambda_0 \big({\textstyle\frac{k}{k_0}}\big)^{\!\!-2}\!\! +\frac{g_0}{4} \Big(C+ {\textstyle \frac{17}{2\pi}}\big(1-\ln g_0^2\big)\Big) \Big[\big({\textstyle\frac{k}{k_0}}\big)^{\!\!-2}\!\!-\big({\textstyle\frac{k}{k_0}}\big)^{\!2}\Big]+{\textstyle \frac{17}{2\pi}} \,g_0 \big({\textstyle\frac{k}{k_0}}\big)^{\!2} \ln \big({\textstyle\frac{k}{k_0}}\big)\:.
 \end{aligned}
\end{equation}
Although the flow cannot be linearized in the familiar way, we observe that it is perfectly well-defined in the vicinity of the {\bf GFP}, exhibiting non-polynomial terms, though. 

\paragraph{(D) Non-Gaussian fixed points.} The system \eqref{glsubsystem} allows for non-Gaussian fixed points as well. This is seen as follows: We can solve the condition $\beta_g(\lambda,g)=0$ for \begin{equation}
g^*(\lambda)= -\frac{1}{8\pi} \frac{N(\lambda)}{P_8/\gamma^2+ P_{10}(\lambda)}\:.
\end{equation}
This solution is substituted into the second fixed point condition $\beta_\lambda(\lambda,g^*(\lambda))=0$. The zeros of this function of $\lambda$ correspond to all non-Gaussian fixed points in the $(\lambda, g)$-plane. Explicitly this condition amounts to the solution of the following equation for $\lambda$:
\begin{equation}\label{gl-FP-cond}
 4\lambda=-\frac{1}{4 \pi}\, g^\ast(\lambda)\bigg[\!12 \ln \!\Big[\frac{\gamma^2-1}{\gamma^2}\Big]^{2} \!\!+ 5\ln(\lambda-1)^{2}\!\!-96\ln m^{2}\!\!-34\ln g^*(\lambda)^{2} \!\!-\ln {\cal N}'\bigg].
\end{equation}
From the asymptotic behavior of this equation, which is linear on the LHS while it is logarithmic on the RHS ($g^\ast(\lambda)$ tends to a constant value for large $|\lambda|$), we expect that there is at least one solution to equation \eqref{gl-FP-cond}. In addition, further solutions can be generated by the variation of $g^\ast(\lambda)$. Especially there will appear solutions in the vicinity of the poles of $g^\ast(\lambda)$. These latter solutions, however, were observed to give rise to fixed points with a tiny basin of attraction, such that they do not at all influence the overall structure of the flow. Therefore these fixed point solutions should not be considered physical.

Typically we find two ``robust'' solutions of eq. \eqref{gl-FP-cond} that can be considered physical. We denote the corresponding fixed points by ${\bf NGFP^1}$ and  ${\bf NGFP^2}$ and discuss their properties separately in the subsequent paragraphs.

\begin{figure}[t]
\centering
{\small
\begin{psfrags}
 \input{Pictures/FP-cond-psfrag.tex}
 \includegraphics[width=0.65\linewidth]{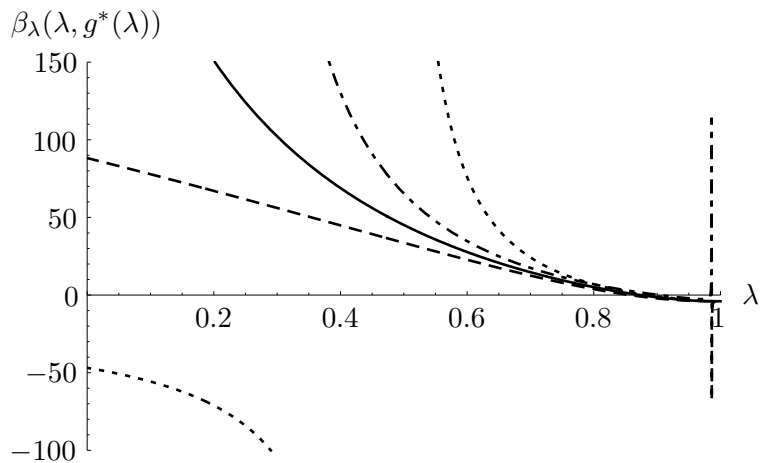}
\end{psfrags}}
\caption{Fixed point condition $\beta_\lambda(\lambda, g^\ast(\lambda))$ of the $(\lambda,g)$-system: At $\gamma=5$ we find a fixed point at $\lambda\approx 0.9$ for all four bases (${\cal B}_1$ solid, ${\cal B}_2$ dashed, ${\cal B}_3$ dot-dashed, ${\cal B}_4$ dotted). In basis ${\cal B}_3$ the function $\beta_\lambda(\lambda, g^\ast(\lambda))$ exhibits a pole close to $\lambda=1$ giving rise to another, however unphysical, fixed point.}
\label{FP-cond-pict}
\end{figure}

\noindent{\bf (i) The fixed point ${\bf NGFP^1}$.} The most stable solution that gives rise to the fixed point ${\bf NGFP^1}$ occurs at $\lambda\approx 0.9$: As the RHS of eq. \eqref{gl-FP-cond} falls to zero at $\lambda=1$ and the LHS grows linearly to 4 at that point, we generically find a solution to the fixed point condition in the interval $\lambda\in [0,1]$. This behavior is depicted for all bases ${\cal B}_i$ and $\gamma=5$ in Fig. \ref{FP-cond-pict}.

For other values of $\gamma$ the situation is very similar. For $\gamma>1$ we generally find that $0.8<\lambda^*<0.9$ while for $\gamma<1$ the FP moves towards smaller $\lambda^*$. Only in basis ${\cal B}_2$ it ceases to exist for $\gamma\lesssim 0.75$.

\begin{figure}[p]
\centering
{\small
\begin{psfrags}
 \input{Pictures/glFPposPlot-psfrag.tex}
 \includegraphics[width=0.65\linewidth]{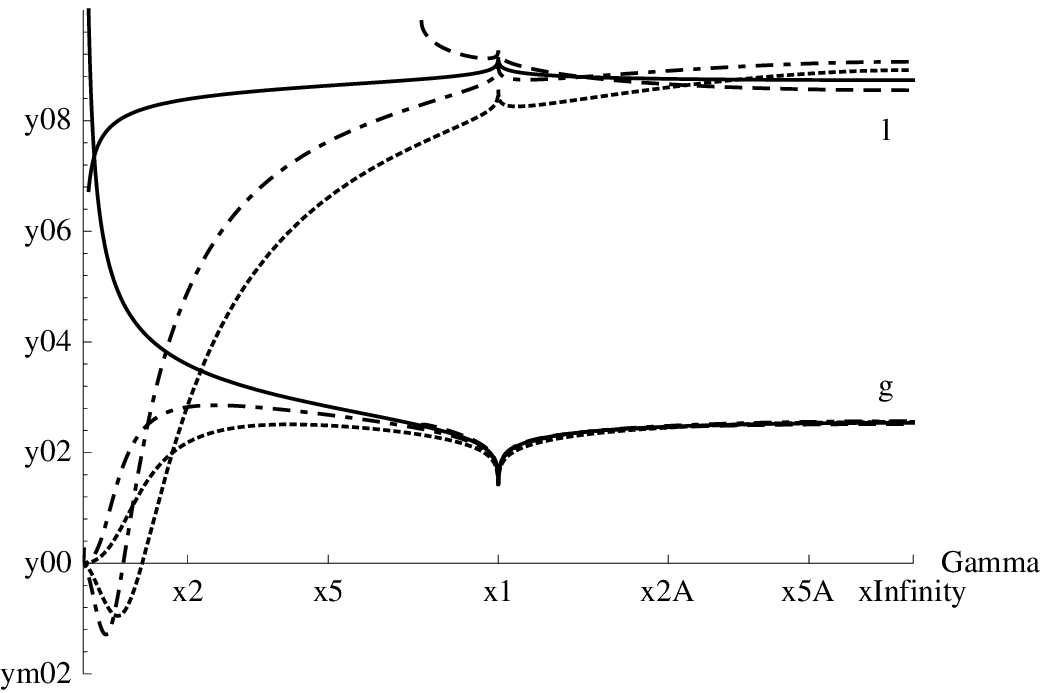}
\end{psfrags}}
\caption{Fixed point position for all four bases (${\cal B}_1$ solid, ${\cal B}_2$ dashed, ${\cal B}_3$ dot-dashed, ${\cal B}_4$ dotted) and $m=1$ as a function of $\gamma$. While for $\gamma>1$ the fixed point position is almost independent of the basis, for small $\gamma$ we find their behavior differing: In basis ${\cal B}_1$ the $g^*$ coordinate diverges, in basis ${\cal B}_2$ the FP ceases to exist at $\gamma\approx 0.75$ and in ${\cal B}_{3,4}$ the FP merges with the {\bf GFP}.}
\label{glFPposPlot}
\end{figure}

In Fig. \ref{glFPposPlot} the fixed point position $(\lambda^\ast, g^\ast)$ is plotted as a function of $\gamma$ for all four bases considered. We find that except for basis ${\cal B}_2$ the fixed point exists in all bases for all values of $\gamma$. For $\gamma>1$ the fixed point lies at almost the same position for all bases. For small $\gamma$, however, the $g^*$ coordinate diverges in basis ${\cal B}_1$, while in the bases ${\cal B}_3$ and ${\cal B}_4$ the fixed point merges with the Gaussian one.

\begin{figure}[p]
\centering
{\small
\begin{psfrags}
 \input{Pictures/glFPCEPlot-psfrag.tex}
 \includegraphics[width=0.75\linewidth]{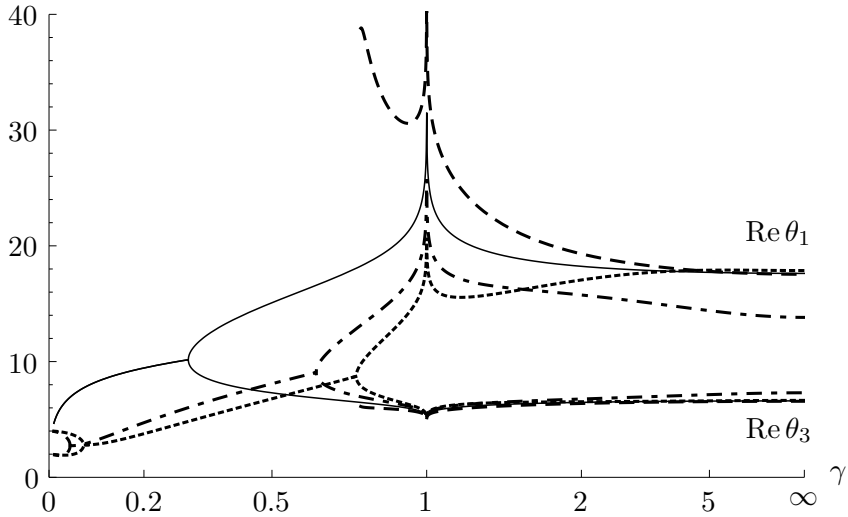}
\end{psfrags}}
\caption{Real part of the critical exponents for all four bases (${\cal B}_1$ solid, ${\cal B}_2$ dashed, ${\cal B}_3$ dot-dashed, ${\cal B}_4$ dotted) and $m=1$ as a function of $\gamma$. }
\label{glFPCEPlot}
\end{figure}

Let us move on and analyze the stability properties of the fixed point ${\bf NGFP^1}$. In Fig. \ref{glFPCEPlot} the real parts of the critical exponents for the four bases are plotted as a function of $\gamma$. At the bifurcation points the critical exponents become real, while in the range of $\gamma$ with a single line only the critical exponents form a complex conjugated pair, whose imaginary part is not depicted. 

We observe that {\it for all bases, and all values of $\gamma$, both critical exponents are positive, \ie the fixed point ${\bf NGFP^1}$ is UV attractive in both directions}. Moreover, the qualitative dependence of the critical exponents on $\gamma$ turns out similar for the different bases: At infinity they start real and we find a peak of the curves close to $\gamma=1$, that probably is an artifact of the singularity of the $\beta$-functions at that point. Shortly after that the critical exponents turn complex, at least for some interval in $\gamma$. The dashed line, corresponding to basis ${\cal B}_2$, stops at $\gamma\approx 0.75$ as the FP vanishes for smaller $\gamma$. Note that the absolute value of the critical exponents is fairly large over the whole range of $\gamma$, which might cast some doubt on the physical significance of this FP. For large $\gamma$, on the other hand, the functions corresponding to 3 out of 4 bases coincide to a remarkably good degree, showing a special robustness of the result in the limit $\gamma\rightarrow \infty$.

\begin{figure}[t]
\centering
{\small
\begin{psfrags}
 \input{Pictures/glFP2Pos-psfrag.tex}
 \resizebox{0.65\linewidth}{!}{\includegraphics{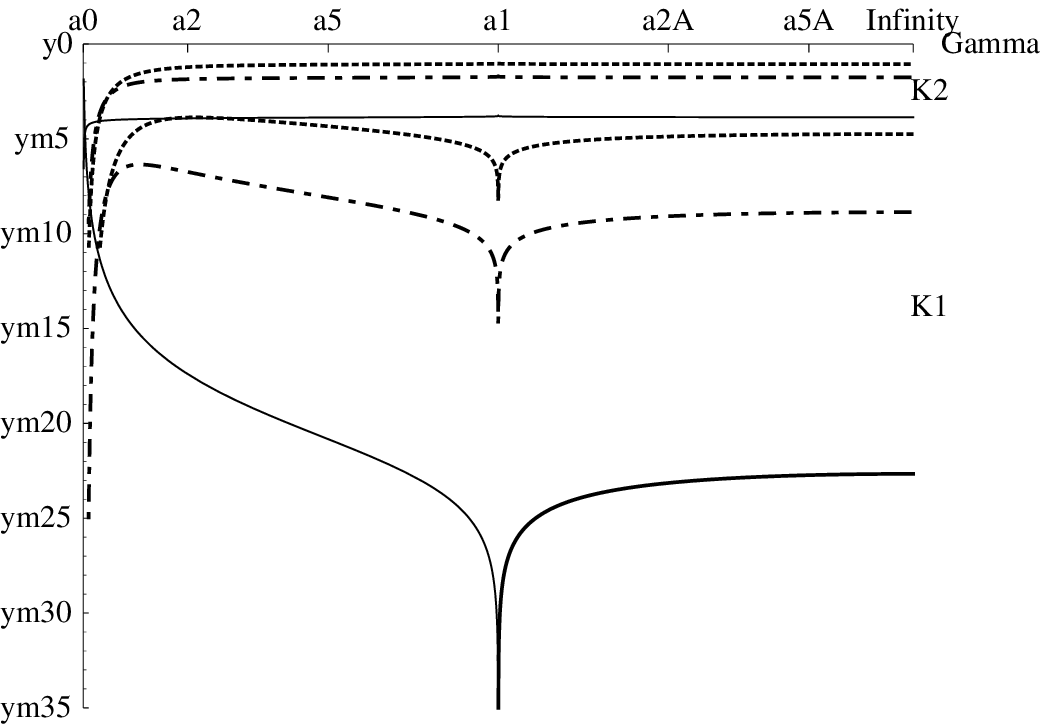}}
\end{psfrags}}
\caption{Fixed point position for the three bases ${\cal B}_1$ (solid), ${\cal B}_3$ (dot-dashed) and ${\cal B}_4$ (dotted) and $m=1$ as a function of $\gamma$. Compared to the first fixed point, the FP position here depends severely on the basis chosen. However, for each basis the $g^*$ coordinate is almost independent of $\gamma$, although $\lambda^*$ varies significantly.}
\label{glFP2Pos}
\end{figure}

\noindent {\bf(ii) The fixed point ${\bf NGFP^2}.$} This FP is only present in the bases ${\cal B}_1$, ${\cal B}_3$ and ${\cal B}_4$, while for basis ${\cal B}_2$ we generally find only the first non-Gaussian fixed point, ${\bf NGFP^1}$. When it exists, ${\bf NGFP^2}$ is located at large negative $\lambda^*$, and the corresponding $g^*$ coordinate typically lies in the range $g^*\in [-4,-1]$.

The fixed point coordinates as functions of $\gamma$ are shown in Fig. \ref{glFP2Pos}. We find that, except for a small region close to $\gamma\approx 0$, that is dominated by the logarithmic divergence, $\lambda^*$ starts off decreasing in value up to $\gamma=1$ where it shows a significant peak, while it stays approximately constant for $\gamma>1$. What catches the eye, however, is that $g^*(\gamma,\lambda^*)$ is constant to a good approximation, although $\lambda^*$ shows such a pronounced variation. Hence, a remarkable compensation in the function $g^*(\lambda^*)$ seems to take place. 

In comparison to ${\bf NGFP^1}$, the position of ${\bf NGFP^2}$ is quite variable, depending on both $\gamma$ and the choice of basis. Also the coordinates take on large absolute values. However, the fixed point position is not a physical observable so that we should not be concerned about these results but first analyze  the absolute value and stability of the critical exponents, which are considered physically more meaningful.

\begin{figure}[t]
\centering
{\small
\begin{psfrags}
 \input{Pictures/glFP2CE-psfrag.tex}
 \includegraphics[width=0.75\linewidth]{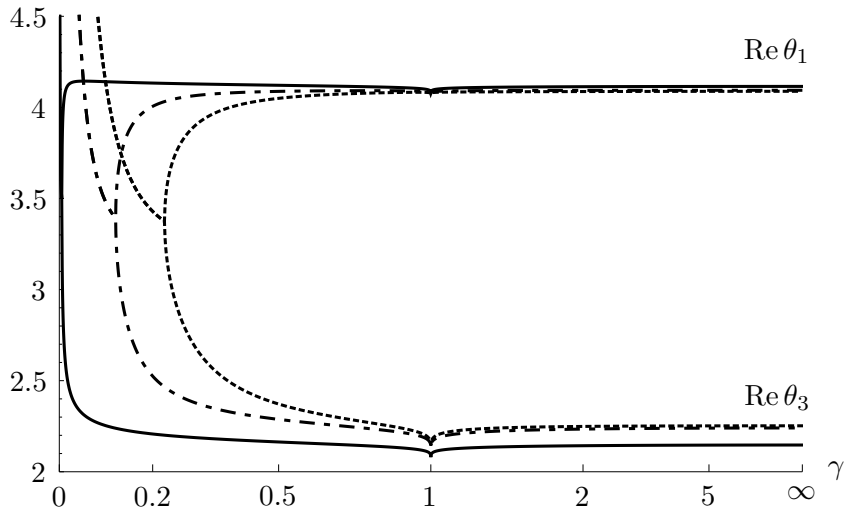}
\end{psfrags}}
\caption{Real part of the critical exponents for the three bases ${\cal B}_1$ (solid), ${\cal B}_3$ (dot-dashed) and ${\cal B}_4$ (dotted), and $m=1$ as a function of $\gamma$. For all $\gamma$ the fixed point is attractive in both directions. The critical exponents are almost constant and independent of the basis in a large part of total range in $\gamma$.}
\label{glFP2CE}
\end{figure}

In Fig. \ref{glFP2CE} the real parts of the critical exponents corresponding to ${\bf NGFP^2}$ are depicted. We observe that the fixed point is UV attractive in both directions, and that the absolute value of the critical exponents lies in a perfectly reasonable range. Moreover, the critical exponents are real, constant and almost independent of the basis chosen for all $\gamma>0.5$. 

Taken together these findings strongly support the physical significance of the second non-Gaussian fixed point, ${\bf NGFP^2}$.

\noindent {\bf (iii) Summary: non-Gaussian fixed points.} We conclude that both non-Gaussian fixed points in principle seem suitable for the asymptotic safety construction. They show the same degree of predictivity as both critical hypersurfaces are two dimensional. In comparison it is difficult to judge which of the two fixed points should be considered most reliable. The first one lies in the positive $(\lambda,g)$-quadrant at a position similar to the one known from metric gravity, but its large critical exponents question its reliability. The second one occurs in the $(\lambda<0,g<0)$-quadrant at large coordinate values, which is hard to reconcile with phenomenology, but shows a remarkable stability of its critical exponents. In the following we will therefore assign the same level of credibility to both fixed points, ${\bf NGFP^1}$ and ${\bf NGFP^2}$, and treat them on the same footing.

The above discussion refers to the case with the dimensionless mass parameter $m$ chosen to $m=1$. As there is no physical mechanism which could generate a second momentum/mass scale besides $k$, a choice of $m\approx1$ is indeed most natural. We tested that qualitatively the situation does not change for other choices of $m\in [0.5,5]$. As in basis ${\cal B}_1$ all $\gamma$-dependence besides the logarithmic term drops out, we can infer that changing $m$ in this case is equivalent to choosing a different $\gamma$: Small $m$ then correspond to $\gamma \ll 1$ and larger $m$ to $\gamma\approx 1$. Qualitatively this correspondence is also found for the other bases. The explicit $m$ dependence of the FP properties will be analyzed in more detail when we discuss the corresponding ``lifts'' of the fixed points in the 3-dimensional truncation.

\begin{figure}[p]
\centering
{\small
\begin{psfrags}
 \input{Pictures/gl02Plane-psfrag.tex}
 \includegraphics[width=0.37\linewidth]{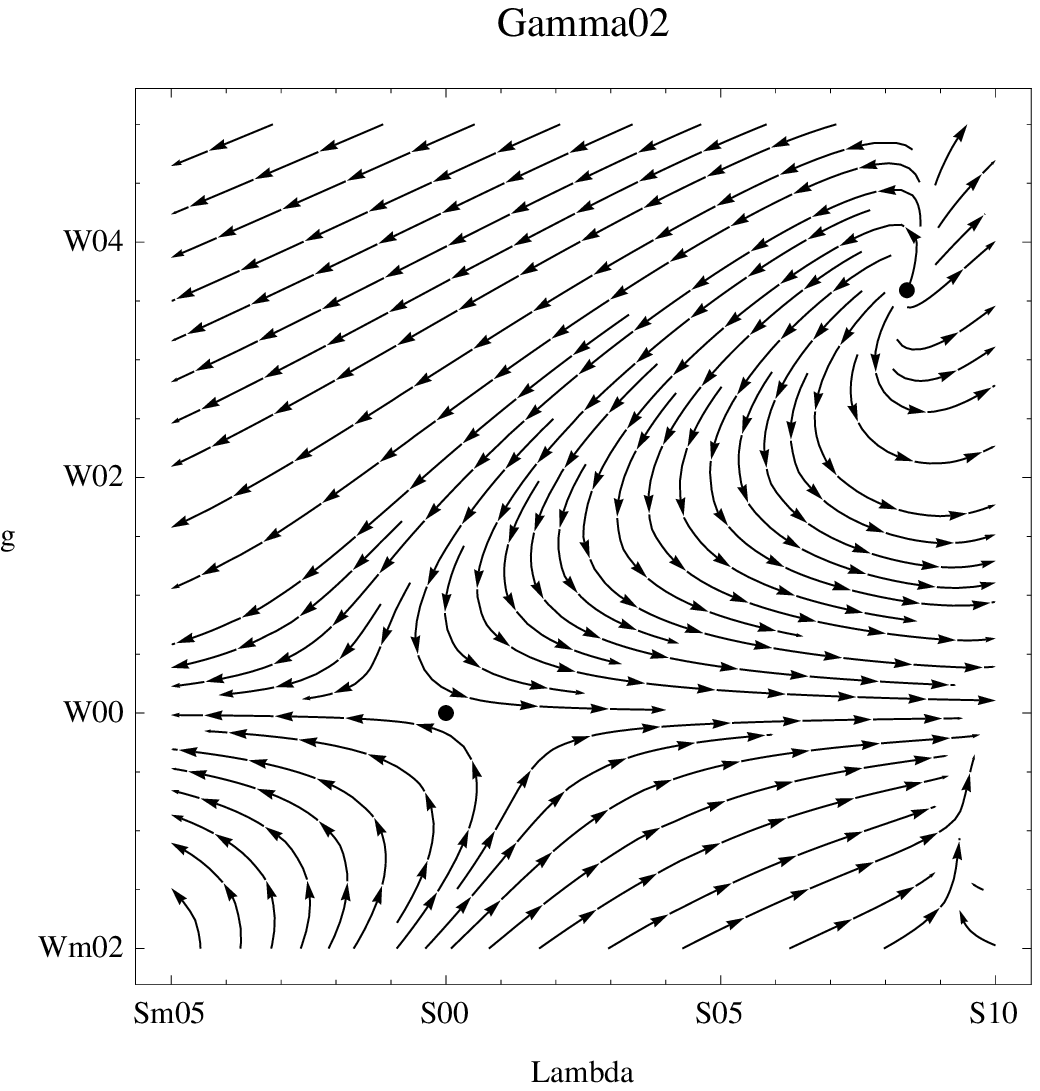}
\end{psfrags}
\qquad
\begin{psfrags}
 \input{Pictures/gl05Plane-psfrag.tex}
 \includegraphics[width=0.37\linewidth]{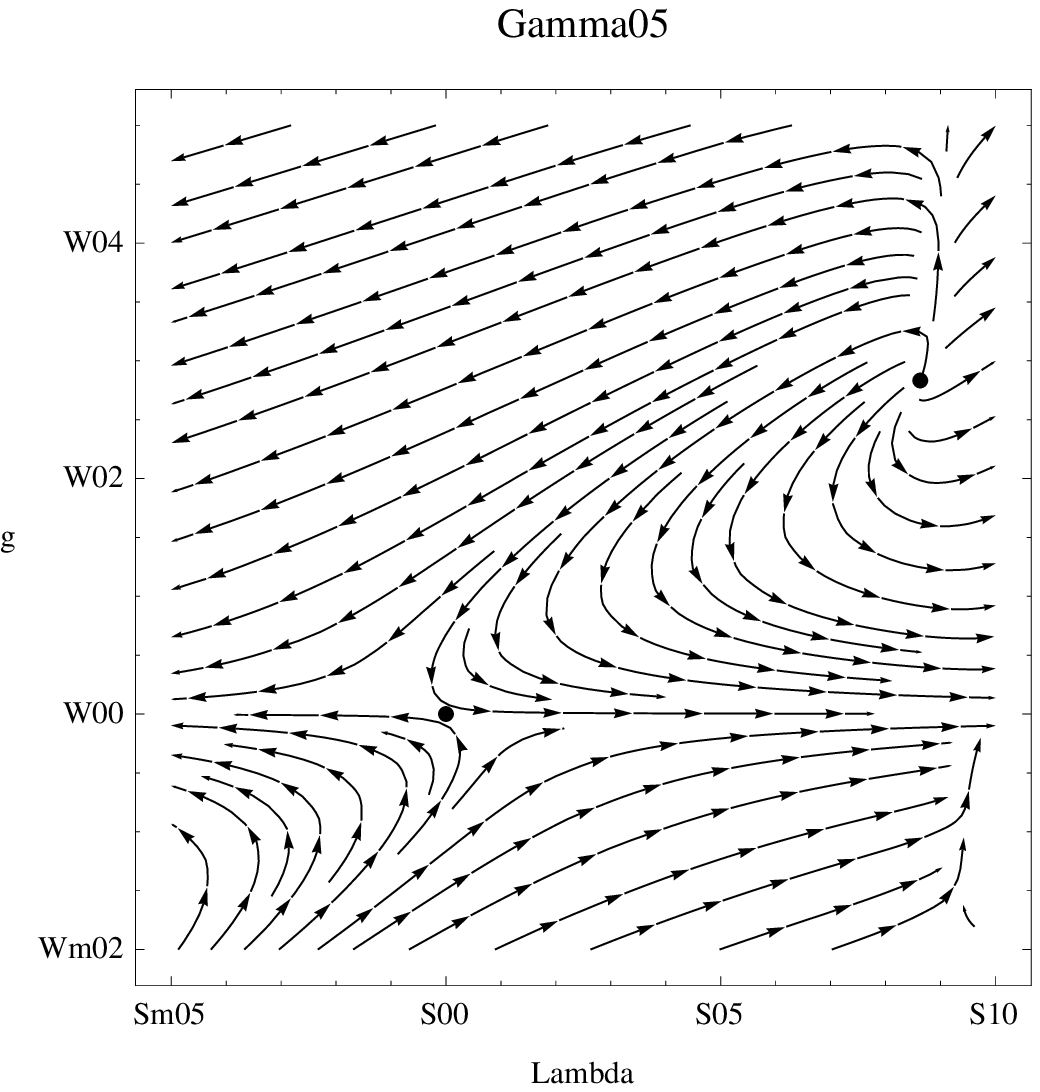}
\end{psfrags}\\[0.5cm]
\begin{psfrags}
 \input{Pictures/gl1Plane-psfrag.tex}
 \includegraphics[width=0.37\linewidth]{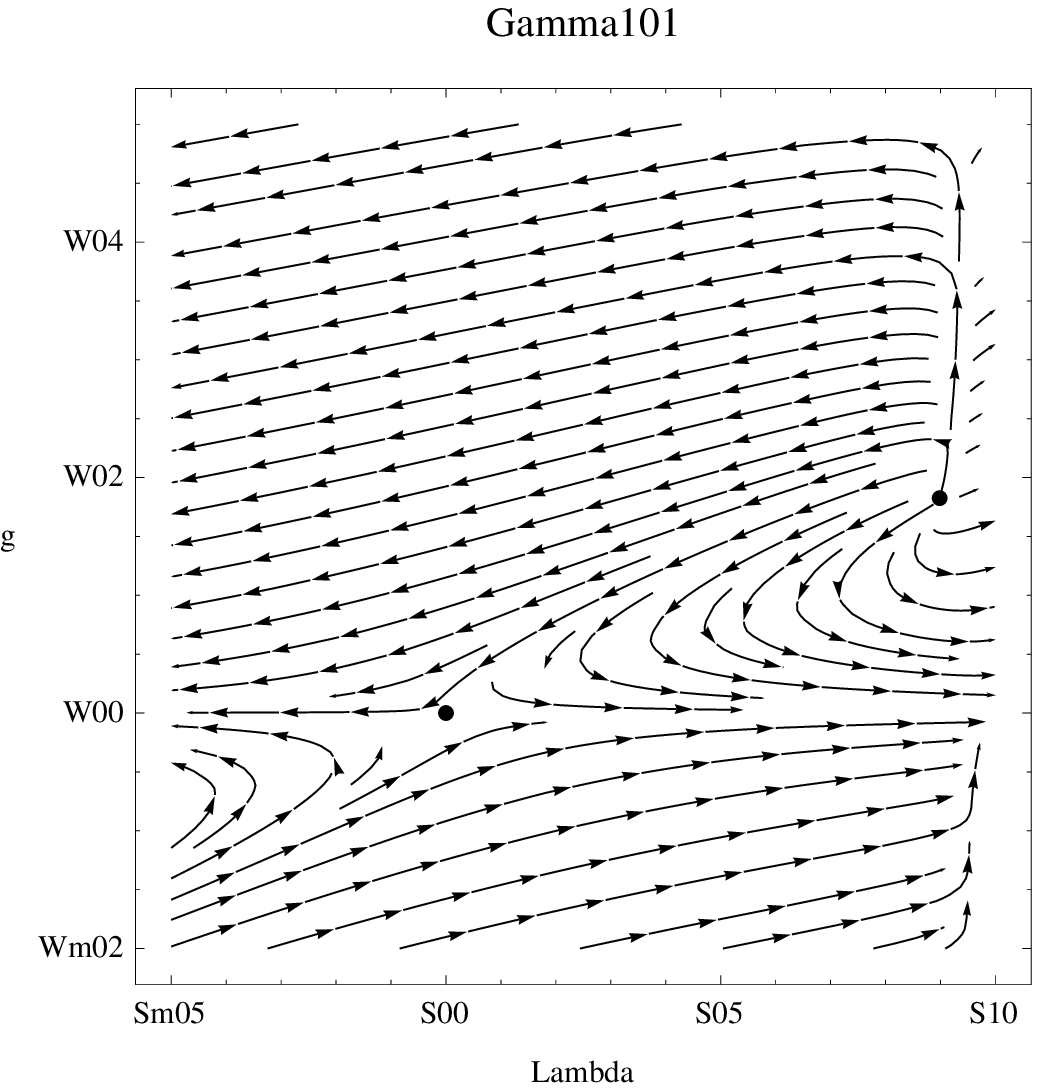}
\end{psfrags}
\qquad
\begin{psfrags}
 \input{Pictures/gl2Plane-psfrag.tex}
 \includegraphics[width=0.37\linewidth]{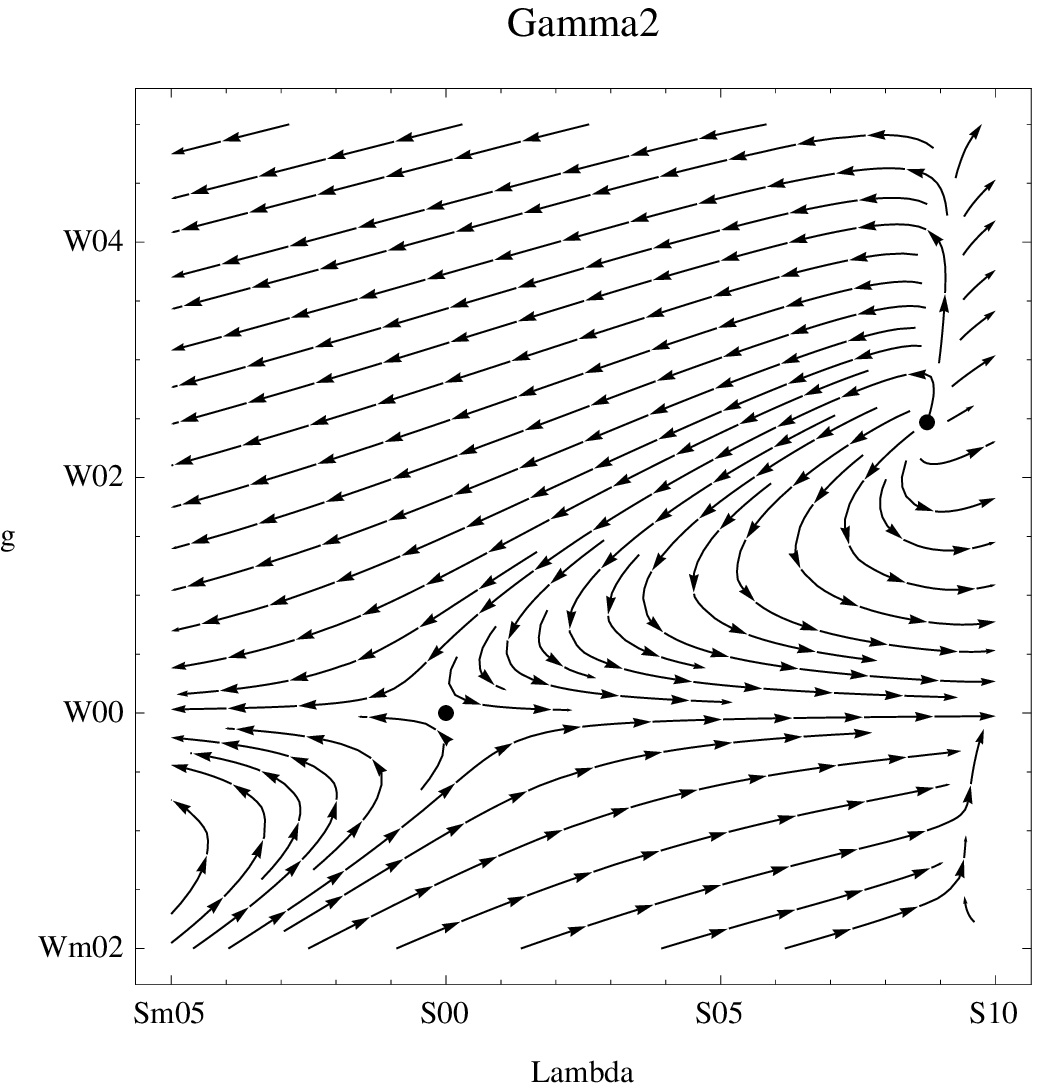}
\end{psfrags}\\[0.5cm]
\begin{psfrags}
 \input{Pictures/gl5Plane-psfrag.tex}
 \includegraphics[width=0.37\linewidth]{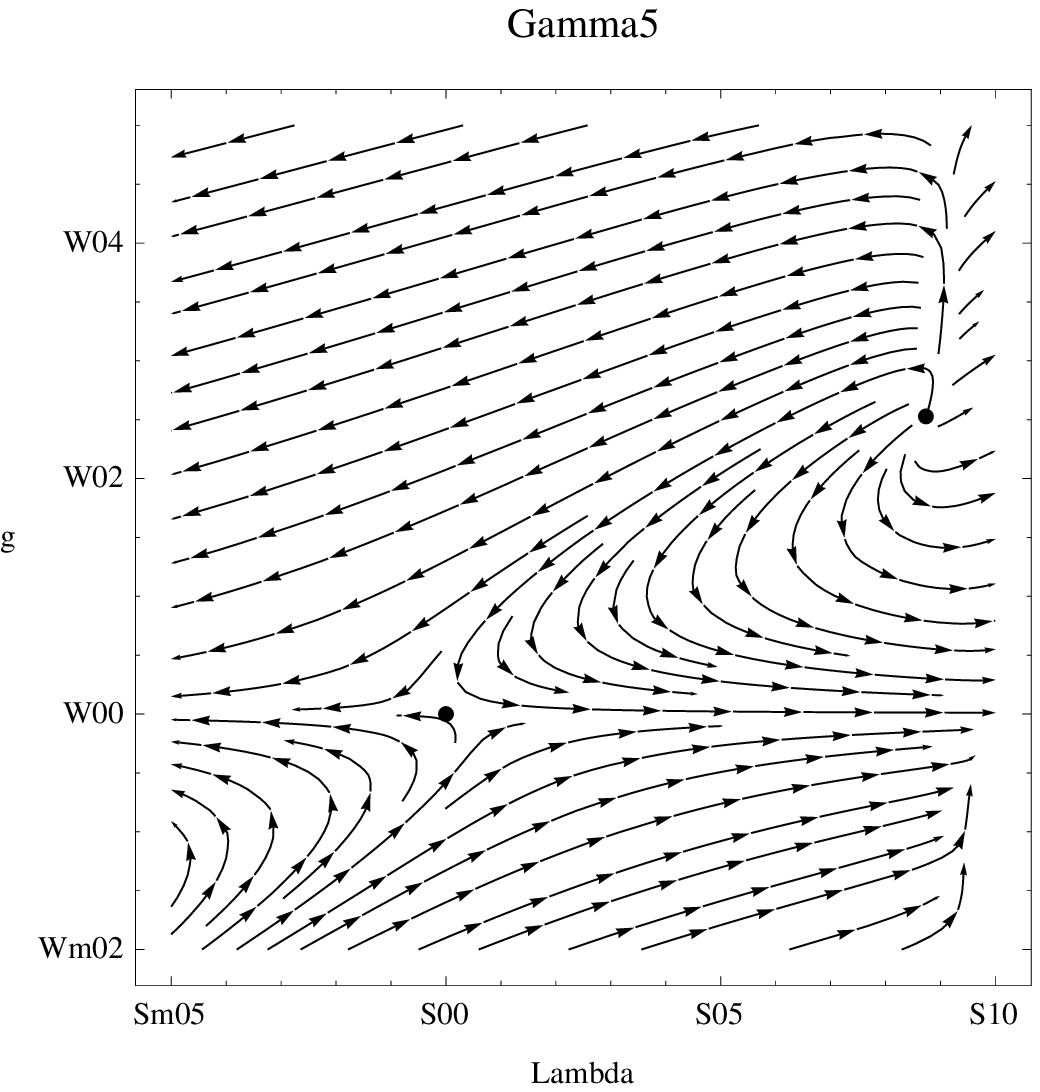}
\end{psfrags}
\qquad
\begin{psfrags}
 \input{Pictures/glinftyPlane-psfrag.tex}
 \includegraphics[width=0.37\linewidth]{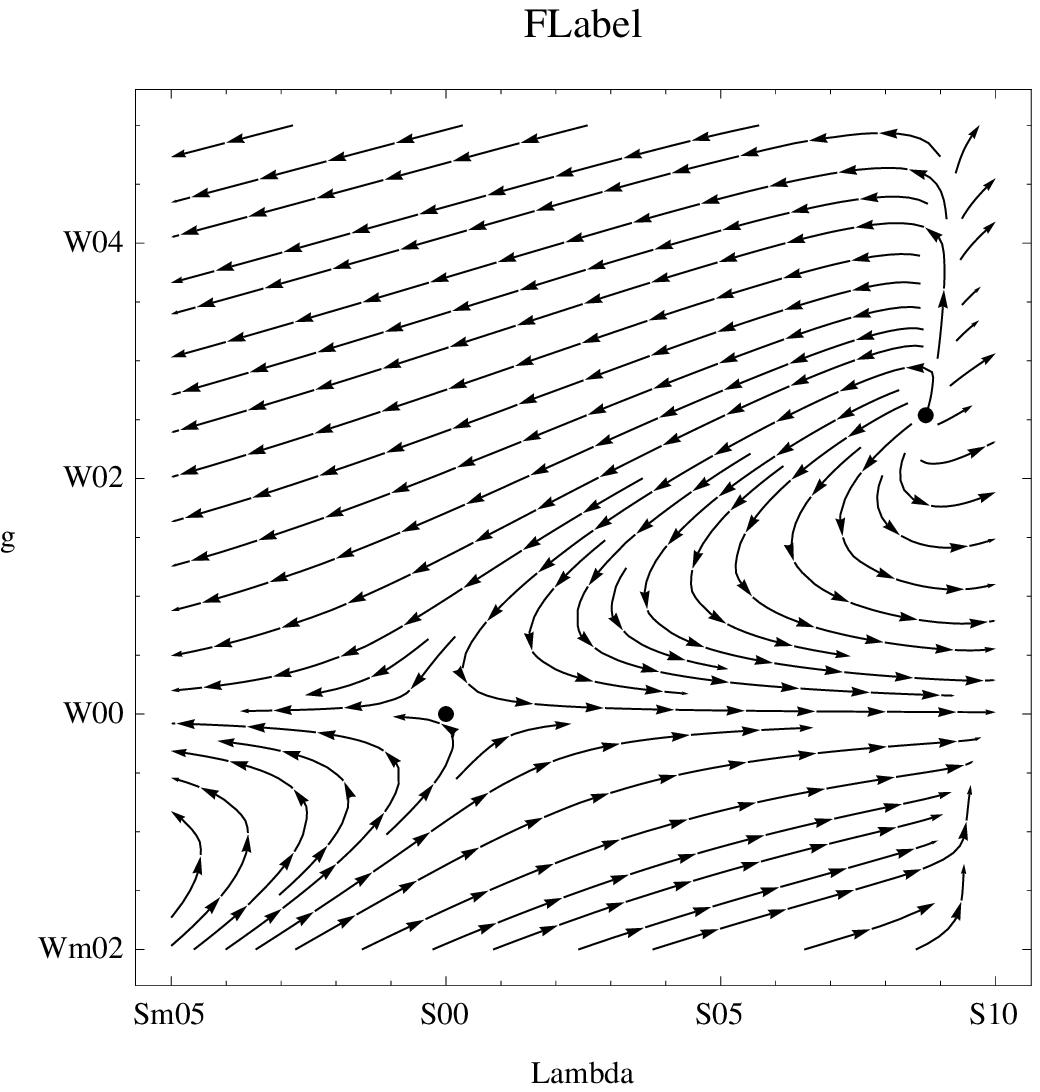}
\end{psfrags}}
\caption{Phase portraits of the $(\lambda, g)$-truncation for different fixed values of $\gamma$ in basis ${\cal B}_1$ focusing on ${\bf NGFP^1}$ in the positive $(\lambda, g)$-quadrant. While the fixed point position changes slightly, the qualitative features of the flow are remarkably similar for all choices of $\gamma$.}
\label{glPlanes1}
\end{figure}

\begin{figure}[p]
\centering
{\small
\begin{psfrags}
 \input{Pictures/gl02Plane2-psfrag.tex}
 \includegraphics[width=0.37\linewidth]{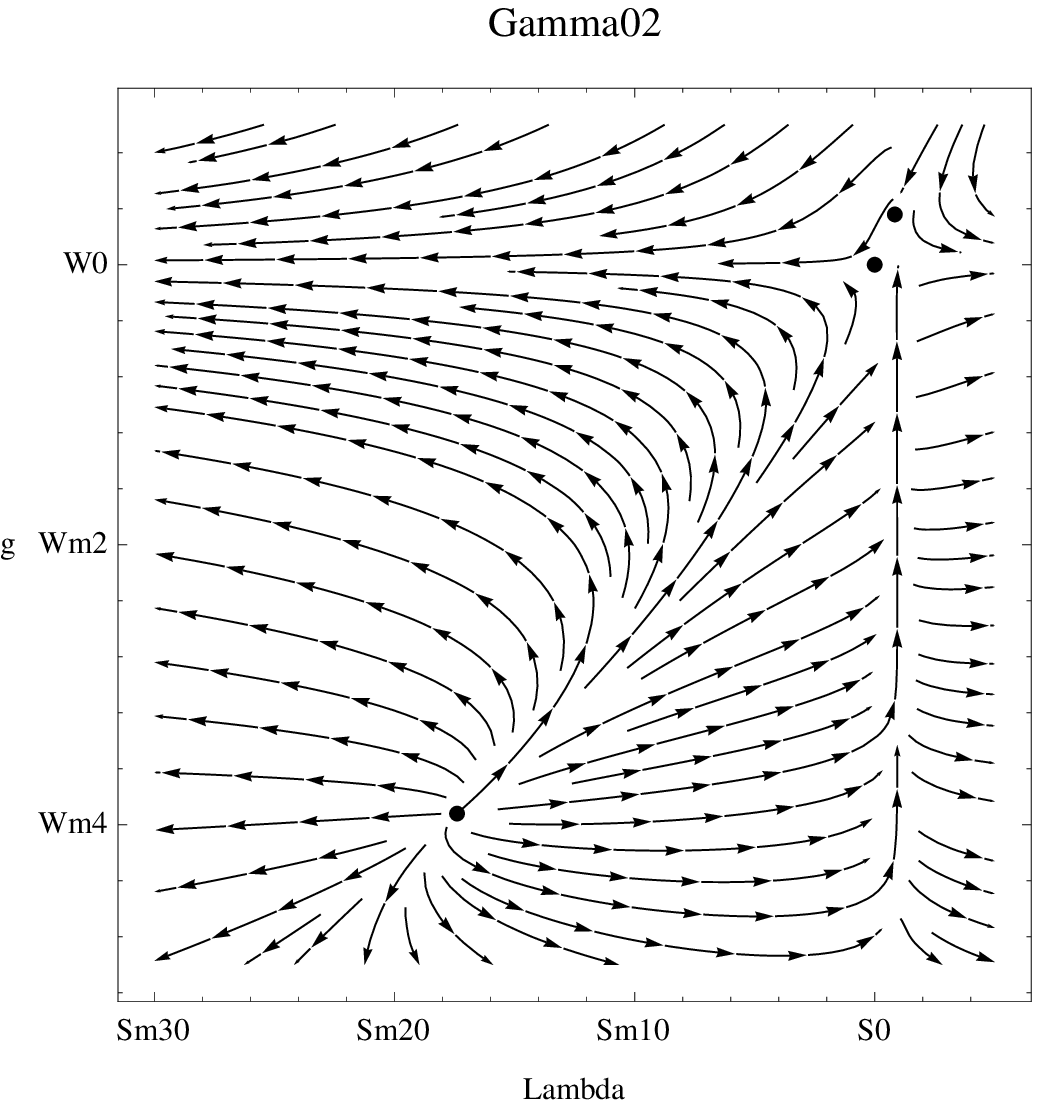}
\end{psfrags}
\qquad
\begin{psfrags}
 \input{Pictures/gl05Plane2-psfrag.tex}
 \includegraphics[width=0.37\linewidth]{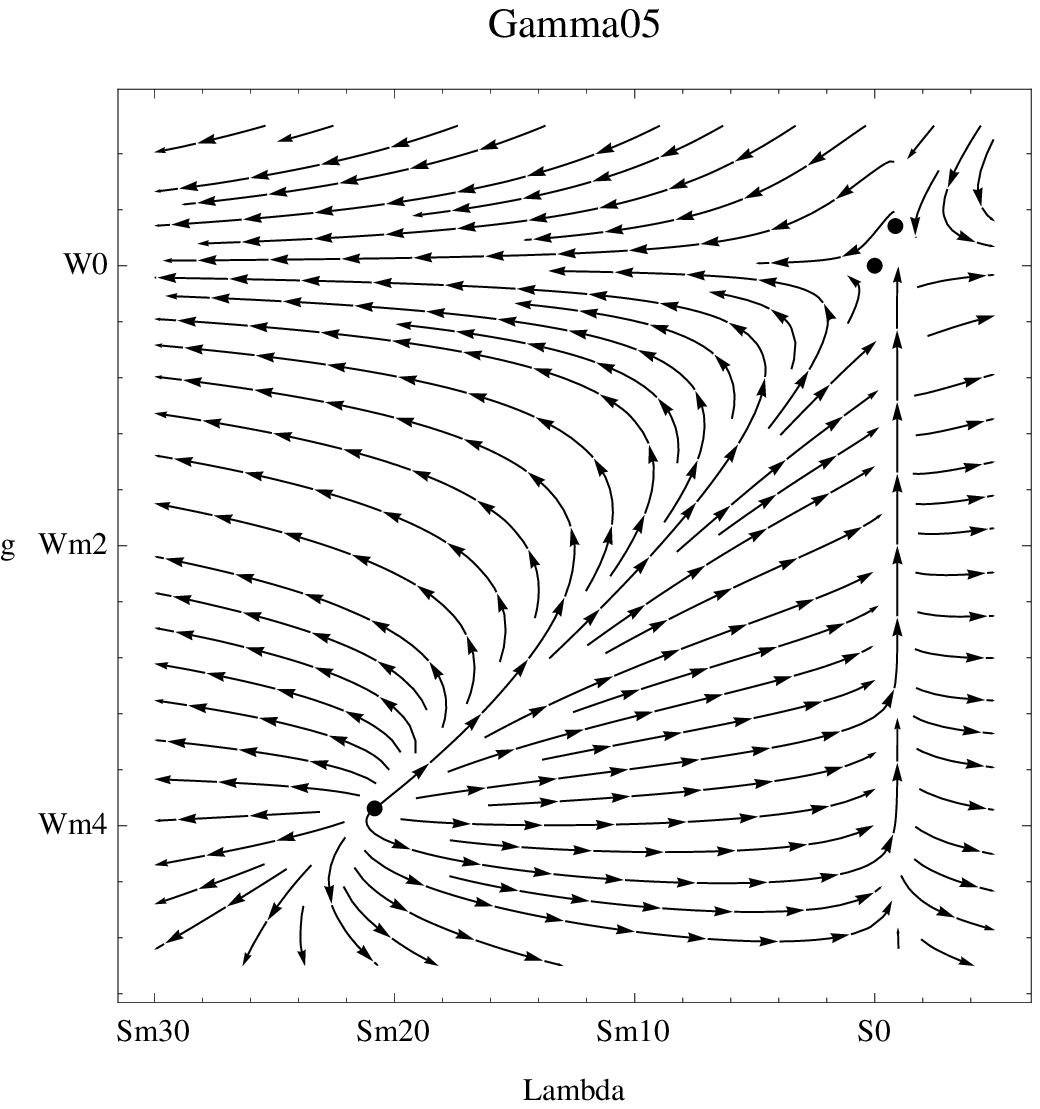}
\end{psfrags}\\[0.5cm]
\begin{psfrags}
 \input{Pictures/gl1Plane2-psfrag.tex}
 \includegraphics[width=0.37\linewidth]{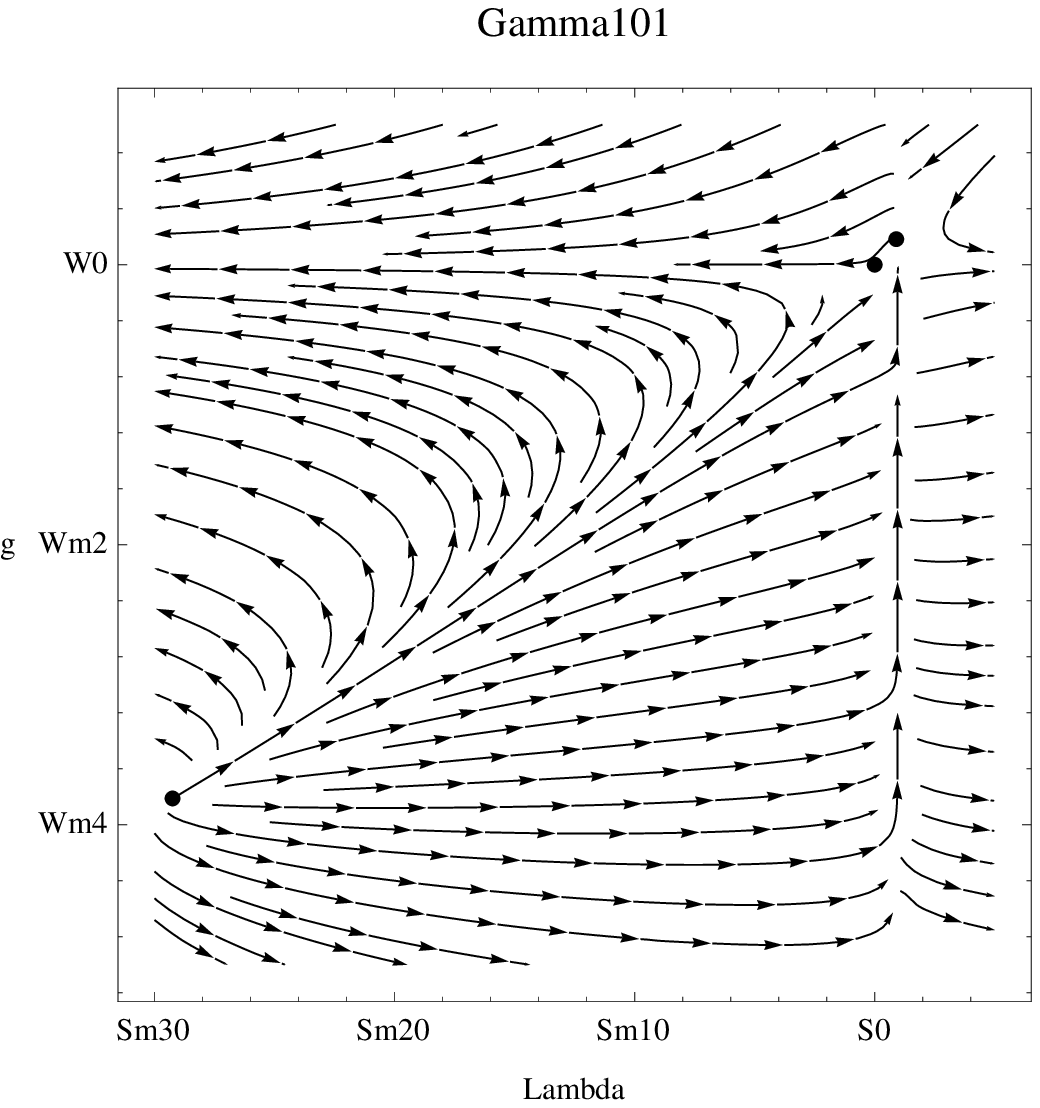}
\end{psfrags}
\qquad
\begin{psfrags}
 \input{Pictures/gl2Plane2-psfrag.tex}
 \includegraphics[width=0.37\linewidth]{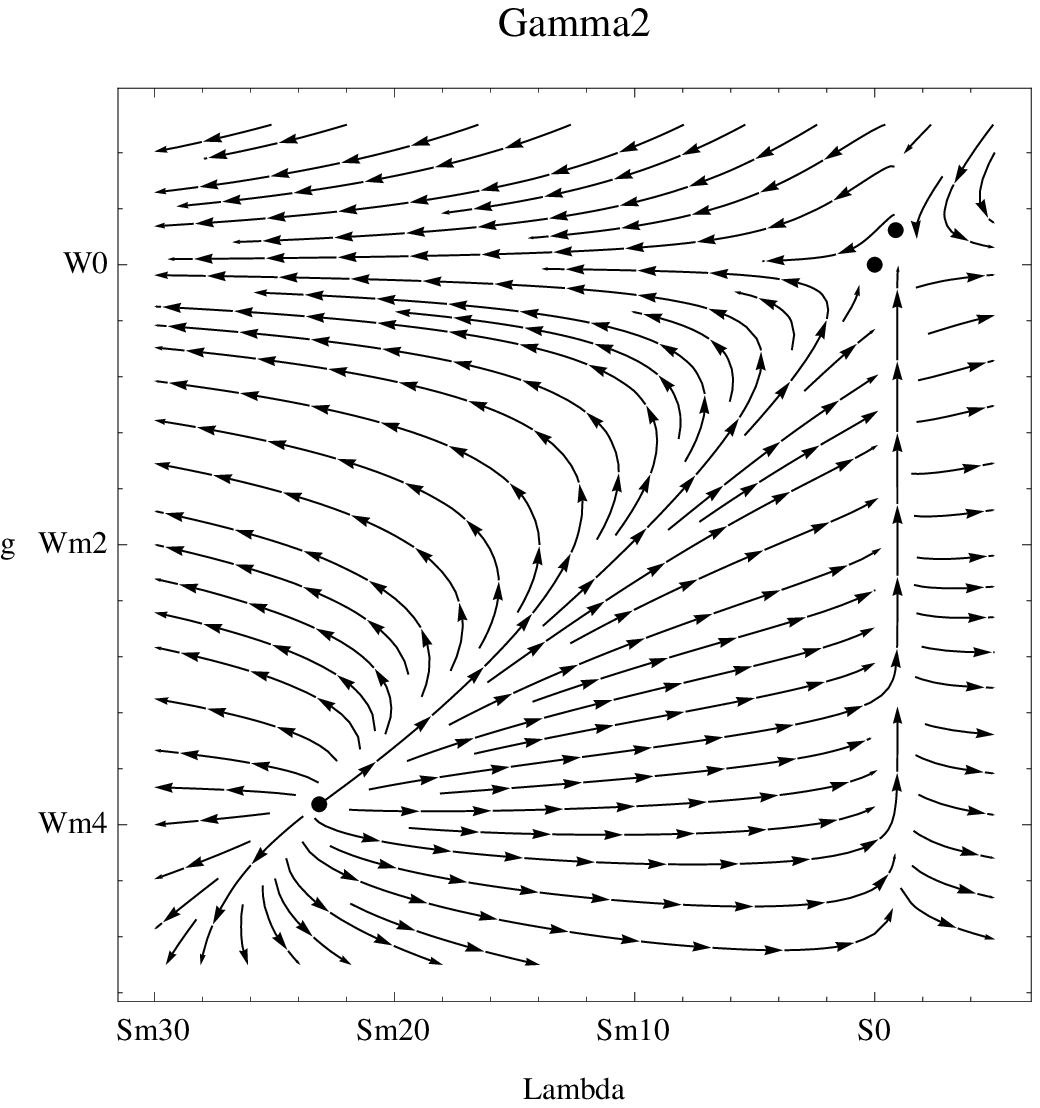}
\end{psfrags}\\[0.5cm]
\begin{psfrags}
 \input{Pictures/gl5Plane2-psfrag.tex}
 \includegraphics[width=0.37\linewidth]{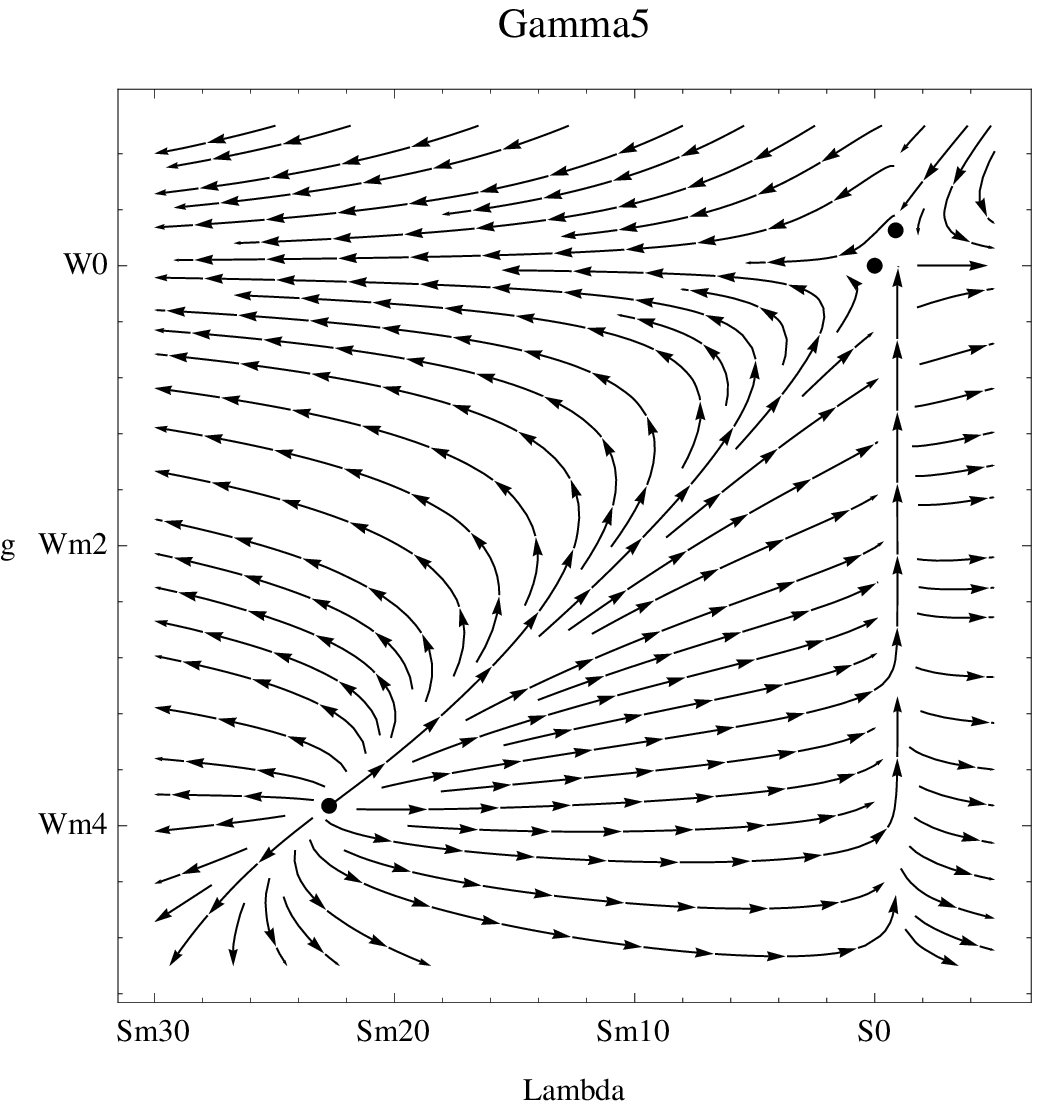}
\end{psfrags}
\qquad
\begin{psfrags}
 \input{Pictures/glinftyPlane2-psfrag.tex}
 \includegraphics[width=0.37\linewidth]{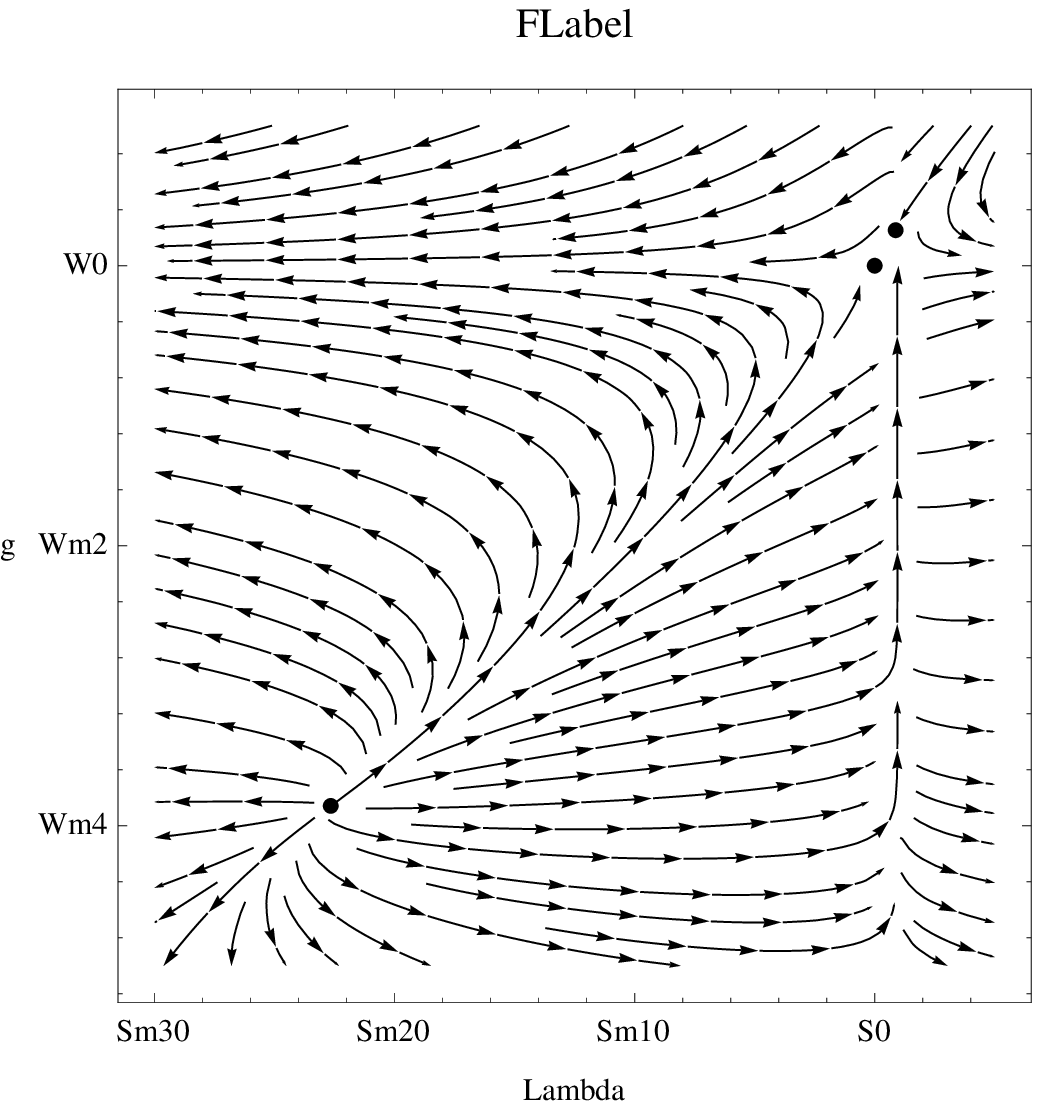}
\end{psfrags}}
\caption{Phase portraits of the $(\lambda, g)$-truncation for different fixed values of $\gamma$ in basis ${\cal B}_1$ focusing on ${\bf NGFP^2}$ in the negative $(\lambda, g)$-quadrant. Again only the fixed point moves, leaving the RG flow qualitatively unchanged for all choices of $\gamma$.}
\label{glPlanes2}
\end{figure}

\paragraph{(E) Phase portrait of the $(\lambda, g)$-truncation.}
Next we discuss the global features of the phase portrait in the $(\lambda,g)$-truncation. As the two non-Gaussian fixed points occur at very different coordinate scales, it is not possible to depict the resulting flow in the vicinity of both FPs equally well in a single diagram. For that reason the two Figs. \ref{glPlanes1} and \ref{glPlanes2} each focus on one of the fixed points and its interplay with the Gaussian FP. The figures are restricted to the ${\cal B}_1$ basis, but the qualitative features of the flow were found to be similar for all bases that show the respective fixed point.

Fig. \ref{glPlanes1} pictures the ${\bf NGFP^1}$ in the positive $(\lambda>0,g>0)$-quadrant. Qualitatively all panels of different $\gamma$ resemble each other, while only the fixed point position moves slightly. 

The $\lambda$-axis, being the critical surface $\mathscr{S}_{\rm UV}$ of the {\bf GFP} is an IR attractive line that cannot be crossed by any trajectory. All points above this axis with $\lambda<1$ lie on trajectories that are asymptotically safe \wrt ${\bf NGFP^1}$. The divergence of the $\beta$-functions at $\lambda=1$ is approached in a controlled way, such that the flow stops on this line.

The similarity of this phase portrait with the one of metric gravity in Einstein-Hilbert truncation is striking. In complete analogy to the metric case a classification of the asymptotically safe RG trajectories is possible: We find the NGFP connected to the {\bf GFP} by a separatrix (type IIa trajectory) that separates the trajectories with a negative IR cosmological constant (type Ia) from those with a positive one (type IIIa). 

However, one significant difference to the metric case should be noted: Following the separatrix from ${\bf NGFP^1}$ at positive $\lambda$ to the IR we find that {\it the cosmological constant turns negative before ending in the ${\bf GFP}$}. In general such a situation occurs if the second eigendirection of stability matrix at the ${\bf GFP}$ points in a direction of negative $\lambda$. Here, due to the logarithmic divergence of the stability matrix at the ${\bf GFP}$, the separatrix even ends up {\it tangent to the $\lambda$-axis} in the ${\bf GFP}$. 

In Fig. \ref{glPlanes2} the second NGFP in the $(\lambda<0,g<0)$-quadrant is depicted. Again, qualitatively all panels of different $\gamma$ resemble each other, but the fixed point position changes. We see that the points in the negative $g$-halfplane with $\lambda<1$ are all attracted to ${\bf NGFP^2}$ in the UV. Also here, we find a separatrix connecting the FP and the ${\bf GFP}$ that separates positive from negative IR cosmological constants. Due to the limiting nature of $\lambda$-axis no trajectory that is asymptotically safe \wrt ${\bf NGFP^2}$ will run to positive $g$ in the IR. Hence, the physical significance of these trajectories is questionable if we expect a positive Newton constant in the IR for phenomenological reasons.

\paragraph{(F) Comparison to the proper-time flow.}
If we compare the results obtained here with the RG study using a proper-time flow equation in \cite{je:longpaper} we find that the fixed point properties do not fully coincide. In \cite{je:longpaper} the $(\lambda, g)$-truncation shows {\it three} non-Gaussian fixed points, all of which occur at $g^*>0$. Two of them have one UV attractive and one repulsive direction, while the third one is attractive in both directions. Thus, only this last one is comparable to our two NGFPs. If we were to compare it with one of our FPs,  we would choose our second fixed point at negative $g$: Both have large absolute values of their $(\lambda^*,g^*)$-coordinates, that are of the same order of magnitude and the critical exponents are similar. In \cite{je:longpaper} they are given as $\{3.4,1.8\}$ while we find in the limit $\gamma\rightarrow\infty$ on average $\{4.1,2.2\}$ which is indeed quite similar at the expected level of accuracy. However, more evidence should be collected before one can reliably identify the fixed points.

\subsection[The $(\gamma,g)$-subsystem]{The $\boldsymbol{(\gamma,g)}$-subsystem}
\paragraph{(A) Differential equations.} The $(\gamma,g)$-truncation, that we are about to discuss next, is based on the two $\beta$-functions \eqref{Beta_g} and \eqref{Beta_gamma}:

\begin{equation}\label{ggsubsystem}
\boxed{ 
\begin{aligned}
  \partial_t g=\beta_g(\lambda,\gamma,g)&=g\left[2+\eta_N(\lambda,\gamma,g)\right],\\
 \partial_t \gamma=\beta_\gamma(\lambda,\gamma,g)&= -\frac{16 \pi g \gamma}{N(\lambda)} \left[P_9(\lambda)+\frac{1}{\gamma^2}P_8(\lambda)+P_{10}(\lambda)\right].
 \end{aligned}}
\end{equation}

\vspace{0.5cm}

{\noindent}Now $\lambda$ is thought of as a fixed external parameter that does not run, and we study the properties of the flow in dependence on its constant value. 

We observe that the system \eqref{ggsubsystem} does not suffer from the limitations due to logarithmic divergences as these only occur in $\beta_\lambda$; nevertheless we again find the $\lambda=1$-barrier of the flow due to the (double) zero of the denominator $N(\lambda)$. In the bases ${\cal B}_2$, ${\cal B}_3$ and ${\cal B}_4$, where the polynomial $P_8(\lambda)$ is non-vanishing an additional divergence at $\gamma=0$ is found. Surprisingly, the limit of chiral gravity, $\gamma=\pm 1$, is perfectly well defined in this two-dimensional truncation.

We find that $\beta_g$ is an even function of $\gamma$, while $\beta_\gamma$ is odd. Together this leads to a symmetry of the flow under $\gamma\mapsto -\gamma$, under which the flow remains unchanged.

\begin{figure}[ht]
\centering
{\small
\begin{psfrags}
 \input{Pictures/ggSketch1-psfrag.tex}
 \subfigure[${\cal B}_1$]{\includegraphics[width=0.45\linewidth]{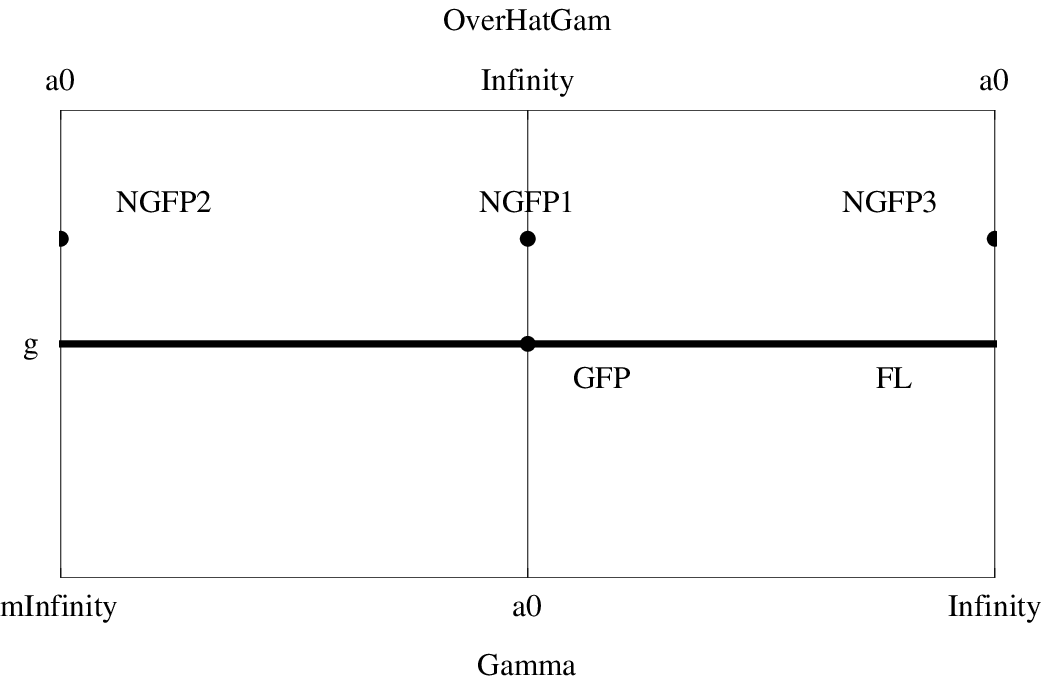}}
\end{psfrags}
\begin{psfrags}
 \input{Pictures/ggSketch234-psfrag.tex}
\subfigure[${\cal B}_2,{\cal B}_3,{\cal B}_4$]{\includegraphics[width=0.45\linewidth]{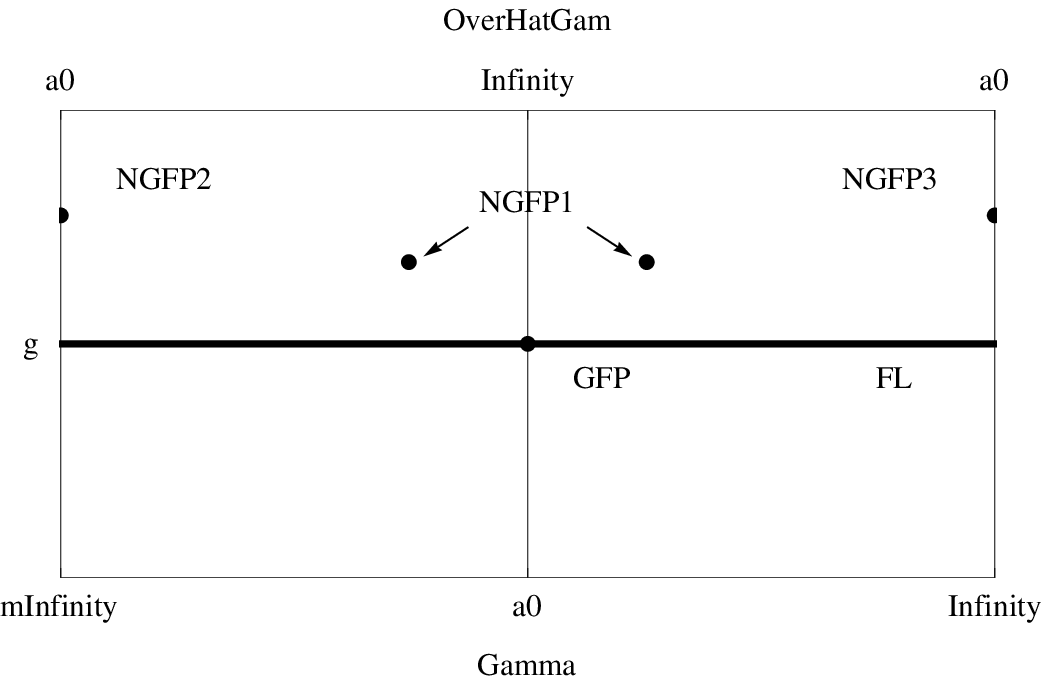}}
\end{psfrags}
}
\caption{Sketch of the fixed point structure of the $(\gamma,g)$-system.}
\label{ggSketch}
\end{figure}

\paragraph{(B) Fixed point structure.}
Let us, first of all, illuminate the general situation in theory space by a schematic plot of the fixed points we are going to discuss and the names given to them (cf. Fig. \ref{ggSketch}). We find, that in the $(\gamma,g)$-subsystem the fixed point structure depends on the basis chosen. For basis ${\cal B}_1$ there is a NGFP at $\gamma=0$ (denoted ${\bf NGFP'_0}$), while in the bases ${\cal B}_{2,3,4}$ we find a pair of NGFPs at finite non-zero $\gamma$ (${\bf NGFP'_{fin}}$). Besides that a second NGFP (${\bf NGFP'_{\infty}}$) and a fixed line at $g=0$ is present in all bases, that includes the Gaussian fixed point ${\bf GFP}$. (The alternative coupling $\hat\gamma$ used in Fig. \ref{ggSketch} will be explained below.)

\paragraph{(C) The Gaussian fixed point.}
The first and most obvious solution of the fixed point condition $(\beta_\gamma,\beta_g)=(0,0)$ is the limit of vanishing Newton constant $g$. It gives rise to a {\bf fixed line} $g=0$ with $\gamma$ arbitrary that contains the {\bf GFP} at $(\gamma,g)=(0,0)$. This fixed line turns out IR attractive (in $g$-direction) for all values of $\lambda$ as the stability matrix along this line is of the form
\begin{equation}
 \mathscr{B}\Big|_{g=0}=\begin{pmatrix}
              2&0\\ \partial_g\beta_\gamma(\lambda,\gamma,g\!=\!0)&0
             \end{pmatrix}.
\end{equation}

For the search of non-Gaussian fixed points we have to distinguish basis ${\cal B}_1$ from the other bases ${\cal B}_2$, ${\cal B}_3$, ${\cal B}_4$ as the absence of $P_8(\lambda)$ changes the mechanism how the non-Gaussian fixed point at finite $\gamma$ is established. For this reason the fixed point ${\bf NGFP'_0}$ which is found in ${\cal B}_1$ exhibits properties that differ from those obtained by taking the limit $P_8\rightarrow 0$ of the fixed point ${\bf NGFP'_{fin}}$ found for the other three bases.

\paragraph{(D) The fixed point ${\bf NGFP'_{fin}}$.} Let us first discuss ${\bf NGFP'_{fin}}$, that exists in the bases ${\cal B}_2$, ${\cal B}_3$ and ${\cal B}_4$. Since $P_8$ is non-zero in this case, the only possibility to satisfy the condition $\beta_\gamma=0$ for $g\neq0$ is that the terms in the square brackets add up to zero (cf. \eqref{ggsubsystem}). Hence the fixed point coordinate $\gamma^*$ is determined by the condition
\begin{equation}\label{ggFPcond1}
 \big(\gamma^*(\lambda)\big)^2= -\frac{P_8(\lambda)}{P_9(\lambda)+P_{10}(\lambda)}\:.
\end{equation}
We observe that the condition can only be satisfied by real $\gamma$ if the right hand side is positive. As $P_8$ takes on a constant value in $(0,0,0)$-gauge for all three bases, sign changes of the RHS only occur at the zeros of $P_9+P_{10}$, which therefore limit the domain of existence of the FP in $\lambda$. At these limiting values of $\lambda$ the fixed point coordinate $\gamma^*$ diverges. Moreover, as $P_9+P_{10}$ does not have any poles, $\big(\gamma^*\big)^2=0$ is only obtained in the limit $\lambda\rightarrow \infty$. For any fixed value of $\lambda$, except for the zeros of $P_9+P_{10}$, we thus find a fixed point solution ${\bf NGFP'_{fin}}$ arising at finite non-zero $\gamma^*$. Note that the solutions to \eqref{ggFPcond1} come in pairs of $\pm \gamma^*$, which is not surprising as the total RG flow is invariant under the transition $\gamma\mapsto -\gamma$.

The corresponding $g^*$-coordinate is found from $\beta_g(\gamma^*(\lambda),g)=0$ resulting in 
\begin{equation}
 g^*(\lambda)=\frac{1}{8 \pi}\frac{N(\lambda)}{P_9(\lambda)}\:.
\end{equation}
Since $N(\lambda)\geq0$ the fixed point lies at positive $g^*$ if $P_9(\lambda)$ is positive.

With analytical expressions for the fixed point coordinates at hand, we can also find an analytical expression for the critical exponents of the fixed point ${\bf NGFP'_{fin}}$. As $\left.\partial_g\beta_\gamma\right|_{{\bf NGFP'_{fin}}}=0$ the stability matrix at the fixed point is triangular, such that one of its eigenvectors points in the direction of the $g$-axis. The corresponding critical exponents are found to 
\begin{equation}
\theta_g=2,\qquad \theta_2= 4\frac{P_9(\lambda)+P_{10}(\lambda)}{P_9(\lambda)}\:.
\end{equation}
Note that the divergence of the critical exponent $\theta_2$ at $P_9(\lambda)=0$ occurs at the same $\lambda$-value as the pole of $g^*(\lambda)$, such that both quantities change sign there. The zeros of $\theta_2$, on the other hand, occur at the poles of $\gamma^*(\lambda)$.

\begin{figure}[tb]
\centering
{\small
\begin{psfrags}
 \input{Pictures/ggNGFPfinpos2-psfrag.tex}
 \includegraphics[width=0.45\linewidth]{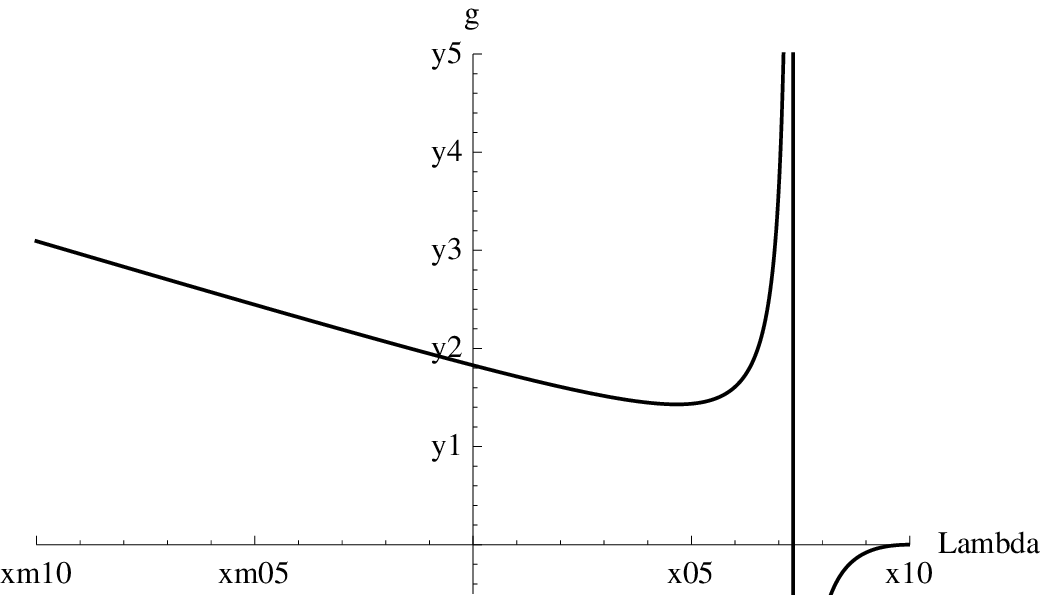}
\end{psfrags}
\begin{psfrags}
 \input{Pictures/ggNGFPfinpos1-psfrag.tex}
 \includegraphics[width=0.45\linewidth]{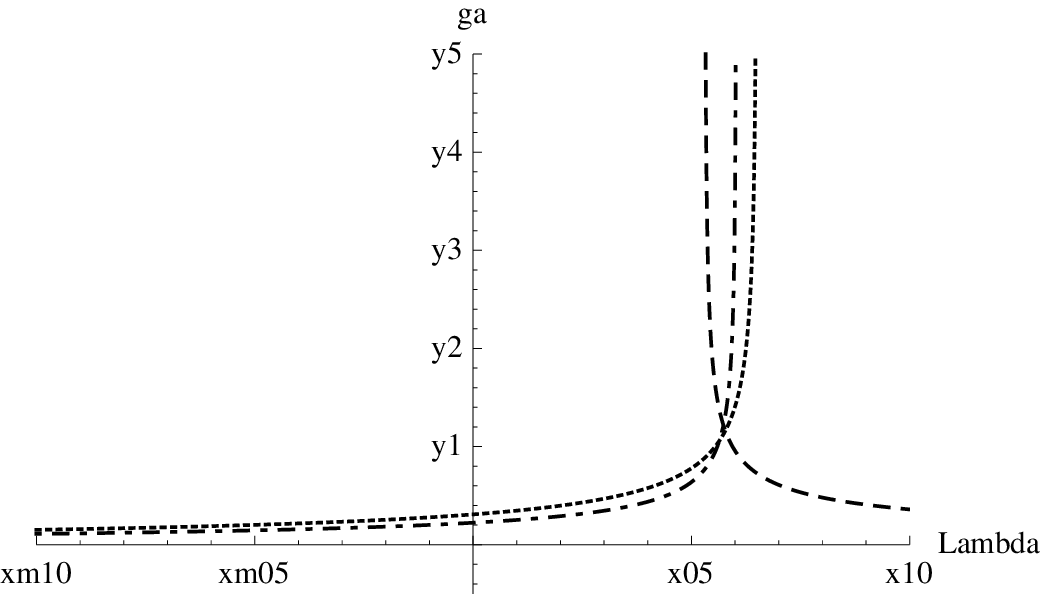}
\end{psfrags}}
\caption{Position of the fixed point ${\bf NGFP'_{fin}}$ as a function of $\lambda$. While $g^*$ only depends on the (in our case fixed) choice of basis in the parity-odd $T^2$ sector, $\gamma^*$ differs for the three bases ${\cal B}_2$ (dashed), ${\cal B}_3$ (dot-dashed) and ${\cal B}_4$ (dotted).}
\label{ggNGFPfinpos}
\end{figure}

In Fig. \ref{ggNGFPfinpos} we plot the coordinates of ${\bf NGFP'_{fin}}$ as functions of $\lambda$ for the three bases ${\cal B}_2$, ${\cal B}_3$ and ${\cal B}_4$. Interestingly, $g^*(\lambda)$ is found basis independent because it only depends on $P_9(\lambda)$, whose explicit form is fixed by our choice of basis in the parity-odd torsion squared subspace. 

We find a positive coordinate $g^*>0$ for all $\lambda\leq 0.733$, and therefore in particular for small $|\lambda|\ll 1$. The domain of existence of the FP differs for the three bases: While for ${\cal B}_3$ and ${\cal B}_4$ the FP exists up to $\lambda\leq0.603$ and $\lambda\leq0.651$, respectively, for ${\cal B}_2$ is ceases to exist for $\lambda\leq 0.53$. Thus, there is a small interval in $\lambda$ in which the FP exists for all three bases. Nonetheless, we can confirm the observation from the $(\lambda,g)$-truncation, that predictions concerning fixed point existence and stability in basis ${\cal B}_2$ differ considerably from the other bases. For that reason we will consider the $\beta$-functions in basis ${\cal B}_2$ as the exception from the rule, which can be traced back to the fact, that all coefficients in $P_{10}(\lambda)$ for this basis have the opposite sign compared to the other bases.

\begin{figure}[tb]
\centering
{\small
\begin{psfrags}
 \input{Pictures/ggNGFPfinCE-psfrag.tex}
 \includegraphics[width=0.65\linewidth]{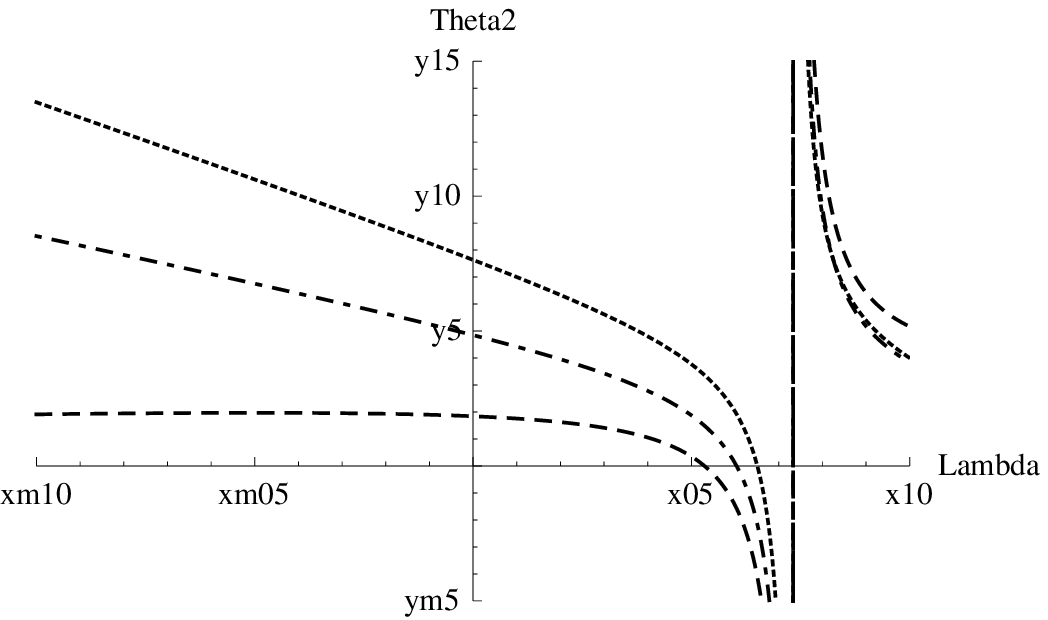}
\end{psfrags}}
\caption{The critical exponent $\theta_2$ of the fixed point ${\bf NGFP'_{fin}}$ as a function of $\lambda$. In the region $-1\leq\lambda\leq 0.5$ the function is positive for all three bases ${\cal B}_2$ (dashed), ${\cal B}_3$ (dot-dashed) and ${\cal B}_4$ (dotted), corresponding to a second UV attractive direction.}
\label{ggNGFPfinCE}
\end{figure}

Considering basis ${\cal B}_2$ less credible, we find that the results of bases ${\cal B}_3$ and ${\cal B}_4$ lie perfectly in line with each other. Both predict a pair of fixed points which lies for $\lambda<0.5$ at very small $|\gamma^*|$. In the range $0.5 < \lambda <0.733$ both coordinates diverge, which is considered a hint that the results in this range become less trustworthy for these large values of the cosmological constant. In Fig. \ref{ggNGFPfinCE} we have plotted the critical exponent $\theta_2$ of the fixed point. For the bases ${\cal B}_3$ and ${\cal B}_4$ we find that it corresponds to a UV attractive direction ($\theta_2>0$) exactly over the whole range in $\lambda$ in which the fixed point exists. 

Taken together with the result $\theta_g=2$ we conclude that for small $\lambda$ both bases predict a pair of fixed points ${\bf NGFP'_{fin}}$ at finite $\gamma$, that is attractive in both directions. The absolute value of $\theta_2$, as a little blemish, turns out fairly large.

\paragraph{(E) The fixed point ${\bf NGFP'_{0}}$.} As already pointed out above, for basis ${\cal B}_1$ we find a different fixed point solution ${\bf NGFP'_{0}}$. With all $P_8$-terms absent in the $\beta$-functions \eqref{ggsubsystem} there is only one non-Gaussian fixed point solution given by
\begin{equation}
 g^*(\lambda)=-\frac{1}{8 \pi}\frac{N(\lambda)}{P_{10}(\lambda)},\qquad \gamma^*=0\:.
\end{equation}
In this case the stability matrix at the fixed point ${\bf NGFP'_{0}}$ is diagonal as also $\partial_\gamma \beta_g=0$. Thus, both critical exponents can be associated with the $g$ and $\gamma$ axes in theory space and we find
\begin{equation}
 \theta_g=2,\qquad \theta_\gamma=-2\, \frac{P_9(\lambda)+P_{10}(\lambda)}{P_{10}(\lambda)}\,.
\end{equation}
We observe that both $g^*$ and $\theta_\gamma$ of this new fixed point ${\bf NGFP'_{0}}$ cannot be obtained by taking the limit $P_8\rightarrow0$ of the corresponding quantities of ${\bf NGFP'_{fin}}$. Thus, in basis ${\cal B}_1$ a truly different fixed point is present.

\begin{figure}[t]
\centering
{\small
\begin{psfrags}
 \input{Pictures/ggNGFP0pos-psfrag.tex}
 \includegraphics[width=0.45\linewidth]{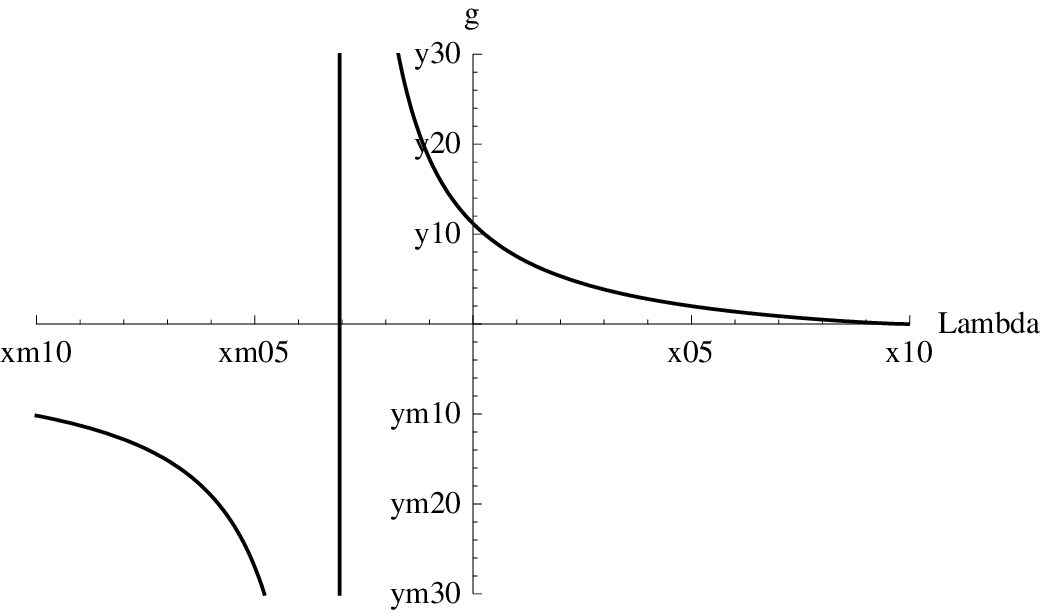}
\end{psfrags}
\begin{psfrags}
 \input{Pictures/ggNGFP0CE-psfrag.tex}
 \includegraphics[width=0.45\linewidth]{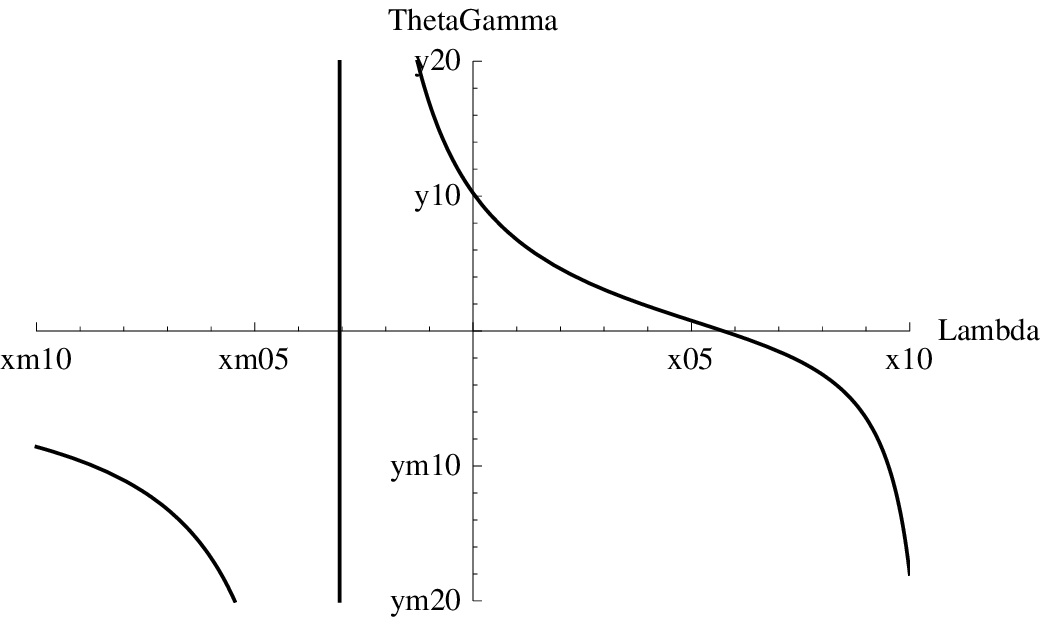}
\end{psfrags}}
\caption{$g^*$-coordinate and critical exponent $\theta_\gamma$ of the fixed point ${\bf NGFP'_{0}}$ that is only present in basis ${\cal B}_1$ as functions of $\lambda$.}
\label{ggNGFP0pos}
\end{figure}

In Fig. \ref{ggNGFP0pos} (left panel) the position of the fixed point is depicted. While $\gamma^*=0$ for all $\lambda$ we find that the fixed point value of the Newton constant is positive for all $ -0.306\leq \lambda\leq 1$, where $g^*$ diverges at the lower boundary. In the right panel we see, that $\theta_\gamma$ diverges at the same point and is positive up to a value of $\lambda \approx 0.573$. 

We conclude that in basis ${\cal B}_1$ the pair of fixed points ${\bf NGFP'_{fin}}$ at finite non-zero $\gamma^*$ gets replaced by ${\bf NGFP'_{0}}$, located at $\gamma^*=0$ for all $\lambda$. Despite this difference both fixed points have in common, that for small $|\lambda|$ they are UV attractive in both directions of the $(\gamma,g)$-plane. 

\paragraph{(F) Coordinate charts and the $\gamma \leftrightarrow 1/\gamma$ duality.}
In order to fully explore the theory space spanned by the Immirzi and the curvature term, we also have to consider its 1-dimensional subspace where the Immirzi term {\it is absent}. This subspace corresponds to the limit $\gamma\rightarrow\infty$ with $g$ arbitrary and is thus not contained in the coordinate chart we have employed so far. 

For this reason it is necessary to cover theory space by {\it two coordinate charts} and to introduce a second coordinate $\hat \gamma$, which in the overlapping region is related to $\gamma$ by $\hat \gamma=\frac{1}{\gamma}$. Hence, similar to the stereographic projection of a sphere $S^1$, the charts overlap at all values of $\gamma$ except for the two points $\gamma=0$ and $\hat\gamma=0$, that are covered by only one of the two charts.

If we analyze the behavior of the $\beta$-functions under this change of coordinates, we are not only enabled to examine the $\gamma\rightarrow\infty$-limit properly, but we also find a most remarkable property of the RG flow in the $(\gamma,g)$-truncation. Under the coordinate change $\gamma\mapsto 1/\hat\gamma$ the $\beta$-functions are transformed according to $\beta_g(\lambda,\gamma,g)\mapsto\beta_g(\lambda,1/\hat\gamma,g)$ and $\beta_\gamma(\lambda,\gamma,g)\mapsto\beta_{\hat\gamma}(\lambda,\hat\gamma,g)=-{\hat\gamma}^2\beta_\gamma(\lambda,1/\hat\gamma,g)$. This gives rise to the explicit form
\begin{equation}\label{Beta_gammaHat}
 \begin{aligned}
  \beta_g(\lambda,\hat{\gamma},g)&=g\left[2+\frac{16 \pi g}{N(\lambda)}\big({\hat\gamma}^2 P_8(\lambda)+P_{10}(\lambda)\big)\right],\\[0.3cm]
  \beta_{\hat\gamma}(\lambda,\hat{\gamma},g)&=\frac{16 \pi g \hat{\gamma}}{N(\lambda)}\left[P_9(\lambda)+\hat{\gamma}^2 P_8(\lambda) +P_{10}(\lambda)\right]\,.
 \end{aligned}
\end{equation}
If we now compare the system \eqref{Beta_gammaHat} with the original one in eq. \eqref{ggsubsystem} that reads
\begin{equation}
 \begin{aligned}
  \beta_g(\lambda,\gamma,g)&=g\left[2+\frac{16 \pi g}{N(\lambda)}\bigg(\frac{1}{\gamma^2} P_8(\lambda)+P_{10}(\lambda)\bigg)\right],\\[0.3cm]
 \beta_\gamma(\lambda,\gamma,g)&= -\frac{16 \pi g \gamma}{N(\lambda)} \left[P_9(\lambda)+\frac{1}{\gamma^2}P_8(\lambda)+P_{10}(\lambda)\right],
 \end{aligned}
\end{equation}
we observe that in the case $P_8(\lambda)=0$ (\ie in basis ${\cal B}_1$) $\beta_g$ is left invariant under the coordinate change and the flow of the Immirzi parameter satisfies
\begin{equation}
\boxed{
\beta_{\hat\gamma}(\lambda,\gamma,g)= -\beta_{\gamma}(\lambda,\gamma,g)\:.}
\end{equation}
Thus, for the {\it same arguments\,} the $\beta$-functions of $\gamma$ and $\hat\gamma$ are equal up to a minus sign. We conclude that, in the case $P_8(\lambda)=0$, {\it the RG flow at large values of the Immirzi parameter $\gamma$ contains the same information as at small values ($1/\gamma$)}. Therefore we refer to the transformation $\gamma\mapsto 1/\gamma$ as a {\it duality map}.

\begin{figure}[tb]
\centering
{\small
\begin{psfrags}
 \input{Pictures/ggNGFPinfpos-psfrag.tex}
 \includegraphics[width=0.45\linewidth]{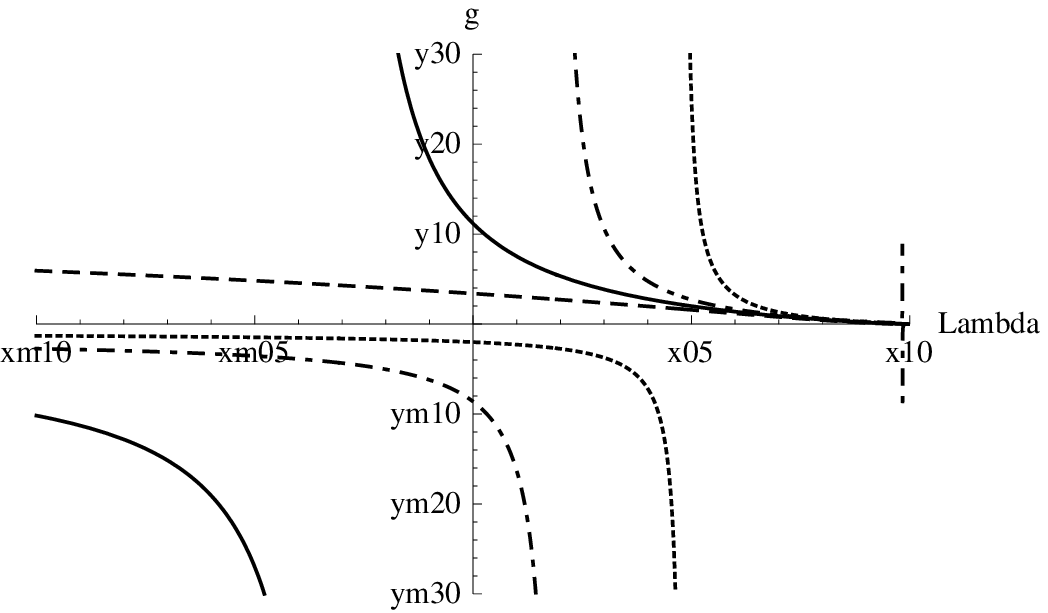}
\end{psfrags}
\begin{psfrags}
 \input{Pictures/ggNGFPinfCE-psfrag.tex}
 \includegraphics[width=0.45\linewidth]{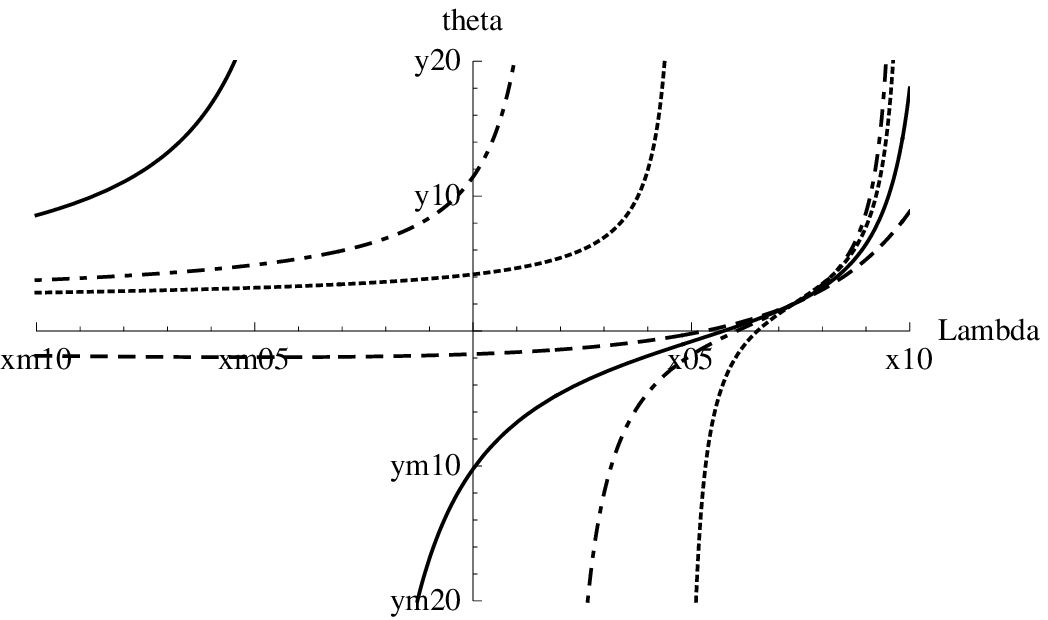}
\end{psfrags}}
\caption{$g^*$-coordinate and critical exponent $\theta_\gamma$ of the fixed point ${\bf NGFP'_{\boldsymbol{\infty}}}$ present in all bases (${\cal B}_1$ solid, ${\cal B}_2$ dashed, ${\cal B}_3$ dot-dashed, ${\cal B}_4$ dotted) as a function of $\lambda$. While the functions for small $\lambda$ differ considerably due to the basis-dependent position of their pole, there is a region of remarkably good agreement of all bases at $\lambda\in[0.7, 0.85]$.}
\label{ggNGFPinfpos}
\end{figure}

\paragraph{(G) The fixed point ${\bf NGFP'_{\boldsymbol{\infty}}}$.} 
Let us now discuss the additional fixed point that arises at $\hat\gamma=0$ and which we refer to as ${\bf NGFP'_{\boldsymbol{\infty}}}$. It is found in all four bases at 
\begin{equation}
 \hat{\gamma}^*=0,\qquad g^*=-\frac{1}{8 \pi}\frac{N(\lambda)}{P_{10}(\lambda)}\:.
\end{equation}
Also at this fixed point the stability matrix is diagonal such that the critical exponents can be associated with the coordinate directions. Explicitly we find for them
\begin{equation}
 \theta_g=2,\qquad \theta_{\hat\gamma}=2 \frac{P_9(\lambda)+P_{10}(\lambda)}{P_{10}(\lambda)}\:.
\end{equation}
Although this fixed point exists in all bases, we find (cf. Fig. \ref{ggNGFPinfpos}) that the predictions concerning its position and attractivity properties are severely basis-dependent for small $|\lambda|$. However there is an interval $\lambda\in [0.7,0.85]$ where the functions $g^*(\lambda)$ and $\theta_{\hat{\gamma}}(\lambda)$ coincide surprisingly well for all four bases. Also the limits at large negative $\lambda$ ($<-10$) are qualitatively similar at least for the bases ${\cal B}_1$, ${\cal B}_3$ and ${\cal B}_4$. Our findings in the previous subsection on the $(\lambda,g)$-truncation suggest already that it is precisely these regions of $\lambda$ where 3-dimensional ``lifts'' of the FPs in the $(\lambda,g)$-truncation will be found. Hence, we conclude that the $(\gamma,g)$-truncation in the limit $\gamma\rightarrow\infty$ is particularly reliable for those constant $\lambda$-values that correspond to the 3-dimensional fixed point values, not only because $\beta_\lambda$ vanishes there, but also due to the enhanced basis-independence.

\paragraph{(H) The interplay of the fixed points.}
A final important observation is that the zeros of $\theta_{\hat\gamma}$ of ${\bf NGFP'_{\boldsymbol{\infty}}}$ coincide exactly with those of $\theta_\gamma$ or $\theta_2$ for ${\bf NGFP'_{0}}$ and ${\bf NGFP'_{fin}}$, respectively, and the functions cross zero in such a way, that the fixed points are antagonistic for all bases and all $\lambda$ as long as both exist, \ie one is UV attractive while the other is repulsive along the second direction. Recall, however, that the zeros of $\theta_2$ are also the points at which ${\bf NGFP'_{fin}}$ ceases to exist, such that in each basis ${\bf NGFP'_{fin}}$ does not change its attractivity properties. 

In basis ${\cal B}_1$, however, this antagonism of the fixed points ${\bf NGFP'_{\boldsymbol{\infty}}}$ and ${\bf NGFP'_{0}}$ is taken to another level: Besides having the same fixed point value $g^*(\lambda)$ we find that the critical exponents satisfy $\theta_\gamma=-\theta_{\hat\gamma}$ for all $\lambda$. This is because for ${\cal B}_1$ the $\beta$-functions in the two coordinate charts are of the same form, except for a sign change in $\beta_\gamma$ compared to $\beta_{\hat\gamma}$. Thus under the duality map $\gamma\mapsto 1/\gamma$ only the $\gamma$ component of flow switches its sign and the fixed points are mapped onto each other. 

Note that for $\lambda=0.573$ we have $\theta_\gamma=0=\theta_{\hat\gamma}$. As this value of $\lambda$ corresponds to the zero of the sum $P_9(\lambda)+P_{10}(\lambda)$, the $\beta$-function $\beta_\gamma(\gamma,g)$ vanishes at this cosmological constant for all $\gamma$ and $g$. Hence, in this special case, the Immirzi parameter is not running and the straight line connecting ${\bf NGFP'_{\boldsymbol{\infty}}}$ and ${\bf NGFP'_{0}}$ at fixed $g^*$ becomes a UV attractive fixed line. The corresponding phase portrait will be discussed below.

\begin{figure}[p]
\begin{minipage}[t]{0.21\linewidth}
\vspace{1.3cm}
$\lambda=0.45$:\\[5.78cm]
$\lambda=0.8$:\\[5.78cm]
$\lambda=-10$:
\end{minipage}
\hfill
\begin{minipage}[t]{0.78\linewidth} 
\vspace{0pt}
\centering
{\small
\begin{psfrags}
 \input{Pictures/ggPhasePortrait045B3-psfrag.tex}
 \includegraphics[width=\linewidth]{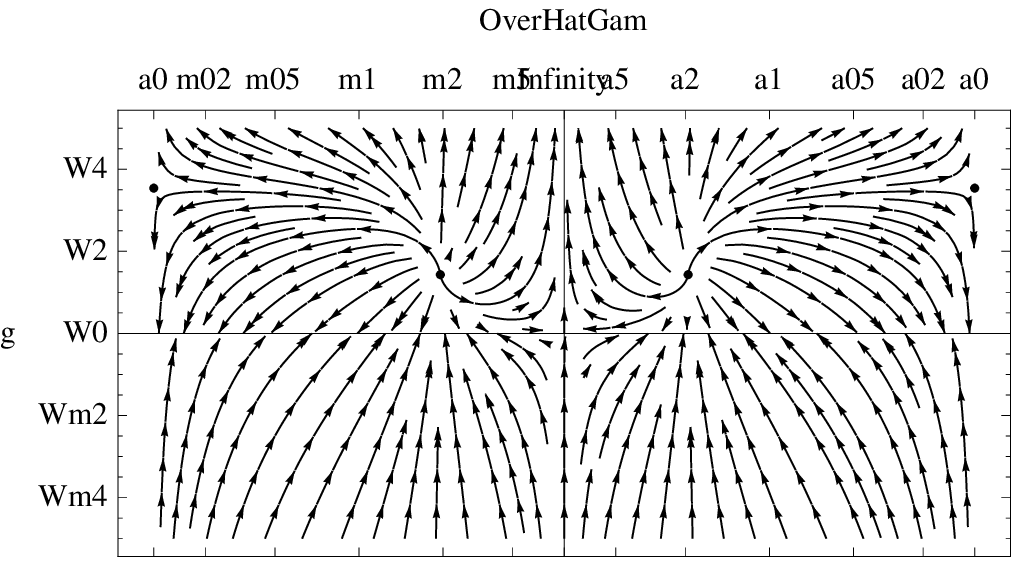}
\end{psfrags}\\[-1cm]
\begin{psfrags}
 \input{Pictures/ggPhasePortrait08B3-psfrag.tex}
 \includegraphics[width=\linewidth]{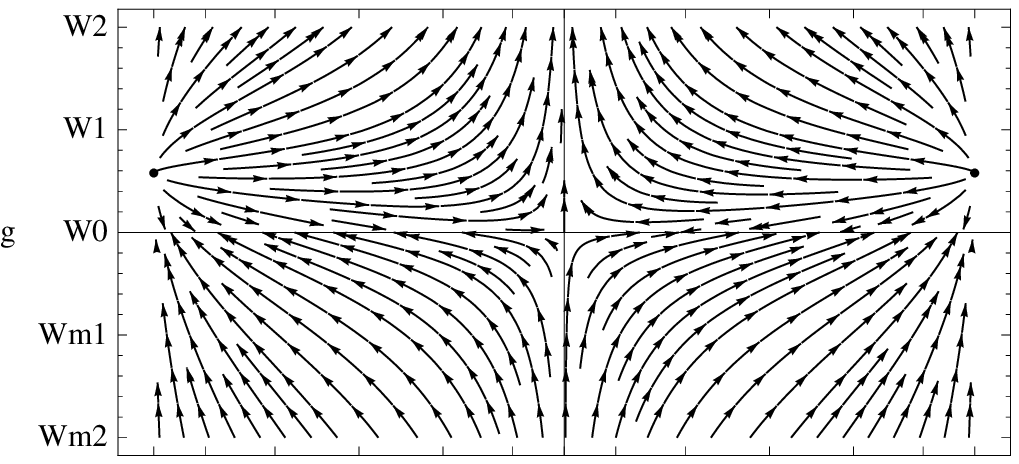}
\end{psfrags}\\[-1cm]
\begin{psfrags}
 \input{Pictures/ggPhasePortrait-10B3-psfrag.tex}
 \includegraphics[width=\linewidth]{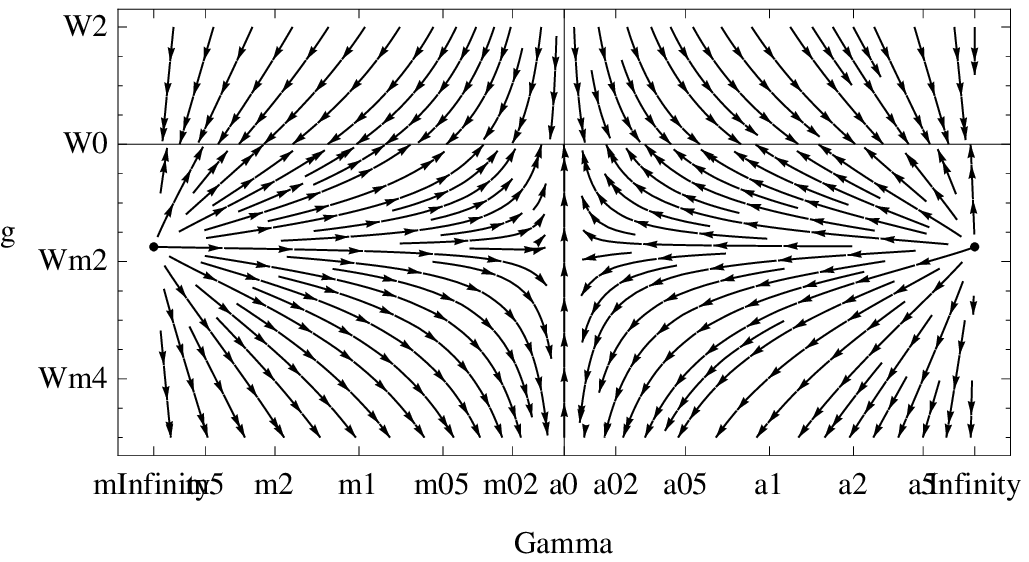}
\end{psfrags}\\[-2cm]
}
\end{minipage}
\caption{Phase portraits of the $(\gamma,g)$-flow in basis ${\cal B}_3$ for various fixed values of $\lambda$. At small $\lambda$ both ${\bf NGFP'_{\boldsymbol{\infty}}}$ and ${\bf NGFP'_{fin}}$ exist. For increasing $\lambda$ ${\bf NGFP'_{fin}}$ moves to larger $|\gamma|$ until it merges with ${\bf NGFP'_{\boldsymbol{\infty}}}$ and ceases to exist.  For large (negative) $\lambda$ the duality found in basis ${\cal B}_1$ is approximately established.} 
\label{ggPhasePortraitB3}
\end{figure}

\paragraph{(I) Phase portraits.}
In Fig. \ref{ggPhasePortraitB3} we have plotted the phase portraits of the $(\gamma,g)$-truncat\-ion in basis ${\cal B}_3$ for three different values of the cosmological constant. The corresponding phase portraits in basis ${\cal B}_4$ look qualitatively similar, while the results for ${\cal B}_2$ are considered less credible (see above). In Fig. \ref{ggPhasePortraitB3} we used an {\it arctangent} rescaling of the $\gamma$-axis in order to compactify it such that the fixed point at $\hat{\gamma}=0$ can be depicted in the same diagram. Clearly, this results in a highly non-linear scale of the $\gamma$-axis.

We observe that ${\bf NGFP'_{\boldsymbol{\infty}}}$ is present in all three panels of Fig. \ref{ggPhasePortraitB3}, and it always gives rise to a complete RG trajectory on the $\hat{\gamma}=0$ line with the fixed point as the UV limit and the $\gamma$-axis as its IR endpoint.

The pair of fixed points ${\bf NGFP'_{fin}}$ only exists in the first panel with $\lambda=0.45$ and it is UV attractive in both directions as discussed above, while ${\bf NGFP'_{\boldsymbol{\infty}}}$ only has one attractive direction in this case. We find a trajectory that connects both non-Gaussian fixed points and one linking ${\bf NGFP'_{fin}}$ with the origin. These trajectories act as separatrices: All trajectories below them are complete: In the UV they approach ${\bf NGFP'_{fin}}$ while in the IR they end on the fixed line $g=0$. Therefore we find an asymptotically safe trajectory for any IR value of the Immirzi parameter. This statement remains true for all phase portraits of the $(\gamma,g)$-truncation we analyzed. 
Note also that the flow in $\gamma$ direction changes sign on the line $\gamma=\gamma^*$ of ${\bf NGFP'_{fin}}$. For that reason each trajectory lies completely either in the region of larger or of smaller $\gamma$ compared to the fixed point value. In the region $|\gamma|<|\gamma^*|$ the Immirzi parameter runs to smaller (absolute) values in the IR, while for $|\gamma|>|\gamma^*|$ it runs to larger values. In the first panel we may assert the latter to be the predominant direction of the $\gamma$-flow as $|\gamma^*|<1$. (In the $g<0$-halfplane the running is just reversed.)

For larger values of $\lambda$ (as \eg $\lambda=0.8$ in the second panel of Fig. \ref{ggPhasePortraitB3}) this is no longer true: The pair of fixed points ${\bf NGFP'_{fin}}$ moves to larger values $|\gamma^*|$ until it merges with ${\bf NGFP'_{\boldsymbol{\infty}}}$ and ceases to exist for even larger values of $\lambda$. By moving the fixed points outwards the predominant direction of the $\gamma$ flow changes until in the whole upper halfplane the $\gamma$ flow points to smaller values in the IR. Then, as depicted in the second panel, all trajectories in the $g>0$-halfplane are asymptotically safe \wrt ${\bf NGFP'_{\boldsymbol{\infty}}}$.

To conclude the discussion of the phase portraits in basis ${\cal B}_3$ we want to point out an interesting mechanism that is present in all bases. As $P_8(\lambda)$ is, for any basis (and any gauge), a polynomial in $\lambda$ whose degree is smaller by 2 compared to the denominator $N(\lambda)$, the corresponding terms in the $\beta$-functions are suppressed quadratically for large $\lambda$, while the $P_9$ terms are suppressed only linearly and the $P_{10}$ terms become constant. Thus, at large $|\lambda|$ the duality of the $\beta$-functions known from basis ${\cal B}_1$ is established approximately in the whole $(\gamma,g)$-plane, except for a narrow strip around the origin of the order $|\gamma|\lesssim1/|\lambda|$. For this reason the third panel of Fig. \ref{ggPhasePortraitB3} is very similar to the phase portraits in basis ${\cal B}_1$, which we will discuss next. The only qualitative difference is that no fixed point on the $g$-axis arises, as the $\gamma$-divergence in the $\beta$-functions persists. 

\begin{figure}[p]
\begin{minipage}[t]{0.21\linewidth}
\vspace{1.3cm}
$\lambda=0$:\\[5.78cm]
$\lambda=0.57$:\\[5.78cm]
$\lambda=0.8$:
\end{minipage}
\hfill
\begin{minipage}[t]{0.78\linewidth} 
\vspace{0pt}
\centering
{\small
\begin{psfrags}
 \input{Pictures/ggPhasePortrait0B1-psfrag.tex}
 \includegraphics[width=\linewidth]{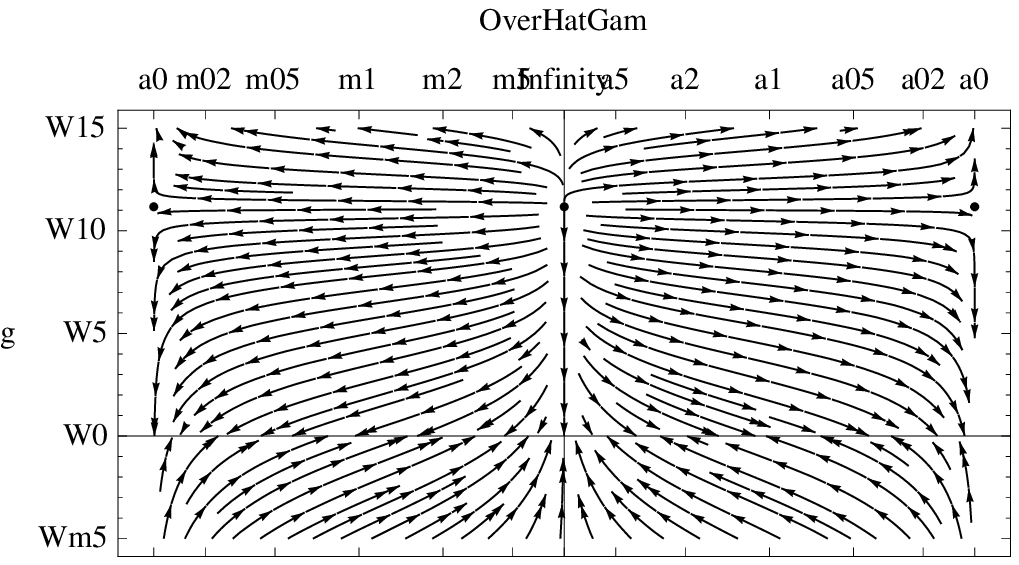}
\end{psfrags}\\[-1cm]
\begin{psfrags}
 \input{Pictures/ggPhasePortrait057B1-psfrag.tex}
 \includegraphics[width=\linewidth]{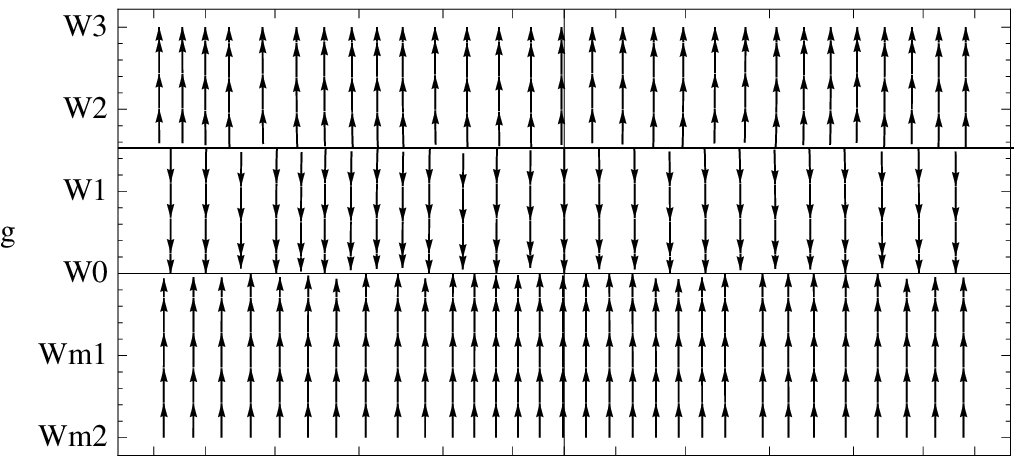}
\end{psfrags}\\[-1cm]
\begin{psfrags}
 \input{Pictures/ggPhasePortrait08B1-psfrag.tex}
 \includegraphics[width=\linewidth]{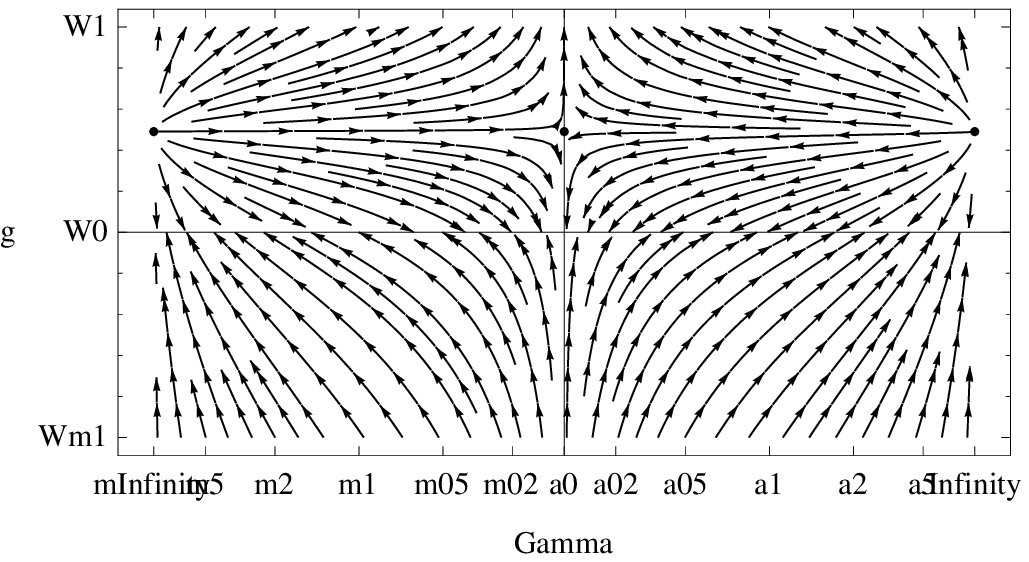}
\end{psfrags}\\[-2cm]
}
\end{minipage}
\caption{Phase portraits of the $(\gamma,g)$-flow in basis ${\cal B}_1$ for various fixed values of $\lambda$. We always find the fixed points ${\bf NGFP'_{\boldsymbol{\infty}}}$ and ${\bf NGFP'_{0}}$, while $\lambda$ determines the direction of the $\gamma$-flow. For $\lambda\approx 0.57$ it vanishes such that all trajectories run vertically.}
\label{ggPhasePortraitB1}
\end{figure}

Let us turn over to the RG flow resulting from basis ${\cal B}_1$, which is depicted in Fig. \ref{ggPhasePortraitB1} for three different values of $\lambda$. We find the two fixed points ${\bf NGFP'_{\boldsymbol{\infty}}}$ and ${\bf NGFP'_{0}}$ as deduced above, which govern the flow in the whole $g>0$-halfplane. Depending on the value of $\lambda$, one of the fixed points has two attractive directions while the other shows only one. This antagonistic behavior prescribes the predominant direction of the $\gamma$-flow. As in basis ${\cal B}_3$ we find for small $\lambda$ ($\lambda=0$, upper panel) that it is directed to larger values of the Immirzi parameter in the IR while for larger values ($\lambda=0.8$, lower panel) it reverses its direction. Different to the case of ${\cal B}_3$ this change does not happen due to a movement of one of the fixed points but only due to a change of their attractivity properties.  For that reason we find a specific value of the cosmological constant ($\lambda \approx 0.57$, central panel) for which the flow of the Immirzi parameter stops completely. In this special case all complete RG trajectories lie between the fixed lines $g=0$ (IR limit) and $g=g^*$ (UV limit). If the exact RG flow showed this behavior we could construct a quantum theory of gravity with a prescribed fixed value of the Immirzi parameter, that would not change under RG transformations.

For all other values of the cosmological constant we find an asymptotically safe trajectory for any prescribed IR value of $\gamma_{\rm IR}$ (with $g_{\rm IR}=0$) which in the UV either runs to $\gamma=0$ or $\hat{\gamma}=0$, \ie to the value of the fixed point with two UV attractive directions. In this case we only find $\gamma$ unrenormalized if we start with the fixed point values $\gamma_{\rm IR}=0$ or $\hat{\gamma}_{\rm IR}=0$ in the IR.

\paragraph{(J) Comparison to the PT flow.}
When we compare the results of this subsection with the corresponding results of the proper-time flow analysis of \cite{je:longpaper} we find indeed a considerable degree of (qualitative) similarity. Besides the fixed line at $g=0$, which is present in both studies, our most stable result is the existence of a NGFP at $\hat{\gamma}=0$, which shows one attractive and one repulsive direction at small $\lambda$, but is attractive in both directions if we prescribe the physical fixed point values $\lambda^*$ of the $(\lambda,g)$-truncation. This qualitative behavior was also found in \cite{je:longpaper}.

Besides that, the proper-time RG study discovered a second NGFP at $\gamma=0$, which shows an antagonistic behavior to the first NGFP that results in a predominant direction of the RG flow of the Immirzi parameter. While this direction coincides with our findings in both bases ${\cal B}_3$ and ${\cal B}_1$, the second fixed point is only found at $\gamma=0$ if we decide for basis ${\cal B}_1$. In this case, however, the analogies between the results of both studies can be extended even further, as we will explain next.

In \cite{je:longpaper} the $\beta$-functions were much more complicated than the present ones. Nevertheless to a good approximation they showed a very simple dependence on the Immirzi parameter $\gamma$ which gave rise to a conjecture concerning their form in an exact treatment, namely that {\it $\beta_g$ is independent of $\gamma$, and $\beta_\gamma$ is proportional to $\gamma$}. In basis ${\cal B}_1$ we find that our $\beta$-functions are exactly of this predicted form (cf. \eqref{ggsubsystem} for $P_8=0$)!

Moreover, the complete halt of the RG running of $\gamma$ is only possible if $\beta_{\gamma}$ shows such a simple $\gamma$ dependence. Although the Immirzi parameter staying unrenormalized under RG transformations appears as a very special case in the above discussion (only in basis ${\cal B}_1$ and only at $\lambda\approx0.57$), one can argue that this behavior should actually be expected for an RG study of the Holst truncation with {\it vanishing} cosmological constant, that properly treats parity-even and -odd contributions on the same footing: 

Since the Immirzi parameter is a relative coupling between the parity-even curvature term and the parity-odd Immirzi term, a vanishing of its flow means that both terms are renormalized exactly the same way and therefore share the Newton constant as their common coupling, while the Immirzi parameter only sets a fixed ($k$-independent) ratio between parity-even and -odd terms in $\Gamma_k$. 

We can only expect such a symmetry to exist between both sorts of terms if we employ a truncation ansatz which exhibits the same symmetry between its scalar and pseudo-scalar constituents and make sure that the method applied to calculate the RG flow respects this symmetry throughout. With this in mind it is clear that our method cannot maintain such a symmetry even if it is present in the original truncation. This is due to the gauge-fixing action (and resulting ghost action) we used, that only contains scalar and no pseudo-scalar terms.

It has been shown in \cite{je:longpaper} that the terms in $\beta_\gamma$ corresponding to those in the square bracket $[P_9(\lambda)+P_{10}(\lambda)]$ in \eqref{ggsubsystem} (for basis ${\cal B}_1$) are proportional to the difference of scalar and pseudo-scalar contributions to the running of $\gamma$. If the symmetry of the truncation was maintained, this bracket would therefore vanish for $\lambda=0$. By choosing $\lambda \neq 0$ we deliberately break the symmetry of the underlying truncation, as the cosmological constant term corresponds to an additional scalar constituent, that has no pseudo-scalar counterpart.

On the other hand it seems quite natural that we can use the value of the cosmological constant to control the amount of scalar contributions to the renormalization of the Immirzi parameter. From this point of view we find that it is possible to effectively restore the symmetry between parity-even and -odd terms in $\Gamma_k$, that was broken by the gauge-fixing procedure, by choosing $\lambda\approx 0.57$. Thus, {\it our result corroborates impressively the conjecture of an inherent symmetry between scalar and pseudo-scalar terms in the $(\gamma,g)$-truncation that was put forward in} \cite{je:longpaper} referring to the simple form of $\beta_\gamma$, as we are able to restore this symmetry for a specific value of $\lambda$ in basis ${\cal B}_1$.

\paragraph{(K) Conclusion.} In summary, the results of this subsection have shown that {\it the RG analysis of the $(\gamma, g)$-truncation using the new \WHlike\ flow equation and the proper-time RG study of the same truncation carried out in \cite{je:longpaper} reinforce each other}. Moreover, among the a priori equally suitable bases ${\cal B}_i$ in theory space, ${\cal B}_1$ is singled out as the one that, first, maximizes the physically significant similarities between both studies and, second, incorporates a symmetry property of the truncation, that leads to a $\beta$-function of the Immirzi parameter of the simple form 
\begin{equation}
\beta_\gamma=g\, \gamma\, f(\lambda),
\end{equation}
which is lost otherwise. For these reasons we shall consider basis ${\cal B}_1$ as being the most reliable one from now on; hence, we will concentrate on this basis in the discussion of the RG flow in the full 3-dimensional truncation covered in the next subsection.

\subsection[The complete $(\lambda,\gamma,g)$-system]{The complete $\boldsymbol{(\lambda,\gamma,g)}$-system}
In this subsection we analyze the RG flow in the complete 3-dimensional coupling constant space of the Holst truncation. As before we will first discuss its fixed point structure for all four bases ${\cal B}_i$ and study the dependence of their properties on the mass parameter $m$ that was held fixed to unity up to now. From the previous subsections we keep in mind that ${\cal B}_1$ amounts to the most physical basis in theory space. Therefore the subsequent analysis of the phase diagram will be restricted to this basis.

The system of flow equations under consideration in this subsection is given by \eqref{Beta_System}. Concerning its global properties we find that the $\gamma\mapsto -\gamma$ symmetry of the flow is preserved as $\beta_\lambda$ is an even function of $\gamma$ as well. Moreover, we observe that the 3-dimensional coupling space is divided by three planes, which no RG trajectory can cross: The $g\!=\!0$-plane, where $\beta_g=0$, the $\gamma\!=\!0$-plane (with $\beta_\gamma=0$) and the $\lambda\!=\!1$-plane, where all three $\beta$-functions diverge. The line $g=0=\lambda$ can, by inspection, be identified as a {\it fixed line} and it corresponds to the GFP known from the $(\lambda,g)$-truncation for all values of $\gamma$. As all trajectories in the half-space $\lambda>1$ are separated from the classical regime close to this Gaussian fixed line, we will not consider this part of the coupling space any further. 

\paragraph{(A) Fixed point structure.}
Since $\beta_{\hat\gamma}$ vanishes in the $\hat\gamma\!=\!0$-plane, setting $\hat\gamma=0$ amounts to a {\it consistent truncation}, \ie trajectories starting in this $(\lambda,g)$-plane will never leave it. Thus, the trajectories of the three dimensional flow in this plane, coincide with the flow of the $(\lambda,g)$-system in the $\gamma\rightarrow\infty$ limit (cf. section \ref{gl-system}). 

From this observation we can already conclude that in the $\hat\gamma\!=\!0$-plane we find two NGFPs, ${\bf NGFP^1_{\boldsymbol{\infty}}}$ and ${\bf NGFP^2_{\boldsymbol{\infty}}}$, that are UV attractive in both directions of the plane, exhibiting the critical exponents found in section \ref{gl-system}. Therefore, we only have to analyze their third critical exponent, that describes the attractivity property in $\hat\gamma$-direction. We conclude that the fixed points ${\bf NGFP'_{\boldsymbol{\infty}}}$ and $\{{\bf NGFP{}^1},{\bf NGFP{}^2}\}$ that we found in the 2-dimensional truncations are the projections of the 3-dimensional fixed points $\{{\bf NGFP{}^1_{\boldsymbol{\infty}}},{\bf NGFP{}^2_{\boldsymbol{\infty}}}\}$ for fixed $\lambda$ and $\gamma$, respectively.

\begin{table}[t]
\centering \renewcommand{\arraystretch}{1.15}
 \begin{tabular}{lccc}\toprule
   \hspace{1cm}   $(\gamma,g)$  & \hspace{0.4cm}${\bf NGFP{}'_0}$\hspace{0.4cm} & \hspace{0.4cm}${\bf NGFP'_{fin}}$\hspace{0.4cm} & \hspace{0.4cm}${\bf NGFP'_{\boldsymbol{\infty}}}$\hspace{0.4cm}\\
   $(\lambda,g)$  & {\footnotesize(${\cal B}_1$)}   & {\footnotesize(${\cal B}_2,\,{\cal B}_3,\,{\cal B}_4$)} & \\ \midrule\\[-0.4cm]
  ${\bf NGFP{}^1}$& ${\bf NGFP{}^1_0}$ ? & - & ${\bf NGFP{}^1_{\boldsymbol{\infty}}}$\\
 ${\bf NGFP{}^2}$ & ${\bf NGFP{}^2_0}$ ? & - & ${\bf NGFP{}^2_{\boldsymbol{\infty}}}$\\ \bottomrule
 \end{tabular}
\flushleft
\vspace{-3.2cm}
{\setlength{\unitlength}{1cm}
\begin{picture}(2,2)(-3.3,0)
 \put(0,2){\line(2,-3){0.7}}
\end{picture}}
\vspace{1cm}
\caption{Overview of the fixed points present in the different truncations: The first row contains the fixed points of the $(\gamma,g)$-truncation, while the first column displays the ones of the $(\lambda,g)$-truncation. The main body of the table contains the names given to the corresponding fixed points in the $(\lambda,\gamma,g)$-truncation, in case they exist.}\label{FixedPointsTable}
\end{table}

Similarly, setting $\gamma\!=\!0$ amounts to another 2-dimensional consistent truncation as $\beta_\gamma$ vanishes in this plane. However, since $\beta_\lambda$ diverges here, this limit eludes further investigation. Nonetheless we want to stress that the mechanism giving rise to $\{{\bf NGFP{}^1},{\bf NGFP{}^2}\}$ works for arbitrarily small $\gamma$, such that in basis ${\cal B}_1$ we should expect the existence of the three dimensional lifts $\{{\bf NGFP{}^1_0},{\bf NGFP{}^2_0}\}$ in the $\gamma\!=\!0$-plane if only the---probably unphysical---logarithmic divergence of $\beta_\lambda$ at $\gamma\rightarrow 0$ could be removed.

In contrast to this situation, the fixed point solutions ${\bf NGFP}'_{\rm fin}$, that were found in bases ${\cal B}_{\{2,3,4\}}$, are lost when the third coupling $\lambda$ is subject to renormalization as well. Thus, in the 3-dimensional truncation for these bases no fixed point at finite $\gamma^\ast$ is found.

We have summarized the above discussion on the different fixed points and their presence in the different truncations in Table \ref{FixedPointsTable}. 

In the next two paragraphs {\bf (i)} and {\bf (ii)}, respectively, we will analyze the fixed points ${\bf NGFP^1_{\boldsymbol{\infty}}}$ and ${\bf NGFP^2_{\boldsymbol{\infty}}}$ and will thereby focus on their third critical exponent, that describes the attractivity property in $\hat\gamma$-direction. In addition we will investigate the dependence of all fixed point properties on the mass parameter $m$, that has been discussed only qualitatively in section \ref{gl-system} so far. In a third paragraph {\bf (iii)}, we will explore an additional mechanism present only in the full three dimensional truncation that in basis ${\cal B}_1$ gives rise to a pair of fixed points at finite $|\gamma^*|$. We will denote this fixed point by ${\bf NGFP_{fin}}$. In the other bases no equivalent to this fixed point is observed.

\begin{figure}[t]
\centering
{\small
\begin{psfrags}
 \input{Pictures/3DSketch-psfrag.tex}
 \includegraphics[width=0.7\linewidth]{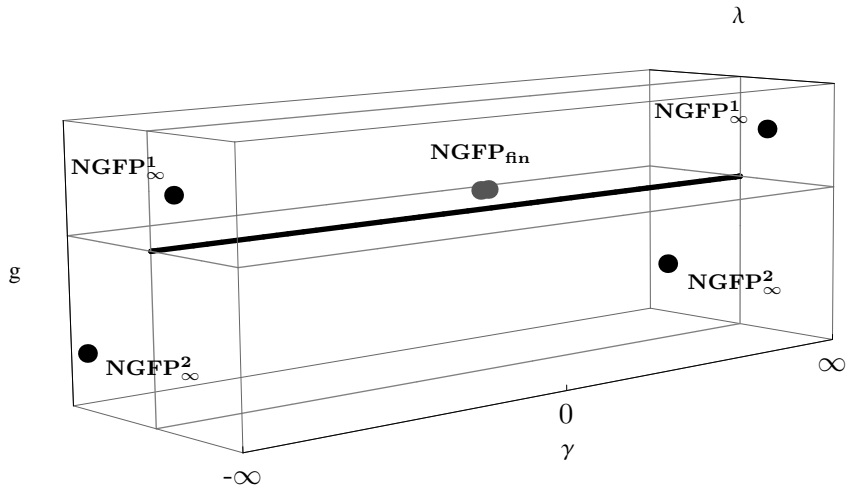}
\end{psfrags}
}
\caption{Sketch of the fixed point structure of the $(\lambda,\gamma,g)$-system. The pair of fixed points ${\bf NGFP_{fin}}$ is shaded in gray as it is only present in basis ${\cal B}_1$, and absent in the other bases.}
\label{3DSketch}
\end{figure}

Before we start to discuss the details of the individual fixed points let us again schematically depict the situation in the 3-dimensional theory space in Fig. \ref{3DSketch}.

\begin{figure}[t]
\centering
{\small
\begin{psfrags}
 \input{Pictures/FP1infPos-psfrag.tex}
 \includegraphics[width=0.65\linewidth]{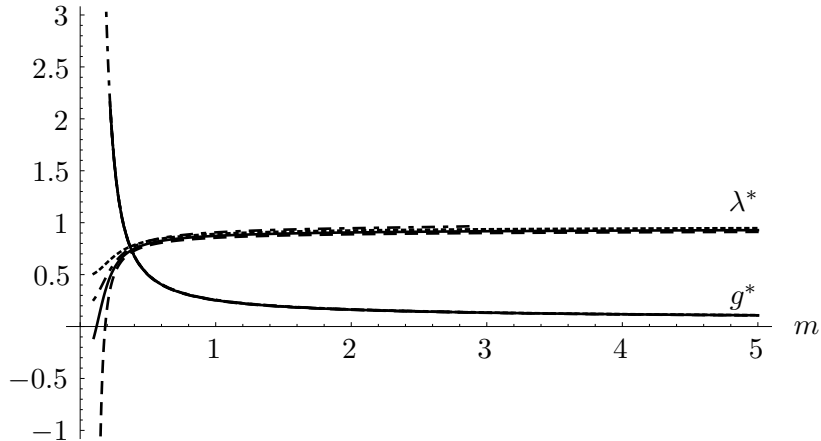}
\end{psfrags}
}
\caption{The fixed point coordinates of ${\bf NGFP^1_{\boldsymbol{\infty}}}$ as functions of $m$. For not to small values of $m$ the curves are almost constant functions and match each other perfectly for the four bases (${\cal B}_1$ solid, ${\cal B}_2$ dashed, ${\cal B}_3$ dot-dashed, ${\cal B}_4$ dotted).}
\label{FP1infPos}
\end{figure}

\begin{figure}[t]
\centering
{\small
\begin{psfrags}
 \input{Pictures/FP1infCE-psfrag.tex}
 \includegraphics[width=0.45\linewidth]{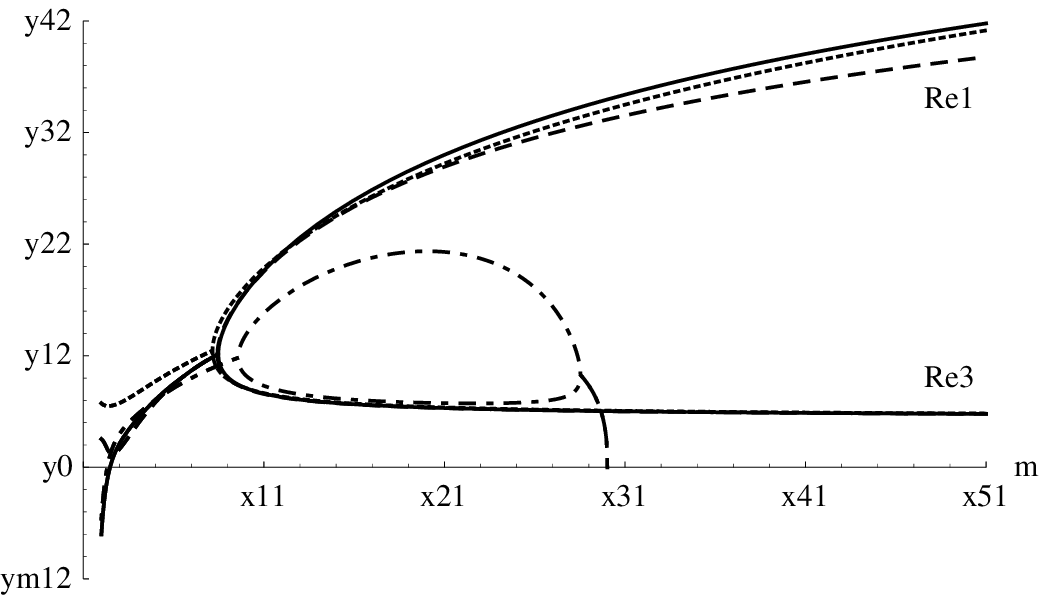}
\end{psfrags}\quad
\begin{psfrags}
 \input{Pictures/FP12infCE-psfrag.tex}
 \includegraphics[width=0.45\linewidth]{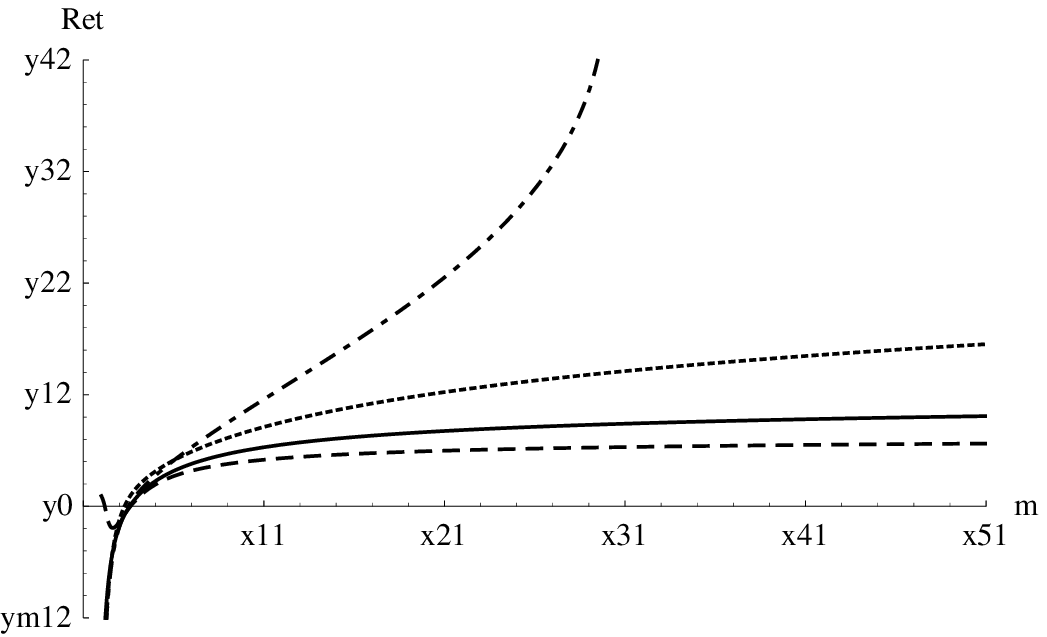}
\end{psfrags}
}
\caption{The critical exponents of ${\bf NGFP^1_{\boldsymbol{\infty}}}$ as functions of $m$. Here the results for the bases ${\cal B}_1$ (solid), ${\cal B}_2$ (dashed) and ${\cal B}_4$ (dotted) are well aligned. While for these bases the FP exists up to large $m$, in basis ${\cal B}_3$ (dot-dashed) it vanishes at $m\approx3$.}
\label{FP1infCE}
\end{figure}

\noindent{\bf (i) The fixed point ${\bf NGFP^1_{\boldsymbol{\infty}}}$.}
The first NGFP in the $\hat\gamma\!=\!0$-plane lies in the $(\lambda>0,g>0)$-quadrant. Its position as a function of the mass parameter $m$ is depicted in Fig. \ref{FP1infPos}. We find the results for the 4 different bases perfectly aligned. Moreover, the position does not depend very much on the value of $m$, at least for $m\gtrsim1$. At small $m$ we find a rapid variation of the FP position with $m$, which seems, in the light of the almost flat curves for larger $m$, unphysical. Thus this is a first indication that our truncation should be trusted in only for $m\gtrsim0.5$.

In Fig. \ref{FP1infCE} we have plotted the corresponding critical exponents. In the left panel we find depicted the real parts of $\theta_1$ and $\theta_3$ that correspond to the eigendirection lying inside the $\hat\gamma\!=\!0$-plane. They start off as a complex pair and turn real at $m\approx 0.8$. While one of the critical exponents becomes unreasonably large the other one stays at moderate values of about 5-8. 

In the right panel of Fig. \ref{FP1infCE} the critical exponent corresponding to the $\hat\gamma$-direction is plotted and we observe that it is increasing with $m$.

For all bases we find that, in the trusted region of $m\gtrsim0.5$, the fixed point is attractive in all three directions. Besides this qualitative similarity found in all bases, the attractivity properties in the bases ${\cal B}_1$, ${\cal B}_2$ and ${\cal B}_4$ are very similar as functions of $m$, even quantitatively. In basis ${\cal B}_3$, on the other hand, the fixed point vanishes at $m\approx 3$ and the critical exponents $\theta_{\hat\gamma}$ gets considerably larger than in the other bases.

\begin{figure}[tb]
\centering
{\small
\begin{psfrags}
 \input{Pictures/FP2infPos-psfrag.tex}
 \includegraphics[width=0.65\linewidth]{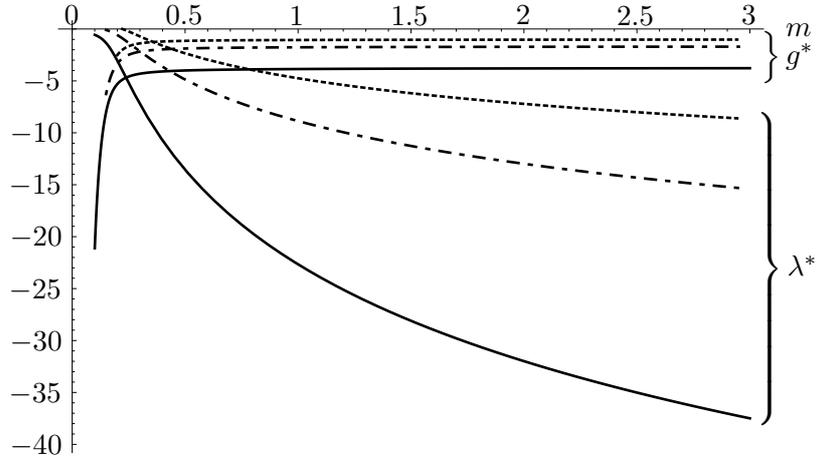}
\end{psfrags}
}
\caption{The fixed point coordinates of ${\bf NGFP^2_{\boldsymbol{\infty}}}$ that exists in the three bases ${\cal B}_1$ (solid), ${\cal B}_3$ (dot-dashed) and ${\cal B}_4$ (dotted) as functions of $m$. While the $\lambda^*$-coordinate varies with $m$ and the choice of basis, for the $g^*$-coordinate this dependence is far less pronounced.}
\label{FP2infPos}
\end{figure}

\begin{figure}[tb]
\centering
{\small
\begin{psfrags}
 \input{Pictures/FP2infCE-psfrag.tex}
 \includegraphics[width=0.45\linewidth]{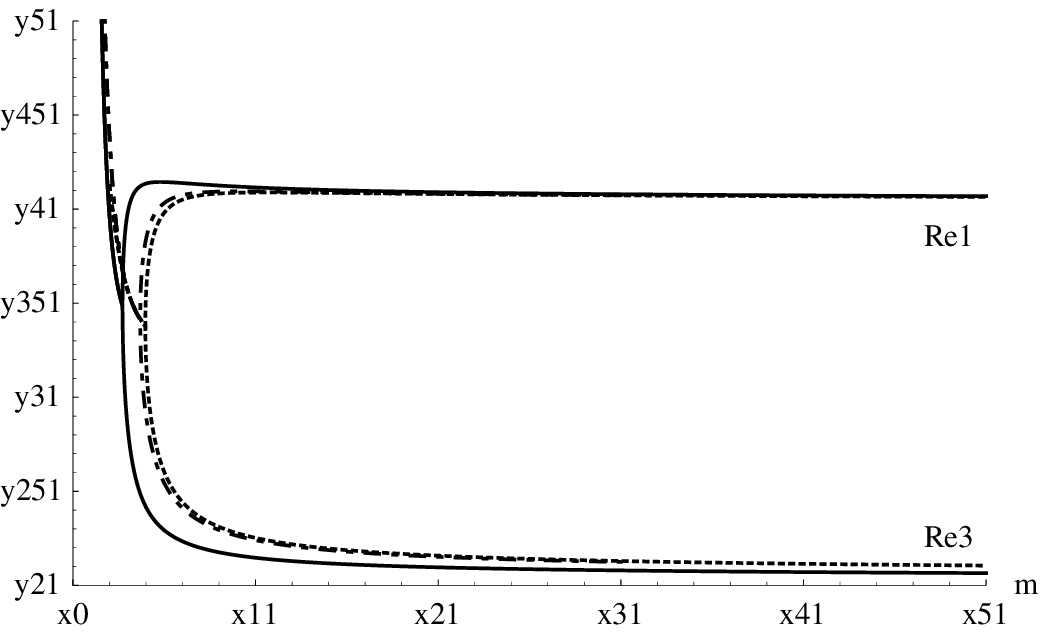}
\end{psfrags}\quad
\begin{psfrags}
 \input{Pictures/FP22infCE-psfrag.tex}
 \includegraphics[width=0.45\linewidth]{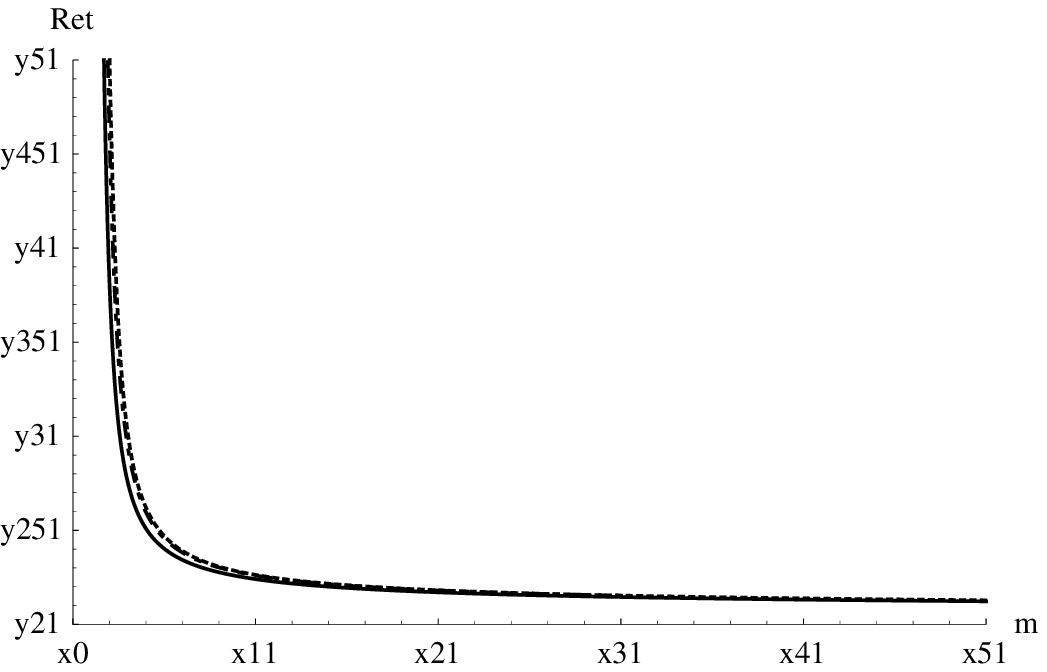}
\end{psfrags}
}
\caption{The critical exponents of ${\bf NGFP^2_{\boldsymbol{\infty}}}$ as functions of $m$. Here the results for the bases ${\cal B}_1$ (solid), ${\cal B}_2$ (dashed) and ${\cal B}_4$ (dotted) are well aligned and are almost constant for $m\geq 0.8$ and stay in a reasonable range.}
\label{FP2infCE}
\end{figure}

\noindent{\bf (ii) The fixed point ${\bf NGFP^2_{\boldsymbol{\infty}}}$.}
In Fig. \ref{FP2infPos} the position of the second fixed point, situated in the $(\lambda<0,g<0)$-quadrant, is shown. Note that this FP does not exist in basis ${\cal B}_2$ such that we are left with three different types of lines in Figs. \ref{FP2infPos} and \ref{FP2infCE}. We find that $\lambda^*$ is decreasing to very large negative values, for increasing $m$, while $g^*$ approaches small absolute values. At small $m$ the $g^*$ coordinate begins to diverge, which we should interpret as the boundary of the trusted region in $m$ which here occurs at $m\approx0.3$.

The corresponding critical exponents are plotted in Fig. \ref{FP2infCE}. We find a very stable prediction of three real and positive critical exponents whose values do not depend on the choice of basis and that show only a slight variation with $m$, for $m\gtrsim 0.7$. Moreover, in contrast to ${\bf NGFP^1_{\boldsymbol{\infty}}}$ all critical exponents take on reasonably small absolute values.

\noindent{\bf (iii) The fixed point ${\bf NGFP_{fin}}$.} 
In basis ${\cal B}_1$, besides the above fixed points in the $\hat\gamma\!=\!0$-plane, an additional NGFP at finite $\gamma$ exists. Its $\lambda^*$ coordinate is given by the condition $\beta_\gamma=0$ and thus corresponds to the zeros of $P_9(\lambda)+P_{10}(\lambda)$ at $\lambda\approx 0.573$ and $\lambda\approx2.918$. As the latter lies behind the $\lambda=1$ divergence of the $\beta$-functions, we will only consider the former. Its corresponding $g^*$-coordinate is found from $\beta_g=0$ to be $g^*=1.525$.

These two coordinates are substituted into $\beta_\lambda(\lambda^*,\gamma,g^*)=0$, and each solution of this equation for $\gamma$ amounts to a fixed point $(\lambda^\ast, \gamma^\ast, g^\ast)$ of the 3-dimensional flow. This last fixed point condition has the structure:
\begin{equation}\label{lastFPcond}
 \beta_\lambda=0= C_1(g^*,\lambda^*,m)+C_2(g^*)\ln\left(\frac{\gamma^2-1}{\gamma^2}\right)^2,
\end{equation}
where $C_1$ and $C_2<0$ are constants that only depend on the other (already fixed) FP coordinates and the mass parameter $m$. From this form we can infer that there is always at least one solution to the condition \eqref{lastFPcond} in the interval $\gamma^2\in (0,1)$, as the logarithm is a smooth function interpolating between $\pm\infty$ for the limits $\gamma\rightarrow 0$ and $\gamma\rightarrow 1$, respectively. We refer to the fixed point corresponding to this solution as ${\bf NGFP_{fin}}$. Whether or not a second solution with $\gamma^2>1$ exists, depends on the value of $m$: For $\gamma>1$ the RHS is a monotonically decreasing function and in the limit $\gamma\rightarrow\infty$ the logarithmic term vanishes. Thus, a second fixed point solution with $|\tilde{\gamma}^*|>1$ exists for those values of $m$ for which $C_1(g^*,\lambda^*,m)<0$.

Note that for the other bases this mechanism is not at work as the dependence of $\beta_g$ and $\beta_\gamma$ on $\gamma$ is different. In this case we can only infer functions $\gamma^*(\lambda)$ and $g^*(\lambda)$ from $\beta_g$ and $\beta_\gamma$. However, the last condition $\beta_\lambda(\lambda,\gamma^*(\lambda),g^*(\lambda))=0$ does not have a solution, such that we do not find a fixed point at finite $\gamma$ in the other bases.

\begin{figure}[t]
\centering
{\small
\begin{psfrags}
 \input{Pictures/FP0Pos-psfrag.tex}
 \includegraphics[width=0.45\linewidth]{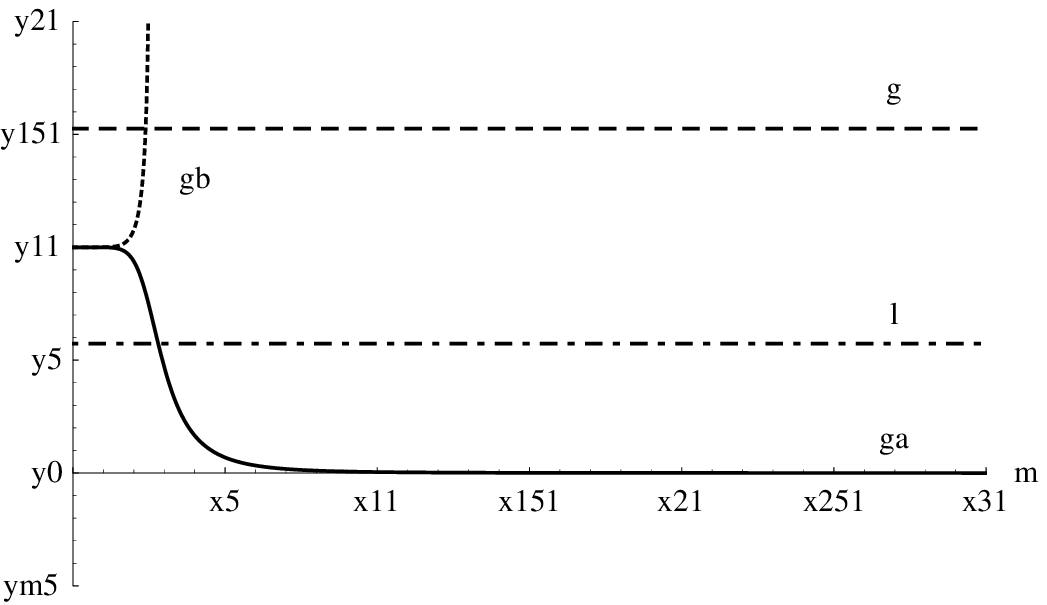}
\end{psfrags}\quad
\begin{psfrags}
 \input{Pictures/FP0CE-psfrag.tex}
 \includegraphics[width=0.45\linewidth]{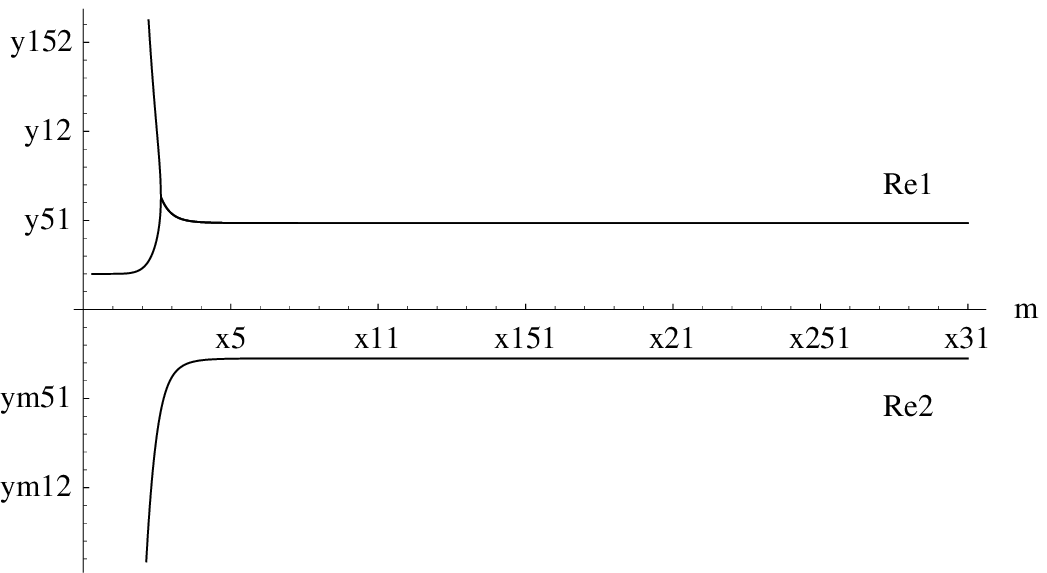}
\end{psfrags}
}
\caption{Position and critical exponents of ${\bf NGFP_{fin}}$ as a function of $m$.
For $m>0.8$ the fixed points have moved to $\gamma^*\approx0$ and the functions are constant from there on. At small $m<0.35$ a second pair of fixed points (with coordinate $\tilde\gamma^*$) exists, which is not further analyzed.}
\label{FP0}
\end{figure}

The exact values of the coordinate $\gamma^*$ and the critical exponents can only be obtained numerically. The resulting functions of $m$ are depicted in Fig. \ref{FP0}. In the left panel we find the fixed values $g^*$ and $\lambda^*$ represented by constant functions, as well as the function $\gamma^*(m)$, corresponding to the solution that exists for all $m$. It starts off at $\gamma^*(0)=\pm1$ and approaches 0 rapidly for $m\gtrsim0.8$. The second solution $\tilde{\gamma}^*$ exists only for $m\lesssim0.35$; as this region of $m$ has been proven to obtain questionable results before, we will not consider this second pair of fixed point solutions any further.

In the right panel of Fig. \ref{FP0} the real parts of the critical exponents corresponding to the fixed point ${\bf NGFP_{fin}}$ are shown. We observe that there is one UV repulsive eigendirection for all $m$.  The other two critical exponents, that are real and positive for small $m$, become a complex conjugated pair at $m\approx 0.4$ with positive real part. Once the fixed point coordinate $\gamma^*$ approaches 0 with increasing $m$, the values of the critical exponents do not change any more.

\begin{figure}[p]
\centering
{\small
%\begin{psfrags}
 %\input{Pictures/Flow3Dposg1-psfrag.tex}
 %\subfigure[]{\includegraphics[width=.83\linewidth]{Pictures/Flow3Dposg1-psfrag.eps}}
%\end{psfrags}\\
\subfigure[]{\includegraphics[width=.81\linewidth]{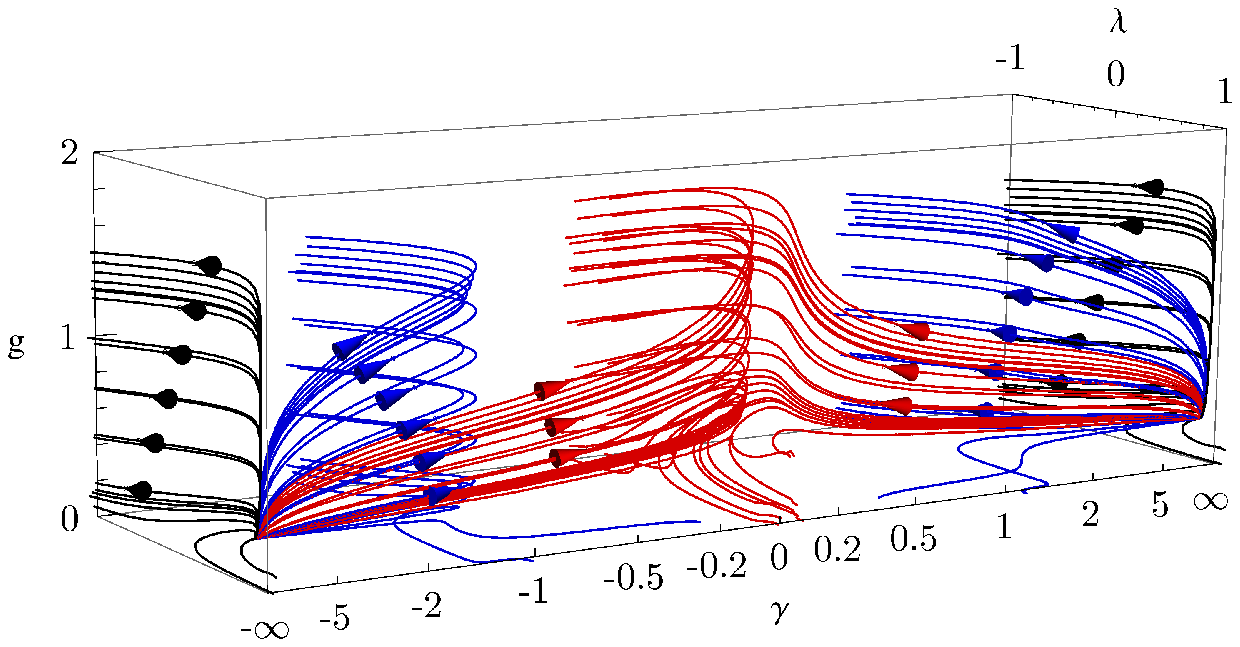}}\\
%\begin{psfrags}
% \input{Pictures/Flow3Dposg2-psfrag.tex}
% \subfigure[]{\includegraphics[width=.83\linewidth]{Pictures/Flow3Dposg2-psfrag.eps}}
%\end{psfrags}\\
\subfigure[]{\includegraphics[width=.81\linewidth]{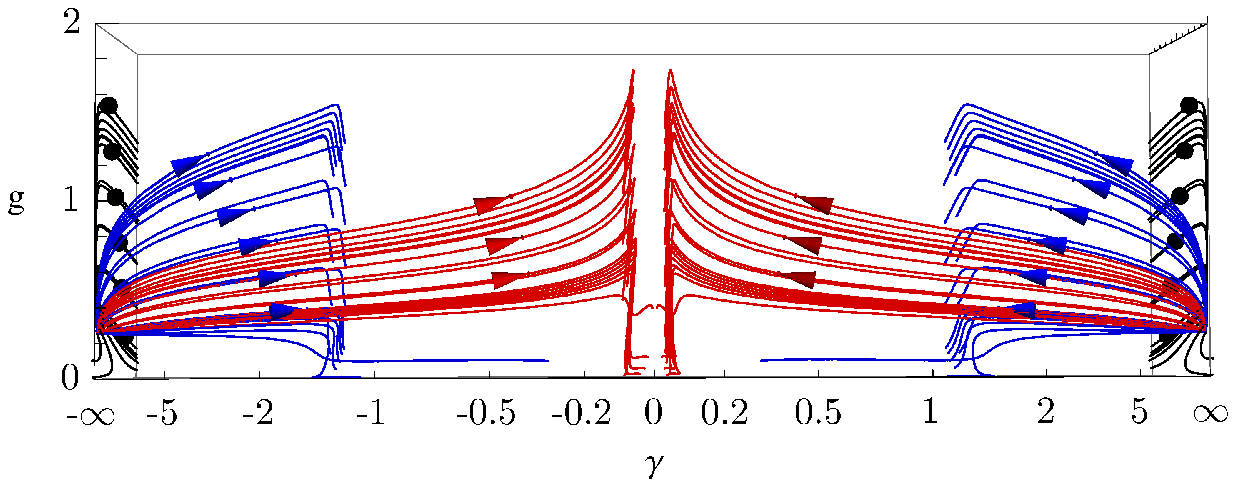}}\\
%\begin{psfrags}
% \input{Pictures/Flow3Dposg3-psfrag.tex}
% \subfigure[]{\includegraphics[width=.83\linewidth]{Pictures/Flow3Dposg3-psfrag.eps}}
%\end{psfrags}
\subfigure[]{\includegraphics[width=.81\linewidth]{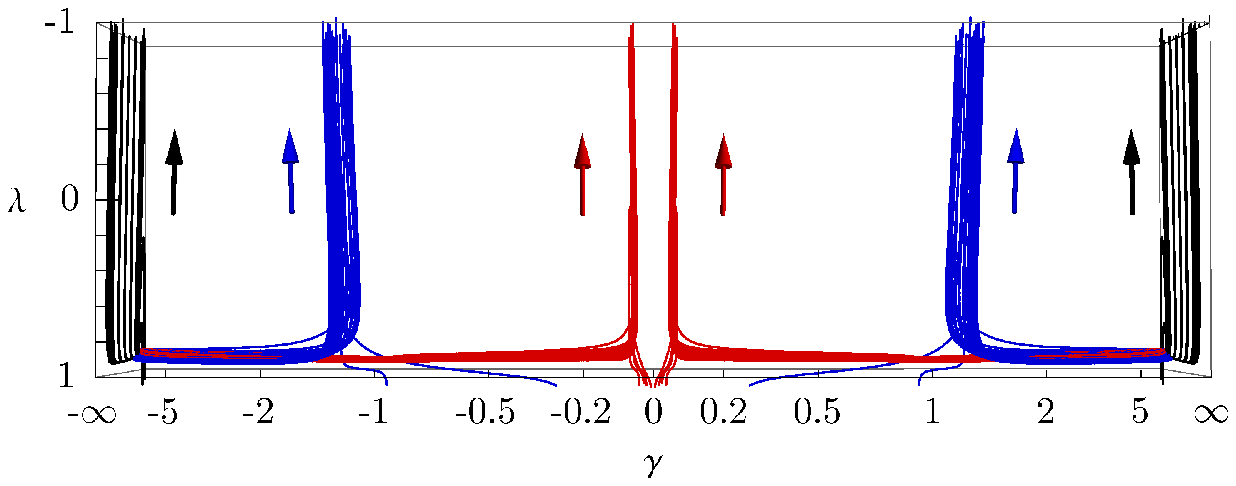}}
}
\caption{RG flow in the part of the 3-dimensional coupling space with $g>0$. All trajectories of the same color pass through points with the same $|\gamma|$: $|\gamma|=0.06$ (red), $|\gamma|=1.38$ (blue) and $|\hat\gamma|=0$ (black). The flow is directed such that all trajectories share the fixed point ${\bf NGFP^1_{\boldsymbol{\infty}}}$ as their UV limit.}
\label{Flow3dposg}
\end{figure}

\begin{figure}[p]
\centering
{\small
%\begin{psfrags}
% \input{Pictures/Flow3Dnegg1-psfrag.tex}
% \subfigure[]{\includegraphics[width=.83\linewidth]{Pictures/Flow3Dnegg1-psfrag.eps}}
%\end{psfrags}\\
\subfigure[]{\includegraphics[width=.81\linewidth]{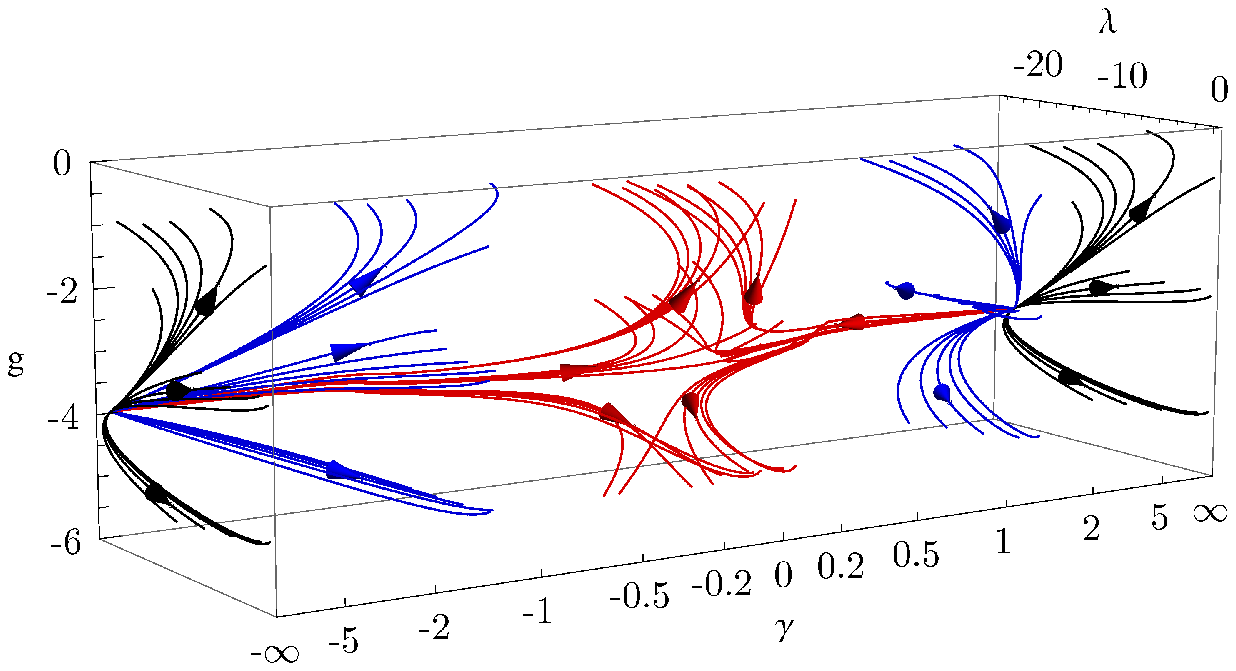}}\\
%\begin{psfrags}
% \input{Pictures/Flow3Dnegg2-psfrag.tex}
% \subfigure[]{\includegraphics[width=.83\linewidth]{Pictures/Flow3Dnegg2-psfrag.eps}}
%\end{psfrags}\\
\subfigure[]{\includegraphics[width=.81\linewidth]{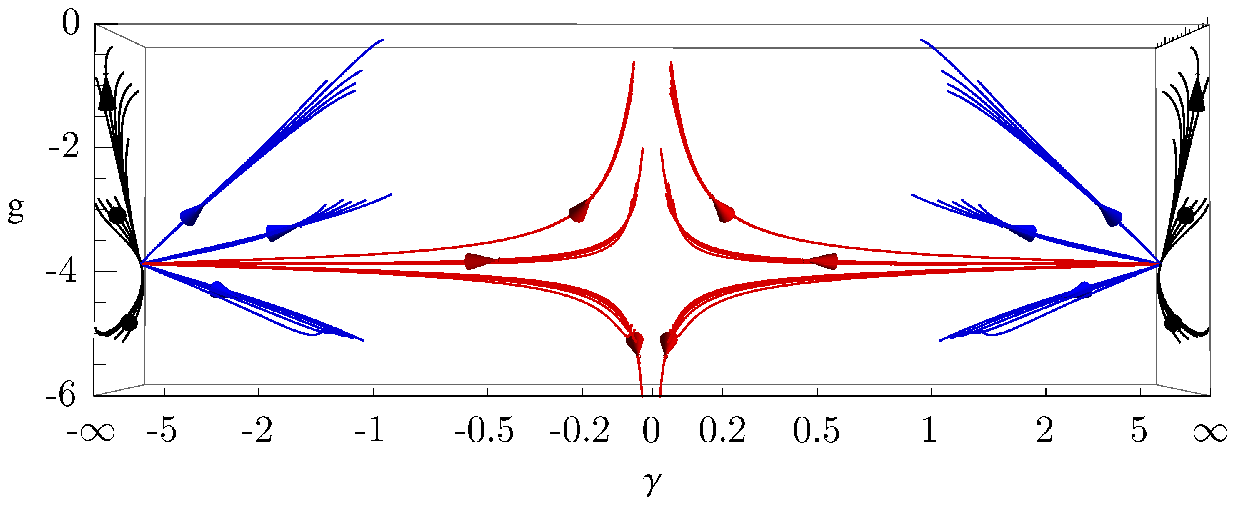}}\\
%\begin{psfrags}
% \input{Pictures/Flow3Dnegg3-psfrag.tex}
% \subfigure[]{\includegraphics[width=.83\linewidth]{Pictures/Flow3Dnegg3-psfrag.eps}}
%\end{psfrags}
\subfigure[]{\includegraphics[width=.81\linewidth]{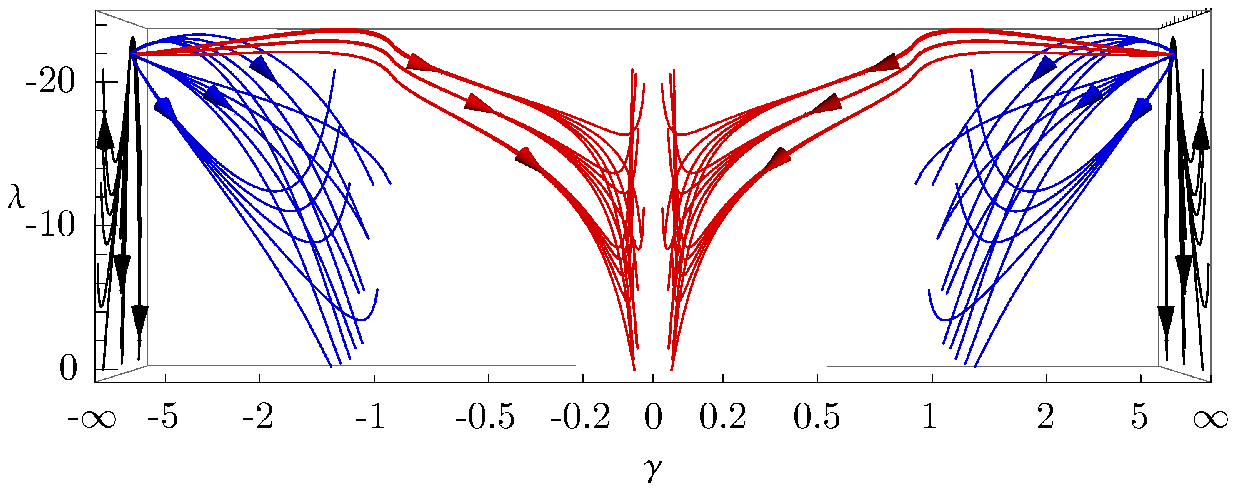}}
}
\caption{RG flow in the part of the 3-dimensional coupling space with $g<0$. All trajectories of the same color pass through points with the same fixed $|\gamma|$: $|\gamma|=0.06$ (red), $|\gamma|=1.38$ (blue) and $|\hat\gamma|=0$ (black). The flow is directed such that all trajectories share the fixed point ${\bf NGFP^2_{\boldsymbol{\infty}}}$ as their UV limit.}
\label{Flow3dnegg}
\end{figure}

\paragraph{(B) The 3D phase portrait.}
The resulting phase portrait of the full 3-dimensional RG flow is depicted in Figs. \ref{Flow3dposg} and \ref{Flow3dnegg}. As it is difficult to visualize a 3-dimensional vector field in every point of space, we have decided to display sets of trajectories, whose starting points lie in planes of fixed $|\gamma|$. The red trajectories start close to the $\gamma\!=\!0$-plane at $\gamma=\pm 0.06$, the blue ones at $\gamma=\pm1.38$ and the black ones lie entirely in the $\hat\gamma\!=\!0$-plane. In addition to the three-dimensional view in (a) we have given a frontal view onto the $(\gamma, g)$-plane in (b) and the $(\lambda,\gamma)$-plane in (c) for a better 3D visualization of the trajectories.

In Fig. \ref{Flow3dposg} we have plotted a region of theory space with $g>0$. The fixed point ${\bf NGFP^1_{\boldsymbol{\infty}}}$ is clearly visible in (a) and (b), lying on the lateral faces of the box that correspond to the $|\gamma|\rightarrow \infty$ limit. All trajectories shown have this fixed point as their UV limit. The second fixed point ${\bf NGFP_{fin}}$ is not visible, as its UV critical hypersurface is only two dimensional.

Note that the flow in the lateral faces, due to the self-consistency of the $\hat\gamma\!=\!0$-truncation, corresponds exactly to the flow depicted in the lower right panel of Fig. \ref{glPlanes1}.

From Fig. \ref{Flow3dposg} (c) we observe that the flow first stays mostly in the $\gamma$-plane of its starting points. Only when $\lambda$ exceeds its fixed point value, the running of $\gamma$ becomes predominant. For that reason (b) looks very similar to the last panel of Fig. \ref{ggPhasePortraitB1} as most of the visible part of the trajectories in this projection lies in the small region $\lambda^*<\lambda<1$. This behavior of the flow can be related to the critical exponents of the fixed point ${\bf NGFP^1_{\boldsymbol{\infty}}}$: The huge value of $\theta_1\approx20$ results in a predominantly fast running of $\lambda$. Only when it has become close to its fixed point value, the other couplings, which are related to critical exponents of much smaller magnitude, are seen to run as well.

Fig. \ref{Flow3dnegg} shows a similar set of plots, only for the $g<0$ part of coupling space. Again we find that the fixed point at $\hat\gamma=0$, ${\bf NGFP^2_{\boldsymbol{\infty}}}$, is dominating the flow. This time the plane $\hat\gamma=0$ (black trajectories) corresponds to the flow depicted in the lower right panel of Fig. \ref{glPlanes2}, but we observe that the other trajectories (blue, red), in contrast to the above case, do not stay in their starting plane of fixed $\gamma$, as the fixed point  ${\bf NGFP^2_{\boldsymbol{\infty}}}$ has three positive critical exponents of the same order of magnitude.

\paragraph{(C) Implication for Asymptotic Safety.} We conclude that in the full Holst truncation both FPs, ${\bf NGFP^1_{\boldsymbol{\infty}}}$ and ${\bf NGFP^2_{\boldsymbol{\infty}}}$, allow for the construction of an asymptotically safe quantum theory and show a basin of attraction that spans the whole part of coupling space with $g>0$ and $g<0$, respectively, that we have analyzed. 

\paragraph{(D) Logarithmic divergences.} Let us finally comment on the influence of the logarithmic divergences in $\beta_\lambda$ on the gross properties of the flow: We have seen from the above examples that the divergence of $\beta_\lambda$ at $\gamma=1$ is crossed by the trajectories smoothly. Indeed this divergence only has a very local effect on the trajectories, that pass the $\gamma\!=\!1$ plane being tangential to it, similarly to the function $x\ln x^2$ crossing $x=0$. Fig. \ref{Flow3dnegg} (c) gives an idea of this behavior by the sharp bend in the red trajectories at $\gamma=1$, while in Fig. \ref{Flow3dposg} (c) the effect is too localized to be visible at the given scale.

Similarly, the fixed point ${\bf NGFP'_0}$ from the $(\gamma, g)$-truncation vanishes due to a logarithmic divergence of $\beta_\lambda$. Again this effect is very local, such that \eg Fig. \ref{Flow3dposg} (b) would not look much different, if an additional fixed point at $\gamma=0$ was present.

If we take into account that the logarithmic contributions to $\beta_\lambda$ correspond to prefactors of a quartic momentum divergence of the path integral, that are most sensitive to all details of the renormalization procedure, we should not take their exact form too seriously. Therefore we should not exclude the existence of one (or more) additional fixed point(s) at $\gamma=0$ with two UV attractive directions in the three dimensional coupling space on the basis of the above results.

\paragraph{(E) Summary and comparison to the PT flow.}
In summary we find that also {\it the 3-dimensional flow generated by the new WH-like equation shows characteristic similarities to the flow obtained in the PT version of the exact FRGE \cite{je:longpaper}}. Besides the existence of a NGFP in the $\hat\gamma\!=\!0$-plane that is UV attractive in all three directions we also find the predominant direction of the $\gamma$-flow towards larger absolute values in the UV that was reported in \cite{je:longpaper}. 

The most prominent differences to the PT study are the exact position of the fixed points in the $\hat\gamma\!=\!0$-plane and the existence of a fixed point at $\gamma=0$. Both features heavily depend on the exact form of the contributions from quartic momentum divergences, that are notoriously unstable under different renormalization procedures. 

Thus, {\it the respective results obtained with the two structurally different flow equations support each other.} In particular this is true for the form of the $\beta$-functions $\beta_g$ and $\beta_\gamma$ as well as most properties of the fixed points in the $\hat\gamma\!=\!0$-plane. Another common feature of both calculations is the {\it absence of a reliable fixed point at finite $\gamma$}. Thus all asymptotically safe trajectories either take on the value $\hat\gamma=0$, corresponding to freely fluctuating torsion, or possibly also $\gamma=0$, where certain torsion components are suppressed completely, in the deep UV. Note that {\it none of these limits can be considered as being equivalent to metric gravity}, since some torsion components fluctuate freely in both cases.

\vspace{0.5cm}
\section{Conclusions}

In this paper we proposed and motivated a new special-purpose functional flow equation which can help us in taming the tremendous algebraic complexity of realistic RG computations in quantum gravity, and which should allow for investigations that are prohibitively difficult by standard methods. The equation descends from the exact FRGE of the EAA by a sequence of specializations and approximations. One of our motivations for the present RG analysis on ${\cal T}_{\rm EC}$ was to assess the viability of the approximations that this ``\WHlike'' flow equation relies on in direct comparison to the proper-time flow study carried out in \cite{je:longpaper}. 

After having analyzed the resulting RG flow in some detail we are now in a position to answer this question to a large part affirmatively. We obtained $\beta$-functions for $g$ and $\gamma$ that are of comparable form to those found in \cite{je:longpaper}, in the sense that they share a similar and simple dependence on the couplings $g$ and $\gamma$, while as a function of $\lambda$ the $\beta$-functions are most complicated.

We saw that the consistency of our truncation based on the Holst action can be optimized by choosing the preferred values $(\aD,\aL,\bD)=(0,0,0)$ of the gauge parameters, which in turn leads to a simplified dependence of the $\beta$-functions on $\lambda$. 

We had to deal with an additional projection ambiguity compared to the corresponding study in \cite{je:longpaper} based on the proper-time FRGE. In a way, this was the price to pay for the purely algebraic character of the new \WHlike\ flow equation: Since we are bound to use constant background fields, the flow of the curvature and the Immirzi term can not be disentangled from the flow of certain torsion squared invariants. We therefore analyzed a set of different projections from the full theory space onto the truncation subspace. They were defined by introducing different bases in theory space. We were then able to show that the similarities to the PT study even grow further if we opt for basis ${\cal B}_1$ which is optimal in this sense.

Disregarding the flow of the cosmological constant we find a system of $\beta$-functions $\{\beta_g,\beta_\gamma\}$ that is of the same structure as the corresponding one in \cite{je:longpaper}. It satisfies exactly the duality property under the mapping $\gamma\mapsto1/\gamma$ of the Immirzi parameter already conjectured in \cite{je:longpaper}. Under this duality transformation the non-Gaussian fixed points ${\bf NGFP_0'}$ at $\gamma=0$ and ${\bf NGFP_\infty'}$ at $\hat\gamma=0$ are mapped onto each other. These fixed points generically show an antagonistic behavior, one being UV attractive in two and the other in one direction, leading to a preferred direction of the $\gamma$-flow, that depends on the fixed value of $\lambda$.

For the special choice of $\lambda\approx 0.57$ it was observed that the $\gamma$-running stops completely. We argued that by picking this value the symmetry between scalar and pseudo-scalar terms in the truncation ansatz, that is spoiled by the gauge fixing and ghost terms, can be effectively restored, leading to an equal renormalization of the curvature and the Immirzi term. In a setting that respects this symmetry throughout, we therefore expect the flow of the Immirzi parameter to vanish, as long as no volume term is present in the (Holst-)\linebreak truncation (\ie for $\lambda=0$).

In this setting where the RG evolution of $\lambda$ is frozen the different conceptions of the Immirzi parameter used in LQG, as a fixed external quantity, and in the RG approach, as a running coupling, are easy to reconcile: In both cases the value of $\gamma$ is not scale dependent and therefore also in the RG setting the different quantum theories can be parametrized by fixed values of the Immirzi parameter. However, the cosmological constant is known to be non-zero, which inevitably leads to a non-trivial running of the Immirzi parameter and in consequence renders a comparison of the role of the Immirzi parameter in LQG and in asymptotically safe gravity more complicated.

At this point we also can make contact to a related perturbative study on the running of the Immirzi parameter. In \cite{bene-speziale,bene-speziale2} a one loop renormalization of the Holst action with vanishing cosmological constant is carried out, keeping track of the powerlike divergences by employing a proper-time cutoff regularization. In this setting the following $\beta$-functions are obtained:

\begin{equation}
 \beta_g= g\Big(2-\frac{17}{3\pi}g\Big),\qquad \beta_\gamma=\frac{4}{3\pi}\,g\, \gamma\:.
\end{equation}
\vspace{0.2cm}

\noindent
They are to be compared with our findings, at $\lambda=0$:

\begin{equation}
 \beta_g= g\Big(2-\frac{9}{16\pi}g\Big),\qquad \beta_\gamma=-\frac{23}{8\pi}\,g\, \gamma\:.
\end{equation}
\vspace{0.2cm}

\noindent
We observe that both calculations lead to $\beta$-functions of the same form. Moreover, the sign of the anomalous dimension $\eta_N$ of the Newton constant coincides for both studies and allows for a fixed point value $g^\ast>0$. The predominant direction of the $\gamma$-flow is found {\it opposite} to our result, however.

Recalling our above discussion one indeed would expect a ``50\,\% chance'' for the sign of $\beta_\gamma$ to (dis-)agree in the two calculations: as a vanishing result $\beta_\gamma=0$ should occur in a perfectly symmetry preserving setting, the sign of the outcome of any calculation that spoils this symmetry will depend on the very details of this calculation (cutoff, gauge fixing, etc.). Unfortunately the perturbative analysis \cite{bene-speziale} is restricted to the case of a vanishing cosmological constant, such that we can not investigate the expected sign change of the $\gamma$-flow there when $\lambda$ is turned on. The main differences, apart from being a one-loop calculation, in the setting of \cite{bene-speziale} compared to our calculation are a different gauge-fixing condition for the diffeomorphisms and the complete neglect of ${\sf O}(4)$-ghost contributions, that may have a crucial influence as was already pointed out in \cite{dhr1, dhr2}. On the technical side also the regularization procedure is different (proper-time cutoff vs. sharp momentum cutoff). Taken together these differences might very well account for the sign difference in the $\gamma$-flow at $\lambda=0$.

When including $\lambda$ as a running coupling into our truncation, the fixed point at $\hat\gamma=0$ gets lifted to the higher dimensional theory space and we find two NGFPs in the $\hat\gamma=0$-plane. The fixed point at $\gamma=0$, however, ceases to exist due to a, probably unphysical, divergence of $\beta_\lambda$ that arises at $\gamma=0$. Thus, the similarities to \cite{je:longpaper} are weaker in the three dimensional truncation, but nonetheless our computation independently supports the existence of NGFPs suitable for the Asymptotic Safety scenario in the $\hat\gamma=0$-plane.

To summarize, we can conclude that the approximations the \WHlike\ flow equation relies on lead only to minor changes of the RG flow as far as its gross topological features are concerned. In addition, with the new setting we were able to identify a preferred choice of gauge parameters, $(\aD,\aL,\bD)=(0,0,0)$, and of the projection scheme, namely to use basis ${\cal B}_1$ in the subspace of torsion squared invariants. Due to the immense reduction in computational effort, which we gain when opting for the \WHlike\ flow equation instead of the PT flow equation, we can only recommend to employ this preferred setting in future RG studies of enlarged truncations that may also include (fermionic) matter, for instance!

\appendix

\section{Classical aspects of torsion}\label{torsion}
To fix our notations and conventions, and to provide some background material for the main part of the paper we summarize in this appendix several classical aspects of spacetimes exhibiting torsion. In {\it subsection (\ref{Torsion:Prelim})} we introduce the torsion tensor and its decomposition into irreducible components. Then we investigate the subspace of theory space spanned by torsion squared monomials and introduce different bases monomials in this space in {\it subsection (\ref{Torsion:Quad_Inv})}. {\it Subsection (\ref{Ho-ActioninTorsion})} contains the Holst action rewritten in several equivalent forms, depending on the metric and various choices of torsion field variables.

\subsection{Preliminaries}\label{Torsion:Prelim}
\paragraph{(A) The Levi-Civita symbol.}
The Levi-Civita symbol $\varepsilon^{abcd}$ denotes the unique totally antisymmetric \Ofour-tensor with $\varepsilon^{0123}=1\ \Rightarrow\varepsilon_{0123}=1$ that is defined in every local frame. The above implication does only hold for a spacetime of Euclidean signature; in the Lorentzian case we pick up a minus sign. For that reason the following basic identities only hold for a Euclidean spacetime:
\begin{equation}
 \begin{aligned}\label{EpsId1}
 \varepsilon^{abcd}\varepsilon_{stuv}&=\delta^{[a}_s\delta^b_t\delta^c_u\delta^{d]}_v\,,\\
  \varepsilon^{abcd}\varepsilon_{stud}&=\delta^{[a}_s\delta^b_t\delta^{c]}_u\,,\\
  \varepsilon^{abcd}\varepsilon_{stcd}&=2 \delta^{[a}_s\delta^{b]}_t\,,\\
  \varepsilon^{abcd}\varepsilon_{sbcd}&=6 \delta^{a}_s\,,\\
  \varepsilon^{abcd}\varepsilon_{abcd}&=24\,.\\
 \end{aligned}
\end{equation}

Throughout this paper the {\it (anti-)symmetrization brackets contain no weighting factor} in our conventions, \ie symbolically we have $[ab]=ab-ba$ and $(ab)=ab+ba$. 

Another useful algebraic identity for the Levi-Civita symbol which we shall need frequently can be derived from the product of three $\varepsilon$-symbols, using the above identities:
\begin{align}
 -\delta^{[e}_a\delta^f_b\delta^{g]}_c\varepsilon_{ghi}^{\ \ \ a}=\varepsilon^{a}_{\ bcd}&\varepsilon^{def}_{\ \ \ g}\varepsilon^{g}_{\ hia}=-\varepsilon^{a}_{\ bcd} \delta^{[d}_h\delta^e_i\delta^{f]}_a\\ \label{EpsId2}
\Leftrightarrow\hspace{3cm}\varepsilon^{[f}_{\ bc[h}\delta^{e]}_{\ i]}&=-\varepsilon^{[f}_{\ hi[b}\delta^{e]}_{\ c]}\:.
\end{align}
Using the Levi-Civita symbol we can express the determinant of the inverse vielbein by
\begin{equation}
 e^{-1}=e_a^{\ 0}e_b^{\ 1}e_c^{\ 2}e_d^{\ 3} \varepsilon^{abcd}\:.
\end{equation}

Multiplying this equation by the vielbein determinant motivates the following definition of the $\varepsilon$-tensor density on the spacetime:
\begin{equation}
 \varepsilon^{\mu\nu\rho\sigma}=e \, e_a^{\ \mu}e_b^{\ \nu}e_c^{\ \rho}e_d^{\ \sigma} \varepsilon^{abcd}\:.
\end{equation}
It obviously inherits the total antisymmetry from the Levi-Civita tensor and satisfies $\varepsilon^{0123}=1$ as well, but it transforms as a tensor {\it density} under diffeomorphisms of spacetime. The $\varepsilon$-tensor density satisfies the identities \eqref{EpsId1} with a factor of $e^2$ on the right hand side as well as \eqref{EpsId2}, both with the (flat) ${\sf O}(4)$-indices substituted by (curved) spacetime ones.

\paragraph{(B) Torsion and contorsion.}
The torsion tensor on an affinely connected spacetime is defined as the vector valued two-form
\begin{equation}\label{DefTorsion}
 T(X,Y)=D_X Y-D_Y X-[X,Y],
\end{equation}
where the brackets denote the Lie-bracket of vector fields. In components we thus find 
\begin{equation}\label{SymmTorsion}
 T^{\lambda}_{\ \mu\nu}=\Gamma^{\lambda}_{[\mu\nu]}\quad\Rightarrow\quad\Gamma^{\lambda}_{\mu\nu}=\frac{1}{2}\big(\Gamma^{\lambda}_{(\mu\nu)}+T^{\lambda}_{\ \mu\nu}\big)\:.
\end{equation}
Using the Ricci condition (metricity of the connection) and \eqref{DefTorsion} one can derive a generalized Koszul formula, that results in an equation relating the general connection and the Levi-Civita connection according to
\begin{equation}\label{KtoTrelation}
 \Gamma^{\lambda}_{\mu\nu}=\big(\Gamma_{\rm \! LC}\big)^{\lambda}_{\mu\nu}+ \frac{1}{2}\big(T^{\lambda}_{\ \mu\nu}-T_{\mu\nu}^{\ \ \lambda}+T_{\nu\ \mu}^{\ \lambda}\big)\equiv\big(\Gamma_{\rm \! LC}\big)^{\lambda}_{\mu\nu}+K^{\lambda}_{\ \mu\nu}\:.
\end{equation}
Here, we have defined the contorsion tensor $K^{\lambda}_{\ \mu\nu}$ as the difference between the Levi-Civita connection $\Gamma_{\rm LC}$ and the metric-compatible connection $\Gamma$ exhibiting torsion $T$. Note that its symmetric part $\Gamma^{\lambda}_{(\mu\nu)}$ does not coincide with $\big(\Gamma_{\rm \! LC}\big)^{\lambda}_{\mu\nu}$.

If we consider the vielbein \ea\ and the spin connection \oab\ as fundamental variables, it is obvious that the Levi-Civita connection, being a function of the metric, can be expressed purely in terms of the vielbein, such that only the contorsion part of the connection depends on \oab:
\begin{equation}
 \Gamma(e,\omega)^{\lambda}_{\mu\nu}=e_{a}^{\ \lambda} \big( \partial_\mu  e^a_{\ \nu} + \omega^{a}_{\ c\mu}e^c_{\ \nu}\big)= \big(\Gamma_{\rm \! LC}(e)\big)^{\lambda}_{\mu\nu}+K(e,\omega)^{\lambda}_{\ \mu\nu}\:.
\end{equation}
If one calculates the Riemann curvature tensor from the above connection $\Gamma(e,\omega)$, one finds that it is related to the field strength tensor $F$ in the following way:
\begin{equation}\label{FRrelation}
 F^{ab}_{\ \ \mu\nu}=e^{a}_{\ \rho} e^{b \sigma}R(e,\omega)^{\ \ \:\rho}_{\mu\nu\ \sigma}
\end{equation}
Moreover, the Riemann tensor can be decomposed into the curvature tensor of the Levi-Civita connection and contorsion terms
\begin{equation}\label{RtoKrelation}
 R(e,\omega)^{\ \ \:\rho}_{\mu\nu\ \sigma}=\big(R_{\rm LC}\big)^{\ \ \:\rho}_{\mu\nu\ \sigma}+D^{\rm LC}_{[\mu}K^{\rho}_{\ \nu]\sigma}+K^{\rho}_{\ [\mu|\tau}K^{\tau}_{\ \nu]\sigma}\:.
\end{equation}

\paragraph{(C) Irreducible components of the torsion tensor.}
In order to identify the independent invariants quadratic in the torsion tensor it is convenient to decompose the torsion tensor into its irreducible components $(T_\mu, S_\mu, q^\lambda_{\ \mu\nu})$. The pertinent orthogonal decomposition reads \cite{Baekler2011a,Shapiro2002}
\begin{equation}\label{T_decomp}
 T^{\lambda}_{\ \mu\nu}= \frac{1}{3}\big(\delta^{\lambda}_{\ \nu}T_{\mu}-\delta^{\lambda}_{\ \mu}T_{\nu} \big)+\frac{1}{6 e}\varepsilon^{\lambda}_{\ \mu\nu\sigma}S^{\sigma} + q^{\lambda}_{\ \mu\nu}
\end{equation}
with $q^{\lambda}_{\ \mu \lambda}=0$, $\varepsilon^{\mu\nu\rho\sigma}q_{\nu\rho\sigma}=0$, and $q^\lambda{}_{\mu\nu}=-q^\lambda{}_{\nu\mu}$. Here, $T_\mu=T^{\lambda}_{\ \mu \lambda}$ is the trace of the torsion tensor and $S^\mu=\frac{1}{e} \varepsilon^{\ \rho\sigma\mu}_{\nu}T^\nu_{\ \rho\sigma}$ is its totally antisymmetric part. The vielbein determinant $e$ in the decomposition ensures that all torsion components defined here transform as genuine tensors under spacetime diffeomorphisms.
 
Let us further investigate the symmetry properties of $q$. First, we note that all contractions of two of its indices vanish due to $q^{\lambda}_{\ \mu \lambda}=0$ and its antisymmetry in the last two indices. Second, as its totally antisymmetric part vanishes, $\varepsilon^{\mu\nu\rho\sigma}q_{\nu\rho\sigma}=0\Leftrightarrow q^{[\nu\rho\sigma]}=0$, we obtain with the antisymmetry in the last two indices
\begin{equation}\label{qProperty}
 0=q^{[\mu\nu\rho]}=q^{[\mu\nu]\rho}-q^{[\mu|\rho|\nu]}-q^{\rho[\nu\mu]}=2\big(q^{[\mu\nu]\rho}+q^{\rho\mu\nu}\big)\quad\Leftrightarrow\quad q^{[\mu\nu]\rho}=-q^{\rho\mu\nu}\:.
\end{equation}

Now it is easy to classify all independent invariants quadratic in the irreducible torsion components $(T,S,q)$:
\paragraph{(i)} For the {\it parity-even} invariants we are left with three possible independent combinations: As $S^{\mu}$ is a pseudo-vector and $T^{\mu}$ as well as $q^{\mu\nu\rho}$ are (true) tensors that can only couple to itself to form a scalar. If we contract $T^{\mu}$ with $q^{\mu\nu\rho}$ the other two indices of $q$ have to be contracted and thus the combination vanishes. In principle, we could also combine $S$ and $q$ using an additional $\varepsilon$-density, but these combinations vanish as $q$ has no totally antisymmetric part. Hence, we are left with the three parity-even invariants
\begin{equation}
\boxed{
 I_1=T^{\mu}T_{\mu},\qquad I_2=S^{\mu}S_{\mu}, \qquad I_3=q^{\mu\nu\rho}q_{\mu\nu\rho}\:.
}
\end{equation}
At first sight one could wonder whether there are additional independent $q^2$-con\-trac\-tions. This is, however, not the case: In total we start with 6 contractions that correspond to the 6 permutations of the indices of the second $q$-factor. The terms with odd permutations are related to the remaining cyclic permutations by the antisymmetry of $q$ in the last two indices. For the cyclic permutations we find with \eqref{qProperty}
\begin{equation}
 \quad q_{\mu\nu\rho}q^{\nu\rho\mu}=\frac{1}{2} q_{\mu\nu\rho}q^{[\nu\rho]\mu}=-\frac{1}{2} q_{\mu\nu\rho}q^{\mu\nu\rho},\quad q_{\mu\nu\rho}q^{\rho\mu\nu}=\frac{1}{2} q_{[\mu\nu]\rho}q^{\rho\mu\nu}=-\frac{1}{2} q_{\rho\mu\nu}q^{\rho\mu\nu}\:,
\end{equation}
such that only $I_3$ remains independent.

\paragraph{(ii)} For the {\it parity-odd} combinations there are only two independent invariants, namely
\begin{equation}
\boxed{
 I_4=S_\mu T^\mu,\qquad I_5=\frac{1}{e}\, \varepsilon_{\alpha\beta\gamma\delta}\,q^{\alpha\beta\mu}\,q^{\gamma\delta}_{\ \ \mu}\:.
}
\end{equation}
The two other $\varepsilon q^2$ combinations one might think of as independent, namely those where either both first indices of the $q$ tensors are contracted or the first index of the first $q$ factor is contracted with the last index of the second factor, are related to $I_5$ according to
\begin{equation}
 \frac{1}{e}\,\varepsilon_{\alpha\beta\gamma\delta}\,q^{\mu\alpha\beta}\,q^{\ \gamma\delta}_{\mu}= 4 I_5,\qquad \frac{1}{e}\,\varepsilon_{\alpha\beta\gamma\delta}\,q^{\mu\alpha\beta}\,q^{\gamma\delta}_{\ \ \mu}= -2 I_5\:.
\end{equation}
These relations can be shown using the identity \eqref{EpsId2} of the $\varepsilon$-symbol.

\subsection{Invariants quadratic in the torsion tensor}\label{Torsion:Quad_Inv}
In this section we discuss all possible field monomials quadratic in the torsion tensor and give their decomposition in the irreducible components introduced above as well as the expressions in terms of the spin connection that remain when the monomials are evaluated for constant background fields $\{\bar{e},\bar\omega\}$.
\paragraph{(A) Parity-even monomials.}
There are three different contractions of the torsion tensor with itself that are not related to each other by its symmetries. As we know already that the space of parity-even monomials is three dimensional and spanned by $I_{\{1,2,3\}}$ we can conclude that they correspond to a different basis of this space. In detail we find that the two bases are related by
\begin{equation}
\boxed{
 \begin{aligned}
  T^{2(+)}_1&= &T^{\mu \nu \rho}T_{\mu \nu \rho}&=\frac{2}{3} I_1 +\frac{1}{6} I_2 + \phantom{\frac{1}{3}}I_3& &\widehat{=}\phantom{-}2\, \bar{\omega}^{abc}\bar{\omega}_{abc}- 2\,\bar{\omega}^{abc}\bar{\omega}_{acb}\\
  T^{2(+)}_2&= &T^{\mu \nu \rho}T_{\nu \mu \rho}&=\frac{1}{3} I_1 -\frac{1}{6} I_2 + \frac{1}{2} I_3 & &\widehat{=}-\phantom{2\,}\bar{\omega}^{abc}\bar{\omega}_{abc}+3\,\bar{\omega}^{abc}\bar{\omega}_{acb}\\
 T^{2(+)}_3&= & T^{\mu \nu}_{\ \ \:\, \mu}T^{\rho}_{\ \:\nu\rho}&= \phantom{\frac{2}{2}}I_1 & &\widehat{=}\phantom{-2} \bar{\omega}^{ab}_{\ \ \: a}\bar{\omega}^{c}_{\ bc}\:.
 \end{aligned}
}
\end{equation}
On the right hand side we have evaluated the monomials on $x$-independent background fields $\{\bar{e},\bar{\omega}\}$. In addition we have incorporated the vielbein into the spin connection, changing its index structure. The explicit vielbein expressions can be reconstructed uniquely.

\paragraph{(B) Parity-odd monomials.}
In the parity-odd sector we find four torsion squared monomials contracted with the $\varepsilon$-symbol. Those four expressions are not linearly independent, as the corresponding subspace of theory space is known to be two dimensional (spanned by $I_4$ and $I_5$); while in the decomposed setting the linear relation between the four monomials becomes obvious, it can also be shown for the undecomposed torsion tensor using the identity \eqref{EpsId2}. In the irreducible component basis the parity-odd torsion squared monomials read
\begin{equation}\label{TsquaredOdd}
\boxed{ 
\begin{aligned}
T^{2(-)}_1&= & \!\!\!\!e^{-1}\varepsilon^{\mu\nu\rho\sigma}T^\tau_{\ \mu\nu}T_{\tau \rho\sigma}&=-\frac{4}{3} I_4+4 I_5 & &\!\!\!\!\widehat{=} \phantom{-2\, \varepsilon_{pqrs}\,\bar{\omega}^{\ pt}_{t} \bar{\omega}^{qrs} + }\;\,4\, \varepsilon_{pqrs}\,\bar{\omega}^{\ pq}_{t} \bar{\omega}^{trs}
\\
T^{2(-)}_2&= & \!\! \!\! e^{-1}\varepsilon^{\rho\ \sigma\tau}_{\ \nu}T^\mu_{\ \mu\rho}T^\nu_{\ \sigma\tau}&=\phantom{-\frac{3}{3}}I_4 & &\!\!\!\!\widehat{=}- 2\, \varepsilon_{pqrs}\,\bar{\omega}^{\ pt}_{t} \bar{\omega}^{qrs}
\\
T^{2(-)}_3 &= & \!\! \!\!e^{-1}\varepsilon^{\mu\nu \ \sigma}_{\ \ \rho}T^\tau_{\ \mu\nu}T^\rho_{\ \tau\sigma}&=\phantom{-}\frac{1}{3}I_4 +2 I_5 & &\!\!\!\!\widehat{=} -2\, \varepsilon_{pqrs}\,\bar{\omega}^{\ pt}_{t} \bar{\omega}^{qrs}+2\, \varepsilon_{pqrs}\,\bar{\omega}^{\ pq}_{t} \bar{\omega}^{trs}
\\  
  T^{2(-)}_4 &= & \!\! \!\!e^{-1}\varepsilon^{\ \ \rho\sigma}_{\mu \nu}T^\mu_{\ \tau\rho}T^{\nu\tau}_{\ \ \sigma}&=-\frac{2}{3} I_4 -\phantom{2}I_5 & &\!\!\!\!\widehat{=}  \phantom{-}2\, \varepsilon_{pqrs}\,\bar{\omega}^{\ pt}_{t} \bar{\omega}^{qrs}-\phantom{2} \varepsilon_{pqrs}\,\bar{\omega}^{\ pq}_{t} \bar{\omega}^{trs}.
 \end{aligned}
}
\end{equation}

{\noindent}As for the torsion tensor there exist four contractions of the spin connection, that are related via \eqref{EpsId2}; in the above we have used in particular

\begin{equation}
\begin{aligned}
 \varepsilon_{pqrs} \bar{\omega}^{pqt}\bar{\omega}^{rs}_{\ \ \:t}&=-4\, \varepsilon_{pqrs} \bar{\omega}^{\ pt}_{t} \bar{\omega}^{qrs}+4 \,\varepsilon_{pqrs} \bar{\omega}^{\ pq}_{t} \bar{\omega}^{trs}
\\
\varepsilon_{pqrs} \bar{\omega}^{pqt} \bar{\omega}^{\ rs}_t&=\phantom{-2\,} \varepsilon_{pqrs} \bar{\omega}^{\ pt}_{t} \bar{\omega}^{qrs}-2 \,\varepsilon_{pqrs} \bar{\omega}^{\ pq}_{t} \bar{\omega}^{trs}
\end{aligned}
\end{equation}

{\noindent}in order to reexpress the right hand side of \eqref{TsquaredOdd} in terms of the other two independent contractions only.

\subsection{Holst action in metric and torsion variables}\label{Ho-ActioninTorsion}
Starting from the Holst action with the tetrad and the spin connection as field variables we now deduce several equivalent actions that depend on the metric and one further independent field: either a general connection $\Gamma$ (allowing for torsion but satisfying the metricity condition), the contorsion $K$, the torsion $T$, or its irreducible components $T,S,q$.

The Holst action in terms of tetrads and the spin connection is given by
\begin{align}
 S_{\rm Ho}&=-\frac{1}{16 \pi G} \int \! {\rm d}^4 x\,  e\bigg[e_a^{\ \mu}e_b^{\ \nu}\bigg(F(e,\omega)^{ab}_{\ \ \mu\nu} - \frac{1}{2\gamma}\varepsilon^{ab}{}_{cd} F(e,\omega)^{cd}{}_{\mu\nu}\bigg) -2\Lambda\bigg]\:.
\intertext{Using \eqref{FRrelation} we can reexpress the field strength $F(e,\omega)$ by a Riemann tensor $R(\Gamma)$, with a connection $\Gamma$ exhibiting torsion, according to}
 S_{\rm Ho}&=-\frac{1}{16 \pi G}\int \!{\rm d}^4 x \, \sqrt{g} \bigg[ R(\Gamma)-\frac{1}{2 \gamma \sqrt{g}}\varepsilon^{\mu\nu}{}_\rho{}^\sigma R(\Gamma)_{\mu\nu}{}^\rho{}_\sigma -2\Lambda\bigg]\:.
\intertext{In a next step the Riemann tensor $R(\Gamma)$ is written in terms of its Levi-Civita counterpart $R_{\rm LC}$ and contorsion terms as given in \eqref{RtoKrelation} resulting in}
 S_{\rm Ho}&= -\frac{1}{16 \pi G}\int \!{\rm d}^4 x \, \sqrt{g} \bigg[R_{\rm LC}+D^{\rm LC}_{[\rho}K^{\rho}_{\ \nu]}{}^{\nu}+K^{\rho}_{\ [\rho|\tau}K^{\tau}_{\ \nu]}{}^{\nu}\nonumber
\\
&\hspace{5.7cm}-\frac{1}{\gamma} \frac{\varepsilon^{\mu\nu}{}_\rho{}^\sigma}{\sqrt{g}}(D^{\rm LC}_{\mu}K^{\rho}_{\ \nu\sigma}+K^{\rho}_{\ \mu\tau}K^{\tau}_{\ \nu\sigma})- 2 \Lambda\bigg]\:.\label{ContorsionAction}
\intertext{Now we can switch from the contorsion variable $K$ to the torsion tensor $T$ using \eqref{KtoTrelation}}
 S_{\rm Ho}&= -\frac{1}{16 \pi G}\!\int \!\!{\rm d}^4 x \, \sqrt{g} \bigg[R_{\rm LC}\!+\! 2 D^{\rm LC}_{\rho}T^{\nu\rho}{}_{\nu}\!+\!\frac{1}{4} T_{\mu\nu\rho}T^{\mu\nu\rho} \!+\!\frac{1}{2} T_{\mu\nu\rho}T^{\nu\mu\rho}\!-\!T^{\mu\nu}{}_{\!\mu}T^{\rho}{}_{\!\nu\rho}\nonumber
\\
&\hspace{4.9cm}-\frac{1}{\gamma} \frac{\varepsilon^{\mu\nu\rho\sigma}}{\sqrt{g}}\bigg(\!\!-\frac{1}{2}D^{\rm LC}_{\mu}T_{ \nu\rho\sigma}+\frac{1}{4}T^{\tau}{}_{\mu\nu}T_{\tau\rho\sigma}\bigg)
- 2 \Lambda\bigg]\:.
\intertext{In a last step we decompose the torsion tensor into its irreducible components \eqref{T_decomp} and obtain}
 S_{\rm Ho}&= -\frac{1}{16 \pi G}\int \!{\rm d}^4 x \, \sqrt{g} \bigg[R_{\rm LC}+2 D^{\rm LC}_{\mu}T^{\mu}-\frac{2}{3} T_\mu T^\mu -\frac{1}{24} S_\mu S^\mu+\frac{1}{2} q_{\mu\nu\rho}q^{\mu\nu\rho}\nonumber
\\
&\hspace{4cm}-\frac{1}{\gamma} \bigg(\frac{1}{2}D^{\rm LC}_\mu S^\mu-\frac{1}{3} T_\mu S^\mu +\frac{\varepsilon^{\mu\nu\rho\sigma}}{\sqrt{g}}q_{\mu\nu}{}^{\tau}q_{\rho\sigma\tau}\bigg) - 2 \Lambda\bigg]\:.
\end{align}
Note that on manifolds with boundary the terms $\propto D^{\rm LC}_{\mu}T^{\mu}$ and $D^{\rm LC}_\mu S^\mu$ would give rise to surface terms.\enlargethispage{0.5cm}

\section{RHS of the FRGE prior to projection}\label{Appendix:UnambiguousRHS}
In this appendix we display the RHS of the \WHlike\ flow equation \eqref{WHlike_konkret2} applied to the Holst truncation \eqref{HoTruncation}.

Expanding the RHS of the FRGE, evaluated for a constant background field configuration, $\{\bar e,\bar\omega\}$, in terms of $\bar{\omega}$ up to second order, we find
\begin{equation}
\begin{aligned}
\partial_t \Gamma_k\bigg|_{\genfrac{}{}{0pt}{}{\bar\omega,\bar e =}{ {\rm const}}}\hspace{-0.3cm}=\! -\frac{k^4}{32 \pi^2}\bigg(\!\ln \frac{(1-\gamma^2)^{24}(1-\lambda)^{12}}{\gamma^{48}g^{68}}-\ln M^{160} \mu^{32}{\cal N}'\bigg)&\!\int\! {\rm d}^4 x\,\bar{e}
\\-\frac{k^2}{1536 \pi^2}\frac{1}{(1-\lambda)^2}\Bigg[\!\bigg(\!\big(70-438 \lambda+373\lambda^2\big)-\frac{20}{\gamma^2}\bigg)&\!\int\! {\rm d}^4 x\, \bar{e}\, \bar{\omega}_{pq}{}^p\bar{\omega}_r{}^{qr}\\
+\bigg(\big(106-174\lambda+43\lambda^2\big)-\frac{20}{\gamma^2}\bigg)&\!\int\! {\rm d}^4 x\, \bar{e}\, \bar{\omega}_{pqr}\bar{\omega}^{pqr}\\
+\bigg(\big(178-186\lambda+43\lambda^2\big)-\frac{20}{\gamma^2}\bigg)&\!\int\! {\rm d}^4 x\, \bar{e}\, \bar{\omega}_{pqr}\bar{\omega}^{prq}\Bigg]\\
-\frac{k^2}{128\pi^2}\frac{23-33\lambda}{(1-\lambda)^2}\,\frac{1}{\gamma} &\!\int\!\! {\rm d}^4 x\, \bar{e}\, \varepsilon_{pqrs}\bar{\omega}_{t}{}^{pt}\bar{\omega}^{qrs}\\
-\frac{5\,k^2}{256\pi^2}\frac{11-15\lambda}{(1-\lambda)^2}\,\frac{1}{\gamma} &\!\int\!\! {\rm d}^4 x\, \bar{e}\, \varepsilon_{pqrs}\bar{\omega}_{t}{}^{pq}\bar{\omega}^{rts}\\
&\hspace{1.8cm}+{\cal O}(\bar\omega^3).
\end{aligned}\label{UnambiguousResult}
\end{equation}
Here ${\cal N}'$ is a pure number with $\ln {\cal N}'\approx 241.42$. In writing down eq. \eqref{UnambiguousResult} we have specialized the general result for the choice of gauge parameters $(\aD,\aL,\bD)=(0,0,0)$, which considerably simplifies the result. The analogous expression for general gauge parameters fills many pages, due to its complicated polynomial structure in $\lambda$ and the gauge parameters.

At this level, the result does not show any ambiguity. However, in order to identify the coefficient functions rhsF and rhsF${}^\ast$ as defined in subsection \ref{EVFLowEq}, we have to define a projection scheme onto the invariants of the Holst truncation. This can be done by specifying a basis in the space of $\bar{\omega}^2$-monomials. Depending on our choice of this basis we are led to different functions rhsF and rhsF${}^\ast$, an effect we refer to as a {\it projection ambiguity}. The explicit expressions for rhsF and rhsF${}^\ast$ projected using the four bases ${\cal B}_i$ ($i=1,\cdots\!,4$) which are given in section \ref{QECG:Results} in fact all have been derived from \eqref{UnambiguousResult}.

\pagestyle{headings}
\bibliographystyle{diss2}
\bibliography{WH-eomega}

%\begin{thebibliography}{99}
%\input{WH-eomega-Bibliography.tex}
%\end{thebibliography}

\end{document}